\documentclass[twocolumn]{aastex701}

\usepackage{hyperref}
\usepackage{multirow}
\usepackage{amsmath}
\usepackage{caption}
\usepackage{longtable}
\usepackage{comment} 

\begin{document}

\title{Magnetar Engines in Broad-lined Type Ic Supernovae and a Unified Picture for Magnetar-powered Stripped-envelope Supernovae}

\correspondingauthor{Jin-Ping Zhu; Bing Zhang}

\author[orcid=0000-0002-9195-4904]{Jin-Ping Zhu}
\affiliation{The Hong Kong Institute for Astronomy and Astrophysics, The University of Hong Kong, Pokfulam Road, Hong Kong, People's Republic of China}
\affiliation{Department of Physics, The University of Hong Kong, Pokfulam Road, Hong Kong, People's Republic of China}
\affiliation{School of Physics and Astronomy, Monash University, Clayton Victoria 3800, Australia}
\affiliation{OzGrav: The ARC Centre of Excellence for Gravitational Wave Discovery, Australia}
\email[show]{\url{jpzhu@hku.hk}}

\author[orcid=0000-0002-9725-2524]{Bing Zhang}
\affiliation{The Hong Kong Institute for Astronomy and Astrophysics, The University of Hong Kong, Pokfulam Road, Hong Kong, People's Republic of China}
\affiliation{Department of Physics, The University of Hong Kong, Pokfulam Road, Hong Kong, People's Republic of China}
\affiliation{The Nevada Center for Astrophysics and Department of Physics and Astronomy, University of Nevada, Las Vegas, NV, 89154, USA}
\email[show]{\url{bzhang1@hku.hk}}

\begin{abstract}

We model the multi-band lightcurves of 80 broad-lined Type Ic supernovae (SNe Ic-BL), including 11 associated with long-duration gamma-ray bursts (lGRBs), using a magnetar engine model with $^{56}$Ni decay. The data are all consistent with a magnetar engine, yielding high-quality fits across the sample. The medians with $1\sigma$ regions for key parameters are initial spin period $P_{\rm{i}}\sim2.03^{+1.81}_{-0.95}\,{\rm{ms}}$, magnetic field $B_{\rm{p}}\sim3.97^{+3.07}_{-1.42}\times10^{15}\,{\rm{G}}$, ejecta mass $M_{\rm{ej}}\sim2.30^{+1.46}_{-1.01}\,M_\odot$, and $^{56}$Ni mass $M_{\rm{Ni}}\sim0.18^{+0.14}_{-0.09}\,M_\odot$, with strong correlations observed for both $M_{\rm{ej}}-P_{\rm{i}}$ (anti-correlation) and $M_{\rm{Ni}}-M_{\rm{ej}}$ (correlation). Comparing the lGRB-associated and non-lGRB-associated samples using fitting parameters, we find no significant difference, although the lGRB-associated sample is slightly brighter, possibly due to observational bias (dimmer ones overshone by afterglow). Relative to ordinary SNe Ic, SNe Ic-BL have similar $^{56}$Ni and ejecta masses, suggesting comparable pre-SN progenitor properties, with differences possibly arising from the presence of a magnetar engine that governs SN dynamics. In comparison with other possible magnetar-powered stripped-envelope SNe (SESNe), including Type Ic superluminous SNe (SLSNe Ic) and fast blue optical transients (FBOTs), we confirm a strong universal $M_{\rm{ej}}-P_{\rm{i}}$ correlation, indicating a common origin. SNe Ic-BL and SLSNe Ic have similar ejecta mass distributions, typically $M_{\rm{ej}}\gtrsim0.5\,M_\odot$, while FBOTs mostly lie below this value. Differences between SNe Ic-BL and SLSNe Ic may arise from magnetar properties, with SN Ic-BL magnetars rotating faster and having stronger fields. Moreover, the $P_{\rm{i}}-B_{\rm{p}}$ distribution of lGRB magnetars largely overlaps with that of SN Ic-BL magnetars. Motivated by binary simulation results, we propose a unified physical classification and progenitor framework for magnetar-powered and ordinary SESNe.

\end{abstract}

\keywords{\uat{Supernovae}{1668}; \uat{High energy astrophysics}{739}; \uat{Gamma-ray bursts}{629}; \uat{Massive stars}{732}  }

\section{Introduction}  \label{sec:Introduction}

Stripped-envelope supernovae (SESNe), including Types IIb, Ib, and Ic SNe (SNe IIb/Ib/Ic), characterized by varying degrees of absence of hydrogen and/or helium features in their spectra \citep[e.g.,][]{Filippenko1997,Modjaz2014,Liu2016,Lyman2016}, are believed to originate from massive stars that have lost part or all of their hydrogen-rich and/or helium envelopes before the explosions \citep[e.g.,][]{Woosley1995,Smartt2009,Eldridge2013,Dessart2020,Aguileradena2023}. Within the SESN population, the diversity in envelope stripping extends to extreme cases such as ultra-stripped SNe (USSNe), typically classified as SNe Ib/Ic, which represent an extreme subclass distinguished by unusually small ejecta masses and are likely produced by progenitors that have undergone extreme mass loss and extensive stripping before explosions \citep{Tauris2013,Tauris2015,Suwa2015}. Moreover, there exists a distinct class of energetic SNe Ic, with significantly broader spectral lines and higher ejecta velocities than those observed in ordinary SESNe Ic, hence the name ``broad-lined Type Ic SNe'' \citep[SNe Ic-BL;][]{Galama1998,Iwamoto1998,Modjaz2016}. Most SNe Ic-BL are discovered as isolated SN events, while only a small fraction are found in association with high-energy transient counterparts, likely powered by relativistic jets from progenitors \citep[$\sim3-30\%$ of SNe Ic-BL may possibly generate relativistic jets;][]{Guetta2007,Ghirlanda2013}. Almost all SNe associated with long gamma-ray bursts (lGRBs) were identified as SNe Ic-BL \citep[see][for reviews]{Woosley2006,Zhang2018}, with famous examples such as GRB\,980425/SN\,1998bw \citep{Galama1998,Iwamoto1998}, GRB\,030329/SN\,2003dh \citep{Hjorth2003}, and GRB\,060218/SN\,2006aj \citep{Campana2006,Pian2006}, except for the ultra-long GRB\,111209A, which is associated with the hydrogen-poor superluminous supernova (SLSN Ic) SN\,2011kl \citep{Greiner2015}. In addition to lGRBs, a few fast X-ray transients (FXTs) recently detected by the {\em Einstein Probe} \citep{Yuan2022,Yuan2025} have also been confirmed to be associated with SNe Ic-BL, such as EP240414a/SN\,2024gsa \citep{Sun2025,Bright2025,Srivastav2025,vanDalen2025}, EP250108a/SN\,2025kg \citep{Li2025,EylesFerris2025,Rastinejad2025,Srinivasaragavan2025}, and EP260321a/SN\,2026gzf \citep{Yuan2026,Chen2026,OConnor2026,Rastinejad2026,MartinCarrillo2026}. The origin of these FXTs remains under active debate, with proposed scenarios including cooling emission of the inner cocoon associated with GRB jets \citep{Zheng2025,Zhu2025,Zheng2026}, weak jet emission \citep{Sun2025,Li2025,Hamidani2025,Hamidani2025EP24041a,Schneider2026}, or shock breakout emission \citep{EylesFerris2025,Srinivasaragavan2025SN2025wkm,Yuan2026}.

The explosion energy of SESNe is generally believed to be supplied by the neutrino-driven mechanism, where $\sim1\%$ of the gravitational binding energy ($\sim10^{53}\,{\rm erg}$) released during core collapse is deposited into the shock via neutrino heating, finally leading to a typical SN explosion energy of $\sim10^{51}\,{\rm erg}$ \citep{Bethe1985,Janka2012,Sukhbold2016}. The explosion energies of SNe Ic-BL inferred from the broad spectral lines can reach up to a few $10^{52}\,{\rm erg}$ \citep[e.g.,][]{Taddia2019,Srinivasaragavan2024}, far exceeding those expected from standard neutrino-driven mechanisms and thus suggesting the presence of additional energy sources. However, the energy sources of SNe Ic-BL remain uncertain. One of the most popular origin models for lGRBs and SNe Ic-BL is the black hole (BH) collapsar scenario, in which a massive, rapidly rotating Wolf-Rayet star undergoes core collapse to form a spinning BH surrounded by an accretion disk with sufficient angular momentum \citep{Woosley1993,Paczynski1998,MacFadyen1999}. Such a BH-disk system can potentially drive powerful relativistic jets via the Blandford-Znajek mechanism \citep{Blandford1977} and simultaneously produce energetic SN Ic-BL explosions \citep[e.g.,][]{MacFadyen2001,Zhang2003,Barkov2008,Gottlieb2024,Shibata2024,Shibata2025}. In addition to the collapsar scenario, the magnetar model has also been extensively proposed, where a newborn, rapidly rotating, strongly magnetized neutron star (a ``millisecond magnetar") powers both lGRBs and SNe Ic-BL by releasing its rotational energy through spin-down processes \citep{Usov1992,Dai1998,Kluzniak1998,Zhang2001,Thompson2004,Zhang2006,Metzger2011,Piro2011,Wang2017,Lin2021,Omand2024,Li2026}. 

Following the core collapse of progenitors, the SN Ic-BL emission is typically thought to be attributed to heating from the radioactive decay of $^{56}$Ni and its daughter $^{56}$Co synthesized in the explosion \citep{Colgate1969}. However, it has been found that the classical ``Arnett-like'' one-dimensional SN model \citep{Arnett1982}, which considers only the radioactive heating of centrally located $^{56}$Ni, fails to reproduce both the peak and the linear decline phase of the SN Ic-BL lightcurves simultaneously. Specifically, the $^{56}$Ni inferred from the linear decline phase (i.e., $^{56}$Ni tail) always predicts a relatively lower peak luminosity than that observed in SNe Ic-BL \cite[e.g.,][]{Sharon2020,Ertl2020,Afsariardchi2021,Woosley2021,Sollerman2022,Rodriguez2023,Rodriguez2024}. To address this issue, \cite{Maeda2003} proposed that the SN Ic-BL ejecta may consist of two components with different velocities and $^{56}$Ni masses, with the outer component shaping the lightcurve near the peak and the dense inner component governing the late-time linear decline. Similarly, \cite{Taddia2019} and \cite{Rastinejad2025} suggested that reproducing the observed lightcurves of SNe Ic-BL requires strong $^{56}$Ni mixing within the ejecta. Beyond adjustments to the geometry and distribution of the ejecta, models with an additional long-lived central energy source have also been actively considered, most commonly a magnetar, to explain the excess peak luminosity relative to that predicted by the $^{56}$Ni tail \citep[e.g.,][]{Ertl2020,Afsariardchi2021,Rodriguez2024}. Using hybrid models that incorporate both a magnetar and $^{56}$Ni, the bolometric lightcurves of SNe Ic-BL, both associated and unassociated with GRBs, have been reproduced with better fits than the $^{56}$Ni-only scenario \citep[e.g.,][]{Cano2016,Wang2016,Wang2017,Afsariardchi2021}. {The magnetars powering SNe Ic-BL are typically inferred to have initial spin periods of $\gtrsim10\,{\rm ms}$, corresponding to initial rotational energies of $\lesssim10^{50}\,{\rm erg}$, with magnetic field strengths of $\sim10^{14}-10^{16}\,{\rm G}$ \citep[see also][for similar magnetar parameters derived by modeling SNe Ic-BL with pure magnetar central engines]{Kumar2024,Kumar2025}. However, this leads to an ``energy puzzle'' for SNe Ic-BL: the measured explosion energies, which can even reach up to a few $10^{52}\,{\rm erg}$, always far exceed the energies provided by both magnetar central engines and neutrino-driven mechanisms. If the measured explosion energies of SNe Ic-BL are primarily supplied by magnetar injection, the initial spin periods must instead be on the order of a few milliseconds\footnote{Additional contributions from other mechanisms cannot be excluded, such as the jittering jets explosion mechanism \citep{Soker2016,Soker2017}.}. This inconsistency indicates that current lightcurve fittings for SNe Ic-BL in the literature may not accurately reflect the underlying magnetar parameters and ejecta properties, with the resulting ``energy puzzle'' potentially arising from at least three factors: (1) fitting the bolometric lightcurves provides limited constraints on the SN parameters, lacking information from ejecta velocities and temperature evolution; (2) the initial ejecta kinetic energy is often treated as a free parameter, although it may, in fact, be set by neutrino-driven mechanisms; (3) treating the initial kinetic energy as a free parameter can lead to models that attribute most of the magnetar rotational energy to the peak emission, whereas in reality most of it is converted into ejecta kinetic energy, thereby underestimating the true initial rotational energy. These considerations highlight the need to investigate how lightcurve fitting of SNe Ic-BL can more accurately constrain their magnetar parameters and ejecta properties.}

Beyond SNe Ic-BL, there also exists a diverse population of extensively studied transients potentially originating from the core-collapse of massive stripped-envelope stars that may be driven by magnetar central engines. A prominent example is the class of SLSNe Ic, a population of intrinsically bright SNe with lightcurve luminosities and total radiated energies that always exceed those of ordinary SNe by a factor of $\sim10-100$ \citep{Qumiby2011,GalYam2012,GalYam2019}. While the extremely high luminosities and radiated energies of SLSNe Ic require a $^{56}$Ni mass far exceeding what is typically produced in ordinary core-collapse SNe, these events can be readily explained by the heating of ejecta from a magnetar central engine \citep{Kasen2010}, making it the most widely accepted scenario to date. Moreover, fast blue optical transients (FBOTs) are a class of fast-evolving, very blue, and luminous transients \citep{Drout2014,Pursiainen2018,Inserra2019} whose progenitors and power sources remain actively debated in the literature. The fact that most FBOTs are observed exclusively in star-forming host galaxies suggests a likely core-collapse origin connected to massive stars \citep{Drout2014,Pursiainen2018,Wiseman2020}. However, similar to SLSNe Ic, the peak luminosities of most FBOTs significantly exceed the maximum achievable luminosity of a standard $^{56}$Ni-powered core-collapse SNe, indicating that an additional power source is required \citep{Inserra2019}. Given that FBOT host galaxies share similar star-formation rates and metallicities with those of SLSNe Ic, SNe Ic-BL and lGRB hosts \citep{Wiseman2020}, FBOTs may be physically related to SLSNe Ic and lGRBs in terms of their progenitors and power sources (i.e., magnetar central engine). Indeed, the magnetar engine model provides a more viable explanation for the observational properties of at least a subset of FBOTs than other alternative scenarios \citep[e.g.,][]{Margutti2019,Wang2019,Liu2022,LiL2024,LiJY2025,Eiden2025,Mukhopadhyay2026}. While statistical studies using the magnetar engine model have been extensively explored for SLSNe Ic \citep[e.g.,][]{Yu2017,Liu2017,Nicholl2017,Chen2023} and FBOTs \citep[e.g.,][]{Liu2022}, a similar comprehensive statistical analysis for SNe Ic-BL has not yet been systematically explored. Addressing this gap is crucial for understanding the potential connections and distinctions among these diverse transients, particularly in terms of their magnetar properties and progenitor origins.

Thanks to the recent wide-field and high-sensitivity transient surveys such as iPTF, ZTF, and LSST, an increasing number of SNe Ic-BL have been discovered \citep[e.g.,][]{Taddia2019,Srinivasaragavan2024}. In this study, we perform a statistical analysis of SNe Ic-BL using the largest sample available from the literature under a magnetar central engine, aiming to investigate the statistical properties of SNe Ic-BL and their magnetars, and to compare them with ordinary SESNe and other magnetar-powered explosions (lGRBs, SLSNe Ic, and FBOTs). The structure of this work is as follows. Section \ref{sec:Model} introduces the magnetar-powered model with $^{56}$Ni contribution. In Section \ref{sec:DataCollection}, we present the observational data and lightcurve fitting. The statistics of the model parameters for SNe Ic-BL, as well as the comparison between SNe Ic-BL associated with lGRBs and those unassociated with lGRBs (abbreviated as GRB-SNe and Non-GRB-SNe, hereafter), are presented in Section \ref{sec:StatisticsSNIcBL}. In Section \ref{sec:StatisticsMagnetarSNe}, we compare the magnetar and explosion parameters of SNe Ic-BL with those of ordinary SESNe and other types of magnetar-powered SESNe. The discussion is presented in Section \ref{sec:Discussion}, where we, in particular, provide a possible unified picture of the phenomenological classification and progenitors of magnetar-powered SESNe and ordinary SESNe. The summary and conclusion are given in Section \ref{sec:Conclusions}. Here, a standard $\Lambda$CDM cosmology of $H_0=67.4\,{\rm km}\,{\rm s}^{-1}\,{\rm Mpc}^{-1}$, $\Omega_{\rm m}=0.315$, and $\Omega_\Lambda=0.685$ \citep{Planck2020} is adopted.

\section{Modeling} \label{sec:Model}

\subsection{Lightcurves of Magnetar-powered SESNe with $^{56}$Ni Contribution}

Here, we adopt a semi-analytical model \citep[e.g.,][]{Kasen2010,Inserra2013,Yu2017} to calculate the emission of magnetar-powered SESNe with $^{56}$Ni contribution. The SN bolometric luminosity can be roughly expressed as \citep{Kasen2010}
\begin{equation}
\label{equ:Bolometric_Lightcurves}
    L_{\rm SN} \approx \frac{cE_{\rm int}}{R\tau}(1-e^{-\tau}),
\end{equation}
where $E_{\rm int}$ is the internal energy of the ejecta, $c$ is the speed of light, $R$ is the radius, and $\tau\approx3\kappa M_{\rm ej}/4\pi R^2$ is the optical depth, with $\kappa\approx0.07\,{\rm cm}^2\,{\rm g}^{-1}$ being the opacity commonly used in modeling SNe Ic-BL \citep[e.g.,][]{Maeda2003,Lyman2016,Lv2018,Taddia2019,Anand2024,Srinivasaragavan2024} and $M_{\rm ej}$ the ejecta mass. The radius $R$ can be given by $R=\int_0^t v_{\rm ej}(t')dt'+R_\star$, where $v_{\rm ej}$ is the ejecta velocity, $t$ is the time, and $R_\star\simeq R_\odot$ denotes a typical radius of a Wolf-Rayet star. For $\tau\gg1$, the above equation reads $L_{\rm SN}=cE_{\rm int}/R\tau$; otherwise, when $\tau\ll1$, it simplifies to $L_{\rm SN}=cE_{\rm int}/R$.

By considering the energy conservation of the ejecta, the evolution of its internal energy $E_{\rm int}$ over time $t$ can be calculated as
\begin{equation}
\label{equ:Energy_Conservation}
    \frac{dE_{\rm int}}{dt} = L_{\rm sd}(t) + L_{\rm rad}(t)- p\frac{dV}{dt} - L_{\rm SN}(t),
\end{equation}
where $L_{\rm sd}$ is the energy injection rate from the magnetar's spin-down, and $L_{\rm rad}$ is the energy injection rate due to the radioactive decay of $^{56}$Ni and $^{56}$Co. Both the spin-down energy injection and $^{56}$Ni are assumed to be centrally located. The term $pdV/dt$ represents the adiabatic cooling of the ejecta, in which the internal energy is used to accelerate the ejecta, where $p=E_{\rm int}/4\pi R^3$ is the pressure and $dV/dt=4\pi R^2v_{\rm ej}$ is the change rate of volume. Thus, one can express the dynamical equation as
\begin{equation}
    \frac{dv_{\rm ej}}{dt} = \frac{4\pi R^2p}{M_{\rm ej}}.
\end{equation}
The initial kinetic energy of the ejecta $E_{\rm kin,i}$ is set by the neutrino-driven mechanism, while the final kinetic energy of the ejecta can be written as $E_{\rm kin,f} = M_{\rm ej}v_{\rm ej,f}^2/2$, where $v_{\rm ej,f}$ is the final characteristic ejecta velocity.

We assume that the newborn magnetar formed after the SN explosion can lose its rotational energy via the magnetic dipole spin-down process, with a luminosity of \citep{Ostriker1971}
\begin{equation}
\label{equ:Magnetar_Injection}
    L_{\rm sd} = \frac{L_{\rm sd,i}}{(1+t/t_{\rm sd})^2},
\end{equation}
with an initial luminosity of $L_{\rm sd,i}\simeq E_{\rm rot,i}/t_{\rm sd}=B_{\rm p}^2R_{\rm mag}^6\Omega_{\rm i}^4/6c^3\simeq10^{49}(B_{\rm p}/10^{15}\,{\rm G})^2(P_{\rm i}/1\,{\rm ms})^{-4}\,{\rm erg}\,{\rm s}^{-1}$ and a spin-down timescale of $t_{\rm sd}=3c^3I_{\rm mag}/B_{\rm p}^2R_{\rm mag}^6\Omega_{\rm i}^2\simeq2\times10^3(B_{\rm p}/10^{15}\,{\rm G})^{-2}(P_{\rm i}/1\,{\rm ms})^2\,{\rm s}$, where $B_{\rm p}$ is the magnetar's polar magnetic field strength, $E_{\rm rot,i}=I_{\rm mag}\Omega_{\rm i}^2/2$ is the initial rotational energy, $\Omega_{\rm i}$ and $P_{\rm i}=2\pi/\Omega_{\rm i}$ are the initial angular frequency and spin period, $I_{\rm mag}\simeq10^{45}\,{\rm g}\,{\rm cm}^2$ is the moment of inertia, and $R_{\rm mag}\simeq10^{\rm 6}\,{\rm cm}$ is the magnetar's radius. The radioactivity luminosity is given by \citep{Nadyozhin1994,Arnett1996}
\begin{equation}
\begin{split}
        L_{\rm rad} = &(6.45e^{-t/\tau_{\rm Ni}}+1.45e^{-t/\tau_{\rm Co}})(1-e^{-At^{-2}})\frac{M_{\rm Ni}}{M_\odot} \\
        &\times10^{43}{\rm erg}\,{\rm s}^{-1},
\end{split}
\end{equation}
where $\tau_{\rm Ni}=8.8\,{\rm d}$ and $\tau_{\rm Co}=111.3\,{\rm d}$ are the $e$-folding mean lifetimes of the nuclei, and $M_{\rm Ni}$ is the nickel mass in solar masses. The item $1-e^{-At^{-2}}$ accounts for the increasing transparency to gamma-ray photons, where $A=3\kappa_\gamma M_{\rm ej}/4\pi v_{\rm ej}^2$ with $\kappa_\gamma$ being the effective gamma-ray opacity for the ejecta \citep[e.g.,][]{Wheeler2015}.

\begin{figure}[t]
\centering
\includegraphics[width=1\linewidth, trim = 72 36 40 60, clip]{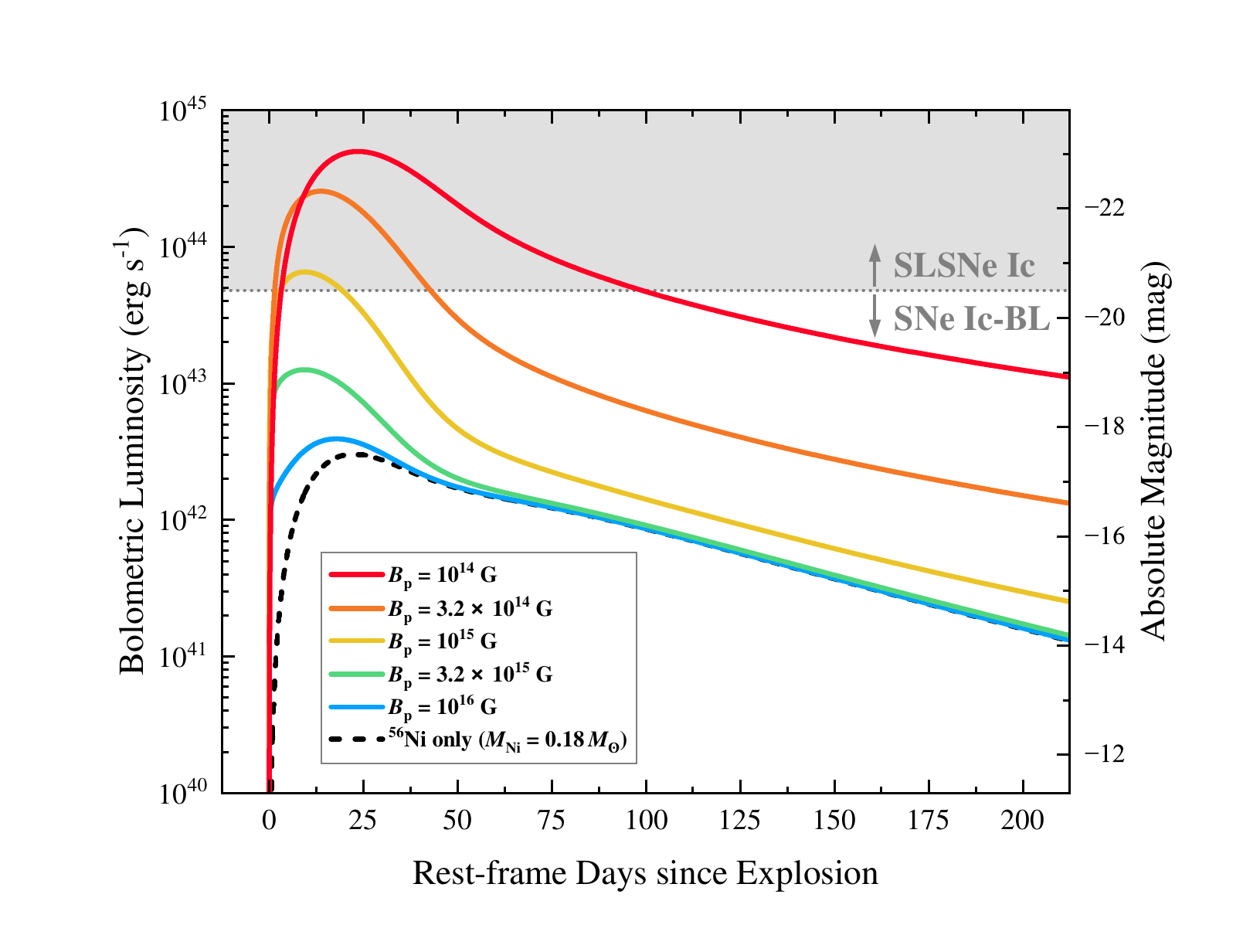}
\caption{Bolometric lightcurves of magnetar-powered SESN with $^{56}$Ni contribution, assuming $P_{\rm i}=2\,{\rm ms}$, $M_{\rm ej}=2.3\,M_\odot$, and $M_{\rm Ni}=0.18\,M_\odot$. The solid red, orange, yellow, green, and blue lines correspond to polar magnetic field strengths of $B_{\rm p}=10^{14}$, $3.2\times10^{14}$, $10^{15}$, $3.2\times10^{15}$, and $10^{16}\,{\rm G}$, respectively. Here, $E_{\rm kin,i}=10^{51}\,{\rm erg}$ and $\kappa_\gamma=0.29\,{\rm cm}^2\,{\rm g}^{-1}$ are adopted. The black dashed line shows the lightcurve of a SN powered only by $^{56}$Ni, with a final kinetic energy comparable to that of the magnetar-powered SESNe. The dashed gray line represents the commonly used peak luminosity threshold to distinguish SLSNe Ic from SNe Ic-BL.  }
\label{fig:Bolometric_Lightcurves}
\end{figure}

As illustrated in Figure \ref{fig:Bolometric_Lightcurves}, we adopt $P_{\rm i}=2\,{\rm ms}$, $M_{\rm ej}=2.3\,M_\odot$, $M_{\rm Ni}=0.18\,M_\odot$, $E_{\rm kin,i}=10^{51}\,{\rm erg}$, and $\kappa_\gamma=0.29\,{\rm cm}^2\,{\rm g}^{-1}$ {(following the median values of inferred parameter distributions reported in Table \ref{tab:Results} in Section \ref{sec:DataCollection} below)}, and evolve Equations (\ref{equ:Bolometric_Lightcurves})-(\ref{equ:Magnetar_Injection}) to explore how different magnetic fields influence the bolometric lightcurves of magnetar-powered SESNe. For a magnetar-powered SESN with $B_{\rm p}\gtrsim10^{14}\,{\rm G}$, the spin-down timescale is $t_{\rm sd}\lesssim9.3(B_{\rm p}/10^{14}\,{\rm G})^{-2}(P_{\rm i}/2\,{\rm ms})^2\,{\rm d}$, always shorter than the ejecta diffusion timescale of $t_{\rm diff}\sim(3\kappa M_{\rm ej}/4\pi cv_{\rm ej,f})^{1/2}\gtrsim14.5(M_{\rm ej}/2.3\,M_\odot)^{3/4}(E
_{\rm kin,f}/6.1\times10^{51}\,{\rm erg})^{-1/2}\,{\rm d}$, indicating that most of the initial magnetar rotational energy is converted into the kinetic energy of the ejecta, where $E_{\rm kin,f} \sim E_{\rm kin,i} + E_{\rm rot,i} = 6.1\times10^{51}\,{\rm erg}$. Furthermore, we ignore the magnetar injection term $L_{\rm sd}$ in Equation (\ref{equ:Energy_Conservation}) and simultaneously set the final kinetic energy to be comparable to that of the magnetar-powered SESNe. The resulting lightcurve (see Figure \ref{fig:Bolometric_Lightcurves}) corresponds to a $^{56}$Ni-powered SN, which can be regarded as an approximation of the $^{56}$Ni contribution in magnetar-powered SESNe.

Figure \ref{fig:Bolometric_Lightcurves} shows that, as the magnetic field decreases (when $B_{\rm p}\gtrsim10^{14}\,{\rm G}$), the resulting longer spin-down timescale allows the initial magnetar rotational energy to be deposited more slowly into the ejecta, leading to more efficient conversion into thermal energy and thus brighter SN emission. In the literature, a peak magnitude threshold of about $M_{\rm peak}\lesssim-20\,{\rm mag}$ to $M_{\rm peak}\lesssim-21\,{\rm mag}$ is empirically used to separate SLSNe Ic from other types of SN transients \citep{GalYam2012,GalYam2019}. As shown in Figure \ref{fig:Bolometric_Lightcurves}, if SLSNe Ic are powered by magnetar engines, the required magnetic field is typically $B_{\rm p}\lesssim10^{15}\,{\rm G}$, in good agreement with the values inferred from observations of SLSN Ic populations \citep[e.g.,][]{Inserra2013,Yu2017,Liu2017,Nicholl2017,Blanchard2020,Chen2023}. In magnetar-powered SLSNe Ic, the luminosities exceed the $^{56}$Ni contribution by over an order of magnitude from peak to late times, indicating that the $^{56}$Ni decay is always negligible in powering their emission. 

When the magnetic field $B_{\rm p}\gtrsim10^{15}\,{\rm G}$, the SN peak luminosities drop below the empirical SLSN Ic threshold, instead falling within the range of typical SNe Ic-BL \citep[e.g.,][]{Taddia2019,Anand2024,Srinivasaragavan2024}, with the $^{56}$Ni decay contributing more significantly at late times. For a stronger magnetic field, the magnetar's contribution to the lightcurve emission decreases, though it always partially contributes to the peak. For a magnetar with $B_{\rm p}\gtrsim10^{15}\,{\rm G}$ the contribution to the late-time lightcurve after $\sim50\,{\rm d}$ is typically secondary. The late-time lightcurve tail becomes entirely $^{56}$Ni-powered when $B_{\rm p}\gtrsim10^{15}\,{\rm G}$. Therefore, it is expected that a newborn magnetar with a few millisecond rotation period and $B_{\rm p}\gtrsim10^{15}\,{\rm G}$ can naturally explain the observed explosion energies of SNe Ic-BL, their late-time $^{56}$Ni tails, and their peak excesses.

\subsection{Photospheric Evolution}

Similar to \cite{Inserra2013,Kashiyama2016,Nicholl2017}, we model the evolution of the photosphere independently of the bolometric lightcurves, tracking its radius, velocity, and temperature. For simplicity, we assume that the SN ejecta can have a broken power-law density profile based on the numerical simulations \citep{Chevalier1989,Matzner1999}, with a flat profile in the inner region and a steep one in the outer region, which can be expressed as follows
\begin{equation}
\label{equ:DensityProfile}
    \rho_{\rm ej}(v,t) = \left\{\begin{aligned}
 \zeta_\rho\frac{M_{\rm ej}}{v_{\rm tr}^3t^3}\left(\frac{v}{v_{\rm tr}} \right)^{-\delta},~v<v_{\rm tr}, \\
 \zeta_\rho\frac{M_{\rm ej}}{v_{\rm tr}^3t^3}\left(\frac{v}{v_{\rm tr}} \right)^{-n},~v\geq v_{\rm tr},
\end{aligned}\right.
\end{equation}
where $v_{\rm tr}\approx\zeta_\nu\sqrt{E_{\rm kin,f}/M_{\rm ej}}$ is the transition velocity, $\zeta_\rho=(n-3)(3-\delta)/4\pi(n-\delta)$, and $\zeta_\nu=\sqrt{2(5-\delta)(n-5)/(n-3)(3-\delta)}$, with the parameters satisfying $\delta < 3$ and $n>5$. Such a broken power-law density profile is consistent with those inferred for SLSNe Ic and SNe Ic-BL in observations \citep{Inserra2013,Nicholl2026}, featuring a dense inner core surrounded by an extended outer envelope. 

The newborn magnetars in SNe Ic-BL may have initial rotation periods of a few milliseconds and magnetic field strengths $B_{\rm p}\gtrsim10^{15}\,{\rm G}$, indicating a spin-down timescale of $t_{\rm sd} \sim 10^1-10^4\,{\rm s}$. Since the spin-down timescale is comparable to the shock-crossing time of the stellar envelope for Wolf-Rayet stars, most of the initial rotational energy can be deposited into the ejecta during this phase, after which the ejecta may evolve toward a self-similar density profile {following Equation (\ref{equ:DensityProfile}), similar to that of ordinary SESNe. Thus, we adopt representative values of $\delta = 1$ and $n = 10$ as our fiducial parameters. The dependence on different density profiles is explored in Section \ref{app_sec:Dependence_Density} of Appendix \ref{app_sec:Dependence}, where we find that different density profiles change the inferred fitting parameters by $\lesssim 20-30\%$.}

The photospheric temperature can be given by
\begin{equation}
    T_{\rm ph}=\max\left[\left(\frac{L_{\rm SN}}{4\pi \sigma_{\rm SB}v_{\rm ph}^2t^2}\right)^{1/4}, T_{\rm floor}\right],
\end{equation}
where $\sigma_{\rm SB}$ is the Stefan-Boltzmann constant, $v_{\rm ph}$ is the photospheric velocity, and $T_{\rm floor}$ is the floor temperature. For $T_{\rm ph}>T_{\rm floor}$, $v_{\rm ph}$ can be determined by solving $\int^{+\infty}_{v_{\rm ph}}\kappa\rho_{\rm ej}(v,t)tdv=2/3$ at different times, while for $T_{\rm ph}=T_{\rm floor}$, $v_{\rm ph}=(L_{\rm SN}/4\pi \sigma_{\rm SB}t^2T_{\rm floor}^4)^{1/2}$. Then, the photospheric radius can be written as $R_{\rm ph}=v_{\rm ph}t$. With the simulated bolometric lightcurve and the evolution of the photospheric radius and temperature, one can derive the monochromatic luminosity as $L_{\nu,\rm SN} = 4\pi R_{\rm ph}^2\cdot\pi B_\nu(T_{\rm ph})$, which is converted to the observed flux density $F_{\nu,\rm SN}=(1+z)L_{\nu,\rm SN}/4\pi D_{\rm L}^2$, and to the AB magnitude via $m_\nu = -2.5\log_{10}(F_{\nu,\rm SN}/3631\,{\rm Jy})$, where $B_\nu$ is the Planck function and $D_{\rm L}$ is the luminosity distance.

\section{Data Collection and Fittings} \label{sec:DataCollection}

{
\begin{deluxetable}{lcccclcccc}
\tablewidth{0pt}
\tabletypesize{\scriptsize}
\tablecaption{Observational Information of SNe Ic-BL. \label{tab:Sample}}
\tablehead{
\colhead{SN Name} & \colhead{$z$} & \colhead{$E(B-V)_{\rm MW}$} & \colhead{$E(B-V)_{\rm host}$} & \colhead{Reference} & \colhead{SN Name} & \colhead{$z$} & \colhead{$E(B-V)_{\rm MW}$} & \colhead{$E(B-V)_{\rm host}$} & \colhead{Reference} \\
\colhead{} & \colhead{} & \colhead{(mag)} & \colhead{(mag)} & \colhead{} & \colhead{} & \colhead{} & \colhead{(mag)} & \colhead{(mag)} & \colhead{} 
}
\startdata
SN\,1998bw$^*$ & 0.0087 & 0.064 & ... & [1] & SN\,2002ap & 0.0022 & 0.071 & 0.008 & [2] \\
SN\,2003dh$^*$ & 0.1685 & 0.025 & ... & [3] & SN\,2003jd & 0.0187 & 0.044 & 0.10 & [4] \\ 
SN\,2003lw$^*$ & 0.1055 & 0.78 & 0.29 & [5] & SN\,2004aw & 0.0163 & 0.10 & 0.35 & [6] \\
SN\,2007ru & 0.01546 & 0.27 & ... & [7] & SN\,2009bb & 0.0099 & 0.098 & 0.482 & [8] \\
SN\,2010bh$^*$ & 0.0592 & 0.117 & 0.061 & [9] & PTF\,10bzf & 0.0498 & 0.012 & ... & [10] \\ 
PTF\,10ciw & 0.1150 & 0.0321 & ... & [11] & PTF\,10gvb & 0.0980 & 0.0212 & ... & [11] \\ 
PTF\,10qts & 0.0907 & 0.029 & ... & [12] & PTF\,10tqv & 0.0795 & 0.061 & ... & [11] \\ 
PTF\,10vgv & 0.0150 & 0.1416 & ... & [11] & PTF\,10aavz & 0.0630 & 0.0479 & 0.08 & [11] \\ 
PTF\,11cmh & 0.1055 & 0.0121 & ... & [11] & PTF\,11lbm & 0.039 & 0.0547 & ... & [11] \\
PTF\,12as & 0.0332 & 0.0213 & 0.25 & [11] & SN\,2013dx$^*$ & 0.145 & 0.038 & ... & [13] \\ 
iPTF\,13alq & 0.0540 & 0.0118 & ... & [11] & SN\,2014ad & 0.01546 & 0.20 & 0.02 & [14] \\ 
iPTF\,14bfu$^*$ & 0.384 & 0.1022 & 0.0578 & [15] & iPTF\,14dby & 0.0736 & 0.0464 & ... & [11] \\ 
iPTF\,15dld & 0.047 & 0.0269 & 0.04 & [11] & iPTF\,15dqg & 0.0577 & 0.053 & ... & [11] \\ 
SN\,2016P & 0.0146 & 0.024 & 0.05 & [16] & SN\,2016coi & 0.0036 & 0.08 & 0.125 & [17] \\ 
SN\,2016jca$^*$ & 0.1475 & 0.0868 & 0.0527 & [18] & SN\,2017iuk$^*$ & 0.0368 & 0.155 & 0.062 &  [19] \\
SN\,2018ie & 0.014 & 0.06 & 0.02 & [20] & SN\,2018bvw & 0.054 & 0.021 & ... & [21] \\ 
SN\,2018ell & 0.0638 & 0.052 & ... & [21] & SN\,2018fip$^*$ & 0.1171 & 0.012 & ... & [22] \\ 
SN\,2018hom & 0.0362 & 0.055 & ... & [21] & SN\,2018kva & 0.043 & 0.027 & ... & [23] \\ 
SN\,2019gwc & 0.038 & 0.012 & ... & [23] & SN\,2019hsx & 0.020652 & 0.042 & 0.31 & [23] \\ 
SN\,2019lci & 0.0292 & 0.067 & 0.118 & [21] & SN\,2019moc & 0.055 & 0.171 & ... & [23] \\ 
SN\,2019pgo & 0.0500 & 0.156 & ... & [21] & SN\,2019qfi & 0.0285 & 0.19 & ... & [21] \\ 
SN\,2020bvc$^*$ & 0.0252 & 0.01 & 0.138 & [21] & SN\,2020hes & 0.0700 & 0.034 & ... & [21] \\ 
SN\,2020hyj & 0.055 & 0.025 & ... & [21] & SN\,2020lao & 0.030814 & 0.045 & ... & [23] \\ 
SN\,2020rfr & 0.0725 & 0.034 & ... & [21] & SN\,2020rph & 0.042 & 0.210 & ... & [23] \\ 
SN\,2020tkx & 0.027 & 0.073 & ... & [21] & SN\,2020abxc & 0.0600 & 0.082 & ... & [21] \\ 
SN\,2020abxl & 0.0815 & 0.111 & ... & [21] & SN\,2020adow & 0.0075 & 0.04 & ... & [21] \\ 
SN\,2021xv & 0.05 & 0.0167 & ... & [21] & SN\,2021bmf & 0.017 & 0.27 & ... & [23] \\ 
SN\,2021fop & 0.077 & 0.029 & ... & [21] & SN\,2021hyz & 0.046 & 0.04 & ... & [21] \\ 
SN\,2021ncn & 0.02461 & 0.046 & ... & [24] & SN\,2021qjv & 0.03803 & 0.0129 & ... & [21] \\ 
SN\,2021too & 0.07 & 0.055 & ... & [23] & SN\,2022cca & 0.042 & 0.019 & ... & [24] \\ 
SN\,2022crr & 0.0188 & 0.134 & ... &  [25] & SN\,2022ofv & 0.0472 & 0.023 & ... & [24] \\ 
SN\,2022wlm & 0.05 & 0.187 & ... & [24] & SN\,2023iwy & 0.030 & 0.075 & ... & [24] \\ 
SN\,2023mee & 0.064 & 0.063 & ... & [24] & SN\,2023qzm & 0.040 & 0.049 & ... & [24] \\ 
SN\,2024tv & 0.048 & 0.112 & ... & [24] & SN\,2024itg & 0.035 & 0.159 & ... & [24] \\ 
SN\,2024rqu & 0.0654 & 0.018 & ... & [24] & SN\,2024aber & 0.07 & 0.020 & ... & [24] \\ 
SN\,2024adml & 0.037 & 0.034 & ... & [24] & SN\,2025op & 0.01133 & 0.027 & ... & [24] \\ 
SN\,2025tt & 0.045 & 0.136 & ... & [24] & SN\,2025cnu & 0.0285 & 0.059 & ... & [24] \\ 
SN\,2025moc & 0.035 & 0.049 & ... & [24] & SN\,2025ocy & 0.02245 & 0.126 & ... & [24] \\ 
SN\,2025shf & 0.0353 & 0.049 & ... & [24] & SN\,2025vaw & 0.019 & 0.049 & ... & [24] \\ 
SN\,2025vef & 0.03 & 0.028 & ... & [24] & SN\,2025wkm$^*$ & 0.1194 & 0.05 & ... & [26] \\
\enddata
\tablecomments{SNe Ic-BL marked with an asterisk ($^*$) denote GRB-SNe. We assume $E(B-V)_{\rm host} = 0$ when the host extinction is unmeasured or negligible. Lightcurve reference: [1] \cite{Clocchiatti2011}; [2] \cite{Foley2003}; [3] \cite{Deng2005}; [4] \cite{Valenti2008}; [5] \cite{Malesani2004}; [6] \cite{Taubenberger2006}; [7] \cite{Sahu2009}; [8] \cite{Pignata2011}; [9] \cite{Cano2011}; [10] \cite{Corsi2011,Taddia2019}; [11] \cite{Taddia2019}; [12] \cite{Walker2014,Taddia2019}; [13] \cite{Toy2016,Volnova2017}; [14] \cite{Sahu2018}; [15] \cite{Cano2015}; [16] \cite{Gangopadhyay2020}; [17] \cite{Prentice2018}; [18] \cite{Cano2017}; [19] \cite{Izzo2019}; A. A. Volnova et al. (in prep); [20] \cite{Prentice2019} ; [21] \cite{Srinivasaragavan2024}; [22] \cite{Rossi2026}; [23] \cite{Anand2024,Srinivasaragavan2024}; [24] ZTF forced-photometry service \citep{Masci2019}; [25] \cite{Mousa2025}; [26] \cite{Srinivasaragavan2025SN2025wkm}.}
\end{deluxetable}
}

In order to precisely constrain the magnetar parameters and ejecta properties of SNe Ic-BL, it is necessary to fit the lightcurve (e.g., bolometric or multi-band) and photospheric evolution (e.g., photospheric radius, temperature, or velocity). Considering that the evolution of the photospheric velocity is highly sensitive to the choice of spectral lines and model assumptions, while multi-band lightcurves provide more reliable information on the evolution of the photospheric radius and temperature, we therefore uniformly fit the multi-band lightcurves of SNe Ic-BL in the following analysis. 

The multi-band lightcurves of spectroscopically confirmed SNe Ic-BL from the literature are collected based on three main criteria: (1) the source must include at least a few data points after $\sim60-70\,{\rm d}$; (2) the source must have at least one data point during the rising phase of the lightcurve; (3) the source should have some data points available in at least two filters around the peak. The criterion (1) ensures that each SN Ic-BL source shows a clear $^{56}$Ni tail, allowing a robust constraint on the $^{56}$Ni mass. The criteria (2) and (3) provide critical early-time and peak coverage of the lightcurve and photospheric evolution, thereby allowing for constraints on the magnetar parameters and ejecta properties. Following these criteria, we collect 80 SNe Ic-BL from the literature reported before {March 7th, 2026,} with their names, redshift $z$, Milky Way extinction $E(B-V)_{\rm MW}$, host extinction $E(B-V)_{\rm host}$, and lightcurve reference listed in Table \ref{tab:Sample}. Among them, 11 SNe Ic-BL are associated with GRBs and/or jet afterglows, which include SN\,1998bw (GRB\,980425), SN\,2003dh (GRB\,030329),  SN\,2003lw (GRB\,031203), SN\,2010bh (GRB\,100316D), SN\,2013dx (GRB\,130702A), iPTF14bfu (GRB\,140606B), SN\,2016jca (GRB\,161219B), SN\,2017iuk (GRB\,171205A), SN\,2018fip (GRB\,180728A), SN\,2020bvc, and SN\,2025wkm (EP\,250827b), whereas no GRB association has been identified for the rest. We note that \cite{Izzo2020,Ho2020} suggested that the early-time bump of SN\,2020bvc could be an off-axis jet afterglow or a cooling emission from a jet-cocoon system. Although no GRB association has been reported, we still classify it as a GRB-SN in the following discussion. 

\begin{figure}[t]
\centering
\includegraphics[width=1\linewidth, trim = 72 34 105 60, clip]{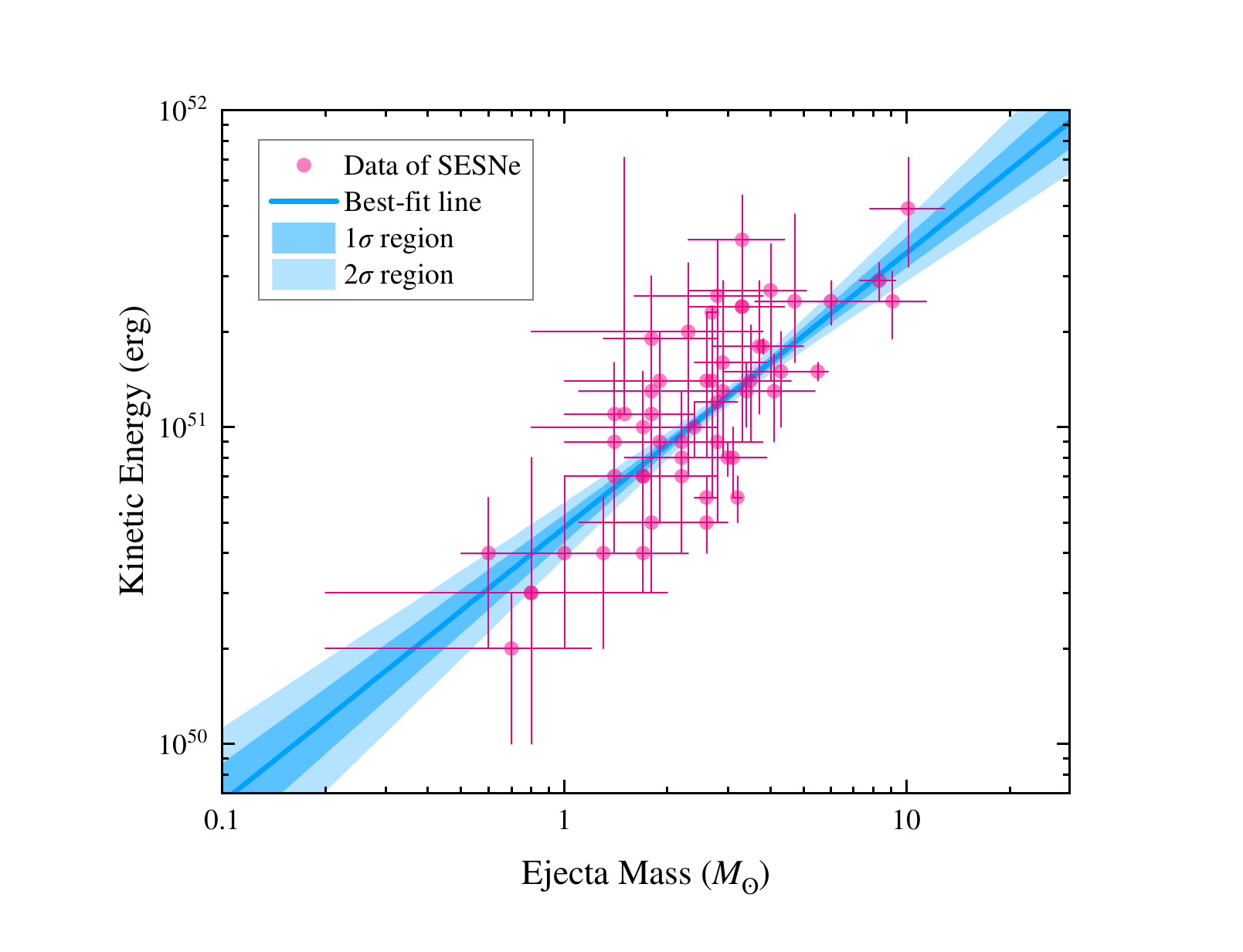}
\caption{Relationship between observed kinetic energy and ejecta mass for ordinary SESNe. The observed data (pink points) are collected from \cite{Lyman2016,Taddia2018}. The best-fit log-log line  and its $1\sigma$ and $2\sigma$ credible region are marked as a blue solid line, a light-blue shaded area, and a darker-blue shaded area, respectively.  }
\label{fig:SESNe_Mej_Ekin}
\end{figure}

We then adopt a Markov Chain Monte Carlo method with the \texttt{emcee}
package \citep{ForemanMackey2013} to fit the multi-band lightcurve of each SN Ic-BL based on the model in Section \ref{sec:Model}. In our fittings, a standard Galactic reddening law with $A_V=3.1$ is considered \citep{Cardelli1989}. In cases where the host-galaxy extinction of an SN Ic-BL is either unreported or measured to be negligible, we assume zero host extinction and only consider Milky Way extinction in our  fittings. Figure \ref{fig:SESNe_Mej_Ekin} shows that the kinetic energies and ejecta masses of ordinary core-collapse SESNe follow an approximately linear relation in log-log space \citep[e.g.,][]{Lyman2016,Taddia2018}, with a fitting line of
\begin{equation}
\label{equ:Neutrino_Energy}
    {E_{\rm kin}/{\rm erg} = (4.82\pm0.44\times10^{50})\times(M_{\rm ej}/M_\odot)^{0.87\pm0.08}.}
\end{equation}
Since the kinetic energies of ordinary SESNe are likely powered primarily by the neutrino-driven mechanism, we take the above equation to represent the initial kinetic energy $E_{\rm kin,i}$ of SN Ic-BL within the magnetar central engine framework. {We do not exclude the possibility that the inferred kinetic energies of ordinary SESNe may be slightly affected by central energy injection. However, as discussed in Section \ref{app_sec:Dependence_E_kin} of Appendix \ref{app_sec:Dependence}, this effect can be neglected provided that the initial rotational energy are much larger than the neutrino-driven initial kinetic energy for each SN Ic-BL. }

There are totally seven free parameters for Non-GRB-SNe in our magnetar-powered model: initial rotation spin $P_{\rm i}$, polar magnetic field strength $B_{\rm p}$, ejecta mass $M_{\rm ej}$, $^{56}$Ni mass $M_{\rm Ni}$, effective gamma-ray opacity $\kappa_\gamma$\footnote{{We note that numerical simulations \citep[e.g.,][]{Colgate1980,Swartz1995} suggested a gamma-ray opacity of $\kappa_\gamma\simeq 0.03-0.05\,{\rm cm}^2\,{\rm g}^{-1}$. However, there is a long-standing issue in the literature that gamma-ray opacity ($\kappa_\gamma \simeq 0.03-0.05\,{\rm cm}^2\,{\rm g}^{-1}$) and the commonly adopted optical opacity ($\kappa \sim 0.07-0.1\,{\rm cm}^2\,{\rm g}^{-1}$) in the literature cannot simultaneously reproduce the lightcurves of SESNe \citep[e.g.,][]{Wheeler2015}. Therefore, following many studies in the literature, we treat the gamma-ray opacity as a free parameter rather than fixing it to a specific value predicted by simulations.}}, floor temperature $T_{\rm floor}$, and time of explosion relative to the first observed data point $t_{\rm first}$. In addition, considering that some SNe Ic-BL may originate from BH collapsars, in which the kinetic energy is provided by the BH collapsar itself and the SN emission arises from $^{56}$Ni decay, the lightcurves of each SN Ic-BL are fitted with the $^{56}$Ni-powered model to compare the fitting quality with that obtained from the magnetar-powered model. Within the $^{56}$Ni-powered framework, we remove $P_{\rm i}$ and $B_{\rm p}$ as model parameters, and instead treat the initial kinetic energy $E_{\rm kin,i}$ as a free parameter. We note that the explosion times of GRB-SNe are set to the detection times of the GRBs, so $t_{\rm first}$ is known for these events. Cocoon cooling emission is expected to decay rapidly \citep[][]{Nakar2017,Piro2018,Zheng2025,Zhu2025} and may not significantly affect the subsequent SN emission. For those GRB-SNe that show an early-time bump likely originating from cocoon cooling emission, the corresponding data points are excluded from the fittings. In contrast, for those with early-time optical afterglow, the flux density is modeled as $F_{\nu,\rm AF}=F_0 t^{-\alpha}\nu^{-\beta}$, where $F_0$ is a normalization constant, $\alpha$ is the temporal decay index, and $\beta$ is the spectral index. Accordingly, three additional fitting parameters are introduced.

{
\startlongtable
\begin{deluxetable*}{lccccccc|c|c}
\tablewidth{0pt}
\tabletypesize{\footnotesize}
\tablecaption{Fitting Results of Derived Parameters for Collected SNe Ic-BL. \label{tab:Results}}
\tablehead{
\colhead{SN Name} & \colhead{$P_{\rm i}$} & \colhead{$B_{\rm p}$} & \colhead{$M_{\rm ej}$} & \colhead{$M_{\rm Ni}$} & \colhead{$\kappa_\gamma$} & \colhead{$T_{\rm floor}$} & \colhead{$t_{\rm first}$} & \colhead{$f_{\rm Ni}$} & \colhead{$\Delta{\rm AIC}$} \\
 & \colhead{(ms)} & \colhead{($10^{15}\,{\rm G}$)} & \colhead{($M_\odot$)} & \colhead{($M_\odot$)} & \colhead{(${\rm cm}^2\,{\rm g}^{-1}$)} & \colhead{($10^3\,{\rm K}$)} &  \colhead{(d)} & \colhead{} & \colhead{}
}
\startdata
SN\,1998bw$^*$ & $2.07^{+0.03}_{-0.03}$ & $2.74^{+0.02}_{-0.02}$ & $2.68^{+0.03}_{-0.02}$ & $0.31^{+0.00}_{-0.00}$ & $0.16^{+0.00}_{-0.00}$ & $4.95^{+0.01}_{-0.01}$ & ... & ${0.11}^{+0.00}_{-0.00}$ & $21979$\\ 
SN\,2002ap & $2.37^{+0.03}_{-0.01}$ & $9.43^{+0.07}_{-0.08}$ & $1.59^{+0.01}_{-0.02}$ & $0.10^{+0.00}_{-0.00}$ & $0.33^{+0.00}_{-0.00}$ & $4.49^{+0.01}_{-0.01}$ & $1.05^{+0.02}_{-0.01}$ & $0.06^{+0.00}_{-0.00}$ & $11644$ \\ 
SN\,2003dh$^*$ & $2.41^{+0.60}_{-0.53}$ & $2.51^{+0.52}_{-0.41}$ & $3.40^{+1.22}_{-0.92}$ & $0.17^{+0.08}_{-0.05}$ & $0.17^{+0.28}_{-0.13}$ & $5.61^{+0.30}_{-0.52}$ & ... & $0.05^{+0.04}_{-0.02}$ & $11$ \\ 
SN\,2003jd & $0.74^{+0.03}_{-0.02}$ & $3.75^{+0.06}_{-0.08}$ & $4.36^{+0.07}_{-0.08}$ & $0.26^{+0.00}_{-0.00}$ & $0.25^{+0.01}_{-0.01}$ & $4.48^{+0.02}_{-0.02}$ & $5.29^{+0.31}_{-0.13}$ & ${0.06}^{+0.00}_{-0.00}$ & $2031$ \\ 
SN\,2003lw$^*$ & $3.45^{+0.07}_{-0.02}$ & $1.48^{+0.01}_{-0.01}$ & $1.93^{+0.03}_{-0.01}$ & $0.49^{+0.00}_{-0.00}$ & $0.64^{+0.07}_{-0.09}$ & $5.32^{+0.03}_{-0.01}$ & ... & ${0.25}^{+0.00}_{-0.00}$ & $2085$ \\ 
SN\,2004aw & $2.02^{+0.06}_{-0.06}$ & $3.98^{+0.09}_{-0.10}$ & $3.08^{+0.10}_{-0.06}$ & $0.23^{+0.00}_{-0.00}$ & $0.28^{+0.04}_{-0.03}$ & $4.41^{+0.02}_{-0.02}$ & $8.19^{+0.27}_{-0.27}$ & ${0.07}^{+0.00}_{-0.00}$ & $1394$ \\ 
SN\,2007ru & $0.84^{+0.01}_{-0.01}$ & $2.77^{+0.01}_{-0.01}$ & $4.96^{+0.04}_{-0.04}$ & $0.19^{+0.00}_{-0.00}$ & $0.23^{+0.01}_{-0.01}$ & $4.50^{+0.01}_{-0.01}$ & $9.01^{+0.02}_{-0.01}$ & ${0.04}^{+0.00}_{-0.00}$ & $3571$ \\  
SN\,2009bb & $1.13^{+0.02}_{-0.02}$ & $5.19^{+0.08}_{-0.08}$ & $2.42^{+0.06}_{-0.04}$ & $0.19^{+0.00}_{-0.00}$ & $0.17^{+0.00}_{-0.00}$ & $4.85^{+0.03}_{-0.03}$ & $1.24^{+0.05}_{-0.03}$ & ${0.08}^{+0.00}_{-0.00}$ & $2884$ \\ 
SN\,2010bh$^*$ & $1.33^{+0.01}_{-0.02}$ & $8.07^{+0.05}_{-0.07}$ & $2.51^{+0.04}_{-0.03}$ & $0.11^{+0.00}_{-0.00}$ & $0.34^{+0.02}_{-0.03}$ & $3.75^{+0.01}_{-0.02}$ & ... & ${0.04}^{+0.00}_{-0.00}$ & $3198$ \\ 
PTF\,10bzf & $3.03^{+0.80}_{-0.34}$ & $3.49^{+0.21}_{-0.32}$ & $2.23^{+0.16}_{-0.21}$ & $0.17^{+0.01}_{-0.01}$ & $0.13^{+0.03}_{-0.02}$ & $6.29^{+0.28}_{-0.28}$ & $5.30^{+0.77}_{-0.40}$ & ${0.08}^{+0.01}_{-0.01}$ & $85$ \\ 
PTF\,10ciw & $2.12^{+0.52}_{-0.78}$ & $4.47^{+0.66}_{-0.43}$ & $1.57^{+0.49}_{-0.23}$ & $0.20^{+0.02}_{-0.02}$ & $0.65^{+0.37}_{-0.23}$ & $5.84^{+0.23}_{-0.18}$ & $2.91^{+0.61}_{-0.73}$ & ${0.13}^{+0.03}_{-0.03}$ & $167$ \\ 
PTF\,10gvb & $1.45^{+0.12}_{-0.11}$ & $3.40^{+0.16}_{-0.20}$ & $2.85^{+0.22}_{-0.15}$ & $0.34^{+0.01}_{-0.02}$ & $0.52^{+0.30}_{-0.14}$ & $4.84^{+0.08}_{-0.08}$ & $3.08^{+0.39}_{-0.28}$ & ${0.12}^{+0.01}_{-0.01}$ & $387$  \\ 
PTF\,10qts & $1.25^{+0.06}_{-0.05}$ & $3.20^{+0.13}_{-0.11}$ & $2.75^{+0.15}_{-0.18}$ & $0.25^{+0.02}_{-0.02}$ & $0.31^{+0.07}_{-0.08}$ & $5.00^{+0.17}_{-0.13}$ & $0.49^{+0.05}_{-0.02}$ & ${0.09}^{+0.01}_{-0.01}$ & $193$ \\ 
PTF\,10tqv & $5.78^{+1.82}_{-0.80}$ & $3.35^{+0.19}_{-0.31}$ & $1.03^{+0.11}_{-0.15}$ & $0.09^{+0.00}_{-0.00}$ & $0.51^{+0.37}_{-0.20}$ & $5.65^{+0.08}_{-0.08}$ & $1.20^{+0.34}_{-0.17}$ & ${0.09}^{+0.01}_{-0.01}$ & $96$ \\ 
PTF\,10vgv & $4.65^{+0.00}_{-0.00}$ & $4.61^{+0.02}_{-0.02}$ & $1.14^{+0.00}_{-0.00}$ & $0.05^{+0.00}_{-0.00}$ & $0.13^{+0.00}_{-0.00}$ & $5.21^{+0.02}_{-0.02}$ & $0.37^{+0.01}_{-0.01}$ & ${0.04}^{+0.00}_{-0.00}$ & $11334$ \\ 
PTF\,10aavz & $1.75^{+0.03}_{-0.08}$ & $2.18^{+0.03}_{-0.02}$ & $3.64^{+0.12}_{-0.07}$ & $0.32^{+0.00}_{-0.00}$ & $0.41^{+0.03}_{-0.03}$ & $5.86^{+0.02}_{-0.02}$ & $3.09^{+0.11}_{-0.13}$ & ${0.09}^{+0.00}_{-0.00}$ & $3723$ \\ 
PTF\,11cmh & $1.86^{+0.11}_{-0.08}$ & $5.32^{+0.16}_{-0.22}$ & $1.62^{+0.08}_{-0.07}$ & $0.20^{+0.01}_{-0.01}$ & $0.33^{+0.09}_{-0.04}$ & $5.79^{+0.06}_{-0.07}$ & $1.51^{+0.21}_{-0.13}$ & ${0.12}^{+0.01}_{-0.01}$ & $449$ \\ 
PTF\,11lbm & $2.17^{+0.05}_{-0.03}$ & $5.71^{+0.06}_{-0.06}$ & $2.10^{+0.03}_{-0.04}$ & $0.08^{+0.00}_{-0.00}$ & $0.63^{+0.27}_{-0.26}$ & $5.15^{+0.02}_{-0.02}$ & $1.46^{+0.09}_{-0.06}$ & ${0.04}^{+0.00}_{-0.00}$ & $770$ \\ 
PTF\,12as & $3.56^{+0.16}_{-0.09}$ & $8.18^{+0.36}_{-0.36}$ & $1.02^{+0.06}_{-0.06}$ & $0.08^{+0.00}_{-0.00}$ & $0.10^{+0.01}_{-0.00}$ & $5.87^{+0.10}_{-0.10}$ & $0.62^{+0.03}_{-0.02}$ & ${0.08}^{+0.01}_{-0.01}$ & $247$ \\ 
SN\,2013dx$^*$ & $2.70^{+0.08}_{-0.12}$ & $2.63^{+0.11}_{-0.07}$ & $1.59^{+0.06}_{-0.06}$ & $0.22^{+0.02}_{-0.01}$ & $0.12^{+0.03}_{-0.03}$ & $5.34^{+0.12}_{-0.12}$ & .. & ${0.14}^{+0.02}_{-0.01}$ & $193$ \\ 
iPTF\,13alq & $1.64^{+0.04}_{-0.06}$ & $2.90^{+0.09}_{-0.04}$ & $3.02^{+0.07}_{-0.04}$ & $0.16^{+0.01}_{-0.00}$ & $0.20^{+0.11}_{-0.03}$ & $4.83^{+0.03}_{-0.02}$ & $9.63^{+0.16}_{-0.38}$ & ${0.05}^{+0.00}_{-0.00}$ & $236$ \\ 
SN\,2014ad & $1.11^{+0.00}_{-0.00}$ & $5.76^{+0.01}_{-0.01}$ & $2.51^{+0.00}_{-0.01}$ & $0.19^{+0.00}_{-0.00}$ & $0.39^{+0.00}_{-0.00}$ & $5.15^{+0.00}_{-0.00}$ & $2.02^{+0.00}_{-0.00}$ & ${0.08}^{+0.00}_{-0.00}$ & $34893$ \\ 
iPTF\,14bfu$^*$ & $1.31^{+0.21}_{-0.23}$ & $2.70^{+0.42}_{-0.24}$ & $3.17^{+0.50}_{-0.51}$ & $0.36^{+0.08}_{-0.06}$ & $0.57^{+0.69}_{-0.30}$ & $6.09^{+0.54}_{-0.61}$ & ... & ${0.11}^{+0.04}_{-0.03}$ & $33$ \\ 
iPTF\,14dby & $3.25^{+0.05}_{-0.10}$ & $5.14^{+0.19}_{-0.29}$ & $2.73^{+0.16}_{-0.09}$ & $0.18^{+0.00}_{-0.01}$ & $0.06^{+0.01}_{-0.01}$ & $4.09^{+0.05}_{-0.10}$ & $0.52^{+0.07}_{-0.04}$ & ${0.07}^{+0.00}_{-0.01}$ & $245$\\ 
iPTF\,15dld & $2.85^{+0.33}_{-0.66}$ & $6.87^{+0.47}_{-0.70}$ & $1.22^{+0.22}_{-0.10}$ & $0.10^{+0.00}_{-0.00}$ & $0.44^{+0.08}_{-0.08}$ & $5.32^{+0.06}_{-0.05}$ & $1.67^{+1.14}_{-0.39}$ & ${0.08}^{+0.01}_{-0.01}$ & $308$ \\ 
iPTF\,15dqg & $0.84^{+0.02}_{-0.02}$ & $2.92^{+0.06}_{-0.07}$ & $3.56^{+0.13}_{-0.10}$ & $0.36^{+0.01}_{-0.01}$ & $0.15^{+0.02}_{-0.01}$ & $5.11^{+0.05}_{-0.05}$ & $1.08^{+0.03}_{-0.02}$ & ${0.10}^{+0.01}_{-0.01}$ & $770$ \\ 
SN\,2016P & $1.61^{+0.07}_{-0.01}$ & $8.58^{+0.11}_{-0.07}$ & $2.00^{+0.04}_{-0.06}$ & $0.09^{+0.00}_{-0.00}$ & $0.29^{+0.01}_{-0.00}$ & $4.69^{+0.02}_{-0.01}$ & $2.74^{+0.08}_{-0.02}$ & ${0.04}^{+0.00}_{-0.00}$ & $11106$ \\ 
SN\,2016coi & $2.16^{+0.00}_{-0.00}$ & $4.30^{+0.00}_{-0.00}$ & $4.13^{+0.00}_{-0.00}$ & $0.13^{+0.00}_{-0.00}$ & $0.15^{+0.00}_{-0.00}$ & $4.12^{+0.00}_{-0.00}$ & $1.31^{+0.00}_{-0.00}$ & ${0.03}^{+0.00}_{-0.00}$ & $176799$ \\ 
SN\,2016jca$^*$ & $1.71^{+0.03}_{-0.07}$ & $3.57^{+0.11}_{-0.08}$ & $2.07^{+0.06}_{-0.08}$ & $0.27^{+0.01}_{-0.01}$ & $0.60^{+0.11}_{-0.16}$ & $5.33^{+0.08}_{-0.07}$ & ... & ${0.13}^{+0.01}_{-0.01}$ & $1021$ \\ 
SN\,2017iuk$^*$ & $1.30^{+0.03}_{-0.05}$ & $4.73^{+0.08}_{-0.12}$ & $3.28^{+0.16}_{-0.09}$ & $0.22^{+0.00}_{-0.00}$ & $0.48^{+0.02}_{-0.03}$ & $4.02^{+0.04}_{-0.03}$ & ... & ${0.07}^{+0.00}_{-0.00}$ & $988$ \\ 
SN\,2018ie & $7.61^{+0.06}_{-0.01}$ & $9.42^{+0.06}_{-0.09}$ & $0.52^{+0.01}_{-0.01}$ & $0.03^{+0.00}_{-0.00}$ & $0.47^{+0.02}_{-0.06}$ & $6.02^{+0.03}_{-0.03}$ & $0.62^{+0.03}_{-0.01}$ & ${0.06}^{+0.00}_{-0.00}$ & $5490$ \\ 
SN\,2018bvw & $2.46^{+0.14}_{-0.12}$ & $5.25^{+0.26}_{-0.37}$ & $1.40^{+0.09}_{-0.05}$ & $0.25^{+0.01}_{-0.01}$ & $0.10^{+0.01}_{-0.01}$ & $5.29^{+0.15}_{-0.12}$ & $0.93^{+0.26}_{-0.22}$ & ${0.18}^{+0.01}_{-0.02}$ & $150$ \\ 
SN\,2018ell & $0.91^{+0.06}_{-0.04}$ & $7.69^{+0.58}_{-0.54}$ & $3.09^{+0.20}_{-0.23}$ & $0.28^{+0.01}_{-0.01}$ & $0.75^{+0.33}_{-0.29}$ & $4.63^{+0.15}_{-0.10}$ & $1.17^{+0.12}_{-0.08}$ & ${0.09}^{+0.01}_{-0.01}$ & $154$\\ 
SN\,2018fip$^*$ & $3.15^{+0.17}_{-0.08}$ & $2.62^{+0.06}_{-0.11}$ & $2.10^{+0.07}_{-0.06}$ & $0.25^{+0.01}_{-0.02}$ & $0.21^{+0.07}_{-0.04}$ & $4.41^{+0.16}_{-0.26}$ & ... & ${0.12}^{+0.01}_{-0.01}$ & $491$ \\ 
SN\,2018hom & $1.02^{+0.03}_{-0.04}$ & $2.65^{+0.10}_{-0.06}$ & $3.30^{+0.07}_{-0.09}$ & $0.30^{+0.01}_{-0.01}$ & $0.38^{+0.08}_{-0.06}$ & $5.46^{+0.04}_{-0.06}$ & $10.24^{+0.31}_{-0.42}$ & ${0.09}^{+0.01}_{-0.00}$ & $431$ \\ 
SN\,2018kva & $2.89^{+0.12}_{-0.15}$ & $3.42^{+0.13}_{-0.10}$ & $2.05^{+0.11}_{-0.10}$ & $0.15^{+0.01}_{-0.01}$ & $0.21^{+0.06}_{-0.03}$ & $5.58^{+0.16}_{-0.16}$ & $0.42^{+0.03}_{-0.03}$ & ${0.07}^{+0.01}_{-0.01}$ & $266$ \\ 
SN\,2019gwc & $2.88^{+0.02}_{-0.03}$ & $4.70^{+0.06}_{-0.03}$ & $1.48^{+0.02}_{-0.02}$ & $0.11^{+0.00}_{-0.00}$ & $0.25^{+0.01}_{-0.01}$ & $5.10^{+0.04}_{-0.03}$ & $0.72^{+0.01}_{-0.01}$ & ${0.08}^{+0.00}_{-0.00}$ & $2477$ \\ 
SN\,2019hsx & $5.85^{+0.38}_{-0.30}$ & $5.65^{+0.13}_{-0.12}$ & $1.27^{+0.05}_{-0.05}$ & $0.07^{+0.00}_{-0.00}$ & $0.59^{+0.46}_{-0.17}$ & $5.34^{+0.10}_{-0.11}$ & $0.55^{+0.10}_{-0.04}$ & ${0.05}^{+0.00}_{-0.00}$ & $389$ \\ 
SN\,2019lci & $3.34^{+0.21}_{-0.11}$ & $3.60^{+0.08}_{-0.07}$ & $1.77^{+0.06}_{-0.09}$ & $0.10^{+0.01}_{-0.01}$ & $0.42^{+0.31}_{-0.16}$ & $5.42^{+0.09}_{-0.07}$ & $1.76^{+0.11}_{-0.08}$ & ${0.06}^{+0.01}_{-0.00}$ & $219$ \\ 
SN\,2019moc & $1.57^{+0.03}_{-0.06}$ & $2.70^{+0.05}_{-0.05}$ & $3.59^{+0.15}_{-0.05}$ & $0.31^{+0.01}_{-0.01}$ & $0.38^{+0.17}_{-0.09}$ & $4.64^{+0.07}_{-0.05}$ & $1.43^{+0.04}_{-0.07}$ & ${0.09}^{+0.00}_{-0.00}$ & $2930$ \\ 
SN\,2019pgo & $3.61^{+0.00}_{-0.00}$ & $3.33^{+0.01}_{-0.01}$ & $1.39^{+0.00}_{-0.00}$ & $0.30^{+0.00}_{-0.00}$ & $0.15^{+0.00}_{-0.00}$ & $5.27^{+0.03}_{-0.02}$ & $0.43^{+0.00}_{-0.00}$ & ${0.21}^{+0.00}_{-0.00}$ & $588$ \\ 
SN\,2019qfi & $1.57^{+0.17}_{-0.11}$ & $7.27^{+0.40}_{-0.49}$ & $2.20^{+0.23}_{-0.18}$ & $0.11^{+0.00}_{-0.01}$ & $0.37^{+0.20}_{-0.10}$ & $4.37^{+0.09}_{-0.07}$ & $1.16^{+0.35}_{-0.18}$ & ${0.05}^{+0.01}_{-0.01}$ & $84$ \\ 
SN\,2020bvc$^*$ & $3.52^{+0.00}_{-0.00}$ & $2.28^{+0.01}_{-0.01}$ & $1.60^{+0.00}_{-0.00}$ & $0.26^{+0.00}_{-0.00}$ & $0.11^{+0.00}_{-0.00}$ & $7.34^{+0.01}_{-0.01}$ & $1.88^{+0.00}_{-0.01}$ & ${0.16}^{+0.00}_{-0.00}$ & $19381$ \\ 
SN\,2020hes & $1.41^{+0.08}_{-0.09}$ & $3.03^{+0.11}_{-0.09}$ & $2.37^{+0.10}_{-0.12}$ & $0.29^{+0.01}_{-0.01}$ & $0.28^{+0.04}_{-0.03}$ & $5.66^{+0.07}_{-0.09}$ & $4.34^{+0.24}_{-0.25}$ & ${0.12}^{+0.01}_{-0.01}$ & $580$ \\ 
SN\,2020hyj & $3.09^{+0.12}_{-0.10}$ & $4.93^{+0.21}_{-0.21}$ & $1.96^{+0.11}_{-0.08}$ & $0.14^{+0.01}_{-0.01}$ & $0.19^{+0.16}_{-0.03}$ & $4.87^{+0.05}_{-0.05}$ & $1.50^{+0.09}_{-0.10}$ & ${0.07}^{+0.01}_{-0.01}$ & $825$ \\ 
SN\,2020lao & $2.01^{+0.03}_{-0.03}$ & $4.97^{+0.05}_{-0.06}$ & $1.85^{+0.05}_{-0.03}$ & $0.13^{+0.00}_{-0.00}$ & $0.30^{+0.02}_{-0.02}$ & $4.90^{+0.02}_{-0.03}$ & $0.41^{+0.01}_{-0.01}$ & ${0.07}^{+0.00}_{-0.00}$ & $3777$ \\ 
SN\,2020rfr & $4.43^{+0.37}_{-0.56}$ & $2.68^{+0.13}_{-0.18}$ & $1.15^{+0.15}_{-0.08}$ & $0.16^{+0.02}_{-0.02}$ & $0.23^{+0.16}_{-0.06}$ & $5.46^{+0.79}_{-0.36}$ & $2.67^{+0.27}_{-0.17}$ & ${0.14}^{+0.02}_{-0.02}$ & $36$ \\ 
SN\,2020rph & $1.70^{+0.12}_{-0.08}$ & $10.88^{+0.77}_{-0.89}$ & $2.40^{+0.13}_{-0.10}$ & $0.30^{+0.00}_{-0.00}$ & $0.15^{+0.01}_{-0.01}$ & $4.53^{+0.05}_{-0.07}$ & $2.74^{+0.30}_{-0.36}$ & ${0.12}^{+0.01}_{-0.01}$ & $232$ \\ 
SN\,2020tkx & $1.13^{+0.05}_{-0.03}$ & $4.65^{+0.11}_{-0.13}$ & $3.53^{+0.11}_{-0.14}$ & $0.16^{+0.01}_{-0.01}$ & $0.54^{+0.12}_{-0.09}$ & $4.28^{+0.04}_{-0.03}$ & $1.66^{+0.14}_{-0.08}$ & ${0.05}^{+0.00}_{-0.00}$ & $617$ \\ 
SN\,2020abxc & $1.57^{+0.17}_{-0.17}$ & $2.39^{+0.10}_{-0.12}$ & $3.38^{+0.42}_{-0.30}$ & $0.31^{+0.02}_{-0.03}$ & $0.34^{+0.38}_{-0.20}$ & $4.56^{+0.26}_{-0.26}$ & $2.68^{+0.28}_{-0.22}$ & ${0.09}^{+0.01}_{-0.02}$ & $94$ \\ 
SN\,2020abxl & $0.73^{+0.04}_{-0.04}$ & $3.36^{+0.27}_{-0.19}$ & $5.17^{+0.35}_{-0.51}$ & $0.44^{+0.02}_{-0.02}$ & $0.47^{+0.59}_{-0.25}$ & $4.52^{+0.27}_{-0.45}$ & $1.03^{+0.15}_{-0.18}$ & ${0.08}^{+0.01}_{-0.01}$ & $159$ \\ 
SN\,2020adow & $6.05^{+0.00}_{-0.00}$ & $5.81^{+0.00}_{-0.00}$ & $0.84^{+0.00}_{-0.00}$ & $0.06^{+0.00}_{-0.00}$ & $0.20^{+0.00}_{-0.00}$ & $5.50^{+0.00}_{-0.01}$ & $2.28^{+0.00}_{-0.00}$ & ${0.07}^{+0.00}_{-0.00}$ & $86213$ \\ 
SN\,2021xv & $4.41^{+0.27}_{-0.19}$ & $3.24^{+0.13}_{-0.09}$ & $0.90^{+0.04}_{-0.06}$ & $0.14^{+0.01}_{-0.01}$ & $0.16^{+0.07}_{-0.03}$ & $5.65^{+0.09}_{-0.14}$ & $0.95^{+0.07}_{-0.06}$ & ${0.15}^{+0.02}_{-0.01}$ & $255$ \\ 
SN\,2021bmf & $2.69^{+0.13}_{-0.18}$ & $2.52^{+0.05}_{-0.04}$ & $3.01^{+0.13}_{-0.11}$ & $0.14^{+0.00}_{-0.00}$ & $0.13^{+0.01}_{-0.01}$ & $5.29^{+0.09}_{-0.09}$ & $2.79^{+0.12}_{-0.23}$ & ${0.05}^{+0.00}_{-0.00}$ & $516$ \\ 
SN\,2021fop & $1.24^{+0.12}_{-0.09}$ & $2.70^{+0.17}_{-0.15}$ & $2.97^{+0.28}_{-0.34}$ & $0.25^{+0.04}_{-0.04}$ & $0.63^{+0.81}_{-0.42}$ & $4.97^{+0.20}_{-0.45}$ & $0.41^{+0.03}_{-0.02}$ & ${0.09}^{+0.02}_{-0.02}$ & $107$ \\ 
SN\,2021hyz & $3.97^{+0.25}_{-0.20}$ & $3.73^{+0.12}_{-0.12}$ & $0.96^{+0.06}_{-0.05}$ & $0.15^{+0.01}_{-0.01}$ & $0.22^{+0.15}_{-0.07}$ & $5.33^{+0.16}_{-0.24}$ & $0.47^{+0.03}_{-0.01}$ & ${0.16}^{+0.01}_{-0.02}$ & $191$ \\ 
SN\,2021ncn & $4.12^{+0.83}_{-0.35}$ & $13.94^{+0.79}_{-0.89}$ & $0.95^{+0.06}_{-0.08}$ & $0.04^{+0.00}_{-0.00}$ & $0.52^{+0.14}_{-0.12}$ & $4.72^{+0.16}_{-0.46}$ & $1.52^{+0.48}_{-0.24}$ & ${0.04}^{+0.01}_{-0.00}$ & $106$ \\ 
SN\,2021qjv & $3.50^{+0.16}_{-0.12}$ & $5.86^{+0.25}_{-0.18}$ & $1.50^{+0.06}_{-0.08}$ & $0.09^{+0.01}_{-0.00}$ & $0.73^{+0.34}_{-0.40}$ & $4.58^{+0.07}_{-0.07}$ & $0.46^{+0.02}_{-0.02}$ & ${0.06}^{+0.01}_{-0.00}$ & $326$ \\ 
SN\,2021too & $0.77^{+0.00}_{-0.00}$ & $1.98^{+0.01}_{-0.01}$ & $7.51^{+0.02}_{-0.01}$ & $0.76^{+0.00}_{-0.00}$ & $0.77^{+0.02}_{-0.02}$ & $4.28^{+0.01}_{-0.01}$ & $0.35^{+0.00}_{-0.01}$ & ${0.10}^{+0.00}_{-0.00}$ & $5140$ \\ 
SN\,2022cca & $2.18^{+0.10}_{-0.12}$ & $3.70^{+0.17}_{-0.13}$ & $2.18^{+0.17}_{-0.15}$ & $0.17^{+0.01}_{-0.01}$ & $0.29^{+0.17}_{-0.08}$ & $5.53^{+0.15}_{-0.16}$ & $1.66^{+0.10}_{-0.12}$ & ${0.08}^{+0.01}_{-0.01}$ & $184$ \\ 
SN\,2022crr & $2.84^{+0.14}_{-0.14}$ & $4.78^{+0.10}_{-0.08}$ & $1.71^{+0.06}_{-0.03}$ & $0.06^{+0.00}_{-0.00}$ & $0.15^{+0.01}_{-0.01}$ & $5.71^{+0.04}_{-0.03}$ & $3.05^{+0.20}_{-0.16}$ & ${0.03}^{+0.00}_{-0.00}$ & $459$\\ 
SN\,2022ofv & $0.75^{+0.03}_{-0.04}$ & $5.76^{+0.16}_{-0.13}$ & $3.65^{+0.22}_{-0.12}$ & $0.15^{+0.00}_{-0.00}$ & $0.35^{+0.05}_{-0.03}$ & $4.17^{+0.04}_{-0.04}$ & $0.85^{+0.03}_{-0.06}$ & ${0.04}^{+0.00}_{-0.00}$ & $946$ \\ 
SN\,2022wlm &  $1.87^{+0.08}_{-0.09}$ & $3.79^{+0.21}_{-0.23}$ & $2.26^{+0.16}_{-0.13}$ & $0.29^{+0.01}_{-0.02}$ & $0.13^{+0.03}_{-0.02}$ & $4.94^{+0.09}_{-0.09}$ & $0.74^{+0.05}_{-0.04}$ & ${0.13}^{+0.01}_{-0.01}$ & $246$ \\ 
SN\,2023iwy & $2.38^{+0.00}_{-0.00}$ & $1.96^{+0.00}_{-0.00}$ & $1.85^{+0.00}_{-0.00}$ & $0.33^{+0.00}_{-0.00}$ & $0.12^{+0.00}_{-0.00}$ & $5.51^{+0.02}_{-0.01}$ & $1.51^{+0.00}_{-0.01}$ & ${0.18}^{+0.00}_{-0.00}$ & $4855$ \\ 
SN\,2023mee & $1.28^{+0.04}_{-0.05}$ & $2.61^{+0.07}_{-0.06}$ & $3.29^{+0.13}_{-0.11}$ & $0.27^{+0.02}_{-0.01}$ & $0.19^{+0.05}_{-0.03}$ & $5.27^{+0.08}_{-0.10}$ & $1.53^{+0.07}_{-0.06}$ & ${0.08}^{+0.01}_{-0.01}$ & $413$ \\ 
SN\,2023qzm & $2.59^{+0.03}_{-0.08}$ & $3.11^{+0.03}_{-0.03}$ & $1.92^{+0.04}_{-0.03}$ & $0.23^{+0.00}_{-0.00}$ & $0.74^{+0.09}_{-0.14}$ & $5.60^{+0.05}_{-0.05}$ & $0.47^{+0.02}_{-0.01}$ & ${0.12}^{+0.00}_{-0.00}$ & $2127$ \\ 
SN\,2024tv & $1.29^{+0.19}_{-0.16}$ & $5.13^{+0.48}_{-0.68}$ & $2.28^{+0.51}_{-0.36}$ & $0.17^{+0.01}_{-0.02}$ & $0.30^{+0.18}_{-0.08}$ & $4.87^{+0.53}_{-0.27}$ & $0.68^{+0.22}_{-0.12}$ & ${0.08}^{+0.01}_{-0.02}$ & $24$ \\ 
SN\,2024itg & $1.28^{+0.06}_{-0.09}$ & $3.16^{+0.10}_{-0.07}$ & $3.59^{+0.21}_{-0.09}$ & $0.24^{+0.01}_{-0.01}$ & $0.27^{+0.04}_{-0.03}$ & $4.91^{+0.06}_{-0.07}$ & $4.30^{+0.21}_{-0.28}$ & ${0.07}^{+0.00}_{-0.00}$ & $693$ \\ 
SN\,2024rqu & $0.77^{+0.06}_{-0.05}$ & $3.76^{+0.16}_{-0.16}$ & $3.99^{+0.34}_{-0.33}$ & $0.14^{+0.02}_{-0.02}$ & $0.67^{+0.71}_{-0.33}$ & $4.42^{+0.12}_{-0.15}$ & $0.56^{+0.04}_{-0.03}$ & ${0.04}^{+0.01}_{-0.01}$ & $410$ \\ 
SN\,2024aber & $1.23^{+0.20}_{-0.13}$ & $3.83^{+0.36}_{-0.37}$ & $3.55^{+0.44}_{-0.57}$ & $0.21^{+0.03}_{-0.03}$ & $0.43^{+0.53}_{-0.26}$ & $4.27^{+0.24}_{-0.37}$ & $0.66^{+0.15}_{-0.08}$ & ${0.06}^{+0.02}_{-0.01}$ & $32$ \\ 
SN\,2024adml & $1.31^{+0.04}_{-0.05}$ & $3.04^{+0.11}_{-0.08}$ & $4.61^{+0.23}_{-0.24}$ & $0.21^{+0.01}_{-0.01}$ & $0.26^{+0.10}_{-0.04}$ & $4.31^{+0.06}_{-0.07}$ & $0.79^{+0.06}_{-0.06}$ & ${0.05}^{+0.00}_{-0.00}$ & $850$ \\ 
SN\,2025op & $4.00^{+0.00}_{-0.00}$ & $4.92^{+0.01}_{-0.01}$ & $1.53^{+0.00}_{-0.00}$ & $0.09^{+0.00}_{-0.00}$ & $0.10^{+0.00}_{-0.00}$ & $4.91^{+0.01}_{-0.01}$ & $0.45^{+0.00}_{-0.00}$ & ${0.06}^{+0.00}_{-0.00}$ & $6144$ \\ 
SN\,2025tt & $2.41^{+0.19}_{-0.16}$ & $3.85^{+0.26}_{-0.30}$ & $2.31^{+0.21}_{-0.15}$ & $0.34^{+0.01}_{-0.02}$ & $0.95^{+0.30}_{-0.24}$ & $4.64^{+0.13}_{-0.10}$ & $1.16^{+0.11}_{-0.12}$ & ${0.15}^{+0.01}_{-0.02}$ & $114$ \\ 
SN\,2025cnu & $2.41^{+0.16}_{-0.10}$ & $8.55^{+0.26}_{-0.35}$ & $2.06^{+0.14}_{-0.07}$ & $0.14^{+0.00}_{-0.00}$ & $0.35^{+0.11}_{-0.07}$ & $3.82^{+0.04}_{-0.06}$ & $0.70^{+0.07}_{-0.05}$ & ${0.07}^{+0.00}_{-0.00}$ & $136$ \\ 
SN\,2025mco & $2.01^{+0.13}_{-0.07}$ & $9.97^{+0.32}_{-0.34}$ & $2.84^{+0.11}_{-0.11}$ & $0.18^{+0.00}_{-0.00}$ & $0.55^{+0.24}_{-0.12}$ & $3.64^{+0.03}_{-0.03}$ & $0.78^{+0.14}_{-0.05}$ & ${0.06}^{+0.00}_{-0.00}$ & $173$ \\ 
SN\,2025ocy & $1.58^{+0.03}_{-0.04}$ & $8.56^{+0.43}_{-0.29}$ & $1.93^{+0.09}_{-0.11}$ & $0.10^{+0.00}_{-0.00}$ & $0.10^{+0.01}_{-0.02}$ & $4.04^{+0.10}_{-0.06}$ & $0.65^{+0.02}_{-0.02}$ & ${0.05}^{+0.00}_{-0.01}$ & $431$ \\ 
SN\,2025shf & $3.89^{+0.04}_{-0.10}$ & $4.52^{+0.07}_{-0.12}$ & $1.10^{+0.04}_{-0.03}$ & $0.14^{+0.00}_{-0.01}$ & $0.12^{+0.01}_{-0.01}$ & $5.05^{+0.06}_{-0.06}$ & $0.43^{+0.01}_{-0.01}$ & ${0.13}^{+0.01}_{-0.01}$ & $677$ \\ 
SN\,2025vaw & $2.19^{+0.03}_{-0.08}$ & $4.04^{+0.11}_{-0.08}$ & $3.67^{+0.12}_{-0.11}$ & $0.13^{+0.01}_{-0.00}$ & $0.39^{+0.19}_{-0.05}$ & $4.14^{+0.05}_{-0.04}$ & $3.66^{+0.05}_{-0.10}$ & ${0.03}^{+0.00}_{-0.00}$ & $820$ \\ 
SN\,2025vef & $1.95^{+0.05}_{-0.07}$ & $7.31^{+0.15}_{-0.12}$ & $2.71^{+0.07}_{-0.08}$ & $0.21^{+0.00}_{-0.00}$ & $0.35^{+0.06}_{-0.05}$ & $3.89^{+0.03}_{-0.02}$ & $0.36^{+0.01}_{-0.01}$ & ${0.08}^{+0.00}_{-0.00}$ & $964$ \\ 
SN\,2025wkm$^*$ & $1.39^{+0.03}_{-0.04}$ & $2.96^{+0.07}_{-0.07}$ & $3.16^{+0.12}_{-0.10}$ & $0.34^{+0.01}_{-0.01}$ & $0.39^{+0.16}_{-0.10}$ & $5.93^{+0.04}_{-0.04}$ & ... & ${0.11}^{+0.01}_{-0.01}$ & $4372$ \\ 
\hline
GRB-SNe & ${2.05}^{+1.64}_{-0.90}$ & ${2.75}^{+1.72}_{-0.48}$ & ${2.44}^{+1.05}_{-0.78}$ & ${0.26}^{+0.12}_{-0.09}$ & ${0.29}^{+0.40}_{-0.18}$ & ${5.26}^{+1.10}_{-1.14}$ & ... & $0.11^{+0.06}_{-0.06}$ & ... \\
Non-GRB-SNe & ${2.03}^{+1.91}_{-0.99}$ & ${4.20}^{+3.06}_{-1.45}$ & ${2.27}^{+1.59}_{-1.05}$ & ${0.17}^{+0.14}_{-0.08}$ & ${0.29}^{+0.29}_{-0.16}$ & ${4.94}^{+0.71}_{-0.68}$ & ${1.17}^{+2.39}_{-0.71}$ & $0.07^{+0.05}_{-0.03}$ & ... \\
All SNe Ic-BL & ${2.03}^{+1.81}_{-0.95}$ & ${3.97}^{+3.07}_{-1.42}$ & ${2.30}^{+1.46}_{-1.01}$ & ${0.18}^{+0.14}_{-0.09}$ & ${0.29}^{+0.30}_{-0.16}$ & ${4.98}^{+0.74}_{-0.74}$ & ... & $0.08^{+0.06}_{-0.03}$ & ... \\
\enddata 
\tablecomments{SNe Ic-BL marked with an asterisk ($^*$) denote GRB-SNe. The distributions of each parameter for the GRB-SN, Non-GRB-SN, and all SN Ic-BL samples are listed in the last three rows, respectively. }
\end{deluxetable*}
}

\begin{figure}[t!]
    \centering
    \includegraphics[width = 1\linewidth , trim = 80 165 100 65, clip]{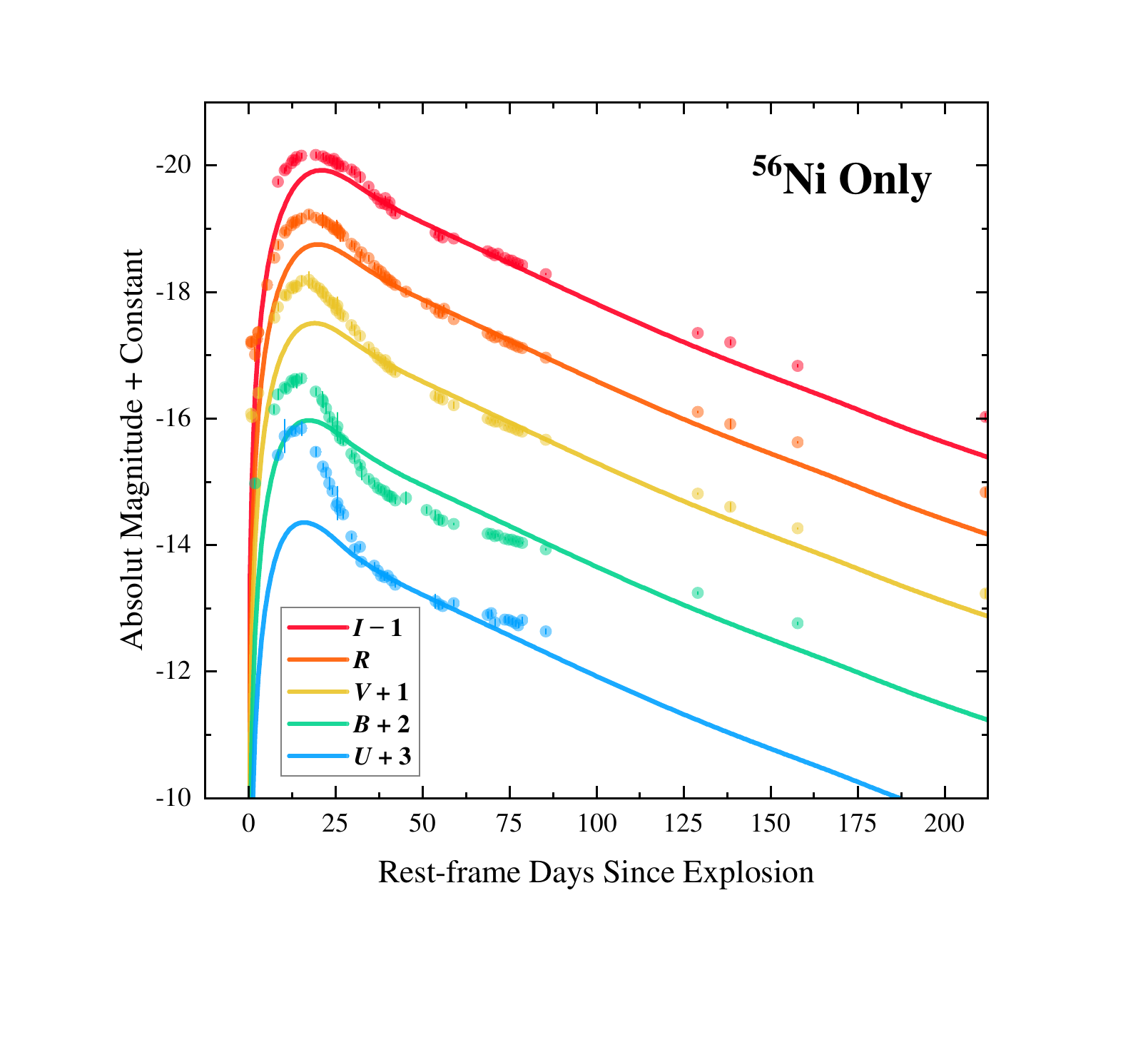}
    \includegraphics[width = 1\linewidth , trim = 80 110 100 65, clip]{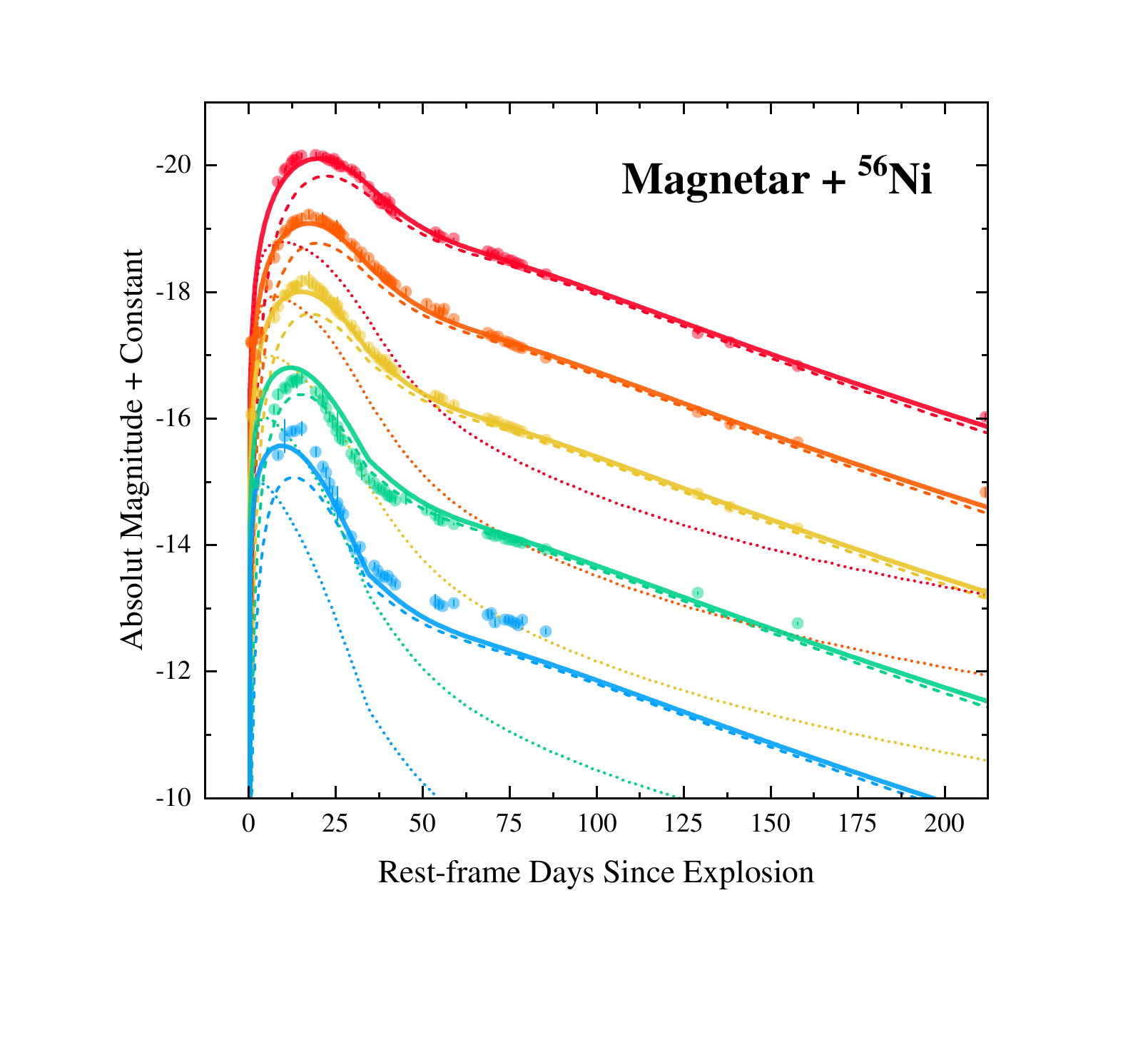}
    \caption{Multi-band observations of SN\,1998bw with best-fit results from the $^{56}$Ni-powered model (top panel) and the magnetar-powered model with $^{56}$Ni contribution (bottom panel). In the magnetar-powered case, the solid lines represent the best-fit multi-band lightcurves, while the contributions from magnetar and $^{56}$Ni emissions are shown as dotted and dashed lines, respectively. The colors correspond to different photometric filters, as indicated in the top panel. }
    \label{fig:Comparsion}
\end{figure}

As an example, the multi-band lightcurves of SN\,1998bw, with the best-fit results from the $^{56}$Ni-powered model and magnetar-powered model, are shown in Figure \ref{fig:Comparsion}. It is clear that the $^{56}$Ni-powered model struggles to fit the multi-band lightcurves of SN\,1998bw, especially near the peaks in each band. When an extra magnetar central engine is included, it can reproduce the peak excess in each band, while the $^{56}$Ni decay can explain the late-time declines. The magnetar-powered model is also strongly favored by the values of Akaike Information Criterion (AIC), {a statistical measure used to compare the relative quality of different models by balacing goodness of fit and model complexity \citep{Akaike1974}. The AIC is defined as ${\rm AIC}=2k-2\ln \mathcal{L}$, where $k$ is the number of free parameters of the model, and $\mathcal{L}$ is the maximum likelihood of the model given the observed data. A lower AIC value indicates that the model is more favorable in explaining the data. } With ${\rm AIC}_{\rm Ni}=24690$ for the $^{56}$Ni-powered model and ${\rm AIC}_{\rm mag}=2711$ for the magnetar-powered model, the resulting $\Delta{\rm AIC}={\rm AIC}_{\rm Ni}-{\rm AIC}_{\rm mag}=21979$ indicates that the magnetar-powered model provides a significantly better explanation for the multi-band lightcurves of SN\,1998bw. 

We list the fitting results (hereafter, all measurements are quoted with the median and $1\sigma$ credible region), the $^{56}$Ni mass fraction $f_{\rm Ni}=M_{\rm Ni}/M_{\rm ej}$, and the $\Delta {\rm AIC}$ values in Table \ref{tab:Results}. Furthermore, the distributions of each parameter for the GRB-SN, Non-GRB-SN, and all SN Ic-BL samples are presented in Table \ref{tab:Results}. The best-fit lightcurves for each SN Ic-BL are shown in Figure \ref{fig:Fits} of Appendix \ref{app_sec:Fits}. As shown, the multi-band lightucurves of all SNe Ic-BL can be well fitted by the magnetar-powered model during both the rising and declining phases. The values of $\Delta{\rm AIC}$ range between $\sim11-2\times10^5$, suggesting that the magnetar-powered model is more strongly favored for explaining the multi-band lightcurves of our collected SN Ic-BL sample compared to the $^{56}$Ni-powered model.

\section{Statistical Properties of Model Parameters in SNe Ic-BL} \label{sec:StatisticsSNIcBL}

Based on the inferred model parameter results presented in Table \ref{tab:Results}, we explore the statistical properties of the magnetar parameters ($P_{\rm i}$ and $B_{\rm p}$; Section \ref{sec:MagnetarParameterSNeIcBL}), explosion parameters ($M_{\rm ej}$ and $P_{\rm i}$; Section \ref{sec:ExplosionParameterSNeIcBL}), and $^{56}$Ni production ($M_{\rm Ni}$ and $f_{\rm Ni}$; Section \ref{sec:NickelProductionSNeIcBL}) in the GRB-SN, Non-GRB-SN, and entire SN Ic-BL samples, and compare our results with recent statistical studies of SNe Ic-BL.  Furthermore, in Section \ref{sec:PopulationEquality}, we assess whether the GRB-SNe and Non-GRB-SNe represent intrinsically different populations using the inferred parameters of our sample. Based on our model fits, we characterize the statistical lightcurve properties of our SN Ic-BL sample in Section \ref{sec:LightcurveSNeIcBL}.

\subsection{Magnetar Parameters} \label{sec:MagnetarParameterSNeIcBL}

\begin{figure}[t!]
    \centering
    \includegraphics[width = 1\linewidth , trim = 33 0 125 80, clip]{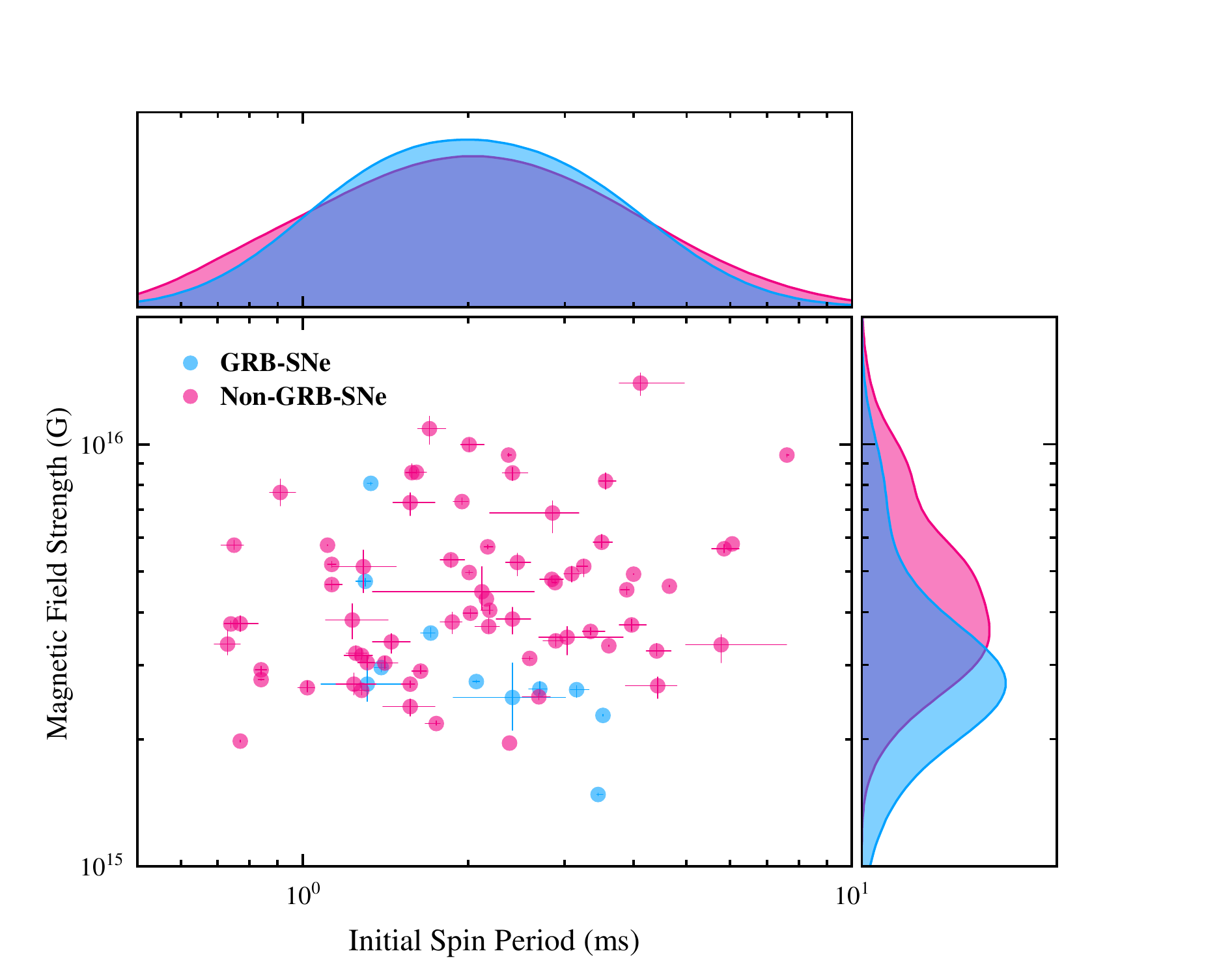}
    \caption{Magnetic field strengths against initial spin periods of GRB-SN (blue) and Non-GRB-SN (purple) magnetars. The top and right panels show the probability density distributions of the initial spin period and magnetic field strength, respectively, derived by the method of kernel density estimation. }
    \label{fig:SNIcBL_Pi_Bp}
\end{figure}

In Figure \ref{fig:SNIcBL_Pi_Bp}, we show the two-dimensional distributions between the initial magnetar spin period $P_{\rm i}$ and magnetic field strength $B_{\rm p}$ for our collected GRB-SN, Non-GRB-SN, and entire SN Ic-BL samples. The $P_{\rm i}$ distributions across these samples are comparable, with medians and $1\sigma$ regions of $2.05^{+1.64}_{-0.90}\,{\rm ms}$, $2.03^{+1.91}_{-0.99}\,{\rm ms}$, and $2.03^{+1.81}_{-0.95}\,{\rm ms}$. In our sample, the fastest-spinning magnetar has an initial spin period of $\sim0.73\,{\rm ms}$, approaching the theoretical lower limit of the Keplerian period of uniformly rotating strange quark stars \citep[e.g.,][]{Haensel2009,Dai2016}. We find that the $B_{\rm p}$ distribution of GRB-SN magnetars is systematically lower than that of Non-GRB-SN magnetars, as indicated by their inferred values of $2.75^{+1.72}_{-0.48}\times10^{15}\,{\rm G}$ and $4.20^{+3.06}_{-1.45}\times10^{15}\,{\rm G}$, respectively. This difference may arise from the observational selection effects. Specifically, on-axis GRBs, which typically occur at cosmological distances and are always associated with very bright afterglows (see Section \ref{sec:SelectionBias} for details), may bias the observed sample toward GRB-SNe powered by magnetars with relatively lower magnetic fields, as these tend to produce brighter SN emission. The $B_{\rm p}$ distribution of the magnetars in the entire SN Ic-BL sample is $3.97^{+3.07}_{-1.42}\times10^{15}\,{\rm G}$.

We then test whether correlations exist in the distribution of $P_{\rm i}$ and $B_{\rm p}$ for the GRB-SN, Non-GRB-SN, and entire SN Ic-BL samples. There is a strong and statistically significant anti-correlation in log-log space for the $P_{\rm i}-B_{\rm p}$ distribution of the GRB-SN sample, with a Pearson correlation coefficient of $r_{\rm P}=-0.718$ and a $p$-value of $0.01$. However, no strong correlation is found in the Non-GRB-SN or entire SN Ic-BL samples, both of which show weak correlations ($r_{\rm P}=0.249$ and $0.149$), with only marginal or no statistical significance ($p$-$\rm value=0.05$ and $0.19$, respectively). 

Since the spin-down timescale of SNe Ic-BL is always much shorter than the ejecta diffusion timescale, a large fraction of the initial magnetar rotational energy is expected to be converted into ejecta kinetic energy, such that the measured kinetic energy provides a reasonable estimate of the initial magnetar rotational energy. Using the \cite{Arnett1982} model powered by $^{56}$Ni decay, \cite{Lyman2016}, \cite{Taddia2019}, and \cite{Srinivasaragavan2024} fitted the bolometric lightcurves and photometric velocities near peak for samples of 8, 27, and 20 SNe Ic-BL, respectively, the majority of which are Non-GRB-SNe. Their inferred final kinetic energy $E_{\rm kin,f}$ distributions are $4.0^{+7.3}_{-1.9}\times10^{51}\,{\rm erg}$, $5.5^{+9.3}_{-4.2}\times10^{51}\,{\rm erg}$, and $2.5^{+5.5}_{-1.7}\times10^{51}\,{\rm erg}$, corresponding to $P_{\rm i}$ distributions of SN Ic-BL magnetars of $2.23^{+0.84}_{-0.90}\,{\rm ms}$, $1.90^{+1.95}_{-0.74}\,{\rm ms}$, and $2.78^{+2.17}_{-1.21}\,{\rm ms}$. Despite being inferred using a different modeling method, the median and $1\sigma$ credible region of the $P_{\rm i}$ distribution for SN Ic-BL magnetars obtained in this work, i.e., $2.03^{+1.81}_{-0.95}\,{\rm ms}$, are broadly consistent with previous estimates, with particularly good agreement with those of \cite{Lyman2016} and \citet{Taddia2019}.

\subsection{Explosion Parameters} \label{sec:ExplosionParameterSNeIcBL}

\begin{figure}[t!]
    \centering
    \includegraphics[width = 1\linewidth , trim = 33 0 125 80, clip]{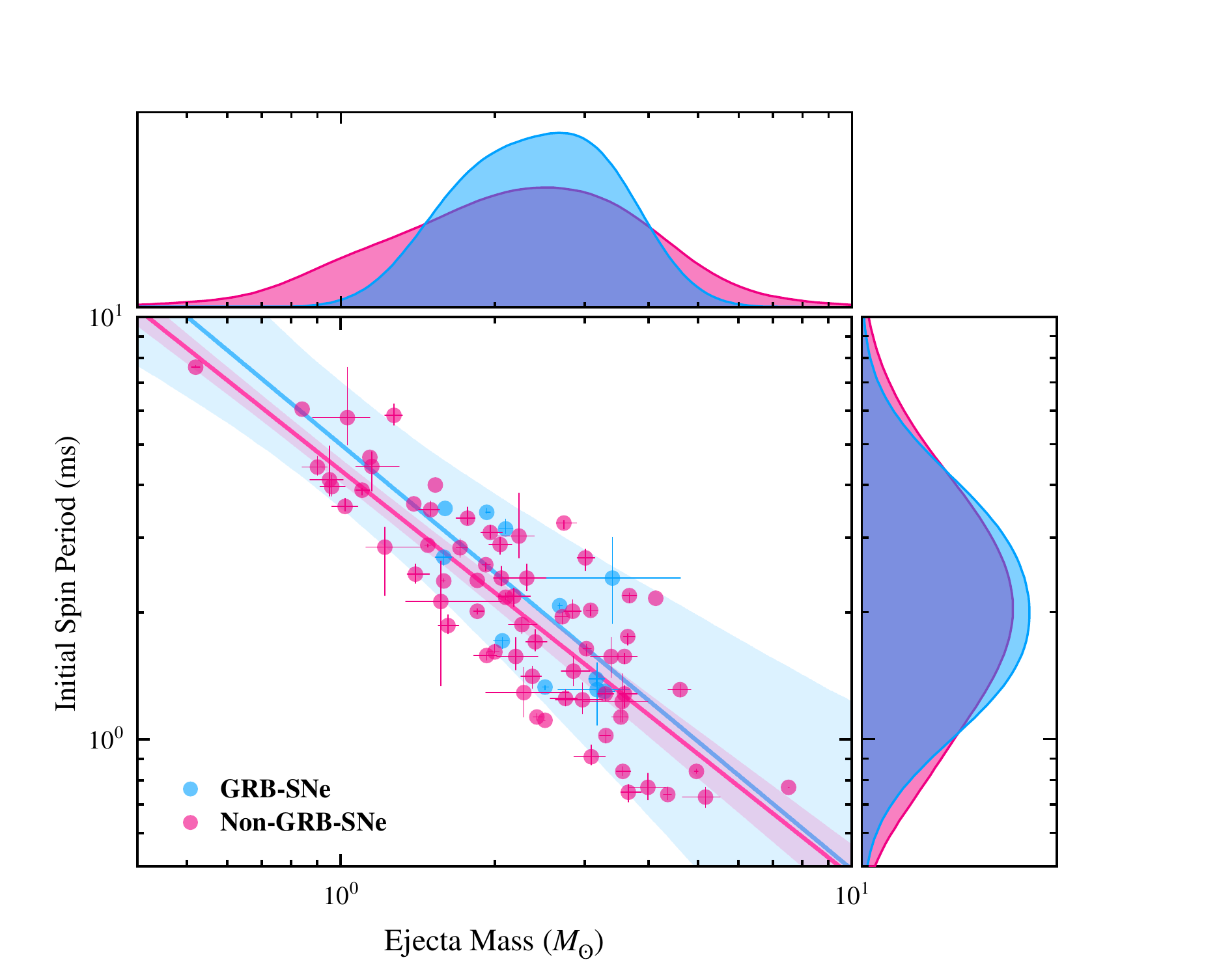}
    \caption{Initial spin periods of magnetars against ejecta masses for GRB-SNe (blue) and Non-GRB-SNe (purple). The best-fit log-log lines and 1$\sigma$ credible regions for GRB-SNe and Non-GRB-SNe are shown as solid lines and shaded areas, respectively. The top and right panels show the probability density distributions of the ejecta mass and initial spin period, respectively. }
    \label{fig:SNIcBL_Mej_Pi}
\end{figure}

As shown in Figure \ref{fig:SNIcBL_Mej_Pi}, the inferred ejecta mass $M_{\rm ej}$ distribution of the GRB-SN sample is slightly larger and more concentrated than that of the Non-GRB-SN sample, with values of $2.44^{+1.05}_{-0.78}\,M_\odot$ and $2.27^{+1.59}_{-1.05}\,M_\odot$, respectively. For the entire SN Ic-BL sample, the inferred $M_{\rm ej}$ distribution is $2.30^{+1.46}_{-1.01}\,M_\odot$, predominantly spanning between $\sim1$ and $4\,M_\odot$. Including the magnetar mass, this implies that the pre-SN mass of SN Ic-BL progenitors is typically in the range of $\sim2.5-5.5\,M_\odot$. 

Since the explosion energy of SNe Ic-BL is primarily supplied by the magnetar rotational energy, the initial spin period $P_{\rm i}$ of the magnetar is expected to be directly linked to the final explosion energy. Figure \ref{fig:SNIcBL_Mej_Pi} shows the two-dimensional distributions between $M_{\rm ej}$ and $P_{\rm i}$ for the GRB-SN and Non-GRB-SN samples, both revealing clear and similar anti-correlations in log-log space. The $M_{\rm ej}-P_{\rm i}$ relationship for the GRB-SN sample can be expressed as
\begin{equation}
    P_{\rm i}/{\rm ms} = 10^{0.70\pm0.15}\times(M_{\rm ej}/M_\odot)^{-1.01\pm0.38},
\end{equation}
with a Pearson correlation coefficient of $r_{\rm P}=-0.709$, indicating a strong anti-correlation. This anti-correlation is much stronger for the Non-GRB-SN sample, with a Pearson coefficient of $r_{\rm P}=-0.856$, which is given by
\begin{equation}
    P_{\rm i}/{\rm ms} = 10^{0.64\pm0.03}\times(M_{\rm ej}/M_\odot)^{-0.96\pm0.07}.
\end{equation}
For the entire SN Ic-BL sample, the $M_{\rm ej}-P_{\rm i}$ anti-correlation in log-log space is
\begin{equation}
    P_{\rm i}/{\rm ms} = 10^{0.64\pm0.03}\times(M_{\rm ej}/M_\odot)^{-0.96\pm0.07}
\end{equation}
with a Pearson coefficient of $r_{\rm P}=-0.845$. The corresponding $p$-values are $0.015$, $\sim10^{-20}$, and $\sim10^{-22}$ for the $M_{\rm ej}-P_{\rm i}$ distributions of the GRB-SN, Non-GRB-SN, and entire SN Ic-BL samples, respectively, indicating that these anti-correlations are statistically significant. {We note that although we set the initial neutrino-driven kinetic energy as a correlation function of $M_{\rm ej}$ in order to avoid unphysical convergence, the inferred $M_{\rm ej}-P_{\rm i}$ anti-correlations for SNe Ic-BL are intrinsic and are not induced by the $M_{\rm ej}-E_{\rm kin,i}$ correlation (i.e., Equation \ref{equ:Neutrino_Energy}) observed in ordinary SESNe. This is because the inferred initial magnetar rotational energy, which dominates the final kinetic energy of SNe Ic-BL, is typically much larger than the initial neutrino-driven kinetic energy, and thus the latter has a negligible impact on the inferred parameters (as also explored in Section \ref{app_sec:Dependence_E_kin} of Appendix \ref{app_sec:Dependence}). }

\citet{Lyman2016}, \citet{Taddia2019}, and \citet{Srinivasaragavan2024} reported that the ejecta mass distributions for SNe Ic-BL are $2.16^{+2.91}_{-0.97}\,M_\odot$, $3.10^{+4.24}_{-1.93}\,M_\odot$ and $1.91^{+2.86}_{-1.01}\,M_\odot$, respectively. These results suggest that the ejecta masses of SNe Ic-BL are not particularly massive and are broadly consistent with the distribution inferred in our work, $2.30^{+1.46}_{-1.01}\,M_\odot$. Their $M_{\rm ej}-P_{\rm i}$ distributions were also found to show a strong anti-correlation in log-log space, with $r_{\rm P}=-0.729$, $-0.788$, and $-0.802$ and corresponding $p$-values of $0.04$, $\sim10^{-6}$, and $2\times10^{-5}$, respectively. By comparison, our sample exhibits a tighter anti-correlation, with $r_{\rm P}=-0.845$ and a corresponding $p$-value of $\sim10^{-22}$, which may be partly attributed to the larger sample size adopted in this work.

\subsection{$^{56}$Ni Production} \label{sec:NickelProductionSNeIcBL}

\begin{figure}[t!]
    \centering
    \includegraphics[width = 1\linewidth , trim = 33 0 125 80, clip]{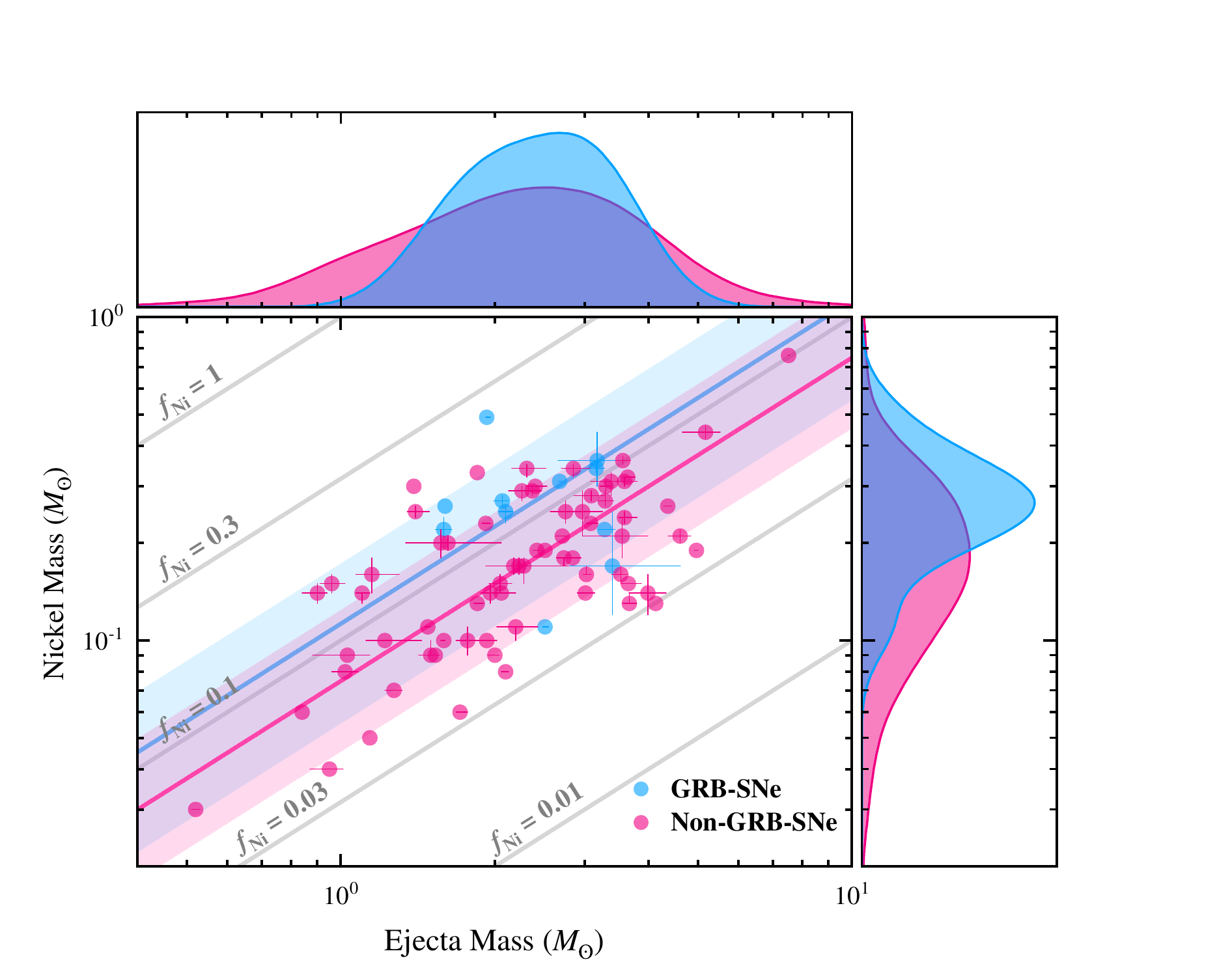}
    \caption{$^{56}$Ni masses against ejecta masses for GRB-SNe (blue) and Non-GRB-SNe (purple). The medians and 1$\sigma$ credible regions of the $^{56}$Ni mass fractions for GRB-SNe and Non-GRB-SNe are shown as colored solid lines and shaded areas, respectively. The top and right panels show the probability density distributions of the ejecta mass and $^{56}$Ni mass, respectively. The $^{56}$Ni mass fractions ranging from $f_{\rm Ni}=0.01$ to $1$ are marked by the gray solid lines. }
    \label{fig:SNIcBL_Mej_MNi}
\end{figure}

We show the distributions between the ejecta mass $M_{\rm ej}$ and $^{56}$Ni mass $M_{\rm Ni}$ for the GRB-SN and Non-GRB-SN samples in Figure \ref{fig:SNIcBL_Mej_MNi}. No clear correlation is found for the $M_{\rm ej}-M_{\rm Ni}$ distribution in log-log space of the GRB-SN sample, with a Pearson correlation coefficient of $r_{\rm P}=-0.109$. In contrast, a strong correlation is observed for the Non-GRB-SN sample as well as for the entire SN Ic-BL sample in log-log space, with $r_{\rm P}=0.684$ and $r_{\rm P}=0.636$, and corresponding $p$-values of $\sim9\times10^{-11}$ and $\sim2\times10^{-10}$, respectively. We find that the GRB-SN sample tends to have a higher $^{56}$Ni production than the Non-GRB-SN sample, with the $M_{\rm Ni}$ distributions of $0.26^{+0.12}_{-0.09}\,M_\odot$ and $0.17^{+0.14}_{-0.08}$, respectively, corresponding to the $^{56}$Ni mass fraction $f_{\rm Ni}$ distributions of $0.11^{+0.06}_{-0.06}$ and $0.07^{+0.05}_{-0.03}$. This difference may still be partly attributable to observational selection effects (see Section \ref{sec:SelectionBias}), since on-axis GRBs are typically very remote and are followed by very bright afterglows, SNe with larger $^{56}$Ni yields, and hence higher luminosities, are more easily detected. For the entire SN Ic-BL sample, the inferred $M_{\rm Ni}$ and $f_{\rm Ni}$ distributions are $0.18^{+0.14}_{-0.09}\,M_\odot$ and $0.08^{+0.06}_{-0.03}$, respectively.

Since \cite{Lyman2016}, \cite{Taddia2019}, and \cite{Srinivasaragavan2024} assumed that the peak luminosity is entirely powered by $^{56}$Ni decay, the resulting $M_{\rm Ni}$ distributions of $0.30^{+0.31}_{-0.15}\,M_\odot$, $ 0.27^{+0.50}_{-0.20}\,M_\odot$, and $0.32^{+0.21}_{-0.18}\,M_\odot$ are systematically higher, by about a factor of two, compared to our results. Consistently, their $f_{\rm Ni}$ values of $f_{\rm Ni} = 0.10^{+0.13}_{-0.05}$, $0.39^{+0.13}_{-0.10}$, and $0.15^{+0.06}_{-0.09}$ are also systematically higher than the distribution inferred in our work. Furthermore, in contrast to our $M_{\rm ej}-M_{\rm Ni}$ distribution, which shows a strong and highly significant correlation, they reported weak to moderate or statistically insignificant correlations, with $r_{\rm P}=0.543$, $0.408$ and $0.011$, and corresponding $p$-values of $0.164$, $0.03$, and $0.965$, respectively. 

\subsection{Testing for Population Equality between GRB-SNe and Non-GRB-SNe} \label{sec:PopulationEquality}

Whether GRB-SNe and Non-GRB-SNe originate from the same or different underlying populations remains an open and actively debated question in the literature. \cite{Japelj2018} found that the distributions of projected offsets from the host centers and metallicities differ between GRB-SNe and Non-GRB-SNe, suggesting a genuine difference in the nature of the two types of explosions. \cite{Modjaz2016} measured the expansion velocities of 11 GRB-SNe and 10 Non-GRB-SNe, finding that GRB-SNe have systematically higher expansion velocities than Non-GRB-SNe. Recent work by \cite{Finneran2025} analyzed a larger spectroscopic sample of GRB-SNe and Non-GRB-SNe, finding that their expansion velocity distributions substantially overlap, and therefore suggesting that GRB-SNe and Non-GRB-SNe are drawn from the same underlying population of events. Based on our inferred model parameters in Table \ref{tab:Results}, we can also test for population equality between GRB-SNe and Non-GRB-SNe.

While the distributions of $B_{\rm p}$ and $M_{\rm Ni}$ differ between the GRB-SN and Non-GRB-SN samples, likely due to selection effects, their $P_{\rm i}$, $M_{\rm ej}$, $\kappa_\gamma$, and $T_{\rm floor}$ share consistent medians and $1\sigma$ regions, as listed in Table \ref{tab:Results}. Given these mixed signals, a systematic test is necessary to explore whether the GRB-SN and Non-GRB-SN samples  originate from fundamentally the same population or not. This leads us to apply the energy distance test \citep{Szekely2007} for the datasets of the two samples.

Consider the null hypothesis that GRB-SNe and Non-GRB-SNe are derived from the same origin, with the multi-dimensional datasets represented by $\mathbf{X}$ and $\mathbf{Y}$, respectively. The energy distance used to test this null hypothesis is given by
\begin{equation}
    \mathcal{E}(\mathbf{X},\mathbf{Y}) = 2 \mathbb{E}\| \mathbf{X} - \mathbf{Y} \| - \mathbb{E}\| \mathbf{X} - \mathbf{X}' \| - \mathbb{E}\| \mathbf{Y} - \mathbf{Y}' \|,
\end{equation}
where $\|\cdot\|$ denotes the Euclidean distance between two datasets, $\mathbb{E}$ represents the expected value, and primed $\mathbf{X}'$ and  $\mathbf{Y}'$ refer to independent random draws from the respective datasets. Then, we use a permutation test to simulate the $p$-value by
\begin{equation}
    p{\text{\rm -value}} = \frac{\sum_{i=1}^{n_{\rm perm}}\mathbf{1}(\mathcal{E}_{{\rm perm,}i}\ge \mathcal{E}) + 1}{n_{\rm perm} + 1},
\end{equation}
where the number of permutations we set is $n_{\rm perm}=10^5$, $\mathcal{E}_{{\rm perm,}i}$ is the energy distance for the $i$-th permutation, and $\mathbf{1}$ is the indicator function. To simulate each permuted distance $\mathcal{E}_{{\rm perm,}i}$, the datasets $\mathbf{X}$ and $\mathbf{Y}$ are combined into a single dataset, which is randomly split into two new datasets, followed by the calculation of the energy distance for the permuted new datasets. 

By including the inferred fitting results of all six physical parameters, except for $t_{\rm first}$, for 11 GRB-SNe and 69 Non-GRB-SNe listed in Table \ref{tab:Results}, the calculated $p$-value is $\sim0.07$. Although this $p$-value is close to the commonly used significance level of $\alpha=0.05$, we cannot fully reject the null hypothesis, and hence, we conclude that there is no significant difference between our collected GRB-SN and Non-GRB-SN samples. 

\subsection{Lightcurve Properties} \label{sec:LightcurveSNeIcBL}

\begin{figure}[t!]
    \centering
    \includegraphics[width = 1\linewidth , trim = 33 67 125 80, clip]{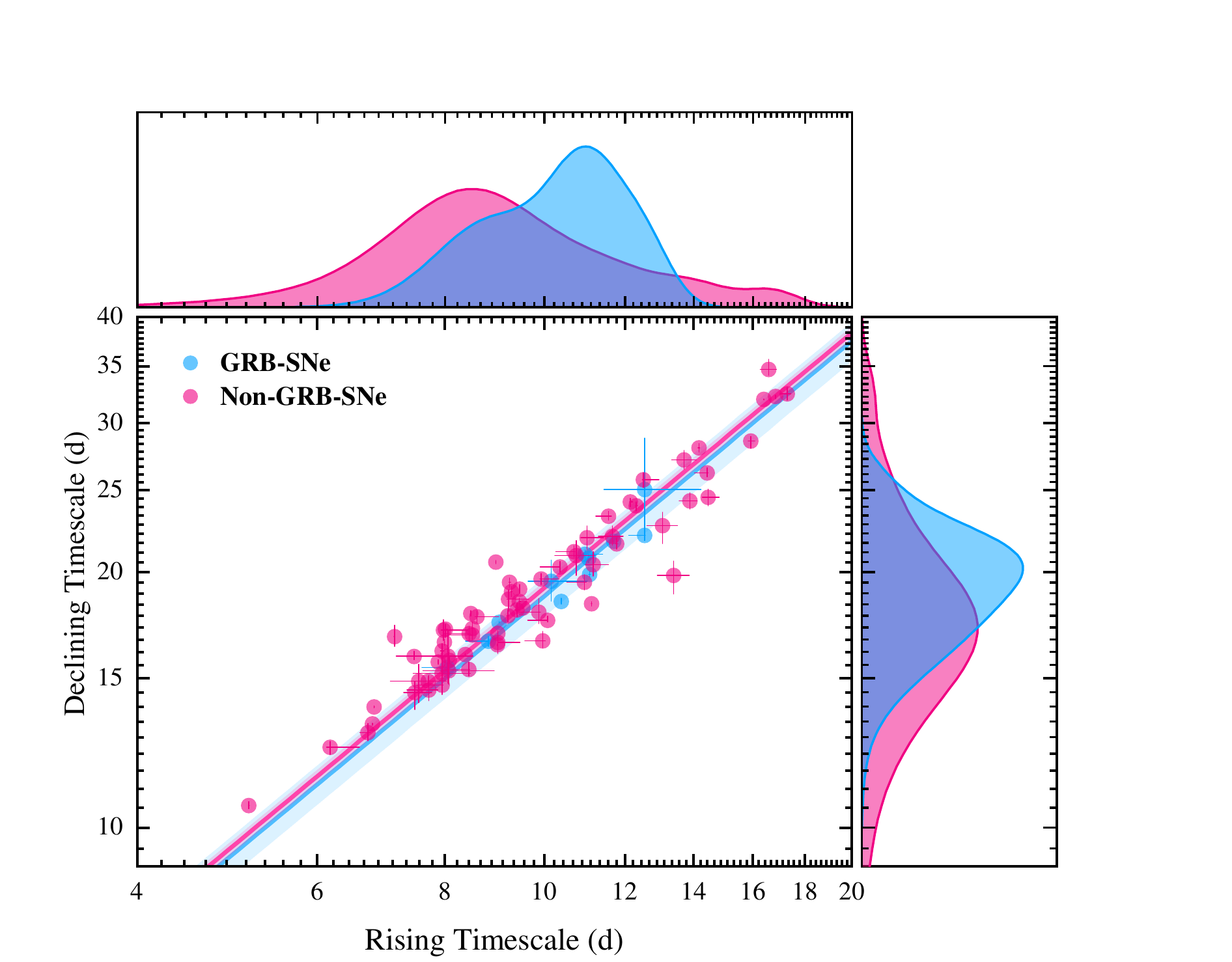}
    \includegraphics[width = 1\linewidth , trim = 33 0 125 233, clip]{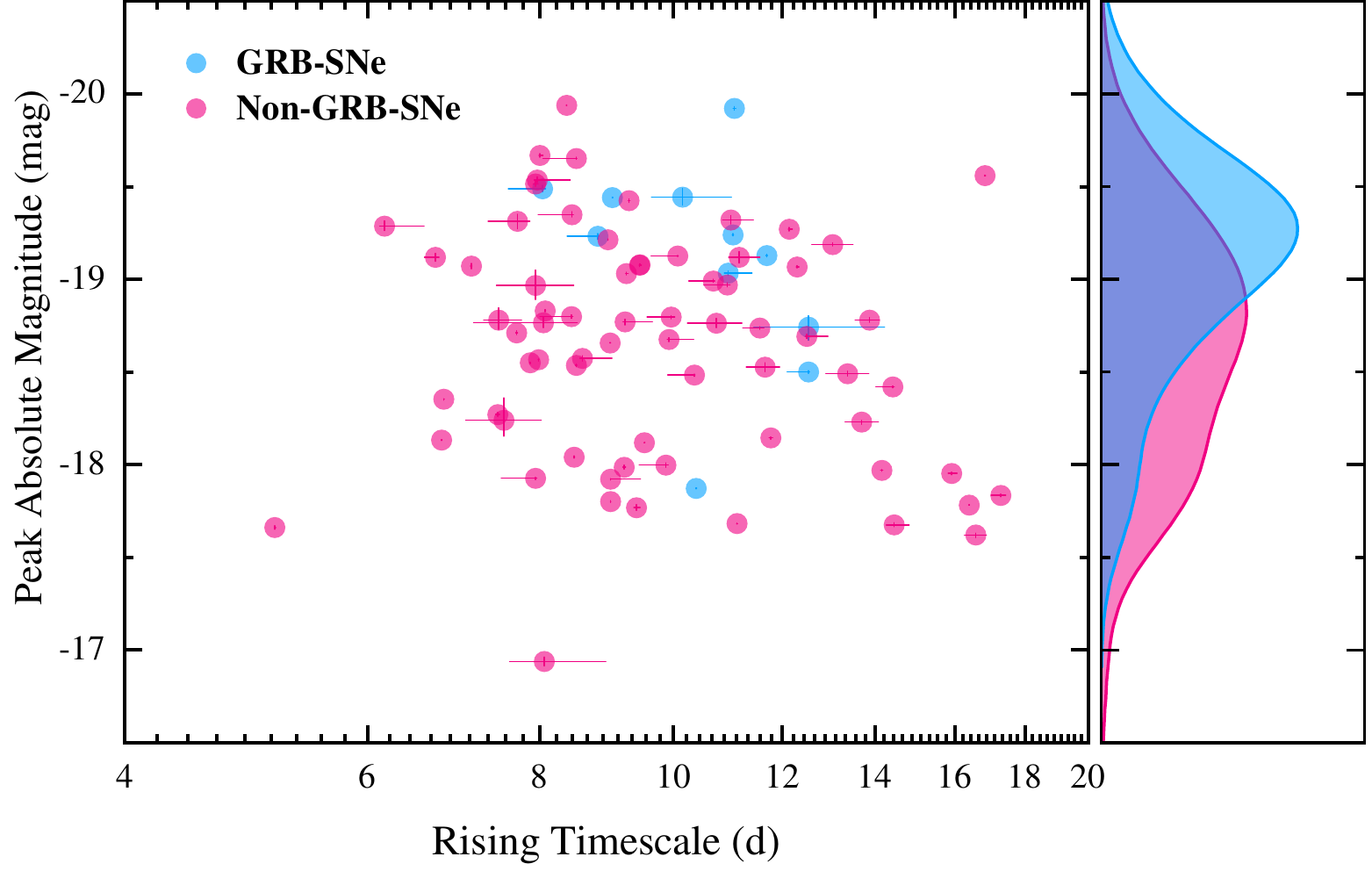}
    \caption{Rising timescales against decline timescales (middle panel) and against peak absolute magnitudes (bottom panel) for GRB-SNe (blue) and Non-GRB-SNe (purple). In the middle panel, the median and $1\sigma$ credible region of the fitted proportional relation $t_{\rm dec}\propto t_{\rm rise}$ for GRB-SNe and Non-GRB-SNe are shown as colored solid lines and shaded areas, respectively. The top panel show the probability density distributions of the rising timescale, while the two right panels show those of the declining timescale and peak absolute magnitude, respectively. }
    \label{fig:SNIcBL_LightcurveProperties}
\end{figure}

Based on the parameter values constrained from the model fittings, we derive the rising and declining timescales ($t_{\rm rise}$ and $t_{\rm dec}$) above half of the peak luminosity for the SNe Ic-BL, as well as the peak absolute magnitudes $M_{\rm peak}$ and the peak bolometric luminosities $L_{\rm peak}$, together with their $1\sigma$ credible regions, which are summarized in Table \ref{tab:Lightcurve}.

We find that the distributions of the rising and declining timescales of our Non-GRB-SNe sample, $t_{\rm rise}=9.41^{+3.77}_{-2.01}\,{\rm d}$ and $t_{\rm dec}=18.25^{+6.23}_{-3.50}\,{\rm d}$, are overall slightly shorter than those of our GRB-SNe sample, $t_{\rm rise}=10.69^{+1.54}_{-1.91}\,{\rm d}$ and $t_{\rm dec}=19.83^{+3.18}_{-3.32}\,{\rm d}$. As shown in the middle panel of Figure \ref{fig:SNIcBL_LightcurveProperties}, both samples show clear and comparable linear $t_{\rm rise}-t_{\rm dec}$ relationships, with the best-fitting results of $t_{\rm dec} = (1.92\pm0.03)t_{\rm rise}$ and $t_{\rm dec} = (1.87\pm0.10)t_{\rm rise}$, respectively. The corresponding Pearson correlation coefficients are $r_{\rm P} = 0.959$ and $r_{\rm P} = 0.958$, with $p$-values of $2\times10^{-38}$ and $3\times10^{-6}$, indicating statistically significant linear correlations. For the entire SN Ic-BL sample, the distributions of the rising and declining timescales are $t_{\rm rise}={9.63}^{+3.30}_{-2.11}\,{\rm d}$ and $t_{\rm dec}={18.51}^{+5.58}_{-3.54}\,{\rm d}$, respectively, and a best-fitting relation is $t_{\rm dec} = (1.91\pm0.03)t_{\rm rise}$ with $r_{\rm P}=0.959$ and a $p$-value of $2\times10^{-44}$. We also show the $t_{\rm rise}-M_{\rm peak}$ distributions in the bottom panel of Figure \ref{fig:SNIcBL_LightcurveProperties}. Our GRB-SN and Non-GRB-SN samples have peak absolute magnitude distributions of $M_{\rm peak}=-19.19^{+0.65}_{-0.41}\,{\rm mag}$ (corresponding peak luminosity distribution of $L_{\rm peak}=14.43^{+6.22}_{-6.84}\times10^{42}\,{\rm erg}\,{\rm s}^{-1}$) and $M_{\rm peak}=-18.67^{+0.77}_{-0.67}\,{\rm mag}$ ($L_{\rm peak}=9.18^{+6.71}_{-5.06}\times10^{42}\,{\rm erg}\,{\rm s}^{-1}$), respectively. Thus, GRB-SNe are brighter than Non-GRB-SNe, with an overall difference of $\sim0.5\,{\rm mag}$ in $M_{\rm peak}$, corresponding to a factor of $\sim1.6$ in peak luminosity, consistent with our discussion in Sections \ref{sec:MagnetarParameterSNeIcBL} and \ref{sec:NickelProductionSNeIcBL}. Moreover, no statistically significant linear correlation is found in the $t_{\rm rise}-M_{\rm peak}$ and hence $t_{\rm dec}-M_{\rm peak}$ distributions for the GRB-SN and Non-GRB-SN samples.

{
\begin{deluxetable}{lcccclcccc}
\tablewidth{0pt}
\tabletypesize{\fontsize{6.5pt}{9pt}\selectfont}
\tablecaption{Lightcurve Properties of SNe Ic-BL. \label{tab:Lightcurve}}
\tablehead{
\colhead{SN Name} & \colhead{$t_{\rm rise}$} & \colhead{$t_{\rm dec}$} & \colhead{$M_{\rm peak}$} & \colhead{$L_{\rm peak}$} & \colhead{SN Name} & \colhead{$t_{\rm rise}$} & \colhead{$t_{\rm dec}$} & \colhead{$M_{\rm peak}$} & \colhead{$L_{\rm peak}$} \\
\colhead{} & \colhead{(day)} & \colhead{(day)} & \colhead{(mag)} & \colhead{($10^{42}\,{\rm erg}\,{\rm s}^{-1}$)} & \colhead{} & \colhead{(day)} & \colhead{(day)} & \colhead{(mag)} & \colhead{($10^{42}\,{\rm erg}\,{\rm s}^{-1}$)}
}
\startdata
SN\,1998bw$^*$ & ${11.69}^{+0.01}_{-0.01}$ & ${21.75}^{+0.14}_{-0.14}$  & ${-19.13}^{+0.01}_{-0.01}$ & ${13.49}^{+0.09}_{-0.09}$ &
SN\,2002ap & ${11.12}^{+0.01}_{-0.01}$ & ${18.36}^{+0.10}_{-0.12}$ & ${-17.68}^{+0.00}_{-0.00}$ & ${3.56}^{+0.01}_{-0.01}$ \\
SN\,2003dh$^*$ & ${12.54}^{+1.70}_{-1.09}$ & ${25.04}^{+3.75}_{-2.68}$ & ${-18.74}^{+0.06}_{-0.07}$ & ${9.45}^{+0.62}_{-0.52}$ &
SN\,2003jd & ${9.46}^{+0.00}_{-0.00}$ & ${19.10}^{+0.25}_{-0.21}$ & ${-19.08}^{+0.00}_{-0.01}$ & ${12.88}^{+0.06}_{-0.06}$ \\
SN\,2003lw$^*$ & ${11.08}^{+0.02}_{-0.03}$ & ${19.89}^{+0.16}_{-0.13}$ &  ${-19.92}^{+0.01}_{-0.01}$ & ${28.04}^{+0.25}_{-0.25}$ &
SN\,2004aw & ${14.43}^{+0.03}_{-0.41}$ & ${26.20}^{+0.36}_{-0.27}$ & ${-18.42}^{+0.01}_{-0.01}$ & ${7.03}^{+0.04}_{-0.04}$ \\
SN\,2007ru & ${8.97}^{+0.00}_{-0.00}$ & ${20.57}^{+0.08}_{-0.09}$ & ${-19.21}^{+0.00}_{-0.00}$ & ${14.60}^{+0.05}_{-0.05}$ &
SN\,2009bb & ${8.08}^{+0.00}_{-0.00}$ & ${15.75}^{+0.16}_{-0.14}$ & ${-18.83}^{+0.01}_{-0.01}$ & ${10.26}^{+0.06}_{-0.06}$ \\
SN\,2010bh$^*$ & ${10.39}^{+0.00}_{-0.00}$ & ${18.48}^{+0.16}_{-0.12}$ & ${-17.87}^{+0.00}_{-0.00}$ & ${4.24}^{+0.01}_{-0.01}$ &
PTF\,10bzf & ${11.66}^{+0.29}_{-0.36}$ & ${22.06}^{+0.63}_{-0.53}$  & ${-18.53}^{+0.02}_{-0.02}$ & ${7.75}^{+0.18}_{-0.16}$ \\
PTF\,10ciw & ${7.94}^{+0.52}_{-0.50}$ & ${15.20}^{+0.89}_{-0.85}$ & ${-18.97}^{+0.08}_{-0.08}$ & ${11.63}^{+0.92}_{-0.78}$ &
PTF\,10gvb & ${11.16}^{+0.39}_{-0.19}$ & ${20.41}^{+0.74}_{-0.61}$  & ${-19.12}^{+0.03}_{-0.03}$ & ${13.38}^{+0.39}_{-0.37}$ \\
PTF\,10qts & ${8.45}^{+0.01}_{-0.46}$ & ${16.93}^{+0.36}_{-0.31}$ & ${-19.35}^{+0.01}_{-0.01}$ & ${16.55}^{+0.22}_{-0.21}$ &
PTF\,10tqv & ${7.47}^{+0.29}_{-0.19}$ & ${14.42}^{+0.45}_{-0.64}$ & ${-18.78}^{+0.05}_{-0.07}$ & ${9.79}^{+0.62}_{-0.45}$ \\
PTF\,10vgv & ${6.82}^{+0.00}_{-0.00}$ & ${13.88}^{+0.04}_{-0.04}$  & ${-18.35}^{+0.00}_{-0.00}$ & ${6.61}^{+0.02}_{-0.02}$ &
PTF\,10aavz & ${12.14}^{+0.06}_{-0.02}$ & ${24.20}^{+0.26}_{-0.31}$ & ${-19.27}^{+0.01}_{-0.01}$ & ${15.39}^{+0.15}_{-0.14}$ \\
PTF\,11cmh & ${8.44}^{+0.02}_{-0.43}$ & ${15.36}^{+0.32}_{-0.28}$  & ${-18.80}^{+0.01}_{-0.01}$ & ${9.97}^{+0.08}_{-0.08}$ &
PTF\,11lbm & ${8.48}^{+0.01}_{-0.01}$ & ${17.89}^{+0.13}_{-0.13}$  & ${-18.04}^{+0.01}_{-0.01}$ & ${4.96}^{+0.03}_{-0.03}$ \\
PTF\,12as & ${7.94}^{+0.02}_{-0.44}$ & ${14.72}^{+0.26}_{-0.21}$  & ${-17.93}^{+0.01}_{-0.01}$ & ${4.46}^{+0.02}_{-0.02}$ &
SN\,2013dx$^*$ & ${8.04}^{+0.04}_{-0.45}$ & ${15.44}^{+0.20}_{-0.19}$  & ${-19.49}^{+0.02}_{-0.02}$ & ${18.82}^{+0.31}_{-0.31}$ \\
iPTF\,13alq & ${9.25}^{+0.02}_{-0.01}$ & ${19.47}^{+0.21}_{-0.27}$  & ${-19.03}^{+0.01}_{-0.01}$ & ${12.34}^{+0.08}_{-0.08}$ &
SN\,2014ad & ${9.00}^{+0.00}_{-0.00}$ & ${16.42}^{+0.03}_{-0.05}$  & ${-18.66}^{+0.00}_{-0.00}$ & ${8.75}^{+0.02}_{-0.01}$ \\
iPTF\,14bfu$^*$ & ${10.16}^{+0.86}_{-0.51}$ & ${19.52}^{+1.15}_{-1.03}$  & ${-19.44}^{+0.05}_{-0.04}$ & ${18.05}^{+0.73}_{-0.74}$ &
iPTF\,14dby & ${17.28}^{+0.14}_{-0.27}$ & ${32.47}^{+0.54}_{-0.53}$  & ${-17.83}^{+0.01}_{-0.01}$ & ${4.10}^{+0.03}_{-0.03}$ \\
iPTF\,15dld & ${7.54}^{+0.48}_{-0.47}$ & ${14.87}^{+0.71}_{-0.84}$  & ${-18.24}^{+0.09}_{-0.12}$ & ${5.95}^{+0.70}_{-0.45}$ &
iPTF\,15dqg & ${8.51}^{+0.00}_{-0.46}$ & ${16.87}^{+0.32}_{-0.22}$ & ${-19.65}^{+0.01}_{-0.01}$ & ${21.86}^{+0.12}_{-0.12}$ \\
SN\,2016P & ${9.01}^{+0.00}_{-0.00}$ & ${16.94}^{+0.12}_{-0.13}$  & ${-17.80}^{+0.01}_{-0.00}$ & ${3.97}^{+0.02}_{-0.02}$ &
SN\,2016coi & ${16.39}^{+0.00}_{-0.00}$ & ${31.98}^{+0.03}_{-0.01}$  & ${-17.78}^{+0.00}_{-0.00}$ & ${3.91}^{+0.00}_{-0.00}$ \\
SN\,2016jca$^*$ & ${8.82}^{+0.01}_{-0.44}$ & ${16.59}^{+0.22}_{-0.23}$  & ${-19.23}^{+0.01}_{-0.02}$ & ${14.86}^{+0.24}_{-0.20}$ &
SN\,2017iuk$^*$ & ${12.53}^{+0.02}_{-0.44}$ & ${22.10}^{+0.43}_{-0.30}$  & ${-18.50}^{+0.01}_{-0.01}$ & ${7.57}^{+0.06}_{-0.05}$ \\
SN\,2018ie & ${5.14}^{+0.01}_{-0.01}$ & ${10.62}^{+0.11}_{-0.10}$ & ${-17.66}^{+0.01}_{-0.01}$ & ${3.49}^{+0.03}_{-0.03}$ &
SN\,2018bvw & ${9.96}^{+0.06}_{-0.39}$ & ${16.60}^{+0.43}_{-0.30}$ & ${-18.80}^{+0.01}_{-0.01}$ & ${9.93}^{+0.08}_{-0.07}$ \\
SN\,2018ell & ${13.38}^{+0.48}_{-0.48}$ & ${19.83}^{+0.79}_{-0.97}$ & ${-18.49}^{+0.01}_{-0.01}$ & ${7.51}^{+0.08}_{-0.09}$ &
SN\,2018fip$^*$ & ${10.96}^{+0.45}_{-0.07}$ & ${21.01}^{+0.23}_{-0.22}$  & ${-19.03}^{+0.01}_{-0.01}$ & ${12.37}^{+0.15}_{-0.13}$ \\
SN\,2018hom & ${8.00}^{+0.04}_{-0.01}$ & ${17.15}^{+0.23}_{-0.38}$ & ${-19.67}^{+0.01}_{-0.01}$ & ${22.20}^{+0.17}_{-0.19}$ &
SN\,2018kva & ${9.93}^{+0.41}_{-0.03}$ & ${19.65}^{+0.29}_{-0.31}$ & ${-18.67}^{+0.01}_{-0.01}$ & ${8.89}^{+0.09}_{-0.09}$ \\
SN\,2019gwc & ${7.87}^{+0.00}_{-0.00}$ & ${15.68}^{+0.10}_{-0.09}$ &  ${-18.55}^{+0.00}_{-0.00}$ & ${7.92}^{+0.03}_{-0.04}$ &
SN\,2019hsx & ${9.41}^{+0.05}_{-0.04}$ & ${18.05}^{+0.32}_{-0.33}$ & ${-17.77}^{+0.01}_{-0.02}$ & ${3.85}^{+0.06}_{-0.05}$ \\
SN\,2019lci & ${8.60}^{+0.43}_{-0.04}$ & ${17.73}^{+0.26}_{-0.29}$ & ${-18.57}^{+0.01}_{-0.01}$ & ${8.10}^{+0.10}_{-0.10}$ &
SN\,2019moc & ${12.30}^{+0.04}_{-0.02}$ & ${23.95}^{+0.27}_{-0.32}$ & ${-19.07}^{+0.01}_{-0.01}$ & ${12.76}^{+0.08}_{-0.07}$ \\
SN\,2019pgo & ${10.08}^{+0.01}_{-0.45}$ & ${17.55}^{+0.20}_{-0.07}$ & ${-19.13}^{+0.00}_{-0.00}$ & ${13.46}^{+0.05}_{-0.04}$ &
SN\,2019qfi & ${9.88}^{+0.04}_{-0.43}$ & ${17.94}^{+0.69}_{-0.54}$  & ${-18.00}^{+0.01}_{-0.01}$ & ${4.76}^{+0.05}_{-0.05}$ \\
SN\,2020bvc$^*$ & ${9.03}^{+0.01}_{-0.00}$ & ${17.46}^{+0.03}_{-0.03}$  & ${-19.44}^{+0.00}_{-0.00}$ & ${18.01}^{+0.03}_{-0.04}$ &
SN\,2020hes & ${7.94}^{+0.04}_{-0.03}$ & ${16.16}^{+0.32}_{-0.40}$  & ${-19.52}^{+0.01}_{-0.01}$ & ${19.28}^{+0.23}_{-0.22}$ \\
SN\,2020hyj & ${11.77}^{+0.03}_{-0.03}$ & ${21.61}^{+0.34}_{-0.31}$  & ${-18.14}^{+0.01}_{-0.01}$ & ${5.45}^{+0.03}_{-0.03}$ &
SN\,2020lao & ${7.99}^{+0.01}_{-0.00}$ & ${16.55}^{+0.13}_{-0.15}$ & ${-18.57}^{+0.01}_{-0.01}$  & ${8.04}^{+0.04}_{-0.05}$\\
SN\,2020rfr & ${7.71}^{+0.16}_{-0.37}$ & ${14.53}^{+0.60}_{-0.42}$ & ${-19.31}^{+0.04}_{-0.04}$ & ${15.99}^{+0.56}_{-0.59}$ &
SN\,2020rph & ${13.70}^{+0.38}_{-0.38}$ & ${27.13}^{+0.70}_{-0.65}$  & ${-18.23}^{+0.01}_{-0.01}$ & ${5.90}^{+0.06}_{-0.06}$ \\
SN\,2020tkx & ${10.36}^{+0.01}_{-0.46}$ & ${20.29}^{+0.31}_{-0.28}$ & ${-18.48}^{+0.01}_{-0.01}$ & ${7.45}^{+0.04}_{-0.04}$ &
SN\,2020abxc & ${11.01}^{+0.42}_{-0.15}$ & ${21.97}^{+0.71}_{-0.67}$  & ${-19.32}^{+0.02}_{-0.02}$ & ${16.11}^{+0.36}_{-0.33}$ \\
SN\,2020abxl & ${13.06}^{+0.44}_{-0.45}$ & ${22.70}^{+0.83}_{-1.07}$ & ${-19.19}^{+0.01}_{-0.01}$ & ${14.26}^{+0.12}_{-0.12}$ &
SN\,2020adow & ${6.79}^{+0.00}_{-0.00}$ & ${13.25}^{+0.02}_{-0.04}$ & ${-18.13}^{+0.00}_{-0.00}$ & ${5.39}^{+0.02}_{-0.01}$ \\
SN\,2021xv & ${6.18}^{+0.42}_{-0.05}$ & ${12.43}^{+0.20}_{-0.24}$ & ${-19.29}^{+0.02}_{-0.03}$ & ${15.63}^{+0.40}_{-0.34}$ &
SN\,2021bmf & ${11.56}^{+0.07}_{-0.33}$ & ${23.30}^{+0.30}_{-0.26}$  & ${-18.74}^{+0.01}_{-0.01}$ & ${9.42}^{+0.06}_{-0.06}$ \\
SN\,2021fop & ${7.97}^{+0.45}_{-0.04}$ & ${17.10}^{+0.48}_{-0.48}$ & ${-19.54}^{+0.02}_{-0.02}$ & ${19.66}^{+0.30}_{-0.28}$ &
SN\,2021hyz & ${6.72}^{+0.04}_{-0.12}$ & ${12.94}^{+0.32}_{-0.27}$ & ${-19.12}^{+0.02}_{-0.02}$ & ${13.37}^{+0.27}_{-0.26}$ \\
SN\,2021ncn & ${8.06}^{+0.87}_{-0.46}$ & ${15.31}^{+0.77}_{-0.54}$ & ${-16.94}^{+0.02}_{-0.02}$ & ${1.80}^{+0.04}_{-0.04}$ &
SN\,2021qjv & ${9.21}^{+0.02}_{-0.02}$ & ${17.78}^{+0.29}_{-0.29}$ & ${-17.99}^{+0.01}_{-0.01}$ & ${4.72}^{+0.05}_{-0.04}$ \\
SN\,2021too & ${16.83}^{+0.01}_{-0.00}$ & ${32.24}^{+0.11}_{-0.18}$ & ${-19.56}^{+0.00}_{-0.00}$ & ${20.09}^{+0.04}_{-0.04}$ &
SN\,2022cca & ${9.23}^{+0.43}_{-0.04}$ & ${18.59}^{+0.49}_{-0.59}$ & ${-18.77}^{+0.02}_{-0.02}$ & ${9.70}^{+0.14}_{-0.14}$ \\
SN\,2022crr & ${7.46}^{+0.02}_{-0.03}$ & ${15.91}^{+0.23}_{-0.13}$ & ${-18.27}^{+0.01}_{-0.01}$ & ${6.13}^{+0.06}_{-0.05}$ &
SN\,2022ofv & ${8.51}^{+0.01}_{-0.01}$ & ${17.18}^{+0.30}_{-0.32}$ & ${-18.54}^{+0.01}_{-0.01}$ & ${7.82}^{+0.06}_{-0.05}$ \\
SN\,2022wlm & ${10.94}^{+0.05}_{-0.42}$ & ${19.46}^{+0.41}_{-0.39}$ & ${-18.97}^{+0.01}_{-0.01}$ & ${11.67}^{+0.12}_{-0.12}$ &
SN\,2023iwy & ${8.37}^{+0.00}_{-0.00}$ & ${16.00}^{+0.04}_{-0.04}$ & ${-19.94}^{+0.00}_{-0.00}$ & ${28.47}^{+0.07}_{-0.07}$ \\
SN\,2023mee & ${9.29}^{+0.02}_{-0.02}$ & ${18.98}^{+0.27}_{-0.25}$ & ${-19.42}^{+0.01}_{-0.01}$ & ${17.73}^{+0.17}_{-0.15}$ &
SN\,2023qzm & ${9.46}^{+0.03}_{-0.01}$ & ${18.48}^{+0.16}_{-0.18}$ & ${-19.08}^{+0.01}_{-0.01}$ & ${12.87}^{+0.08}_{-0.09}$ \\
SN\,2024tv & ${8.05}^{+0.47}_{-0.89}$ & ${15.93}^{+1.17}_{-1.13}$ & ${-18.77}^{+0.03}_{-0.03}$ & ${9.67}^{+0.26}_{-0.24}$ &
SN\,2024itg & ${10.69}^{+0.03}_{-0.43}$ & ${21.15}^{+0.33}_{-0.29}$ & ${-18.99}^{+0.01}_{-0.01}$ & ${11.90}^{+0.10}_{-0.09}$ \\
SN\,2024rqu & ${7.14}^{+0.01}_{-0.01}$ & ${16.79}^{+0.52}_{-0.44}$ & ${-19.07}^{+0.01}_{-0.01}$ & ${12.82}^{+0.15}_{-0.17}$ &
SN\,2024aber & ${10.75}^{+0.47}_{-0.50}$ & ${20.93}^{+0.86}_{-1.09}$ & ${-18.76}^{+0.02}_{-0.03}$ & ${9.64}^{+0.23}_{-0.20}$ \\
SN\,2024adml & ${12.50}^{+0.45}_{-0.02}$ & ${25.71}^{+0.48}_{-0.54}$ & ${-18.69}^{+0.01}_{-0.01}$ & ${9.02}^{+0.10}_{-0.09}$ &
SN\,2025op & ${9.53}^{+0.00}_{-0.00}$ & ${18.15}^{+0.03}_{-0.03}$ & ${-18.12}^{+0.00}_{-0.00}$ & ${5.33}^{+0.00}_{-0.00}$ \\
SN\,2025tt & ${13.89}^{+0.22}_{-0.33}$ & ${24.28}^{+0.54}_{-0.54}$ & ${-18.78}^{+0.01}_{-0.01}$ & ${9.80}^{+0.13}_{-0.11}$ &
SN\,2025cnu & ${14.46}^{+0.37}_{-0.19}$ & ${24.51}^{+0.56}_{-0.55}$  & ${-17.68}^{+0.01}_{-0.01}$ & ${3.54}^{+0.04}_{-0.03}$ \\
SN\,2025mco & ${16.58}^{+0.29}_{-0.32}$ & ${34.70}^{+0.94}_{-0.56}$  & ${-17.62}^{+0.01}_{-0.01}$ & ${3.37}^{+0.03}_{-0.04}$ &
SN\,2025ocy & ${9.01}^{+0.46}_{-0.00}$ & ${16.52}^{+0.30}_{-0.49}$ & ${-17.92}^{+0.00}_{-0.00}$ & ${4.44}^{+0.02}_{-0.02}$ \\
SN\,2025shf & ${7.70}^{+0.01}_{-0.01}$ & ${14.88}^{+0.20}_{-0.22}$ & ${-18.71}^{+0.01}_{-0.01}$ & ${9.19}^{+0.11}_{-0.09}$ &
SN\,2025vaw & ${14.17}^{+0.02}_{-0.03}$ & ${28.03}^{+0.34}_{-0.32}$ & ${-17.97}^{+0.00}_{-0.00}$ & ${4.64}^{+0.02}_{-0.02}$ \\
SN\,2025vef & ${15.92}^{+0.14}_{-0.09}$ & ${28.56}^{+0.37}_{-0.55}$ & ${-17.95}^{+0.01}_{-0.01}$ & ${4.57}^{+0.03}_{-0.03}$ &
SN\,2025wkm$^*$ & ${11.05}^{+0.01}_{-0.01}$ & ${-19.24}^{+0.01}_{-0.01}$ &  ${20.76}^{+0.25}_{-0.23}$ & ${14.95}^{+0.08}_{-0.08}$ \\
\hline
GRB-SNe & $10.69^{+1.54}_{-1.91}$ & $19.83^{+3.18}_{-3.32}$ & $-19.19^{+0.65}_{-0.41}$ & $14.43^{+6.22}_{-6.84}$ \\
Non-GRB-SNe & $9.41^{+3.77}_{-2.01}$ & $18.25^{+6.23}_{-3.50}$ & $-18.67^{+0.77}_{-0.67}$ & $9.18^{+6.71}_{-5.06}$ \\
All SNe Ic-BL & ${9.63}^{+3.30}_{-2.11}$ & ${18.51}^{+5.58}_{-3.54}$ & $-18.75^{+0.80}_{-0.66}$ & $9.81^{+7.06}_{-5.45}$ \\
\enddata
\tablecomments{SNe Ic-BL marked with an asterisk ($^*$) denote GRB-SNe. The distributions of each parameter for the GRB-SN, Non-GRB-SN, and entire SN Ic-BL samples are listed in the last three rows, respectively. }
\end{deluxetable}
}

\section{Comparison of SNe Ic-BL with Other Magnetar-powered and Ordinary SESNe} \label{sec:StatisticsMagnetarSNe}

We then compare the statistical properties of the model parameters derived for SNe Ic-BL with those of various transients reported in the literature. Specifically, this includes comparisons of the magnetar parameters (Section \ref{sec:MagnetarParametersAll}), explosion parameters (Section \ref{sec:ExplosionParametersAll}), and lightcurve properties (Section \ref{sec:LightcurveAll}) of SNe Ic-BL with those of other magnetar-powered SESNe, as well as the $^{56}$Ni production of SNe Ic-BL in our sample compared to that of ordinary SNe Ic (Section \ref{sec:NickelProductionAll}).

\subsection{Magnetar Parameters in Magnetar-powered SESNe} \label{sec:MagnetarParametersAll}

\begin{figure}[t!]
    \centering
    \includegraphics[width = 1\linewidth , trim = 33 0 125 80, clip]{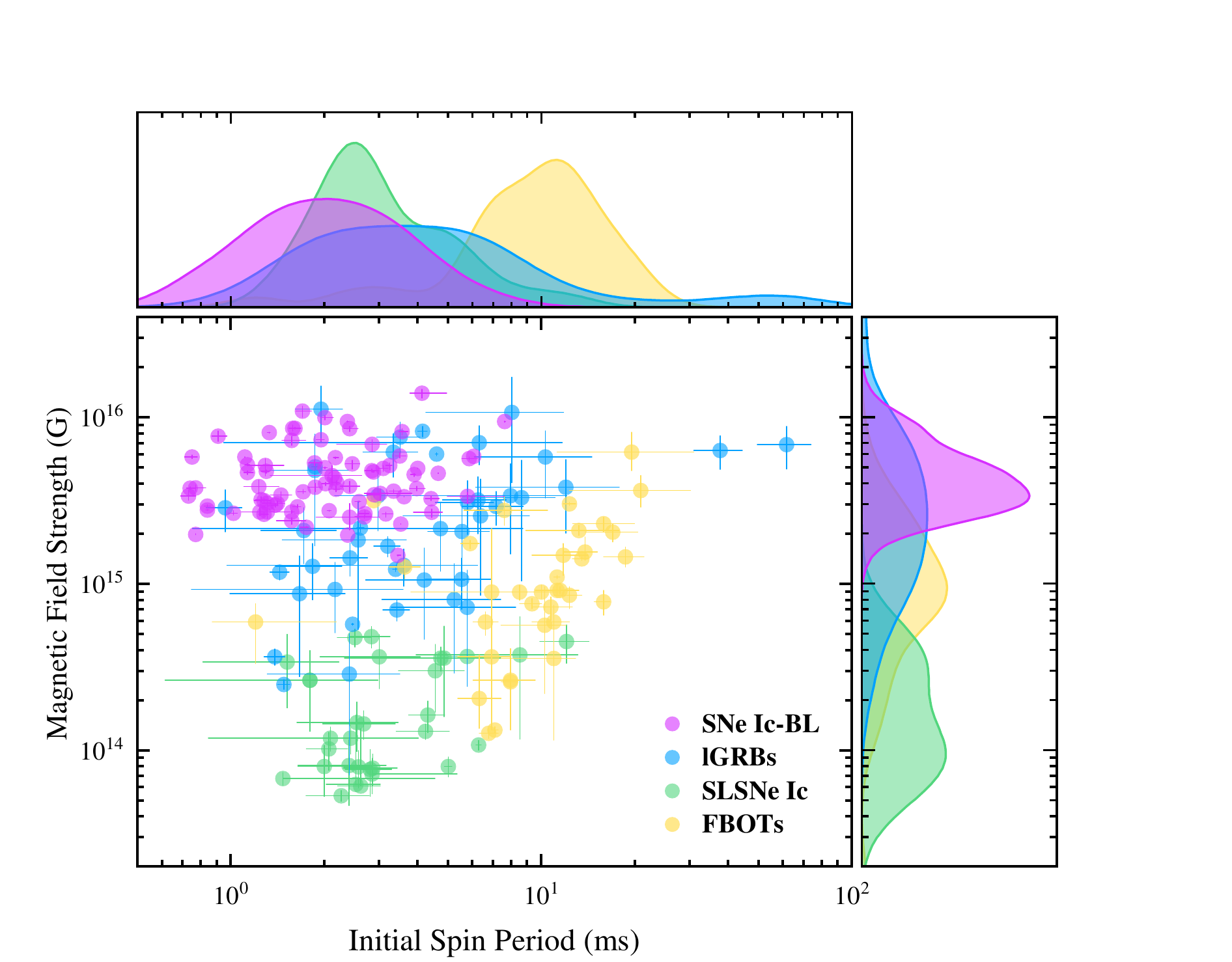}
    \caption{Similar to Figure \ref{fig:SNIcBL_Pi_Bp}, but for magnetic field strengths against initial spin periods of SNe Ic-BL (violet), lGRBs \citep[blue; ][]{Lv2014}, SLSNe Ic \citep[green;][]{Yu2017}, and FBOTs \citep[yellow;][]{Liu2022,Hu2023}. }
    \label{fig:Magnetar_Pi_Bp}
\end{figure}

A shallow decay phase or plateau commonly observed in the early X-ray afterglows of GRBs \citep{Zhang2006, Nousek2006, Obrien2006} is readily interpreted as energy injection from a millisecond magnetar \citep{Zhang2006}, enabling estimates of magnetar parameters. We first compare the $P_{\rm i}-B_{\rm p}$ distribution of SN Ic-BL magnetars in our sample with that of lGRB magnetars derived by \cite{Lv2014} from modeling such X-ray plateaus, as shown in Figure \ref{fig:Magnetar_Pi_Bp}. We find that both classes may harbor rapidly rotating magnetars with extremely high magnetic fields (typically on the order of $10^{15}\,{\rm G}$), with their $P_{\rm i}-B_{\rm p}$ distributions largely overlapping, although the distribution of our SN Ic-BL magnetar sample is not identical to that reported for lGRB magnetars in \cite{Lv2014}. The lGRB magnetars have $P_{\rm i}$ and $B_{\rm p}$ distributions of $3.88^{+6.14}_{-2.17}\,{\rm ms}$ and $2.45^{+5.90}_{-1.74}\times10^{15}\,{\rm G}$, respectively. In comparison, the SN Ic-BL magnetars in our sample have a median initial spin period of $\sim2\,{\rm ms}$, with a $1\sigma$ upper limit of $\sim4\,{\rm ms}$, indicating generally faster rotation than the inferred lGRB magnetars. Moreover, the magnetic field strengths of SNe Ic-BL magnetars are overall higher, with a lower boundary of $\sim10^{15}\,\mathrm{G}$, significantly exceeding that of lGRB magnetars ($\sim3\times10^{14}\,\mathrm{G}$). An energy test, as introduced in Section \ref{sec:PopulationEquality}, applied to the $P_{\rm i}-B_{\rm p}$ distributions yields a $p$-value of $\sim10^{-4}$, indicating that the SN Ic-BL and lGRB magnetars are statistically unlikely to be drawn from the same parent distribution. However, it should be noted that the magnetar parameters inferred for lGRBs are subject to non-negligible uncertainties. As emphasized by \cite{Lv2014}, the derived $P_{\rm i}$ and $B_{\rm p}$ correspond to the magnetar properties via the shallow decay phase or plateau, while the early prompt GRB emission is expected to consume a fraction of the magnetar’s rotational and magnetic energy, implying that the true $P_{\rm i}$ may be shorter and the $B_{\rm p}$ stronger than the inferred values. In addition, several poorly constrained parameters, such as microphysical parameters, jet beaming, and circumburst environment, are fixed in their analysis, while different assumptions could systematically affect the inferred $P_{\rm i}-B_{\rm p}$ distribution. Taking these uncertainties into account, we cannot exclude the possibility that the magnetar properties inferred for SNe Ic-BL and lGRBs are consistent with a similar underlying population.

The $P_{\rm i}$ and $B_{\rm p}$ distributions of SLSN Ic magnetars reported by \cite{Yu2017} are $2.81^{+2.52}_{-0.92}\,{\rm ms}$ and $1.46^{+2.66}_{-0.85}\times10^{14}\,{\rm G}$, respectively, with their relationship shown in Figure \ref{fig:Magnetar_Pi_Bp}. The resulting $P_{\rm i}-B_{\rm p}$ distribution is broadly consistent with those inferred from other studies \citep{Liu2017, Nicholl2017, Chen2023}. Our results indicate an overall trend in which SN Ic-BL magnetars rotate faster and have stronger magnetic fields than SLSN Ic magnetars at the time of their formation. In particular, the magnetic field strengths of SN Ic-BL magnetars are overall higher than those of SLSN Ic magnetars, with an apparent gap between the lower boundary of the SN Ic-BL magnetar distribution at $\sim10^{15}\,{\rm G}$ and the upper boundary of the SLSN Ic magnetar distribution at $\sim4\times10^{14}\,{\rm G}$. This gap may result from three effects: (1) the contribution of $^{56}$Ni decay is often neglected in lightcurve modeling of SLSNe Ic; for relatively faint SLSNe Ic, the $^{56}$Ni decay could still make a non-negligible contribution to both the early- and late-time emission, potentially leading to an underestimation of $B_{\rm p}$; (2) bolometric lightcurve fitting for relatively faint SLSNe Ic that may actually host strong-$B_{\rm p}$ magnetars may underestimate the magnetar contribution to the ejecta kinetic energy, resulting in underestimated $E_{\rm rot,i}$ (i.e., overestimated $P_{\rm i}$) and $B_{\rm p}$; (3) selection effects related to sample construction: most SLSNe Ic in \cite{Yu2017} have peak absolute magnitudes $\lesssim-21\,{\rm mag}$ a historically adopted luminosity threshold for SLSNe \citep{GalYam2012}, whereas the majority of identified SNe Ic-BL are fainter than $\gtrsim-20\,{\rm mag}$, which may artificially exclude some events in the intermediate luminosity range and thus contribute to the observed gap.

\cite{Liu2022} found that FBOTs can host magnetars with $P_{\rm i}$ following a distribution of  $9.81^{+5.95}_{-4.31}\,{\rm ms}$ and $B_{\rm p}$ following a distribution of $9.3^{+15.2}_{-6.4}\times10^{14}\,{\rm G}$. As shown in Figure \ref{fig:Magnetar_Pi_Bp}, while a few FBOT magnetars show relatively rapid rotation, the sample as a whole appears to rotate more slowly than magnetars powering SNe Ic-BL and SLSNe. As a comparison, the $1\sigma$ lower boundary of the FBOT $P_{\rm i}$ distribution ($\sim5\,{\rm ms}$) is comparable to the $1\sigma$ upper boundaries inferred for SNe Ic-BL and SLSN magnetars. The $B_{\rm p}$ distribution of FBOT magnetars spans a wide range, from $\sim10^{14}-10^{16}\,{\rm G}$, covering the $B_{\rm p}$ ranges inferred for SNe Ic-BL and SLSN magnetars.

\subsection{Explosion Parameters in Magnetar-powered SESNe}  \label{sec:ExplosionParametersAll}

\begin{figure}[t!]
    \centering
    \includegraphics[width = 1\linewidth , trim = 33 0 125 80, clip]{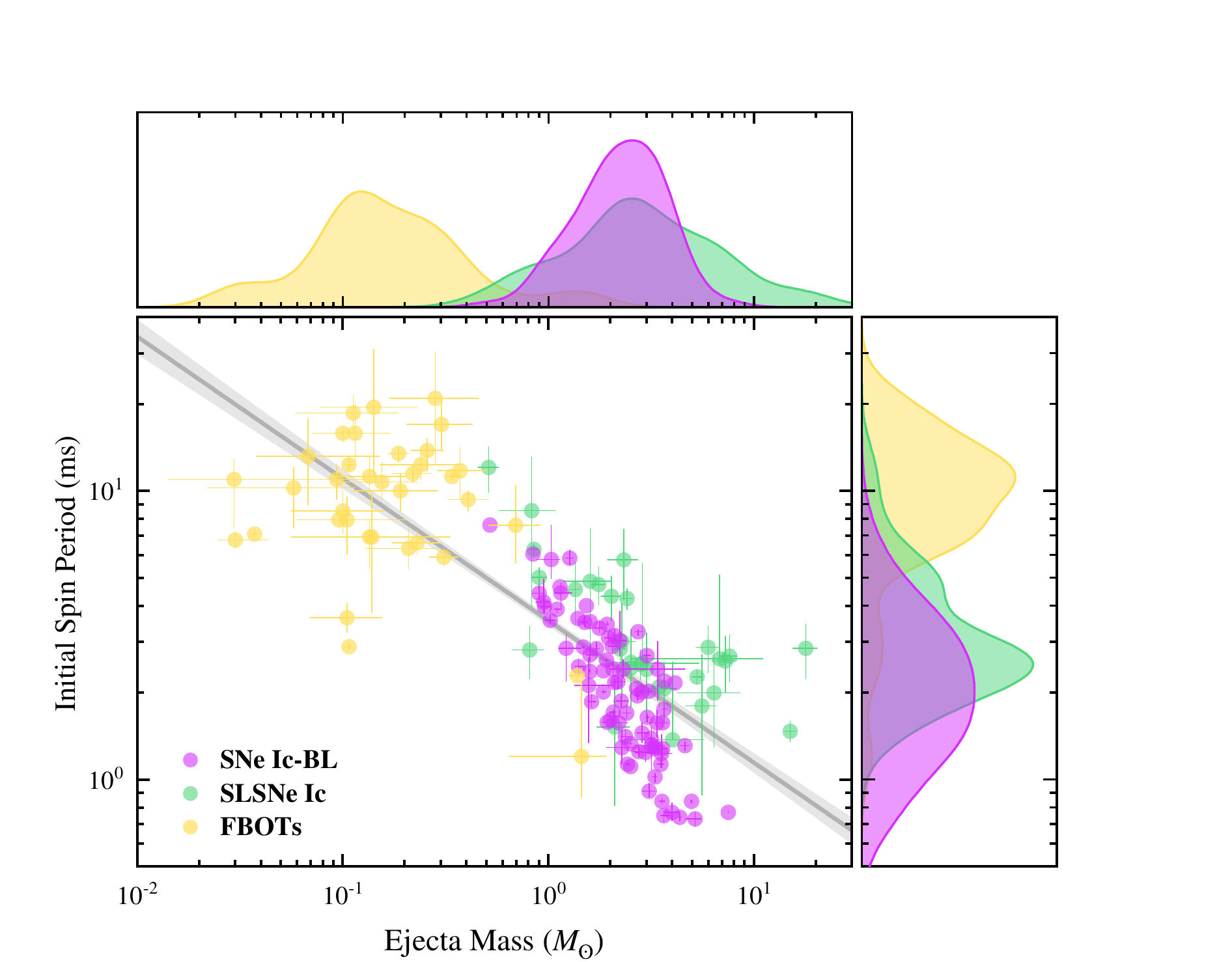}
    \caption{Similar to Figure \ref{fig:SNIcBL_Mej_Pi}, but showing initial spin periods of magnetars against ejecta masses for SNe Ic-BL (violet), SLSNe Ic \citep[green;][]{Yu2017}, and FBOTs \citep[yellow;][]{Liu2022,Hu2023}. The gray solid line and shaded area represent the best-fit log-log universal correlation and $1\sigma$ credible region across the three transient samples, respectively.  }
    \label{fig:Magnetar_Mej_Pi}
\end{figure}

Figure \ref{fig:Magnetar_Mej_Pi} shows the $M_{\rm ej}-P_{\rm i}$ distributions in log-log space for SNe Ic-BL in our sample, along with SLSNe Ic from \cite{Yu2017} and FBOTs from \cite{Liu2022,Hu2023}. While a universal $M_{\rm ej}-P_{\rm i}$ anti-correlation across these diverse transients was originally identified by \cite{Liu2022}, their $P_{\rm i}$ values for SNe Ic-BL were simply derived by assuming that the initial magnetar rotational energy is primarily converted into the observed kinetic energy. Here, by inferring $P_{\rm i}$ of SNe Ic-BL via direct lightcurve fitting, we also independently confirm this universal $M_{\rm ej}-P_{\rm i}$ anti-correlation in log-log space across SNe Ic-BL, SLSNe Ic, and FBOTs, as follows:
\begin{equation}
\label{equ:UniversalRelationship}
    P_{\rm i}/{\rm ms} = 10^{0.55\pm0.02}\times(M_{\rm ej}/M_\odot)^{-0.49\pm0.03}.
\end{equation}
As illustrated in Figure \ref{fig:Magnetar_Mej_Pi}, this $M_{\rm ej}-P_{\rm ej}$ anti-correlation is very strong and statistically robust, with a Pearson correlation coefficient of $r_{\rm P}=-0.806$ and a $p$-value of $\sim7\times10^{-35}$. This anti-correlation suggests that SNe Ic-BL, SLSNe, and FBOTs may originate from a common physical source.

Despite the presence of this universal correlation shared by SNe Ic-BL, SLSNe, and FBOTs, we find that their $M_{\rm ej}-P_{\rm i}$ distributions do not fully overlap, but instead occupy distinct regions of the parameter space. The $M_{\rm ej}$ distributions of SNe Ic-BL and SLSNe Ic are broadly similar, being $2.30^{+1.46}_{-1.01}\,M_\odot$ and $2.81^{+2.52}_{-0.95}\,M_\odot$, respectively, although SLSNe Ic potentially have a slightly higher fraction of events involving larger ejecta masses. As described in Section \ref{sec:MagnetarParametersAll}, the $P_{\rm i}$ distributions of SN Ic-BL and SLSN Ic magnetars are ${2.03}^{+1.81}_{-0.95}\,{\rm ms}$ and $2.81^{+2.52}_{-0.92}\,{\rm ms}$, corresponding to the initial magnetar rotational energy $E_{\rm rot,i}$ of $4.9^{+12.2}_{-3.5}\times10^{51}\,{\rm erg}$ and $2.54^{+3.27}_{-1.83}\times10^{51}\,{\rm erg}$, respectively. While the two populations have comparable ejecta mass ranges and thus initial neutrino-driven explosion energies, SN Ic-BL magnetars inject more rotational energy into the ejecta, resulting in observed explosion energies that are roughly twice those of SLSNe Ic.  When compared with SLSNe Ic and SNe Ic-BL, FBOTs occupy a more distinct region of the $M_{\rm ej}-P_{\rm i}$ distribution, with systematically slower-spinning magnetars and lower ejecta masses of $0.15^{+0.22}_{-0.08}\,M_\odot$. The ejecta mass boundary between FBOTs and the other two populations is $\sim0.5\,M_\odot$. Although SNe Ic-BL, SLSNe Ic, and FBOTs may arise from a broadly common physical origin revealed by the universal correlation, the systematic differences in their $M_{\rm ej}-P_{\rm i}$ distributions likely reflect subtle but important variations in their progenitor properties. We discuss the connections and differences of their progenitors in more detail in Section \ref{sec:Discussion}.

\subsection{Lightcurve Properties of Magnetar-powered SESNe} \label{sec:LightcurveAll}

\begin{figure}[t!]
    \centering
    \includegraphics[width = 1\linewidth , trim = 33 67 125 80, clip]{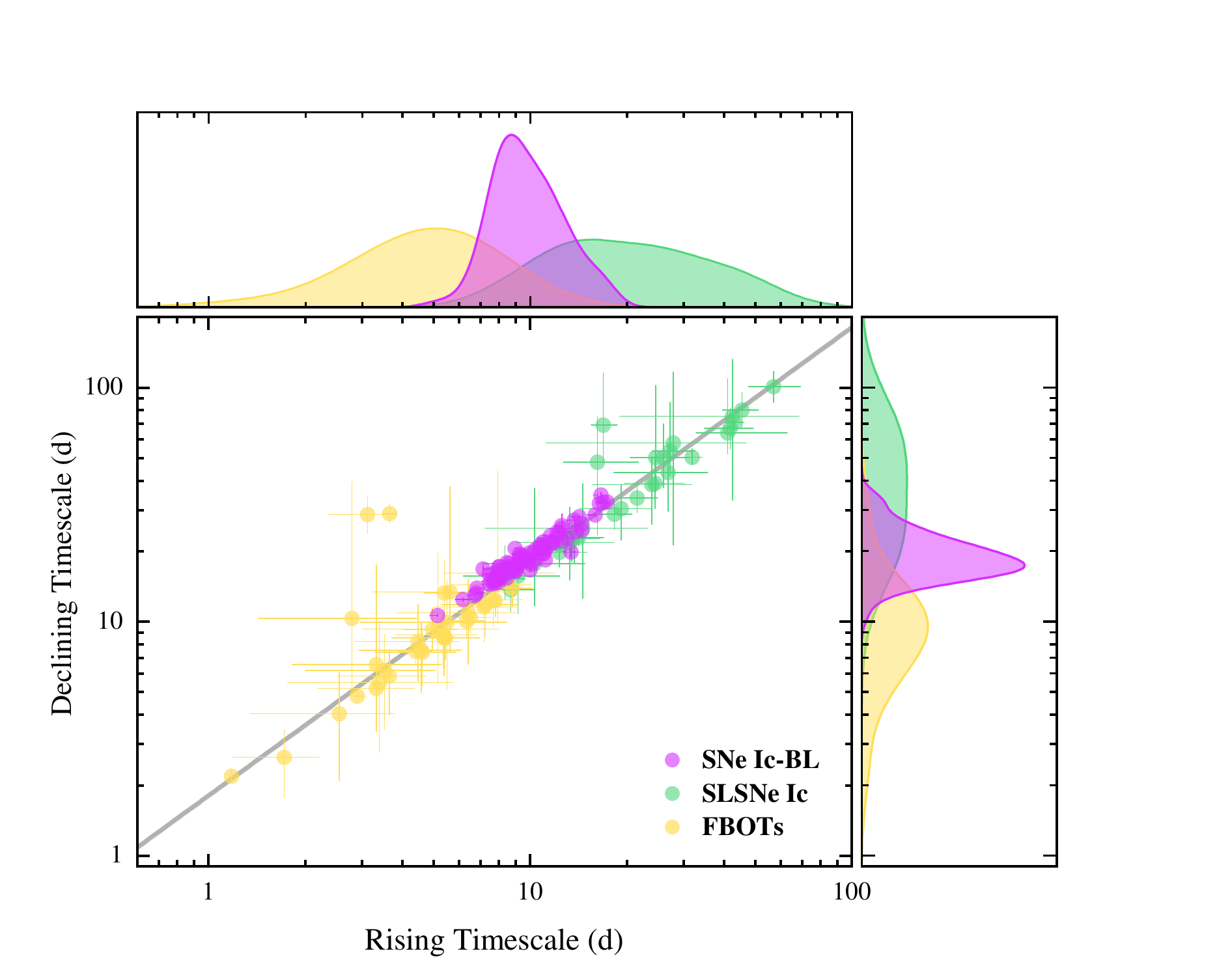}
    \includegraphics[width = 1\linewidth , trim = 33 0 125 233, clip]{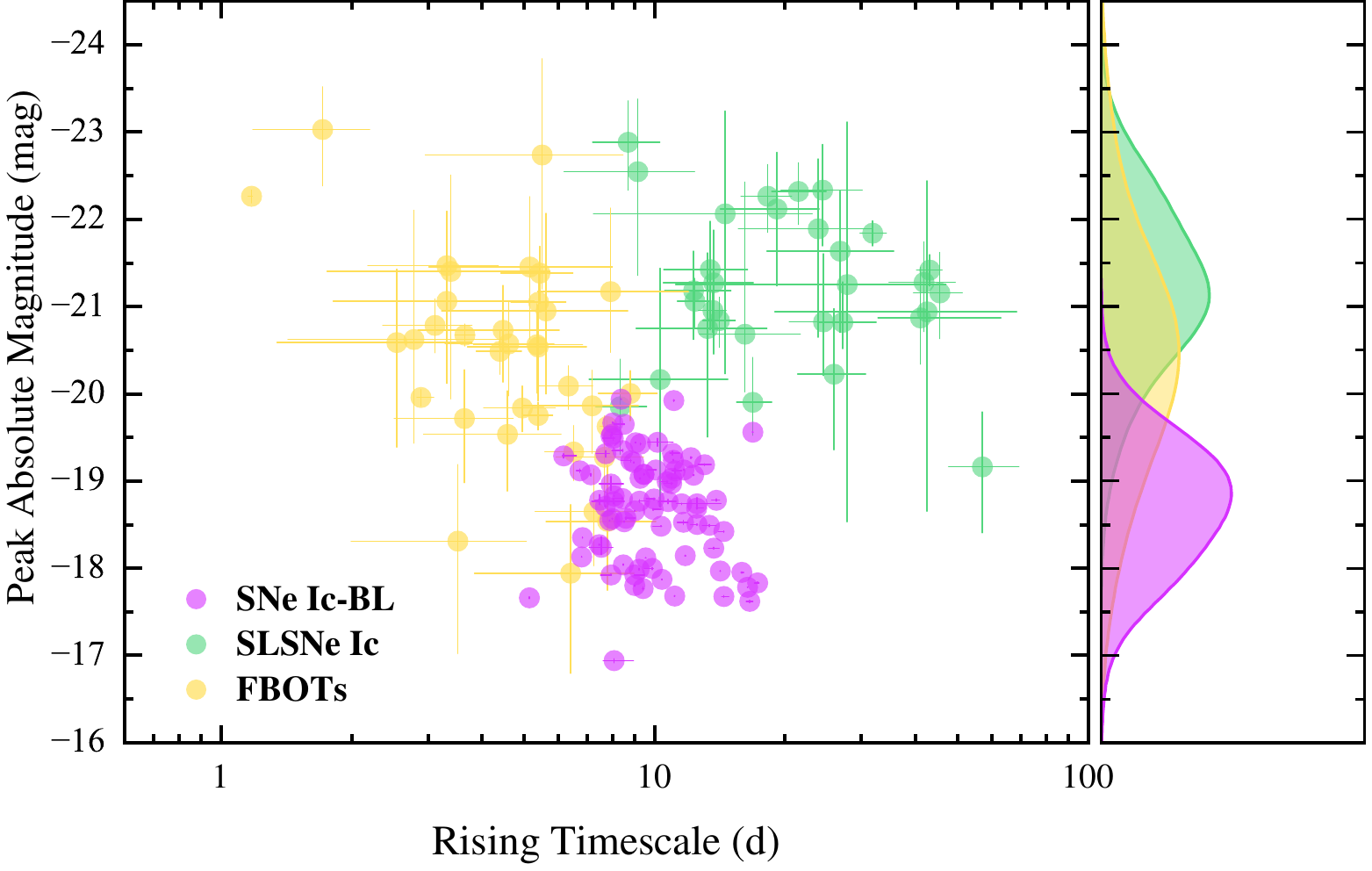}
    \caption{Similar to Figure \ref{fig:SNIcBL_LightcurveProperties}, but showing rising timescales against declining timescales (middle panel) and against peak absolute magnitudes (bottom panel) for SNe Ic-BL (violet), SLSNe Ic \citep[green;][]{Yu2017}, and FBOTs \citep[yellow;][]{Liu2022,Hu2023}. In the middle panel, the gray solid line and shaded area represent the best-fit log-log universal correlation and $1\sigma$ credible region across the three transient samples, respectively.  }
    \label{fig:Magnetar_LightcurveProperties}
\end{figure}

The parameters $t_{\rm rise}$, $t_{\rm dec}$, and $M_{\rm peak}$ describe the basic morphology of transient lightcurves and can be directly measured from observational data, making them useful for classifying transients. We therefore include the $t_{\rm rise}-t_{\rm dec}$ and $t_{\rm rise}-M_{\rm peak}$ distributions of the SLSN Ic and FBOT samples \citep{Yu2017,Liu2022,Hu2023} in Figure \ref{fig:Magnetar_LightcurveProperties} for comparison with those of our SN Ic-BL sample listed in Table \ref{tab:Lightcurve}. 

For the SLSN Ic sample, the distributions of the rising and declining timescales are $t_{\rm rise}=20.4^{+18.6}_{-10.2}\,{\rm d}$, $38.4^{+30.6}_{-22.5}\,{\rm d}$, respectively. Correspondingly, the FBOT sample yields $t_{\rm rise}=4.95^{+2.37}_{-2.22}\,{\rm d}$, $t_{\rm dec}=9.32^{+4.69}_{-4.24}\,{\rm d}$, respectively. The durations of our SN Ic-BL sample lie between those of the SLSN Ic and FBOT samples, with $t_{\rm rise}={9.63}^{+3.29}_{-5.40}\,{\rm d}$ and $t_{\rm dec}={18.52}^{+5.57}_{-3.55}\,{\rm d}$. Despite comparable ejecta mass distributions in the SN Ic-BL and SLSN Ic samples (see Sections \ref{sec:MagnetarParametersAll} and \ref{sec:ExplosionParametersAll}), the durations of SLSNe Ic are generally much longer than those of SNe Ic-BL. This is because the SLSN Ic magnetars generally rotate more slowly, injecting less rotational energy into the ejecta, while a larger fraction of the magnetar energy can instead be converted into radiation. In the middle panel of Figure \ref{fig:Magnetar_LightcurveProperties}, we find that the SN Ic-BL, SLSN Ic, and FBOT samples follow a universal strong correlation between $t_{\rm rise}$ and $t_{\rm dec}$, given by $t_{\rm dec} = (1.81\pm0.06)t_{\rm rise}$, with $r_{\rm P} = 0.950$ and a $p$-value of $1.5\times10^{-74}$. For comparison, the correlation derived from the SN Ic-BL sample alone is slightly steeper, $t_{\rm dec}=(1.91\pm0.03)t_{\rm rise}$. This difference suggests that the lightcurves of SLSNe Ic and FBOTs tend to decline faster relative to their rising timescales than those of SN Ic-BL. One possible explanation is that the decline phase of SN Ic-BL lightcurves may receive an additional contribution from the radioactive decay of $^{56}$Ni, which can moderately extend the post-peak evolution.

The SLSN Ic and FBOT samples have peak absolute magnitude (peak luminosity) distributions of $M_{\rm peak}=-21.20^{+0.86}_{-0.99}\,{\rm mag}$ ($L_{\rm peak}=9.1^{+13.5}_{-5.0}\times10^{43}\,{\rm erg}\,{\rm s}^{-1}$) and $M_{\rm peak}=-20.42^{+1.27}_{-1.16}\,{\rm mag}$ ($L_{\rm peak}=4.43^{+8.47}_{-3.05}\times10^{43}\,{\rm erg}\,{\rm s}^{-1}$), respectively. Overall, both populations are systematically brighter than SNe Ic-BL, which have $M_{\rm peak}=-18.75^{+0.80}_{-0.66}\,{\rm mag}$ ($L_{\rm peak}=9.81^{+7.06}_{-5.45}\times10^{42}\,{\rm erg}\,{\rm s}^{-1}$). We find that the $t_{\rm rise}-M_{\rm peak}$ distributions of these three types of transients occupy three distinct regions, as displayed in the bottom panel of Figure \ref{fig:Magnetar_LightcurveProperties}. The SN Ic-BL sample shows a relatively narrow distribution of $t_{\rm rise}$, typically within $\sim7-20\,{\rm d}$, with $M_{\rm peak}$ mainly ranging from $\sim-17$ to $\sim-20\,{\rm mag}$. In contrast, the SLSN Ic sample shows a much broader range of $t_{\rm rise}$, which are $\sim8-60\,{\rm d}$, and are systematically brighter, with $M_{\rm peak}\lesssim-20\,{\rm mag}$. FBOTs are characterized by much shorter $t_{\rm rise}$ of $\sim1-8\,{\rm d}$ and are also predominantly brighter than $M_{\rm peak}\lesssim-20\,{\rm mag}$, although a small fraction of events are fainter and partially overlap with the SN Ic-BL population. These distinct distributions suggest that the $t_{\rm rise}-M_{\rm peak}$ plane can serve as a useful diagnostic for distinguishing between these transients, and may provide a simple observational criterion for the early classification of newly discovered events in time-domain surveys.

\subsection{$^{56}$Ni Production in SNe Ic-BL and Ordinary SNe Ic} \label{sec:NickelProductionAll}

\begin{figure}[t!]
    \centering
    \includegraphics[width = 1\linewidth , trim = 33 0 125 80, clip]{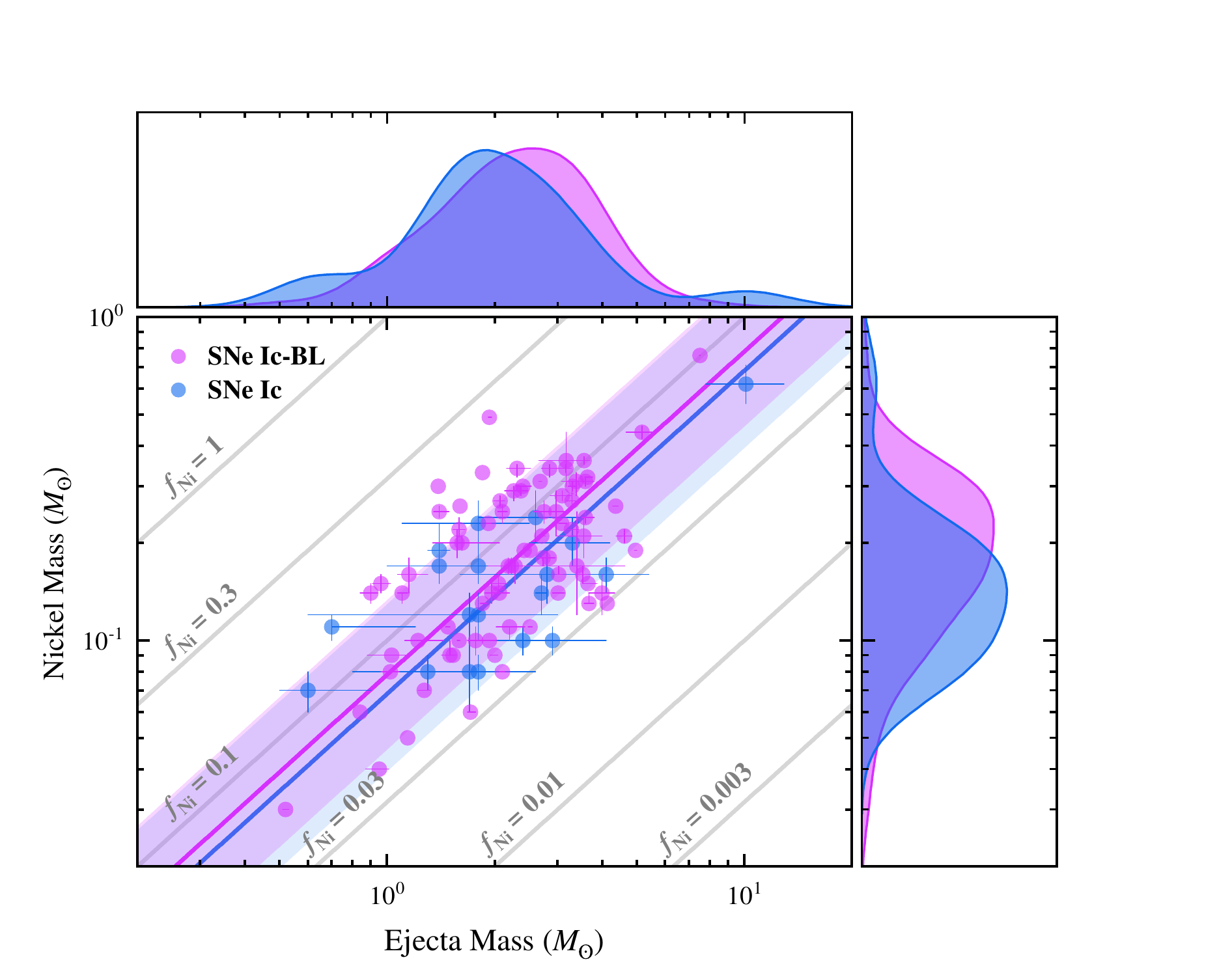}
    \caption{Similar to Figure \ref{fig:SNIcBL_Mej_MNi}, but showing $^{56}$Ni masses against ejecta masses for SNe Ic-BL (violet) and ordinary SNe Ic \citep[darked blue][]{Lyman2016,Taddia2018}.  }
    \label{fig:SNIc_Mej_MNi}
\end{figure}

In Figure \ref{fig:SNIc_Mej_MNi}, we compare the $M_{\rm ej}-M_{\rm Ni}$ distribution of SNe Ic-BL in our sample with that of ordinary SNe Ic from \cite{Lyman2016,Taddia2018}, as both classes are expected to lack hydrogen and helium envelopes before the explosion. Relative to SNe Ic-BL, which have $M_{\rm ej}$ and $M_{\rm Ni}$ distributions of $2.30^{+1.46}_{-1.01}\,M_\odot$ and $0.18^{+0.14}_{-0.09}\,M_\odot$, respectively, SNe Ic tend to have slightly smaller, but basically comparable values, with their $M_{\rm ej}$ and $M_{\rm Ni}$ distributions being $1.99^{+1.54}_{-0.87}\,M_\odot$ and $0.14^{+0.10}_{-0.06}\,M_\odot$. The $M_{\rm ej}-M_{\rm Ni}$ distributions of SNe Ic-BL and SNe Ic also show similar correlations, with comparable $^{56}$Ni mass fractions of $f_{\rm Ni} = 0.08^{+0.06}_{-0.03}$ and $0.07^{+0.06}_{-0.03}$, respectively. 

We then apply an energy test introduced in Section \ref{sec:PopulationEquality} to compare the $M_{\rm ej}-M_{\rm Ni}$ distributions between SNe Ic-BL and SNe Ic. The resulting $p$-value is $\sim0.10$, indicating that there is no statistically significant difference between these distributions. This further suggests that SNe Ic-BL and SNe Ic may originate from progenitors with broadly similar pre-SN properties, while their differences are likely driven by the presence of a central engine after core collapse, which in turn depends on the progenitor’s rotation. More specifically, SNe Ic-BL progenitors may retain substantially higher angular momentum prior to their explosions, enabling the collapsing inner core to produce a rapidly rotating magnetar via angular momentum conservation. Conversely, SNe Ic may give rise to more slowly rotating NSs or magnetars, whose spin-down process plays a negligible role in shaping the explosion dynamics.

\section{Discussion} \label{sec:Discussion}

\subsection{A Unified Picture of Magnetar-powered and Ordinary SESNe}

In this section, we combine the results presented above to explore the connections and distinctions among SNe Ic-BL, lGRBs, SLSNe Ic, FBOTs, and ordinary SESNe. Building on current theoretical progenitor models, we further propose a possible unified progenitor scenario underlying these transients.

\subsubsection{A Physical Classification of Magnetar-powered and Ordinary SNe} \label{sec:PhysicalClassification}

Based on the dominant energy source driving the explosion, SESNe can be broadly classified into two classes: ``{\em magnetar-powered SESNe}'' and ``{\em neutrino-driven SESNe}.'' The former class includes SNe Ic-BL, SLSNe Ic, and FBOTs, while the latter mainly corresponds to ordinary SESNe. However, the role of the magnetar differs among these subclasses.

\begin{figure*}[t!]
    \centering
    \includegraphics[width = 0.49\linewidth , trim = 73 105 100 65, clip]{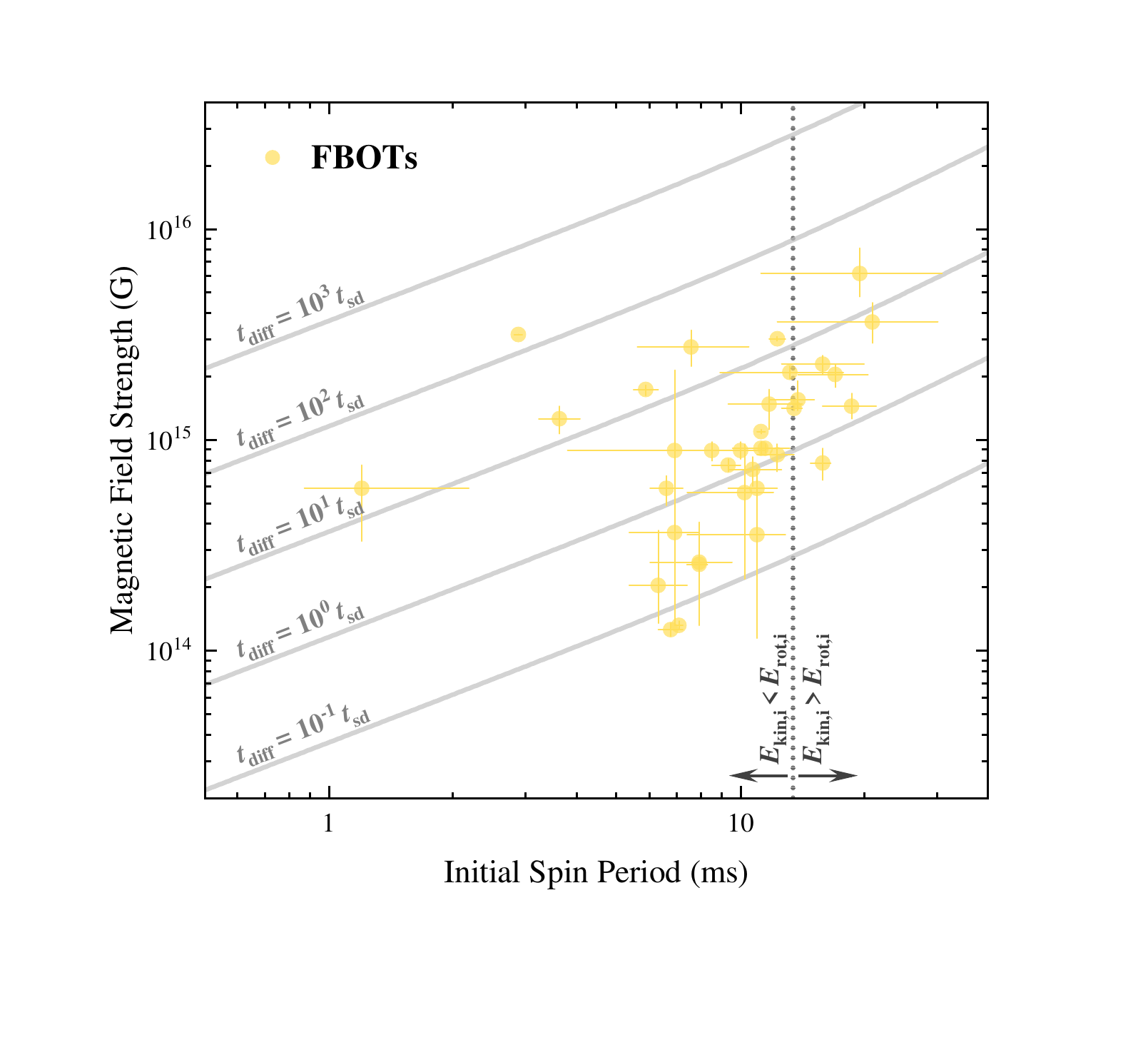}
    \includegraphics[width = 0.49\linewidth , trim = 68 105 105 65, clip]{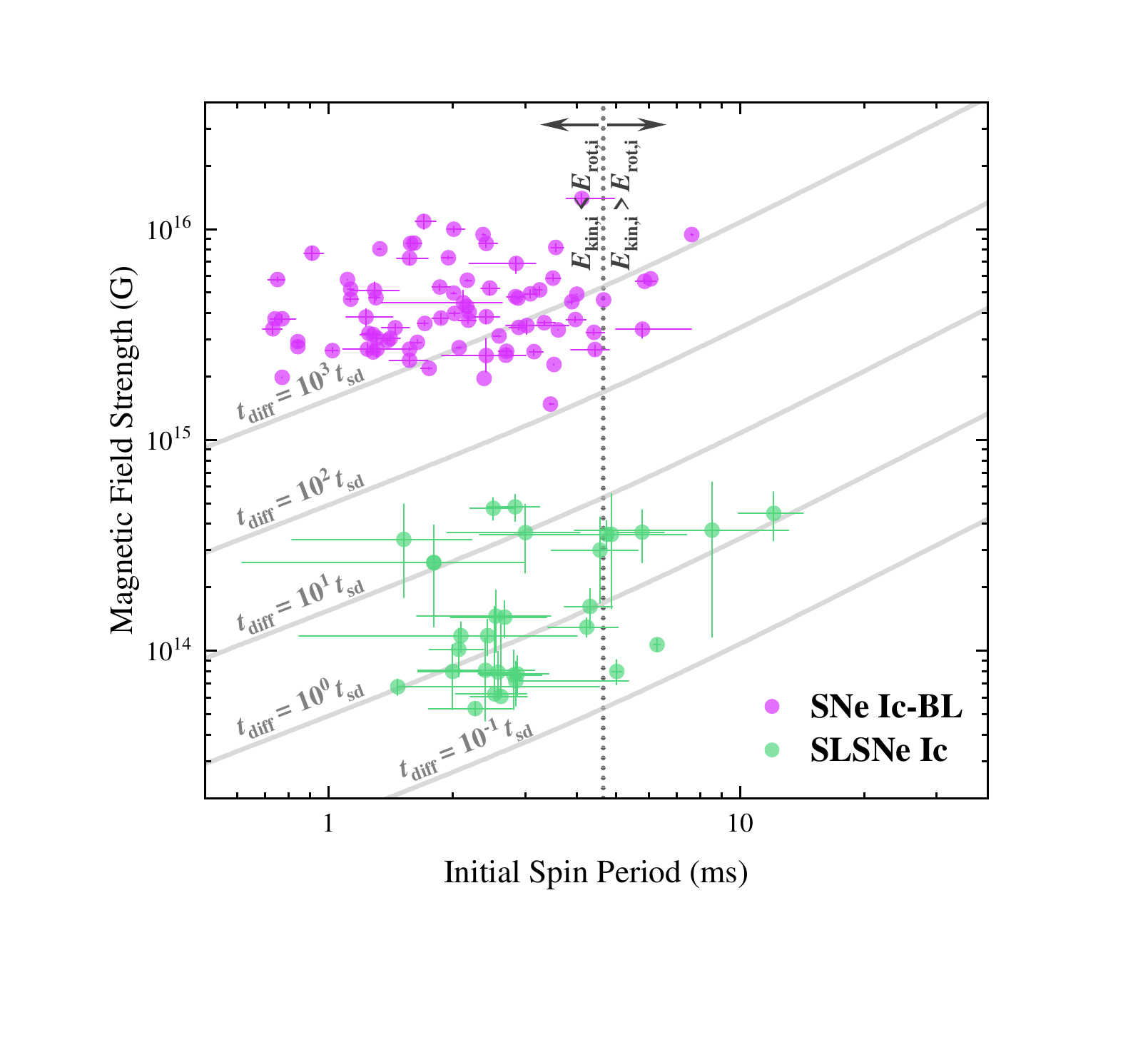}
    \caption{Ratio of the diffusion timescale to the spin-down timescale is shown as gray solid lines corresponding to $t_{\rm diff}/t_{\rm sd} = 10^{-1}$, $10^{0}$, $10^{1}$, $10^{2}$, and $10^{3}$ in the parameter space of initial spin period and magnetic field strength. $t_{\rm diff}$ is calculated with $M_{\rm ej}=0.2\,M_\odot$ and $E_{\rm kin,i}\sim1.1\times10^{50}\,{\rm erg}$ (left panel), and $M_{\rm ej}=2\,M_\odot$ and $E_{\rm kin,i}\sim9.3\times10^{50}\,{\rm erg}$ (right panel). The distributions of initial spin period and magnetic field strength for FBOTs, SNe Ic-BL, and SLSNe Ic the same as those used in Figure \ref{fig:Magnetar_Pi_Bp}. The gray dotted lines mark the boundary defined by $E_{\rm kin,i} = E_{\rm rot,i}$, separating the regions with $E_{\rm kin,i} < E_{\rm rot,i}$ on the left and $E_{\rm kin,i} > E_{\rm rot,i}$ on the right.}
    \label{fig:DiffusionTimescaleSpindownTimescale}
\end{figure*}

Although SNe Ic-BL, SLSNe Ic, and FBOTs are observationally classified as distinct types of transients, the existence of a universal $M_{\rm ej}-P_{\rm i}$ correlation may indicate that they share a common underlying origin. In Sections \ref{sec:MagnetarParametersAll} and \ref{sec:ExplosionParametersAll}, we find that (1) compared to SLSNe Ic and SNe Ic-BL, FBOTs preferentially occupy the region with $M_{\rm ej} \lesssim 1\,M_\odot$ and $P_{\rm i} \gtrsim 5\,\mathrm{ms}$; (2) the magnetic field strengths show systematic differences: FBOT magnetars span a broad $B_{\rm p}$ range of $\sim10^{14}-10^{16}\,\mathrm{G}$, SNe Ic-BL magnetars are typically confined to $\sim10^{15}-10^{16}\,\mathrm{G}$, and SLSNe Ic magnetars always have $\lesssim10^{15}\,{\rm G}$.

Since the spin-down timescale $t_{\rm sd}$ is primarily determined by $P_{\rm i}$ and $B_{\rm p}$, while the diffusion timescale $t_{\rm diff}$ depends on $M_{\rm ej}$ and $P_{\rm i}$, the observational diversity among SNe Ic-BL, SLSNe Ic, and FBOTs essentially reflects differences in the magnetar energy injection rate relative to the photon diffusion timescale. Thus, in Figure \ref{fig:DiffusionTimescaleSpindownTimescale}, we show the ratios of $t_{\rm diff}/t_{\rm sd}$ in the parameter space of $P_{\rm i}$ and $B_{\rm p}$. For illustrative purposes, we adopt two representative ejecta masses: $M_{\rm ej} = 0.2\,M_\odot$ for FBOTs and $M_{\rm ej} = 2\,M_\odot$ for SNe Ic-BL and SLSNe Ic, corresponding to initial neutrino-driven kinetic energies of $\sim 1.1\times10^{50}\,{\rm erg}$ and $\sim9.3\times 10^{50}\,{\rm erg}$, respectively, as calculated from Equation (\ref{equ:Neutrino_Energy}). We note that the former neutrino-driven energy is consistent with that expected for USSNe \citep[e.g.,][]{Suwa2015,Sawada2022}, whereas the latter corresponds to a typical value for ordinary core-collapse SNe \citep[e.g.,][]{Subkhbold2016}. To estimate the ejecta velocity when calculating $t_{\rm diff}$, we approximately assume that the final kinetic energy satisfies $E_{\rm kin,f}\sim E_{\rm kin,i}+E_{\rm rot,i}$. We note, however, that $E_{\rm rot,i}$ can only be partially converted into kinetic energy when $t_{\rm diff} \sim t_{\rm sd}$.

As shown in Figure \ref{fig:DiffusionTimescaleSpindownTimescale}, the magnetar central engine plays a dominant role in the explosions of SNe Ic-BL, SLSNe Ic, and FBOTs, since the initial magnetar rotational energy $E_{\rm rot,i}$ is typically comparable to or significantly exceeds the initial neutrino-driven kinetic energy $E_{\rm kin,i}$. For FBOTs, $t_{\rm diff}$ is generally comparable to $t_{\rm sd}$, with $0.1\lesssim t_{\rm diff}/t_{\rm sd}\lesssim100$ for most sources. For more massive SNe, a larger $M_{\rm ej}$ generally implies a longer $t_{\rm diff}$. Meanwhile, according to the universal $M_{\rm ej}-P_{\rm i}$ anti-correlation we derive in Equation (\ref{equ:UniversalRelationship}), a larger $M_{\rm ej}$ typically corresponds to a smaller $P_{\rm i}$. In this case, if $t_{\rm diff}$ remains of the same order as $t_{\rm sd} \propto B_{\rm p}^{-2} P_{\rm i}^{2}$, this would require a relatively weaker $B_{\rm p}$. Similar to FBOTs, as shown in the right panel of Figure \ref{fig:DiffusionTimescaleSpindownTimescale}, $t_{\rm diff}$ and $t_{\rm sd}$ are also comparable for SLSNe Ic, with $0.1\lesssim t_{\rm diff}/t_{\rm sd} \lesssim 100$. However, unlike FBOT magnetars, whose $B_{\rm p}$ spans a relatively broad range, the inferred $B_{\rm p}$ values of SLSN Ic magnetars are more narrowly distributed and are generally $\lesssim10^{15}\,{\rm G}$. Although the distributions of $P_{\rm i}$, $B_{\rm p}$, and $M_{\rm ej}$ differ between SLSNe Ic and FBOTs, for most sources $t_{\rm diff}$ roughly matches $t_{\rm sd}$. This implies that the initial rotational energy significantly contributes to the SN peak emission, making both classes extremely bright, with the main difference likely being the longer duration of SLSNe Ic. This motivates us to classify them as ``{\em radiatively magnetar-powered SESNe}.'' 

For SNe Ic-BL, the $B_{\rm p}$ values are systematically higher than those of SLSNe Ic, and for most sources $t_{\rm diff}/t_{\rm sd} \sim 10^3$, with all sources having $t_{\rm diff}/t_{\rm sd} \gtrsim10^2$. As a result, the magnetar energy primarily contributes to the ejecta kinetic energy, with only a minor fraction powering the peak luminosity. We therefore classify SNe Ic-BL as ``{\em mechanically magnetar-powered SESNe}.'' Furthermore, given that nearly all lGRBs are observed to be associated with SNe Ic-BL rather than SLSNe Ic, the inferred ultra-high magnetic fields ($B_{\rm p}\gtrsim10^{15}\,{\rm G}$) of newborn millisecond magnetars may play a critical role in the formation of relativistic jets. Such characteristics have been widely suggested in the literature. This picture is broadly consistent with previous studies suggesting that millisecond magnetars with ultra-high magnetic fields are required for launching relativistic jets \citep[e.g.,][]{Usov1992,Kluzniak1998,Zhang2001,Thompson2004,Metzger2011,Zhang2018}.

Finally, newborn NSs with initial rotational energies lower than the neutrino-driven explosion energy may form in ordinary SESNe, in which case the NS rotational energy does not significantly affect the ejecta dynamics. However, the spin-down of NSs may still shape the peak emission and even the late-time tail. Recent statistical and numerical studies increasingly suggest that the emissions of ordinary SESNe, including USSNe, cannot be fully explained by $^{56}$Ni decay alone and instead require an additional energy source, plausibly provided by spin-down magnetars or even pulsars with $P_{\rm i}\gtrsim10\,{\rm ms}$ \citep[e.g.,][]{Li2015,Kashiyama2016,Sharon2020,Ertl2020,Afsariardchi2021,Woosley2021,Sawada2022,Sollerman2022,Rodriguez2023,Rodriguez2024}. These considerations suggest that ordinary SESNe may be ``{\em magnetar/pulsar-aided SESNe}.'' In Figure  \ref{fig:Illustration}, we illustrate our consideration of a preliminary possible physical classification of various SESNe within a unified magnetar framework.

\subsubsection{A Unified Progenitor Model of Magnetar-powered and Ordinary SESNe}

\begin{figure*}[t!]
    \centering
    \includegraphics[width = 1\linewidth , trim = 0 0 0 0, clip]{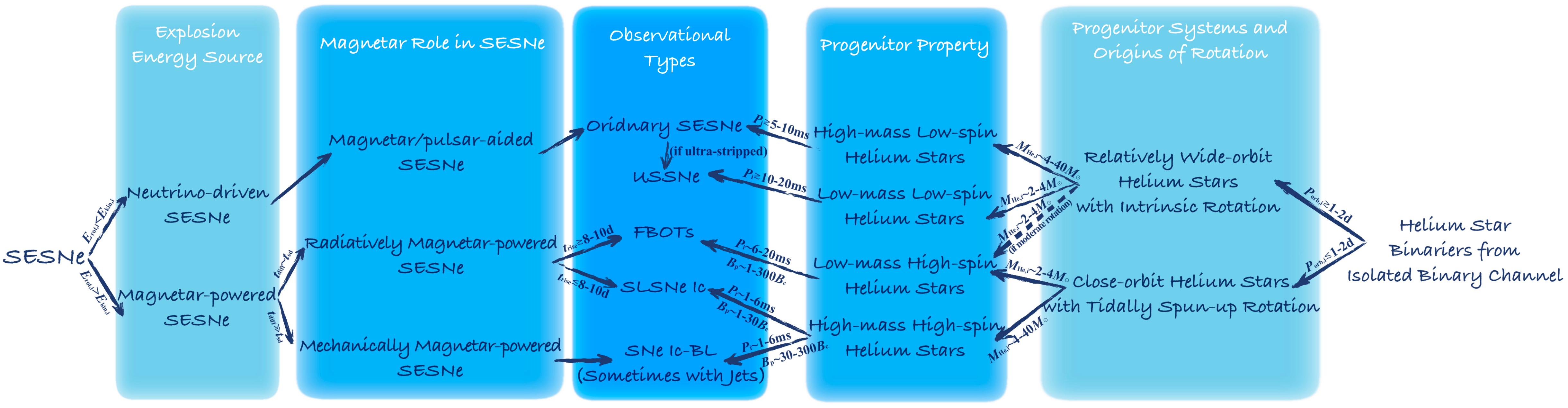}
    \caption{A possible unified picture of magnetar-powered and ordinary SESNe. $B_{\rm c}=4.4\times10^{13}\,{\rm G}$ represents the critical field of the electron Landau quantization.}
    \label{fig:Illustration}
\end{figure*}

Binary stripping through binary mass transfer, including stable Roche-lobe overflow (RLOF) or common-envelope evolution (CEE), as well as stellar wind mass loss \citep[e.g.,][]{Podsiadlowski1992,Nomoto1995,Wellstein1999,Yoon2010,Yoon2017,Eldridge2013,Smith2014,Fang2019,Sun2023}, can produce helium stars spanning a broad mass range. At the high-mass end, the helium stars are observationally classified as Wolf-Rayet stars and are thought to give rise to ordinary SESNe. In the case of low-mass helium stars, late-stage expansion before the explosion can induce an additional phase of Case BB/BC mass transfer through interaction with the companion. This process leads to even more extensive envelope stripping of the helium stars, finally resulting in USSNe with ejecta masses of only a few to several $0.1\,M_\odot$, and leaving behind binary NS systems that serve as important gravitational-wave sources \citep{Tauris2013,Tauris2015,Tauris2017,Suwa2015,Moriya2017,Jiang2021,Wu2022}. 

The origin and formation channels of SNe Ic-BL, SLSNe Ic, and FBOTs remain uncertain and are still under active debate. A magnetar central engine generally requires the pre-SN progenitors to have extremely rapid rotation. A key issue is how a massive star before the SN explosion can attain and retain a rapidly rotating core that eventually produces a fast-spinning central engine, while simultaneously losing its hydrogen and helium envelopes. The evolution of massive stars can be strongly affected by magnetic torques from the Tayler-Spruit dynamo \citep{Spruit2002,Fuller2019}, which can significantly alter the angular momentum transport processes through wind-driven mass loss to spin down stellar cores, making it challenging for single-star models to meet both requirements. One of the most popular progenitor scenarios involves either single main-sequence stars born with rapid rotation \citep[e.g.,][]{Yoon2005,Yoon2006,Woosley2006,AguileraDena2018,Song2023} or stars in binary systems that are spun up through mass transfer, tidal interactions, or stellar mergers \citep[e.g.,][]{Cantiello2007,Eldridge2011,deMink2013,Mandel2016,Riley2021,Ghodla2023}, undergoing quasi-chemically homogeneous evolution (CHE) and directly evolving into fast-spinning Wolf-Rayet stars. This CHE channel can only occur in very low-metallicity environments, which is predicted to occur only at metallicities below $Z<0.3\,Z_\odot$ \citep{Woosley2006}. However, most observed host metallicities of SN Ic-BL, SLSN Ic, and FBOT populations are higher than those expected for the CHE channel, with up to $\lesssim50\%$ of the events found in super-solar environments \citep{Kruhler2015,Nicholl2017SN2017egm,Japelj2018,Modjaz2020,Wiseman2020}, thereby challenging the role of the CHE channel as the sole and primary formation scenario for these transients. 

Alternatively, the classical isolated binary channel provides another set of promising scenarios for producing fast-spinning Wolf-Rayet stars, either through the merger of Wolf-Rayet stars/helium cores \citep{Fryer2005,Kinugawa2017} or via tidal interactions in close binaries that spin up the Wolf-Rayet stars with their companions \citep{Izzard2004,vandenHeuvel2007,Detmers2008,Bogomazov2009,Chrimes2020,Bavera2022,Fuller2022,Hu2023,Hu2026}. The former merger pathway is thought to occur in low-metallicity environments at high redshift, where weaker stellar winds enhance the possibility of mergers and reduce angular momentum loss from the merged Wolf-Rayet stars. However, the contribution of the merger scenario is currently uncertain, but it is likely to be small. For example, \citet{Kinugawa2017} suggested that it can account for the lGRB rate density, which represents only a small fraction of the total rate densities of magnetar-powered SESNe. In contrast, close-orbit Wolf-Rayet stars formed through stable RLOF or CEE channels can retain angular momentum through tidal interactions with their companions across a wide range of metallicities, finally producing millisecond magnetars upon core collapse. The tidal scenario may represent the dominant formation channel for these various types of magnetar-powered SESNe, as supported by the comparable $M_{\rm ej}-P_{\rm i}$ correlations, event rate densities, and metallicity environments inferred from both observations and simulations \citep{Hu2023}. 

Thus, both ordinary and magnetar-powered SESNe are expected to originate from the same isolated helium-star binary progenitors formed through stable RLOF or CEE channels. Indeed, this is supported by statistical analysis of the progenitors of these transients, which reveals several notable similarities: (1) ordinary SESNe (i.e., SNe IIb/Ib/Ic) show broadly similar explosion properties, including ejecta masses, $^{56}$Ni masses, $^{56}$Ni production efficiencies, and kinetic energies \citep[e.g.,][]{Lyman2016,Taddia2018}; (2) the explosion properties of SNe Ic and SNe Ic-BL exhibit comparable $^{56}$Ni masses, ejecta masses, and $^{56}$Ni production efficiencies (as concluded in Section \ref{sec:NickelProductionAll}); (3) the progenitor masses of SNe Ic-BL are comparable to those inferred for SLSNe Ic (as concluded in Section \ref{sec:ExplosionParametersAll}); (4) the host-galaxy properties of ordinary SESNe are broadly similar to those of magnetar-powered transients \citep[e.g.,][]{Chrimes2020}. In addition, several studies provided both direct and indirect observational evidence that main-sequence companions may have existed for some SLSN Ic and SN Ic-BL progenitors \citep[e.g.,][]{Moriya2015,Zhu2024,Zhu2025}, as well as for some ordinary SESNe \citep[e.g.,][]{Maud2004,Maud2016,Cao2013,Ryder2018,Fox2022,Sun2022}. Taken together, these similarities suggest that these different classes of transients may share a common binary evolutionary origin, with the primary difference between ordinary and magnetar-powered SESNe possibly arising from the presence or absence of a rapidly rotating magnetar capable of powering the explosion.

\begin{figure*}[t!]
    \centering
    \includegraphics[width = 0.3523\linewidth , trim = 73 105 105 65, clip]{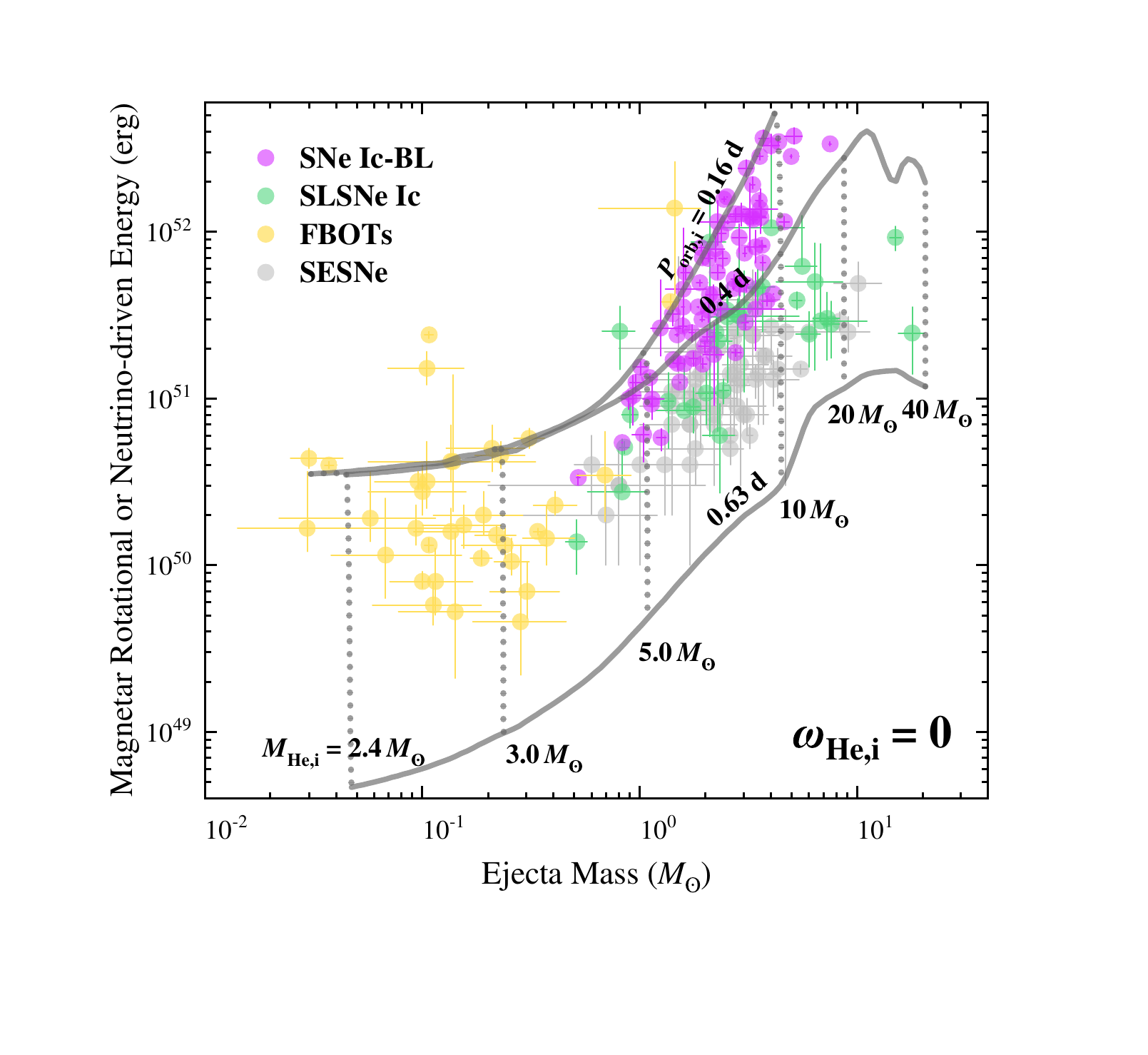}
    \includegraphics[width = 0.315\linewidth , trim = 136 105 105 65, clip]{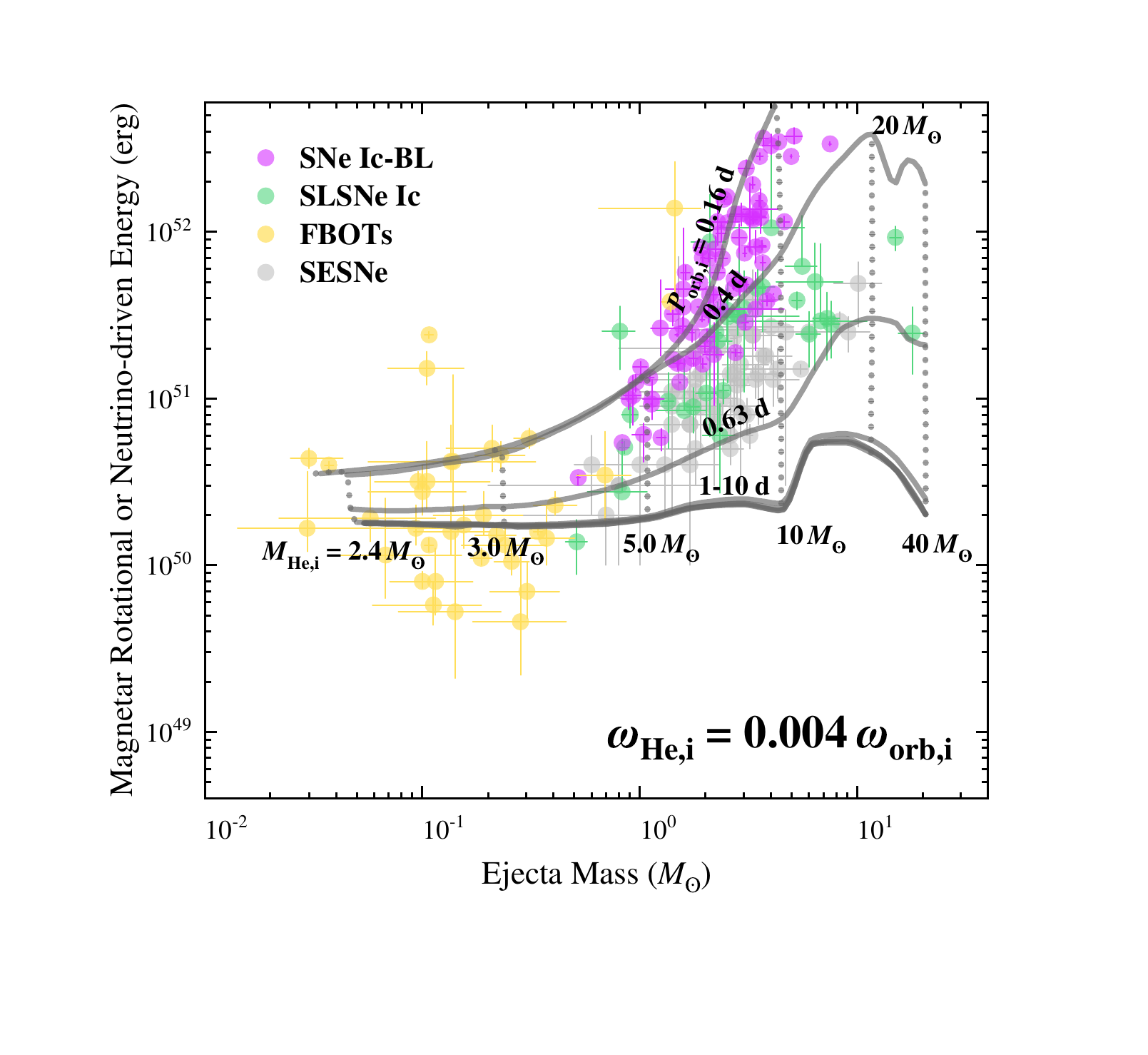}
    \includegraphics[width = 0.315\linewidth , trim = 136 105 105 65, clip]{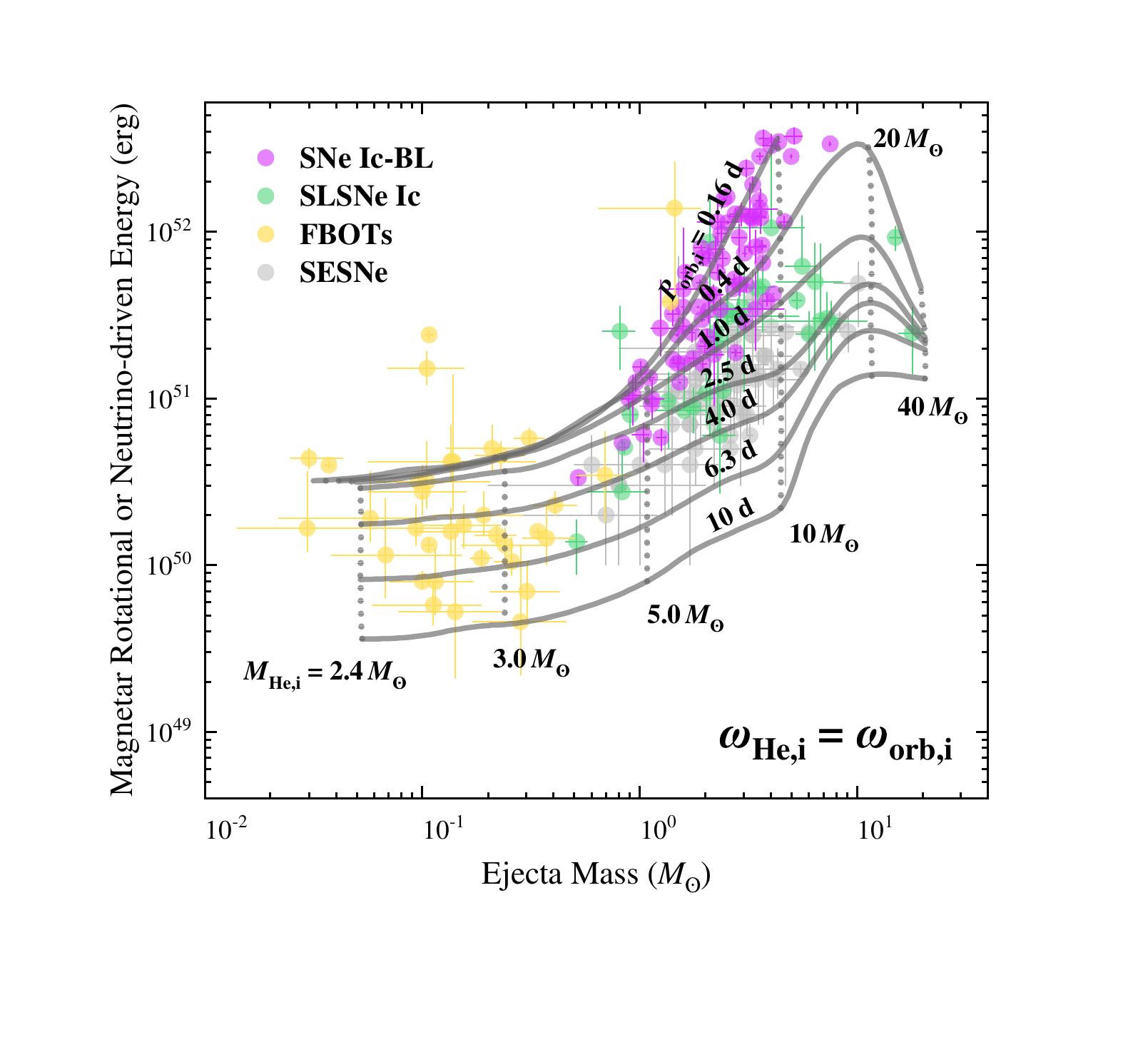}
    \caption{Initial magnetar rotational energies against ejecta masses for SNe Ic-BL (violet points), SLSNe Ic \citep[green points;][]{Yu2017}, and FBOTs \citep[yellow points;][]{Liu2022,Hu2023}. The solid lines show simulations of helium stars in close binary systems \citep{Hu2023} with different initial orbital periods ($0.16-10\,{\rm d}$), while the dotted lines show simulations with different initial helium star masses ($2.4-40\,M_\odot$), as labeled. The left, middle, and right panels correspond to different assumptions for the initial rotation of helium stars: non-rotating, solar-like surface rotation, and tidal synchronization, respectively. The gray points, taken from \cite{Lyman2016,Taddia2018}, correspond to the measured neutrino-driven explosion energies against ejecta masses for ordinary SESNe.}
    \label{fig:MESA}
\end{figure*}

In binary systems, the rotation of newborn NS is closely connected to the rotation of the helium-star progenitor, which is sensitive to the initial conditions and the binary interaction with the companion. It is determined by many parameters, among which the key ones include the initial rotation of the zero-age helium star formed through stable RLOF or CEE channels, the initial helium star mass, the initial orbital period, the companion mass, the metallicity, and the angular momentum transport mechanism \citep{Fuller2022,Hu2023}. When the helium star progenitors have comparable rotation rates, more massive helium stars tend to have larger cores. As a result, more gravitational energy can convert into the NS rotational energy, naturally leading to a $M_{\rm ej}-E_{\rm rot,i}$ correlation. As shown in Figure \ref{fig:MESA}, we directly compare the $M_{\rm ej}-E_{\rm rot,i}$ correlation derived from the $M_{\rm ej}-P_{\rm i}$ anti-correlation of SNe Ic-BL, SLSNe Ic, and FBOTs (Figure \ref{fig:Magnetar_Mej_Pi}) with the simulated $M_{\rm ej}-E_{\rm rot,i}$ results reported in \cite{Hu2023}. Among the various models explored in their work, we adopt one representative parameter set with initial helium star masses of $M_{\rm He,i}\sim2.4-40\,M_\odot$, initial orbital periods of $P_{\rm orb,i}\sim0.16-10\,{\rm d}$, a point companion mass of $1.4\,M_\odot$, a metallicity of $Z\sim0.3\,Z_\odot$, and the standard Tayler-Spruit dynamo model for angular momentum transport \citep{Spruit2002}. We refer the reader to \cite{Hu2023} for more details on the methodology of their models and simulations. Because the initial rotation of the helium star is uncertain, we adopt three different models. In the first two models, we assign a modest initial spin to all helium stars for illustrative purposes, setting the initial surface angular velocity to $\omega_{\rm He,i}=0$ and $\omega_{\rm He,i}=0.004\,\omega_{\rm crit}$, the latter being comparable to the solar rotation rate, where $\omega_{\rm crit}$ is the critical surface angular velocity. In the third model, we assume tidal synchronization, $\omega_{\rm He,i}=\omega_{\rm orb,i}$ with $\omega_{\rm orb,i}=2\pi/P_{\rm orb,i}$ being the initial orbital angular velocity, which may better represent helium stars formed via stable RLOF \citep[e.g.,][]{Schurmann2022}.

As shown in Figure \ref{fig:MESA}, the simulations can broadly reproduce the observed universal $M_{\rm ej}-E_{\rm rot,i}$ correlation in log-log space. We find that SNe Ic-BL and SLSNe Ic mainly originate from helium stars with $M_{\rm He,i}\sim4-40\,M_\odot$ in binaries within $P_{\rm orb,i}\lesssim1-2\,{\rm d}$. In general, tidal interactions are expected to be significant for such short orbital periods \citep[e.g.,][]{Fuller2022,Hu2023}, suggesting that the progenitor rotation of SNe Ic-BL and SLSNe Ic may be largely regulated by tidal interactions. More specifically, since the magnetars inferred in SNe Ic-BL tend to rotate faster than those in SLSNe Ic, the progenitor binaries of SNe Ic-BL may have closer orbits. For ordinary SESNe, although the newborn magnetar/pulsar may rotate very slowly, with rotational energy potentially much lower than the neutrino-driven explosion energy, its spin-down process may still contribute partially to the SN emission. As illustrated in Figure \ref{fig:MESA}, this typically requires wider pre-SN binary systems with orbital periods longer than $P_{\rm orb,i}\gtrsim1-2\,{\rm d}$. In such systems, the rotation of the magnetar/pulsar is more likely determined by the initial rotation of the helium star rather than by tidal interactions. Moreover, we find that the majority ($\sim80\%$) of FBOTs may originate from low-mass ($M_{\rm He,i}\sim2-4\,M_\odot$) helium stars in close binaries. These low-mass helium stars are expected to undergo Case BB/BC mass transfer and further lose their remaining envelopes before exploding as core-collapse or electron-capture SNe. In this scenario, FBOTs may correspond to magnetar-powered USSNe. It is also worth noting that, as shown in Figure \ref{fig:MESA}, if low-mass helium stars have negligible initial rotation, the final magnetar spin is primarily determined by tidal interactions, consistent with the inferred progenitor rotation for SNe Ic-BL and SLSNe Ic. However, if helium stars possess modest initial rotation comparable to the solar rotation rate, the resulting magnetar spin-down can still power FBOT emission. In contrast, producing $^{56}$Ni-powered USSNe requires helium stars with extremely slow initial rotation. 

In Figure \ref{fig:Illustration}, we combine the progenitor scenarios discussed above with the observational characteristics of SESNe summarized in Section \ref{sec:PhysicalClassification} to present a unified picture of the origins of different SESN subclasses.

\subsection{Afterglow Contamination and Observational Bias in GRB-SNe} \label{sec:SelectionBias}

\begin{figure}[t!]
    \centering
    \includegraphics[width = 1\linewidth , trim = 73 105 95 65, clip]{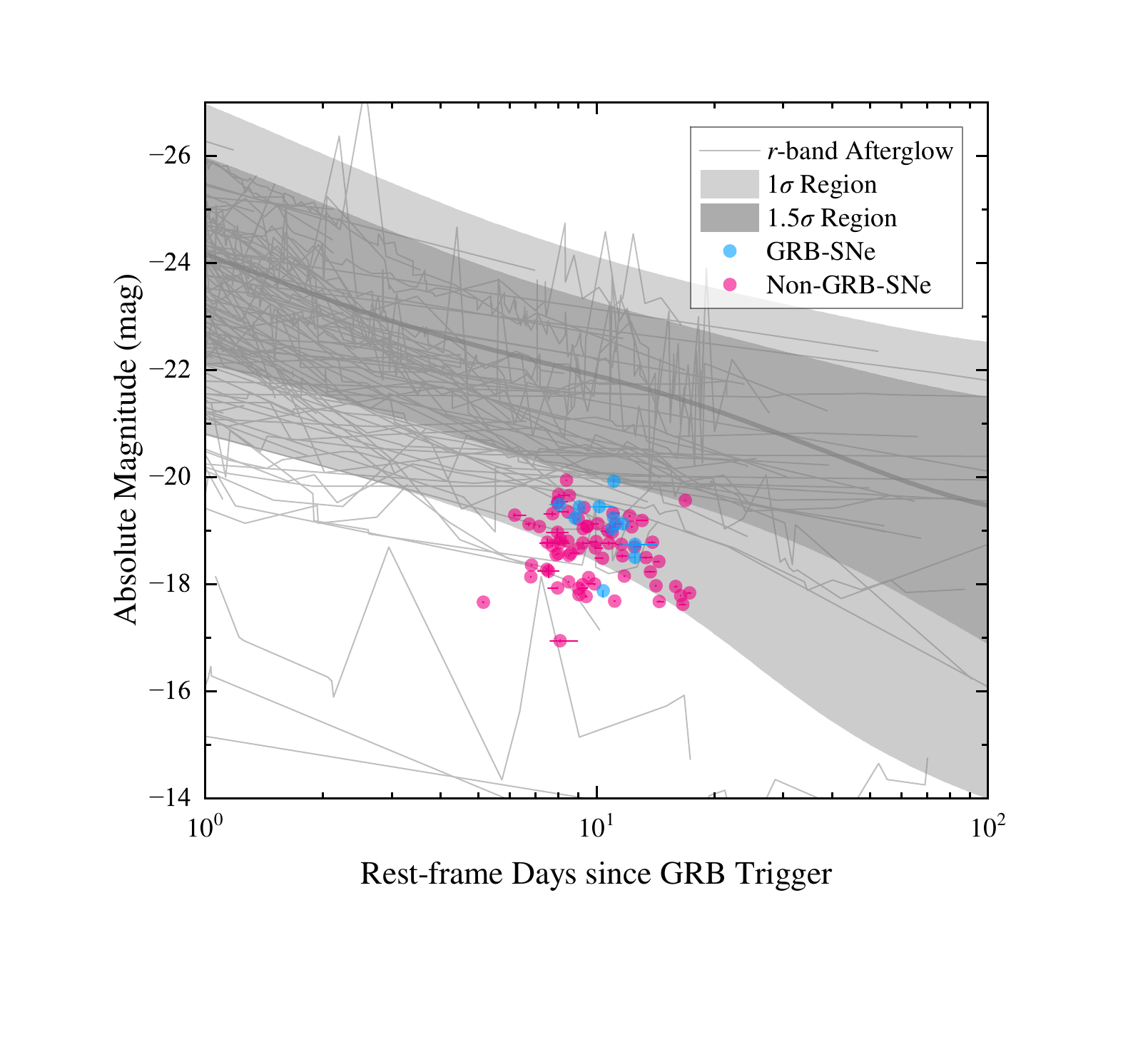}
    \caption{Comparison of the $t_{\rm rise}$–$M_{\rm peak}$ distributions of GRB-SNe (blue points) and Non-GRB-SNe (purple points) with the observed $r$-band afterglow lightcurves of cosmological lGRBs \citep[gray solid lines;][]{Dainotti2024}. The dark gray and light gray shaded regions represent the $1\sigma$ and $1.5\sigma$ ($\sim90\%$) credible regions of the observed afterglow lightcurves, respectively. }
    \label{fig:AfterglowLightcurve}
\end{figure}

In order to perform a preliminary test of whether lGRB afterglows can bias the detection of GRB-SNe toward brighter events,  as found in Sections \ref{sec:MagnetarParameterSNeIcBL}, \ref{sec:NickelProductionSNeIcBL}, and \ref{sec:LightcurveSNeIcBL}, we present in Figure \ref{fig:AfterglowLightcurve} the $r$-band lightcurves of cosmological lGRB afterglows, together with the peak absolute magnitude distributions of our GRB-SN and Non-GRB-SN samples (from Figure \ref{fig:SNIcBL_LightcurveProperties}). The $r$-band afterglow data are taken from \cite{Dainotti2024}, corrected for Galactic extinction and $k$-correction. Considering that the early rising phase of SNe Ic-BL approximately follows ${\propto}t^2$, the rising timescale can be roughly treated as the peak timescale. 

Because lGRBs are usually discovered at cosmological distances, mainly at redshifts ranging from a few 0.1 to several (compared with our Non-GRB-SN sample, which are mostly at redshifts of a few 0.01), only the intrinsically very bright late-time afterglows can be detected, and the observed afterglows are generally very bright near the peak timescales of SNe Ic-BL. Figure \ref{fig:AfterglowLightcurve} shows that a large fraction ($\sim85\%$) of observed afterglows exceed the brightness of both GRB-SNe and Non-GRB-SNe around the peak timescales of $\sim5-20\,{\rm d}$. In fact, for most GRB-SNe, the peak brightness falls above the lower limit of the $1.5\sigma$ ($\sim90\%$) credible region of the afterglow lightcurves. As a result, the SN emission often does not clearly emerge from the afterglow lightcurve around the SN peak and at late times, making it difficult to detect. Thus, only GRB-SNe that are intrinsically bright enough to be detectable above their relatively faint afterglows can be observed. This combination of large distances and bright afterglows therefore biases the observed sample toward intrinsically brighter GRB-SNe.

\subsection{Magnetar Engine for lGRBs in GRB-SNe}

Within a magnetar central engine framework, our analysis in Section \ref{sec:StatisticsSNIcBL} suggests that there is no significant difference between GRB-SNe and Non-GRB-SNe within our sample. This indicates that magnetar engines play an important role in the formation of lGRB jets in SNe Ic-BL, with the jet energy plausibly powered by magnetar spin-down processes. 

Whether a relativistic lGRB jet can be successfully launched and accelerated may depend on the generalized magnetization parameter $\sigma_0$, which quantifies the baryon loading and energy composition of the outflow. As studied by \cite{Metzger2011}, shortly after the SN explosion, the proto-magnetar is sufficiently hot to drive a baryon-loaded neutrino-driven wind, resulting in a heavily mass-loaded outflow that is too baryon-rich to be accelerated to relativistic velocities. As the proto-magnetar cools and the neutrino luminosity declines, the baryon loading is progressively reduced, and $\sigma_0$ of the outflows is increased. In this phase, the outflow can transition to a high-$\sigma$ regime with an average magnetization of $\sigma_{\rm ave}\sim10^2-10^3$ and become Poynting-flux dominated, enabling the launch and acceleration of a relativistic jet. At large scales, the jet can be collimated via its interaction with the stellar envelope and may ultimately break out from the progenitor star \citep[e.g.,][]{Komissarov2007,Bromberg2014}. Subsequent magnetic dissipation processes, such as reconnection, can convert magnetic energy into kinetic energy and non-thermal particles at a characteristic dissipation radius, ultimately powering the prompt emission of lGRBs.

Furthermore, the spin-down power of a magnetar successfully explains energy injection in the afterglow phase, including the X-ray shallow decay phase and the plateau \citep[e.g.,][]{Dai1998,Zhang2001,Zhang2006,Troja2007,Lv2014} that follows in many lGRBs. Thus, magnetars are plausibly essential for powering lGRB jets, their afterglow phases, as well as the emission and dynamics of associated SNe Ic-BL.

\subsection{Origin of Post-peak Lightcurve Bumps}

Detailed systematic studies have revealed that post-peak bumps commonly exist in SLSNe Ic, with $\sim30\%$ to $80\%$ displaying these features \citep[e.g.,][]{Hosseinzadeh2022,Chen2023}. However, in our sample, all SNe Ic-BL lightcurves appear smooth after the peak. In the literature, only two SNe Ic-BL, i.e., SN\,2022xff \citep{Kuncarayakti2023} and SN\,2024xzc \citep{ODwyer2025}, have been reported to identify significant late-time bumps, suggesting that such features are rarely observed in SNe Ic-BL. If SLSNe Ic and SNe Ic-BL share similar central engine models, progenitor properties, and formation channels, one might expect their lightcurves to show comparable origin-related variability. The apparent lack of post-peak bumps in most SNe Ic-BL therefore provides an opportunity to further explore the physical origin of these features.

As one possible explanation, the bumps observed in SLSNe Ic may originate from extrinsic factors, specifically the interaction between the SN ejecta and the circumstellar medium (CSM) expelled before the SN explosion \citep{Chevalier2011,Chatzopoulos2012,Chatzopoulos2013}, a scenario that is commonly invoked in the literature \citep[e.g.,][]{WangSQ2016,Nicholl2016,Liu2018,Gomez2021,West2023,Lin2023,Kumar2026}. Indeed, the bump observed in SN\,2022xxf, accompanied by contemporaneous narrow emission lines, has been interpreted as potentially arising from ejecta-CSM interaction \citep{Kuncarayakti2023}. Given the broadly similar progenitor properties and formation channels inferred for SLSNe Ic and SNe Ic-BL, one might expect that if such bumps in SLSNe Ic are predominantly interaction-powered, a comparable fraction (i.e., $\sim30\%-80\%$) of SNe Ic-BL would also show significant bumps. Nevertheless, current observations do not appear to support such a high rate among SNe Ic-BL. 

Alternatively, the bumps may originate from intrinsic mechanisms related to the magnetar central engine, such as the interaction between the magnetar wind and the companion star \citep[i.e., the so-called ``magnetar-star binary engine'';][]{Zhu2024}, intermittent episodes of enhanced energy release from the magnetar \citep{YuLi2017,Dong2023}, precession-induced modulation of the magnetar \citep{Zhang2025,Farah2026}, or variations in the thermalization efficiency of the injected energy \citep{Vurm2021,Moriya2022}. Since the magnetar always dominates the SLSN Ic emission, variations in its energy injection after the SN peak can be directly reflected in the lightcurve. Conversely, because the magnetar only partially powers the peak emission of SNe Ic-BL, variations in the energy injection are likely smoothed out by diffusion and therefore produce only limited effects on the lightcurve. At later times, when the emission is dominated by the $^{56}$Ni decay, the impact of magnetar variability is further reduced. Therefore, if the bumps are primarily driven by intrinsic engine variability, this framework may potentially explain why a significant fraction of SLSNe Ic show bumps, whereas such features are rarely observed in SNe Ic-BL. In this case, variations in the magnetar central engine in SNe Ic-BL would need to be identified through other observational signatures. For example, the magnetar-star binary engine proposed by \citet{Zhu2024} predicts that the magnetar wind nebula can evaporate the companion main-sequence star and accelerate the evaporated stellar material, which finally collides with the inner layer of the ejecta. This process may produce late-time broad-lined H-rich emission features, as observed in the late-time spectrum of EP250108a/SN\,2025kg \citep{Zhu2025}.

\section{Summary and Conclusion} \label{sec:Conclusions}

In this paper, we developed a magnetar central engine model including $^{56}$Ni decay, in which $^{56}$Ni decay partially contributes to the peak emission and powers the late-time tail, while magnetar injection accounts for the excess luminosity beyond that expected from $^{56}$Ni decay alone. Using this model, we fit the multi-band lightcurves of 80 SNe Ic-BL from the literature, including 11 GRB-SNe and 69 Non-GRB-SNe. In addition, we compare them with ordinary SESNe and other possible magnetar-powered SESNe in terms of the properties of magnetars, explosions, $^{56}$Ni production, and lightcurves, and combine these results with binary simulations to propose a unified framework for their physical classification and progenitors. 

Our fittings reveal that the magnetar engine model with $^{56}$Ni decay can successfully reproduce the lightcurves of all SNe Ic-BL, performing significantly better than the pure $^{56}$Ni-powered model. The inferred medians with $1\sigma$ credible regions of the key parameters are the initial spin period $P_{\rm{i}}\sim2.03^{+1.81}_{-0.95}\,{\rm{ms}}$, magnetic field strength $B_{\rm{p}}\sim3.97^{+3.07}_{-1.42}\times10^{15}\,{\rm{G}}$, ejecta mass $M_{\rm{ej}}\sim2.30^{+1.46}_{-1.01}\,M_\odot$, and $^{56}$Ni mass $M_{\rm{Ni}}\sim0.18^{+0.14}_{-0.09}\,M_\odot$. We find a very strong and statistically significant anti-correlation between $P_{\rm i}$ and $M_{\rm ej}$ following $P_{\rm i}\propto{M}_{\rm ej}^{-0.95\pm0.07}$, while $M_{\rm Ni}$ is strongly correlated with $M_{\rm ej}$, corresponding to a $^{56}$Ni production efficiency of $f_{\rm Ni}\sim0.08^{+0.06}_{-0.03}$. Based on our fittings, the lightcurve properties of our SN Ic-BL sample are the rising timescale $t_{\rm rise}\sim9.63^{+3.30}_{-2.11}\,{\rm d}$, declining timescale $t_{\rm dec}\sim18.51^{+5.58}_{-3.54}\,{\rm d}$, and peak absolute magnitude $M_{\rm peak}\sim-18.75^{+0.80}_{-0.66}\,{\rm mag}$ (corresponding to peak luminosity $L_{\rm peak}\sim9.81^{+7.06}_{-5.45}\times10^{42}\,{\rm erg}\,{\rm s}^{-1}$), with a strong correlation between $t_{\rm rise}$ and $t_{\rm dec}$ given by $t_{\rm dec} = (1.91\pm0.03)t_{\rm rise}$. The inferred peak luminosity of the GRB-SN sample is systematically higher than that of the Non-GRB-SN sample, typically by a factor of $\sim1.6$. This difference is also reflected in the inferred magnetar properties, with GRB-SNe generally hosting magnetars with slightly lower magnetic fields and higher $^{56}$Ni production efficiencies. However, this difference is unlikely to arise from intrinsic physical distinctions between the two populations. Indeed, using the energy test based on the inferred fitting parameters, we find no statistically significant difference between the GRB-SN and Non-GRB-SN samples. Instead, it may be caused by observational biases, since GRB-SNe are typically discovered at larger distances in follow-up observations of lGRBs, while the bright afterglows may outshine the associated SN emission, making intrinsically fainter GRB-SNe difficult to detect. Furthermore, our statistical analysis of GRB-SNe using the magnetar model suggests that the production of lGRBs associated with many GRB-SNe is plausibly powered by a central magnetar, with the energy extracted from its spin-down. The GRB jet is possibly produced via a relativistic jet launched by magnetically driven winds, whose magnetization increases with time as baryon loading decreases \citep{Metzger2011,Zhang2018}. These results may help to provide a renewed perspective on how magnetar-driven lGRBs are produced.

We find no significant difference between SNe Ic-BL and ordinary SNe Ic regarding $^{56}$Ni mass, ejecta mass, and $^{56}$Ni production efficiency, with comparable distributions in the two samples. This suggests that their pre-SN progenitors may be very similar, with the main difference possibly arising from whether a millisecond magnetar can be formed upon core collapse to significantly affect the SN explosion. By comparing SNe Ic-BL with other possible magnetar-powered SESNe, including SLSNe Ic and FBOTs, in terms of their magnetar parameters and explosion properties, we explore a potential intrinsic connection among these transients. A very strong universal anti-correlation $P_{\rm i}\propto M_{\rm ej}^{-0.49\pm0.03}$ is confirmed, suggesting that these different types of transients may share a common origin. However, compared with FBOTs, which typically have small ejecta masses ($M_{\rm ej}\lesssim0.5\,M_\odot$) and magnetars with $P_{\rm i}\sim10\,{\rm ms}$, SNe Ic-BL and SLSNe Ic generally have larger ejecta masses ($M_{\rm ej}\gtrsim0.5\,M_\odot$) and faster-spinning magnetars with $P_{\rm i}$ mainly of a few milliseconds. Despite their similar ejecta mass distributions, the SN Ic-BL magnetars are expected to rotate faster than the SLSN Ic magnetars, with median $P_{\rm i}$ of $\sim2\,{\rm ms}$ and $\sim3\,{\rm ms}$, respectively. Their $B_{\rm p}$ distributions appear to be more clearly separated. FBOTs have $B_{\rm p}$ distributions in the range of $\sim10^{14}-10^{16}\,{\rm G}$ with a median value of $\sim9\times10^{14}\,{\rm G}$, whereas the $B_{\rm p}$ distributions of SNe Ic-BL and SLSNe Ic are roughly divided around $\sim10^{15}\,{\rm G}$, spanning $\sim4\times10^{13}-10^{15}\,{\rm G}$ and $10^{15}-10^{16}\,{\rm G}$, respectively. Moreover, the parameters of lGRB magnetars can be constrained from the X-ray plateaus, with the inferred $P_{\rm i}-B_{\rm p}$ distribution largely overlapping that of SN Ic-BL magnetars. These results suggest that, although these transients may share a common origin, differences in magnetar parameters and ejecta masses likely lead to their diverse observational properties. Indeed, their $t_{\rm rise}-M_{\rm peak}$ distributions form three distinct yet contiguous regions in parameter space.

According to the dominant explosion energy source, we classify SESNe into ``{\em magnetar-powered SESNe}'', which include SNe Ic-BL, SLSNe Ic, and FBOTs, and ``{\em neutrino-driven SESNe}'', which mainly correspond to ordinary SESNe. For SNe Ic-BL, SLSNe Ic, and FBOTs, the inferred initial magnetar rotational energies are usually larger than the neutrino-driven explosion energy and can therefore dominate the SN explosion. However, their roles differ among these subclasses. In SLSNe Ic and FBOTs, the magnetar spin-down timescale is comparable to the photon diffusion timescale, so the magnetar dominates the SN emission. We therefore refer to these events as ``{\em radiatively magnetar-powered SESNe}.'' In contrast, the spin-down timescale of the magnetar in SNe Ic-BL is much shorter than the diffusion timescale. As a result, most of the rotational energy is injected into the ejecta and mainly contributes to the ejecta kinetic energy, while only a small fraction contributes to the peak luminosity. We therefore classify SNe Ic-BL as ``{\em mechanically magnetar-powered SESNe}.'' For ordinary neutrino-driven SESNe, the rotational energy of the magnetar or pulsar is typically much smaller than the neutrino-driven explosion energy. Consequently, it does not significantly affect the ejecta dynamics but may only contribute partially to the SN radiation. In this sense, ordinary SESNe are also described as ``{\em magnetar/pulsar-aided SESNe}''. Both magnetar-powered and ordinary SESNe may originate from binary systems formed through isolated binary evolution. For magnetar-powered SESNe, the helium star progenitors can form close binaries with orbital periods shorter than $\sim1-2\,{\rm d}$ after CEE or stable FLOF. In such systems, the helium stars can be efficiently spun up through tidal interactions with their companions, finally producing rapidly rotating magnetars that significantly affect the SN explosion. More specifically, SNe Ic-BL and SLSNe Ic likely originate from helium stars with masses of $\gtrsim4-40\,M_\odot$ in close binaries, with the progenitors of SNe Ic-BL possibly having  tighter orbits. FBOTs may originate from low-mass helium stars with $M_{\rm He,i}\sim2-4\,M_\odot$ in close binaries, which may correspond to magnetar-powered USSNe. The observational diversity among magnetar-powered SESNe may then arise from differences in the progenitor masses and orbital configurations of these binaries. For ordinary SESNe, the progenitor binaries likely have relatively wider orbits, so that tidal interactions are inefficient in spinning up the helium stars. In this case, the rotation of the resulting magnetar or pulsar is more likely determined by the intrinsic rotation of the helium star after CEE or stable RLOF. We therefore link the observed properties of SESNe to their progenitor systems and propose a possible unified picture connecting magnetar-powered and ordinary SESNe.

\begin{acknowledgments}

We thank {an anonymous referee for constructive suggestions,} Gokul Srinivasaragavan, Sheng Yang, Brad Cenko, and Jesper Sollerman for sharing the data on a subset of the SNe Ic-BL sample analyzed in this paper. We thank Jim Fuller, Ryosuke Hirai, Rui-Chong Hu, Yacheng Kang, Liang-Duan Liu, Wenbin Lu, Hou-Jun L$\ddot{\rm{u}}$, Ilya Mandel, {Noam Soker, J. Craig Wheeler,} Sheng Yang, Yi-Han Iris Yin, and Yun-Wei Yu for their helpful comments. We acknowledge the Research Talent Hub for ITF project (RTH-ITF) (Project No. PiH/270/25GS) from the Innovation and Technology Commission of Hong Kong SAR. The ZTF forced-photometry service was funded under the Heising-Simons Foundation grant \#12540303 (PI: Graham).
\end{acknowledgments}

\begin{contribution}

Jin-Ping Zhu initiated the project, collected the data, and performed the fitting analysis. Jin-Ping Zhu and Bing Zhang discussed the results and completed the manuscript.

\end{contribution}

\software{\texttt{emcee} \citep{ForemanMackey2013}; ZTF forced-photometry service \citep{Masci2019}; \texttt{grblc} \citep{Dainotti2024}}

\clearpage
\appendix
\restartappendixnumbering

\section{Corner Plots of Posteriors}

\begin{figure*}
    \centering
    \includegraphics[width = 0.95\linewidth]{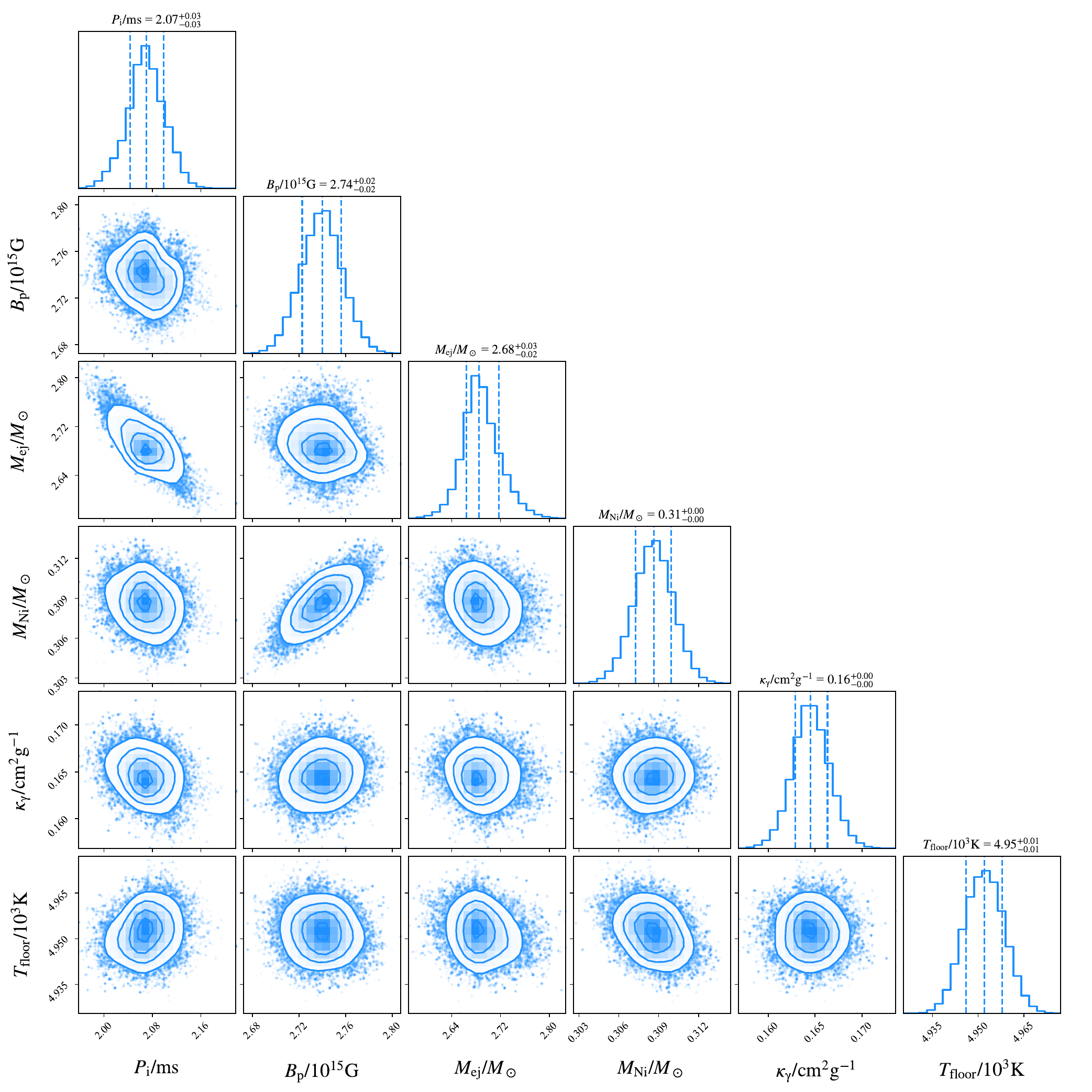}
    \caption{Example posteriors for magnetar-powered model fit to SN\,1998bw. Medians and $1\sigma$ credible regions are labeled. }
    \label{fig:Corner}
\end{figure*}

As an example, we show in Figure \ref{fig:Corner} the posterior distributions of the magnetar-powered model parameters for SN\,1998bw, with the corresponding magnetar-powered lightcurve fit presented in Figure \ref{fig:Comparsion}, to illustrate the convergence of our lightcurve fitting. The posteriors are well behaved and unimodal, indicating good convergence of the Markov Chain Monte Carlo sampling.

\section{Dependence of Multi-band Lightcurves on Model Assumptions} \label{app_sec:Dependence}

\subsection{Density Profile} \label{app_sec:Dependence_Density}

{
\begin{deluxetable*}{l|c|cccccccc}
\tablewidth{0pt}
\tabletypesize{\footnotesize}
\tablecaption{{Example Fitting Results of Derived Parameters with Different Density Models} \label{tab:Results_Density_Profile}}
\tablehead{
\colhead{SN Name} & \colhead{Density Model} & \colhead{$P_{\rm i}$} & \colhead{$B_{\rm p}$} & \colhead{$M_{\rm ej}$} & \colhead{$M_{\rm Ni}$} & \colhead{$\kappa_\gamma$} & \colhead{$T_{\rm floor}$} & \colhead{$t_{\rm first}$} & \colhead{$\Delta{\rm AIC}_{\rm fid}$} \\
\colhead{} & \colhead{} & \colhead{(ms)} & \colhead{($10^{15}\,{\rm G}$)} & \colhead{($M_\odot$)} & \colhead{($M_\odot$)} & \colhead{(${\rm cm}^2\,{\rm g}^{-1}$)} & \colhead{($10^3\,{\rm K}$)} &  \colhead{(d)} & \colhead{} 
}
\startdata
\multirow{4}{*}{SN\,1998bw} & $n=10$ (Fiducial) & $2.07^{+0.03}_{-0.03}$ & $2.74^{+0.02}_{-0.02}$ & $2.68^{+0.03}_{-0.02}$ & $0.31^{+0.00}_{-0.00}$ & $0.16^{+0.00}_{-0.00}$ & $4.95^{+0.01}_{-0.01}$ & ... & $0$ \\ 
& $n=8$ & $1.97^{+0.02}_{-0.02}$ & $2.78^{+0.01}_{-0.02}$ & $2.80^{+0.02}_{-0.02}$ & $0.31^{+0.00}_{-0.00}$ & $0.17^{+0.00}_{-0.00}$ & $4.95^{+0.01}_{-0.01}$ & ... & $150$ \\ 
& $n=6$ & $1.53^{+0.02}_{-0.02}$ & $2.95^{+0.01}_{-0.02}$ & $3.28^{+0.05}_{-0.02}$ & $0.32^{+0.00}_{-0.00}$ & $0.18^{+0.00}_{-0.00}$ & $4.95^{+0.01}_{-0.01}$ & ... & $534$ \\ 
& Uniform & $1.56^{+0.02}_{-0.03}$ & $2.75^{+0.02}_{-0.02}$ & $3.21^{+0.05}_{-0.03}$ & $0.31^{+0.00}_{-0.00}$ & $0.19^{+0.00}_{-0.00}$ & $4.95^{+0.01}_{-0.01}$ & ... & $-199$ \\ 
\hline
\multirow{4}{*}{SN\,2002ap} & $n=10$ (Fiducial) & $2.37^{+0.03}_{-0.01}$ & $9.43^{+0.07}_{-0.08}$ & $1.59^{+0.01}_{-0.02}$ & $0.10^{+0.00}_{-0.00}$ & $0.33^{+0.01}_{-0.00}$ & $4.49^{+0.01}_{-0.01}$ & $1.05^{+0.02}_{-0.01}$ & $0$ \\ 
& $n=8$ & $2.18^{+0.02}_{-0.02}$ & $10.03^{+0.13}_{-0.20}$ & $1.60^{+0.03}_{-0.02}$ & $0.10^{+0.00}_{-0.00}$ & $0.36^{+0.01}_{-0.01}$ & $4.50^{+0.01}_{-0.01}$ & $0.86^{+0.01}_{-0.01}$ & $33$ \\ 
& $n=6$ & $1.72^{+0.02}_{-0.01}$ & $10.09^{+0.21}_{-0.14}$ & $1.90^{+0.02}_{-0.03}$ & $0.10^{+0.00}_{-0.00}$ & $0.40^{+0.01}_{-0.01}$ & $4.51^{+0.01}_{-0.01}$ & $0.55^{+0.01}_{-0.01}$ & $391$ \\ 
& Uniform & $2.15^{+0.01}_{-0.02}$ & $13.01^{+0.10}_{-0.27}$ & $1.38^{+0.03}_{-0.01}$ & $0.13^{+0.00}_{-0.00}$ & $0.40^{+0.00}_{-0.01}$ & $4.47^{+0.01}_{-0.01}$ & $2.08^{+0.02}_{-0.01}$ & $162$\\ 
\hline
\multirow{4}{*}{SN\,2003dh} & $n=10$ (Fiducial) & $2.41^{+0.60}_{-0.53}$ & $2.51^{+0.52}_{-0.41}$ & $3.40^{+1.22}_{-0.92}$ & $0.17^{+0.08}_{-0.05}$ & $0.17^{+0.28}_{-0.13}$ & $5.61^{+0.30}_{-0.52}$ & ...  & $0$ \\ 
& $n=8$ & $2.34^{+0.56}_{-0.57}$ & $2.51^{+0.47}_{-0.42}$ & $3.40^{+1.31}_{-0.73}$ & $0.17^{+0.07}_{-0.05}$ & $0.21^{+0.30}_{-0.15}$ & $5.59^{+0.31}_{-0.50}$ & ... & $0$ \\ 
& $n=6$ & $2.41^{+0.41}_{-0.49}$ & $2.65^{+0.74}_{-0.31}$ & $3.40^{+1.05}_{-0.78}$ & $0.18^{+0.08}_{-0.05}$ & $0.24^{+0.31}_{-0.16}$ & $5.60^{+0.28}_{-0.43}$ & ... & $-1$ \\ 
& Uniform & $1.70^{+0.69}_{-0.56}$ & $3.01^{+1.28}_{-0.51}$ & $3.40^{+1.89}_{-0.86}$ & $0.21^{+0.11}_{-0.05}$ & $0.15^{+0.23}_{-0.09}$ & $5.74^{+0.30}_{-0.26}$ & ... & $0$ \\ 
\hline
\multirow{4}{*}{SN\,2003jd} & $n=10$ (Fiducial) & $0.74^{+0.03}_{-0.02}$ & $3.75^{+0.06}_{-0.08}$ & $4.36^{+0.07}_{-0.08}$ & $0.26^{+0.00}_{-0.00}$ & $0.25^{+0.01}_{-0.01}$ & $4.48^{+0.02}_{-0.02}$ & $5.29^{+0.31}_{-0.13}$ & $0$ \\ 
& $n=8$ & $0.68^{+0.03}_{-0.02}$ & $3.88^{+0.08}_{-0.07}$ & $4.49^{+0.07}_{-0.08}$ & $0.26^{+0.00}_{-0.00}$ & $0.27^{+0.01}_{-0.01}$ & $4.48^{+0.02}_{-0.02}$ & $4.98^{+0.22}_{-0.19}$ & $-2$ \\ 
& $n=6$ & $0.51^{+0.01}_{-0.02}$ & $4.29^{+0.07}_{-0.06}$ & $5.06^{+0.10}_{-0.07}$ & $0.28^{+0.00}_{-0.00}$ & $0.31^{+0.02}_{-0.01}$ & $4.50^{+0.02}_{-0.02}$ & $4.14^{+0.11}_{-0.24}$ & $-17$ \\ 
& Uniform & $1.05^{+0.02}_{-0.04}$ & $2.91^{+0.07}_{-0.04}$ & $4.19^{+0.10}_{-0.08}$ & $0.25^{+0.00}_{-0.00}$ & $0.16^{+0.01}_{-0.01}$ & $4.42^{+0.02}_{-0.02}$ & $9.23^{+0.13}_{-0.43}$ & $-1$ \\ 
\hline
\multirow{4}{*}{SN\,2003lw} & $n=10$ (Fiducial) &  $3.45^{+0.07}_{-0.02}$ & $1.48^{+0.01}_{-0.01}$ & $1.93^{+0.03}_{-0.01}$ & $0.49^{+0.00}_{-0.00}$ & $0.64^{+0.07}_{-0.09}$ & $5.23^{+0.03}_{-0.01}$ & ... & $0$ \\ 
& $n=8$ & $3.42^{+0.10}_{-0.03}$ & $1.46^{+0.02}_{-0.02}$ & $1.92^{+0.04}_{-0.01}$ & $0.49^{+0.00}_{-0.01}$ & $0.59^{+0.15}_{-0.05}$ & $5.29^{+0.03}_{-0.01}$ & ... & $98$ \\ 
& $n=6$ & $2.82^{+0.10}_{-0.09}$ & $1.50^{+0.02}_{-0.03}$ & $2.18^{+0.05}_{-0.07}$ & $0.50^{+0.01}_{-0.01}$ & $0.70^{+0.16}_{-0.22}$ & $5.25^{+0.03}_{-0.04}$ & ... & $129$ \\ 
& Uniform & $2.22^{+0.08}_{-0.15}$ & $1.67^{+0.05}_{-0.03}$ & $2.55^{+0.17}_{-0.07}$ & $0.50^{+0.00}_{-0.00}$ & $0.57^{+0.23}_{-0.11}$ & $5.40^{+0.04}_{-0.03}$ & ... & $643$ \\ 
\hline
\multirow{4}{*}{SN\,2004aw} & $n=10$ (Fiducial) & $2.02^{+0.06}_{-0.06}$ & $3.98^{+0.09}_{-0.10}$ & $3.08^{+0.10}_{-0.06}$ & $0.23^{+0.00}_{-0.00}$ & $0.28^{+0.04}_{-0.03}$ & $4.41^{+0.02}_{-0.02}$ & $8.19^{+0.27}_{-0.27}$ & $0$ \\ 
& $n=8$ & $1.88^{+0.09}_{-0.05}$ & $4.12^{+0.09}_{-0.12}$ & $3.14^{+0.08}_{-0.08}$ & $0.23^{+0.00}_{-0.00}$ & $0.30^{+0.03}_{-0.03}$ & $4.42^{+0.02}_{-0.02}$ & $7.89^{+0.41}_{-0.18}$ & $33$ \\ 
& $n=6$ & $1.43^{+0.05}_{-0.04}$ & $4.57^{+0.13}_{-0.10}$ & $3.51^{+0.09}_{-0.10}$ & $0.24^{+0.00}_{-0.00}$ & $0.37^{+0.05}_{-0.04}$ & $4.45^{+0.02}_{-0.02}$ & $7.41^{+0.37}_{-0.20}$ & $142$ \\ 
& Uniform & $1.60^{+0.09}_{-0.05}$ & $3.73^{+0.10}_{-0.16}$ & $3.90^{+0.15}_{-0.11}$ & $0.23^{+0.00}_{-0.00}$ & $0.26^{+0.04}_{-0.03}$ & $4.40^{+0.02}_{-0.02}$ & $9.61^{+0.68}_{-0.16}$ & $-70$ \\ 
\hline
\multirow{4}{*}{SN\,2007ru} & $n=10$ (Fiducial) &  $0.84^{+0.01}_{-0.01}$ & $2.77^{+0.01}_{-0.01}$ & $4.96^{+0.04}_{-0.04}$ & $0.19^{+0.00}_{-0.00}$ & $0.23^{+0.01}_{-0.01}$ & $4.50^{+0.01}_{-0.01}$ & $9.01^{+0.02}_{-0.01}$ & $0$ \\  
& $n=8$ & $0.78^{+0.01}_{-0.01}$ & $2.81^{+0.01}_{-0.01}$ & $5.17^{+0.04}_{-0.05}$ & $0.19^{+0.00}_{-0.00}$ & $0.22^{+0.01}_{-0.01}$ & $4.50^{+0.01}_{-0.01}$ & $9.01^{+0.02}_{-0.01}$ & $128$ \\ 
& $n=6$ & $0.60^{+0.01}_{-0.01}$ & $2.93^{+0.02}_{-0.01}$ & $6.16^{+0.04}_{-0.07}$ & $0.20^{+0.00}_{-0.00}$ & $0.24^{+0.01}_{-0.01}$ & $4.51^{+0.01}_{-0.01}$ & $9.01^{+0.02}_{-0.01}$ & $459$ \\ 
& Uniform & $0.67^{+0.01}_{-0.01}$ & $2.85^{+0.01}_{-0.01}$ & $5.76^{+0.06}_{-0.06}$ & $0.20^{+0.00}_{-0.00}$ & $0.24^{+0.01}_{-0.01}$ & $4.50^{+0.01}_{-0.01}$ & $9.01^{+0.02}_{-0.01}$ & $-555$ \\ 
\hline
\multirow{4}{*}{SN\,2009bb} & $n=10$ (Fiducial) &   $1.13^{+0.02}_{-0.02}$ & $5.19^{+0.08}_{-0.08}$ & $2.42^{+0.06}_{-0.04}$ & $0.19^{+0.00}_{-0.00}$ & $0.17^{+0.00}_{-0.00}$ & $4.85^{+0.03}_{-0.03}$ & $1.24^{+0.05}_{-0.03}$ & $0$ \\ 
& $n=8$ & $1.05^{+0.02}_{-0.02}$ & $5.40^{+0.07}_{-0.09}$ & $2.48^{+0.07}_{-0.04}$ & $0.20^{+0.00}_{-0.00}$ & $0.18^{+0.01}_{-0.00}$ & $4.84^{+0.03}_{-0.03}$ & $1.05^{+0.03}_{-0.04}$ & $6$ \\ 
& $n=6$ & $0.81^{+0.02}_{-0.02}$ & $6.01^{+0.14}_{-0.11}$ & $2.75^{+0.05}_{-0.07}$ & $0.21^{+0.00}_{-0.00}$ & $0.21^{+0.01}_{-0.01}$ & $4.83^{+0.03}_{-0.03}$ & $0.71^{+0.03}_{-0.02}$ & $11$\\ 
& Uniform & $1.08^{+0.03}_{-0.03}$ & $4.96^{+0.09}_{-0.08}$ & $2.57^{+0.07}_{-0.08}$ & $0.20^{+0.01}_{-0.00}$ & $0.15^{+0.00}_{-0.00}$ & $4.84^{+0.04}_{-0.03}$ & $2.21^{+0.06}_{-0.07}$ & $2$ \\ 
\hline
\multirow{4}{*}{SN\,2010bh} & $n=10$ (Fiducial) &   $1.33^{+0.01}_{-0.02}$ & $8.07^{+0.05}_{-0.07}$ & $2.51^{+0.04}_{-0.03}$ & $0.11^{+0.00}_{-0.00}$ & $0.34^{+0.02}_{-0.03}$ & $3.75^{+0.01}_{-0.02}$ & ... & $0$ \\ 
& $n=8$ & $1.29^{+0.01}_{-0.02}$ & $8.31^{+0.06}_{-0.07}$ & $2.51^{+0.04}_{-0.03}$ & $0.11^{+0.00}_{-0.00}$ & $0.36^{+0.03}_{-0.03}$ & $3.75^{+0.01}_{-0.02}$ & ... & $-53$\\ 
& $n=6$ & $1.05^{+0.01}_{-0.02}$ & $9.61^{+0.13}_{-0.14}$ & $2.58^{+0.05}_{-0.04}$ & $0.12^{+0.00}_{-0.00}$ & $0.42^{+0.03}_{-0.03}$ & $3.71^{+0.02}_{-0.01}$ & ... & $-15$ \\ 
& Uniform & $1.01^{+0.01}_{-0.02}$ & $11.90^{+0.27}_{-0.20}$ & $2.23^{+0.05}_{-0.04}$ & $0.15^{+0.00}_{-0.00}$ & $0.12^{+0.00}_{-0.00}$ & $4.05^{+0.02}_{-0.02}$ & ... & $716$ \\ 
\hline
\multirow{4}{*}{PTF\,10bzf} & $n=10$ (Fiducial) & $3.03^{+0.80}_{-0.34}$ & $3.49^{+0.21}_{-0.32}$ & $2.23^{+0.16}_{-0.21}$ & $0.17^{+0.01}_{-0.01}$ & $0.13^{+0.03}_{-0.02}$ & $6.29^{+0.28}_{-0.28}$ & $5.30^{+0.77}_{-0.40}$ & $0$ \\ 
& $n=8$ & $2.76^{+0.76}_{-0.28}$ & $3.63^{+0.23}_{-0.33}$ & $2.31^{+0.14}_{-0.23}$ & $0.17^{+0.01}_{-0.01}$ & $0.14^{+0.03}_{-0.02}$ & $6.29^{+0.29}_{-0.32}$ & $4.84^{+0.76}_{-0.37}$ & $1$ \\ 
& $n=6$ & $2.32^{+0.57}_{-0.29}$ & $3.83^{+0.28}_{-0.30}$ & $2.50^{+0.20}_{-0.24}$ & $0.17^{+0.01}_{-0.01}$ & $0.16^{+0.03}_{-0.03}$ & $6.29^{+0.33}_{-0.29}$ & $4.25^{+0.66}_{-0.37}$ & $2$ \\ 
& Uniform & $1.90^{+0.36}_{-0.19}$ & $3.61^{+0.22}_{-0.28}$ & $2.98^{+0.28}_{-0.20}$ & $0.18^{+0.01}_{-0.01}$ & $0.15^{+0.04}_{-0.02}$ & $6.28^{+0.23}_{-0.30}$ & $6.21^{+0.77}_{-0.36}$ & $1$ \\ 
\enddata 
\tablecomments{{We define $\Delta \mathrm{AIC}_{\rm fid}$ as the difference between the AIC values of models with different density profiles and those of the fiducial model, defined as the broken-power-law model with $n=10$. }}
\end{deluxetable*}
}

{We present in Table \ref{tab:Results_Density_Profile} the fitting results for several density models, including broken-power-law models with $n=10$ (adopted as the fiducial density model), $8$, and $6$, in which we fix $\delta = 1$, as well as a uniform density model with $v_{\rm ph} \simeq v_{\rm ej}$, using the first 10 discovered events in our sample. In the broken-power-law density model, $n=10$ is typically associated with the outer ejecta structure shaped by shock breakout in the stellar envelope, while smaller values of $n$ have been reported in numerical simulations of magnetar-driven shocks in SN ejecta \citep[e.g.,][]{Suzuki2019,Suzuki2021}. The uniform-density model is frequently employed in the literature to fit multi-band lightcurves of magnetar-powered SNe \citep[e.g.,][]{Nicholl2017,Liu2022}.  }

{We find that different density models primarily affect the inferred values of $P_{\rm i}$, $B_{\rm p}$, and $M_{\rm ej}$. For these 10 example events, we find that the fiducial model always yields the largest $P_{\rm i}$, with other models reducing it by up to $\sim30\%$. In contrast, it produces the lowest $B_{\rm p}$, while the other models increase it by a similar level of up to $\sim30\%$. Furthermore, $M_{\rm ej}$ increases by up to $\sim20\%$ as $n$ decreases by adopting the broken-power-law density profile. The $M_{\rm ej}$ values derived from the uniform-density model also differ from those of the fiducial model by $\sim 20\%$. Overall, we find that the choice of density profile has a relatively modest impact on the inferred values of $P_{\rm i}$, $B_{\rm p}$, and $M_{\rm ej}$.  }

{For the broken-power-law model with different values of $n$, we find that, except for SN\,2003jd and SN\,2010bh, the fiducial model with $n=10$ generally provides the best fit for 4 events and a comparably good description for another 4 events in the SN Ic-BL sample. Thus, among different values of $n$, the fiducial model with $n=10$ provides a better overall description for a larger fraction of the sample. In comparison, the uniform-density model yields a better fit for only 3 SNe Ic-BL, while it also performs worse for 3 events. However, it is worth noting that the assumption $v_{\rm ph} \simeq v_{\rm ej}$ in the uniform-density model is generally unable to reproduce the observed photospheric velocity evolution \citep[e.g.,][]{Nicholl2026}. Therefore, we adopt the $n=10$ model as our fiducial fitting model throughout the main text.}

\subsection{Initial Kinetic Energy} \label{app_sec:Dependence_E_kin}

{
\begin{deluxetable*}{l|c|ccccccccc}
\tablewidth{0pt}
\tabletypesize{\scriptsize}
\tablecaption{{Fitting Results of Derived Parameters for SN\,1998bw and SN\,2002ap with Different Initial Kinetic Energies} \label{tab:Results_Initial_Kinetic_Energy}}
\tablehead{
\colhead{SN Name} & \colhead{$E_{\rm kin,i}$} & \colhead{$P_{\rm i}$} & \colhead{$E_{\rm rot,i}$} & \colhead{$B_{\rm p}$} & \colhead{$M_{\rm ej}$} & \colhead{$M_{\rm Ni}$} & \colhead{$\kappa_\gamma$} & \colhead{$T_{\rm floor}$} & \colhead{$t_{\rm first}$} & \colhead{$\Delta{\rm AIC}_{\rm fid}$} \\
\colhead{} & \colhead{$(10^{51}\,{\rm erg})$} & \colhead{(ms)} & \colhead{$(10^{51}\,{\rm erg})$} & \colhead{($10^{15}\,{\rm G}$)} & \colhead{($M_\odot$)} & \colhead{($M_\odot$)} & \colhead{(${\rm cm}^2\,{\rm g}^{-1}$)} & \colhead{($10^3\,{\rm K}$)} &  \colhead{(d)} & \colhead{} 
}
\startdata
\multirow{6}{*}{SN\,1998bw} & $1.14$ (Fiducial) & $2.07^{+0.03}_{-0.03}$ & $4.67^{+0.14}_{-0.10}$ & $2.74^{+0.02}_{-0.02}$ & $2.68^{+0.03}_{-0.02}$ & $0.31^{+0.00}_{-0.00}$ & $0.16^{+0.00}_{-0.00}$ & $4.95^{+0.01}_{-0.01}$ & ... & $0$ \\ 
& $0.1$ & $1.88^{+0.03}_{-0.02}$ & $5.66^{+0.18}_{-0.12}$ & $2.73^{+0.02}_{-0.01}$ & $2.69^{+0.03}_{-0.03}$ & $0.31^{+0.00}_{-0.00}$ & $0.16^{+0.00}_{-0.00}$ & $4.95^{+0.01}_{-0.01}$ & ... & $0$ \\ 
& $0.32$ & $1.91^{+0.02}_{-0.01}$ & $5.48^{+0.12}_{-0.06}$ & $2.74^{+0.02}_{-0.02}$ & $2.69^{+0.02}_{-0.03}$ & $0.31^{+0.00}_{-0.00}$ & $0.16^{+0.00}_{-0.00}$ & $4.95^{+0.01}_{-0.01}$ & ... & $0$ \\ 
& $1$ & $2.03^{+0.02}_{-0.02}$ & $4.85^{+0.10}_{-0.10}$ & $2.74^{+0.01}_{-0.01}$ & $2.70^{+0.02}_{-0.03}$ & $0.31^{+0.00}_{-0.00}$ & $0.16^{+0.00}_{-0.00}$ & $4.95^{+0.01}_{-0.01}$ & ... & $0$ \\ 
& $3.2$ & $2.80^{+0.13}_{-0.06}$ & $2.55^{+0.26}_{-0.11}$ & $2.67^{+0.02}_{-0.02}$ & $2.66^{+0.03}_{-0.05}$ & $0.31^{+0.00}_{-0.00}$ & $0.16^{+0.00}_{-0.00}$ & $4.95^{+0.01}_{-0.01}$ & ... & $-22$ \\ 
& $10$ & $13.46^{+0.18}_{-0.16}$ & $0.22^{+0.01}_{-0.01}$ & $1.61^{+0.02}_{-0.02}$ & $2.69^{+0.02}_{-0.02}$ & $0.28^{+0.00}_{-0.00}$ & $0.28^{+0.00}_{-0.00}$ & $4.96^{+0.01}_{-0.01}$ & ... & $-376$ \\ 
\hline
\multirow{6}{*}{SN\,2002ap} & $0.72$ (Fiducial) & $2.37^{+0.03}_{-0.01}$ & $3.56^{+0.03}_{-0.06}$ & $9.43^{+0.07}_{-0.08}$ & $1.59^{+0.01}_{-0.02}$ & $0.10^{+0.00}_{-0.00}$ & $0.33^{+0.01}_{-0.00}$ & $4.49^{+0.01}_{-0.01}$ & $1.05^{+0.02}_{-0.01}$ & $0$ \\ 
& $0.1$ & $2.17^{+0.02}_{-0.02}$ & $4.25^{+0.08}_{-0.08}$ & $9.64^{+0.29}_{-0.25}$ & $1.56^{+0.04}_{-0.04}$ & $0.10^{+0.00}_{-0.00}$ & $0.34^{+0.01}_{-0.01}$ & $4.49^{+0.01}_{-0.01}$ & $1.06^{+0.02}_{-0.02}$ & $0$ \\ 
& $0.32$ & $2.25^{+0.02}_{-0.03}$ & $3.95^{+0.11}_{-0.07}$ & $9.58^{+0.39}_{-0.15}$ & $1.56^{+0.03}_{-0.04}$  & $0.10^{+0.00}_{-0.00}$ & $0.34^{+0.01}_{-0.01}$ & $4.49^{+0.01}_{-0.01}$ & $1.06^{+0.02}_{-0.02}$ & $0$  \\ 
& $1$ & $2.47^{+0.04}_{-0.03}$ & $3.28^{+0.09}_{-0.06}$ & $9.66^{+0.21}_{-0.18}$ & $1.56^{+0.03}_{-0.04}$ & $0.10^{+0.00}_{-0.00}$ & $0.34^{+0.01}_{-0.01}$ & $4.49^{+0.01}_{-0.01}$ & $1.06^{+0.02}_{-0.02}$ & $0$ \\ 
& $3.2$ & $12.85^{+0.03}_{-0.08}$ & $0.24^{+0.00}_{-0.00}$ & $6.61^{+0.04}_{-0.05}$ & $1.46^{+0.01}_{-0.01}$ & $0.09^{+0.00}_{-0.00}$ & $0.33^{+0.00}_{-0.00}$ & $4.53^{+0.01}_{-0.01}$ & $1.52^{+0.01}_{-0.01}$ & $-473$\\ 
& $10$ & $15.45^{+0.04}_{-0.27}$ & $0.17^{+0.00}_{-0.01}$ & $4.00^{+0.08}_{-0.01}$ & $2.68^{+0.01}_{-0.02}$ & $0.06^{+0.00}_{-0.00}$ & $0.60^{+0.01}_{-0.02}$ & $4.45^{+0.01}_{-0.01}$ & $1.23^{+0.01}_{-0.02}$ & $-58$ \\ 
\enddata 
\tablecomments{{We define $\Delta \mathrm{AIC}_{\rm fid}$ as the difference between the AIC of models with different initial kinetic energies and that of the fiducial model based on the inferred relationship between initial kinetic energy and ejecta mass for ordinary SESNe (i.e., Equation \ref{equ:Neutrino_Energy}). }}
\end{deluxetable*}
}

{As examples, we fit the multi-band lightcurves of SN\,1998bw and SN\,2002ap, representing a GRB-associated SN and a Non-GRB-SN, respectively, under different initial kinetic energies, with the fitting results listed in Table \ref{tab:Results_Initial_Kinetic_Energy}. We find that when $E_{\rm rot,i}\gg E_{\rm kin,i}$, the fitting results of all parameters are almost insensitive to $E_{\rm kin,i}$. In this regime, the sum of the $E_{\rm rot,i}$ and $E_{\rm kin,i}$ are nearly constant, and the goodness of fit also shows little variation with $E_{\rm kin,i}$. When $E_{\rm kin,i}\gg E_{\rm rot,i}$ and $E_{\rm kin,i}$ is significantly larger than the typical neutrino-driven explosion energy, the inferred magnetar spin period becomes $\gtrsim10\,{\rm ms}$, corresponding to a slowly rotating magnetar whose rotational energy only contributes to the lightcurve emission. Although the fit quality may improve in this regime, the fitting parameters are physically implausible as we discussed in Section \ref{sec:Introduction}.}

\section{Multi-band Lightcurve Fits of SNe Ic-BL with Magnetar-powered Model} \label{app_sec:Fits}

{In Figure \ref{fig:Fits}, we present the multi-band light curves and their magnetar-powered model fits for our SN Ic-BL sample.}

\begin{figure*}
    \centering
    \includegraphics[width = 0.32\linewidth , trim = 80 65 93 35, clip]{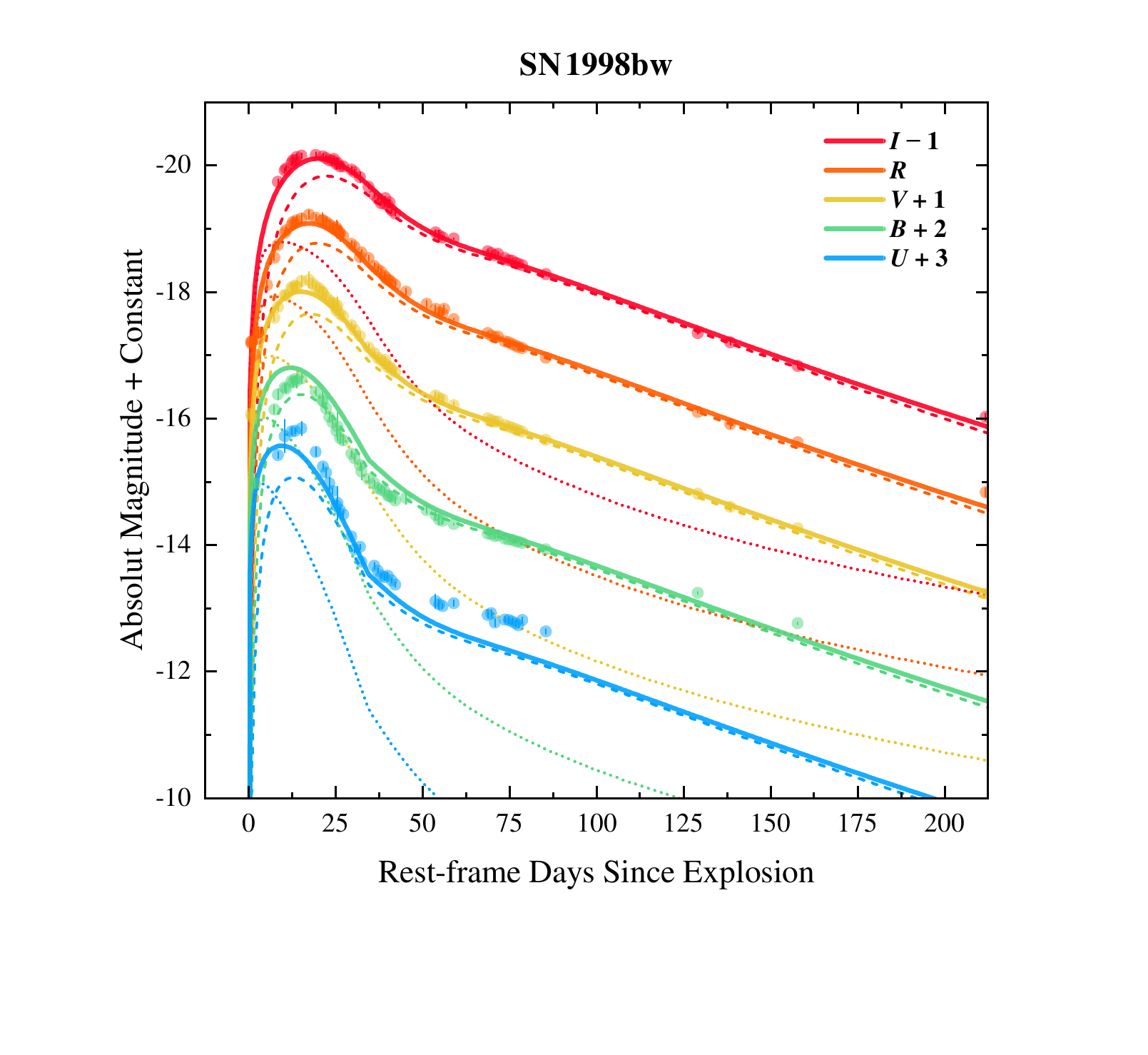}
    \includegraphics[width = 0.32\linewidth , trim = 80 65 93 35, clip]{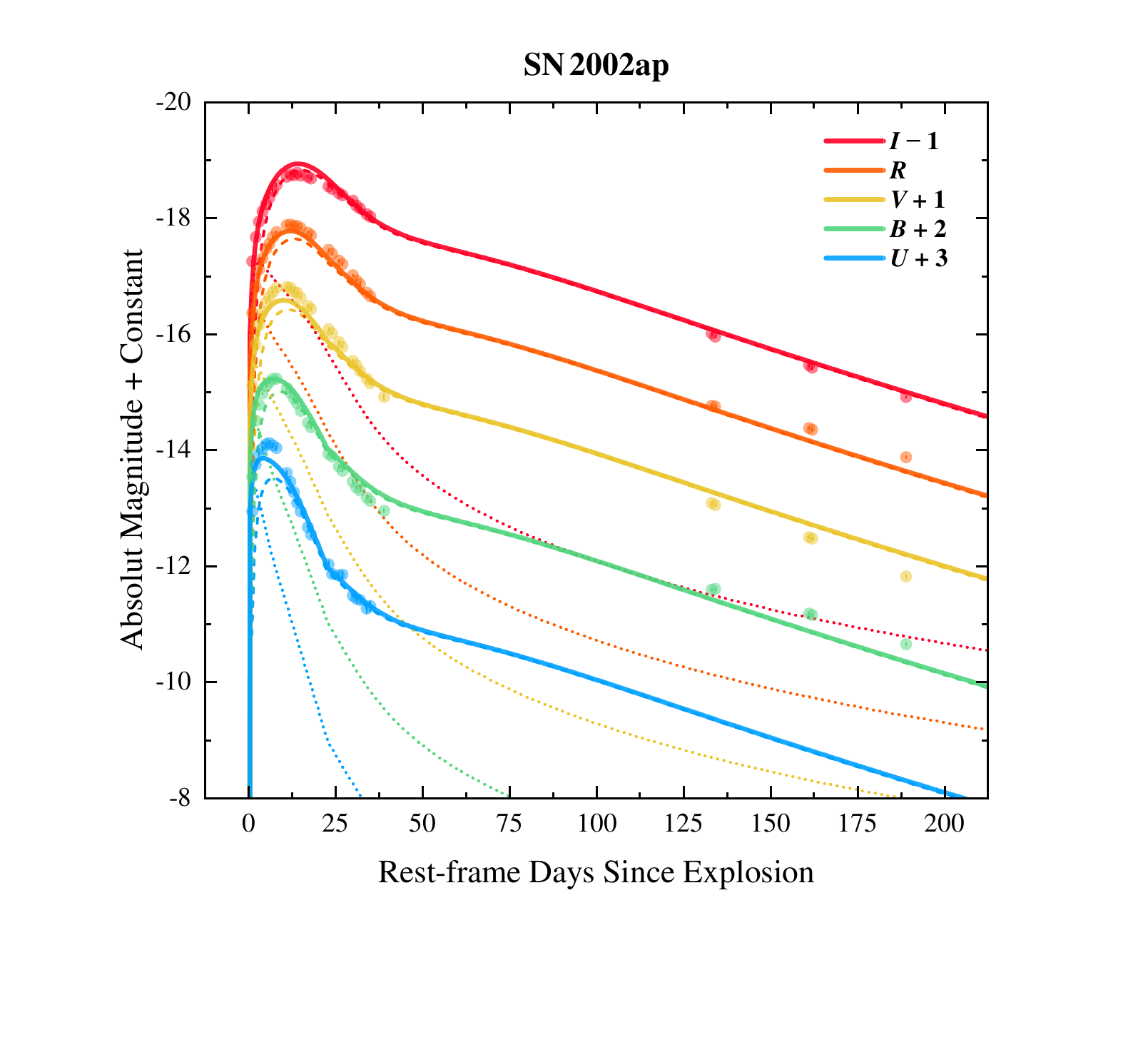}
    \includegraphics[width = 0.32\linewidth , trim = 80 65 93 35, clip]{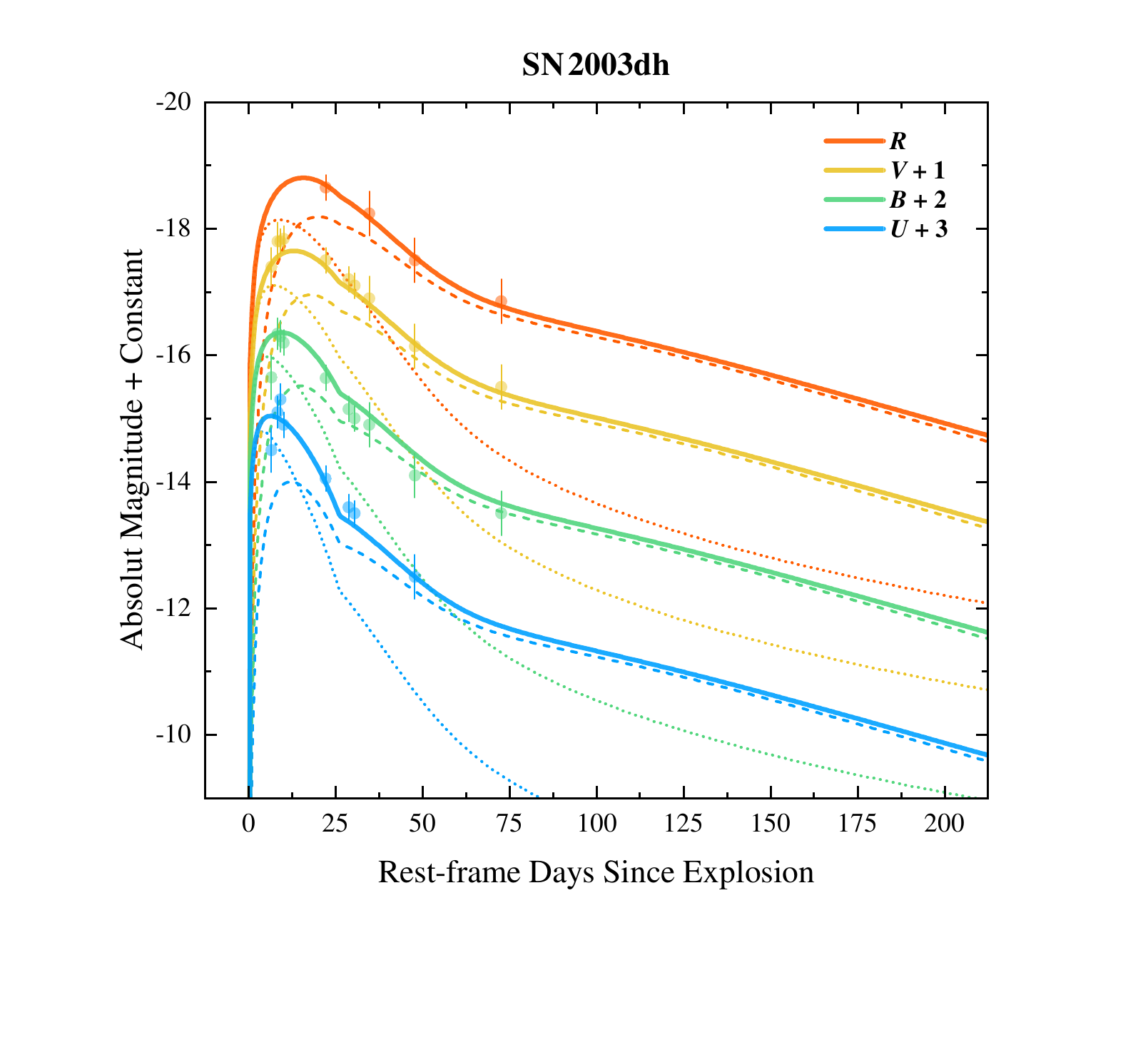}
    \includegraphics[width = 0.32\linewidth , trim = 80 65 93 35, clip]{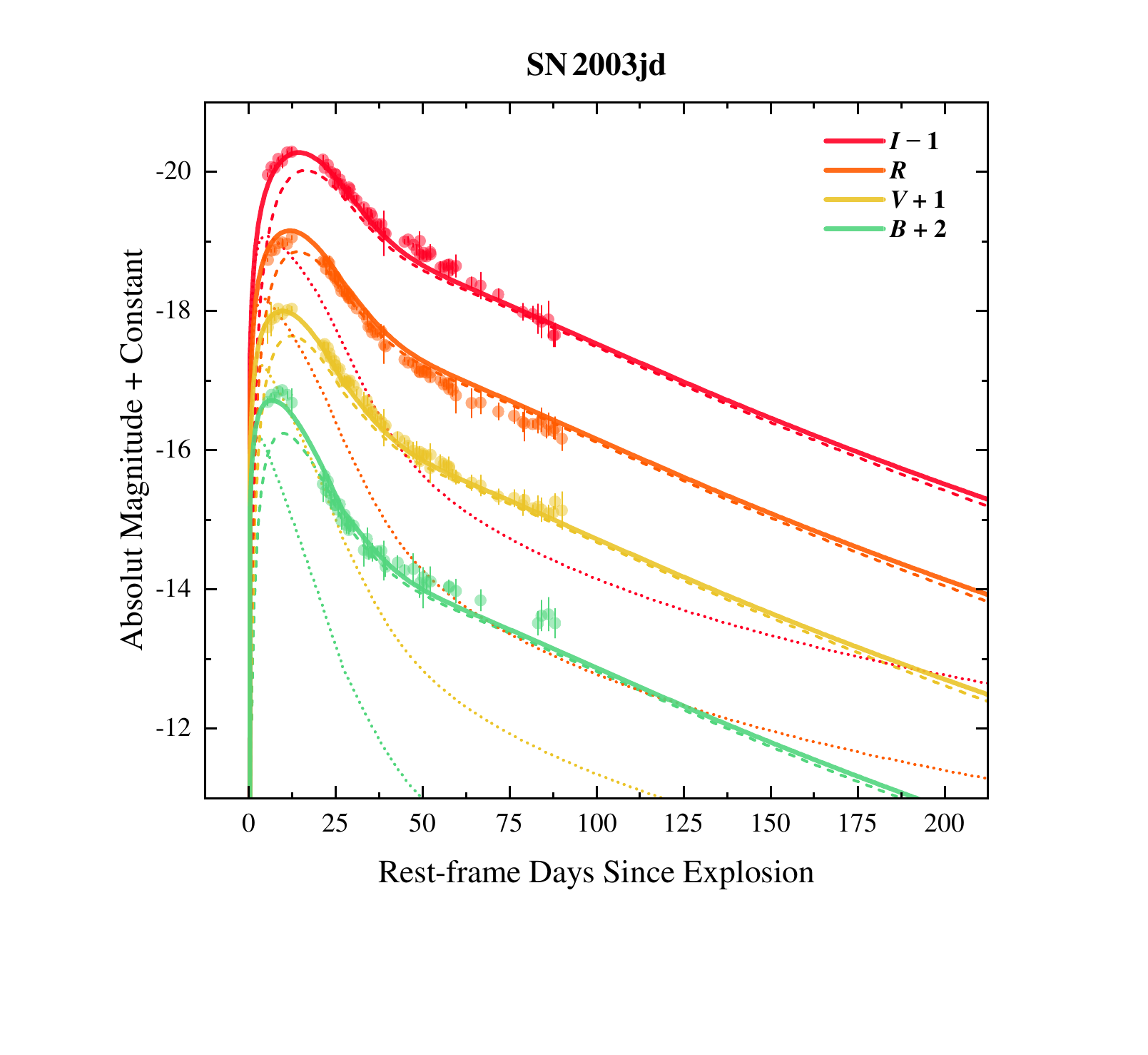}
    \includegraphics[width = 0.32\linewidth , trim = 80 65 93 35, clip]{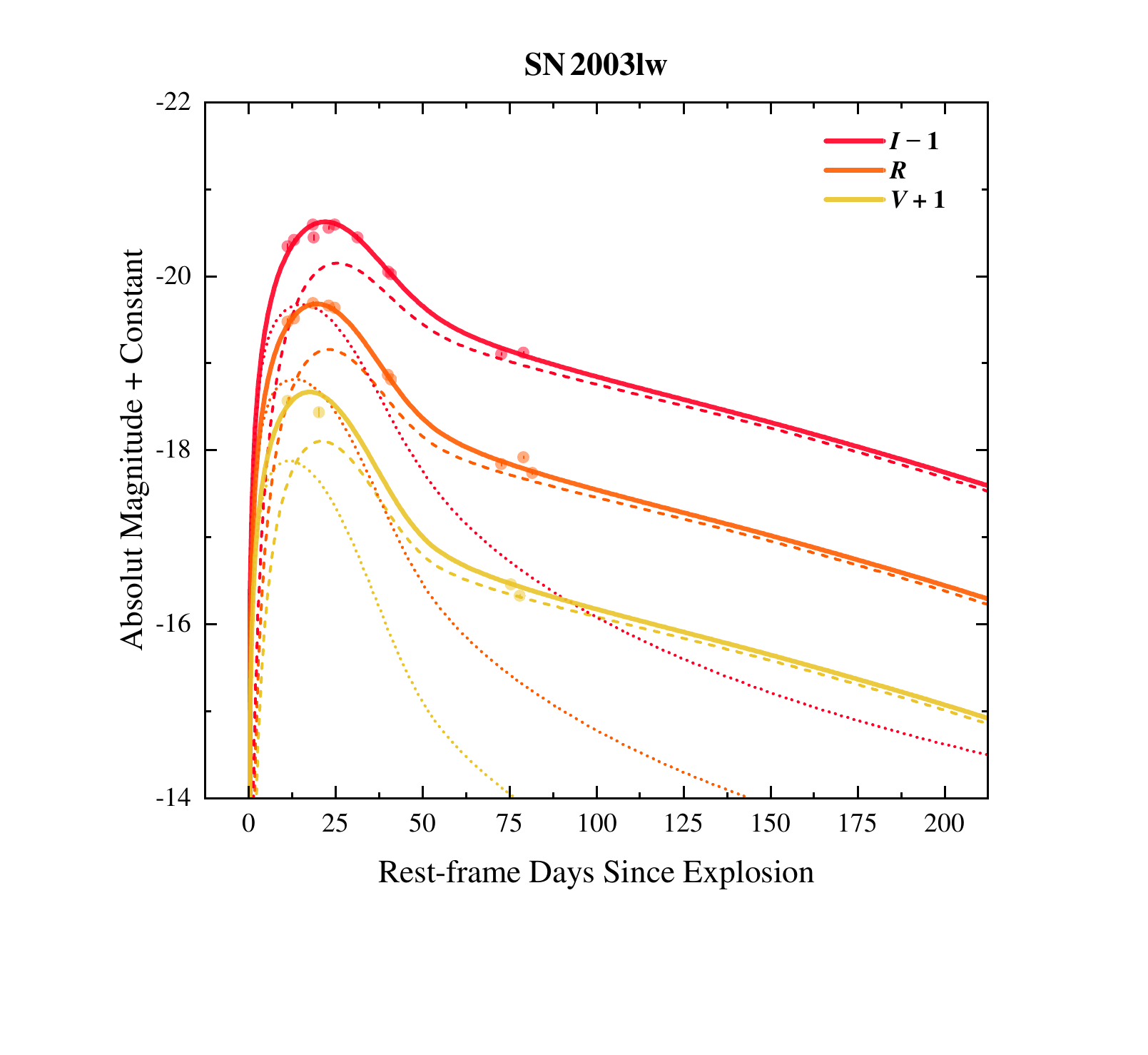}
    \includegraphics[width = 0.32\linewidth , trim = 80 65 93 35, clip]{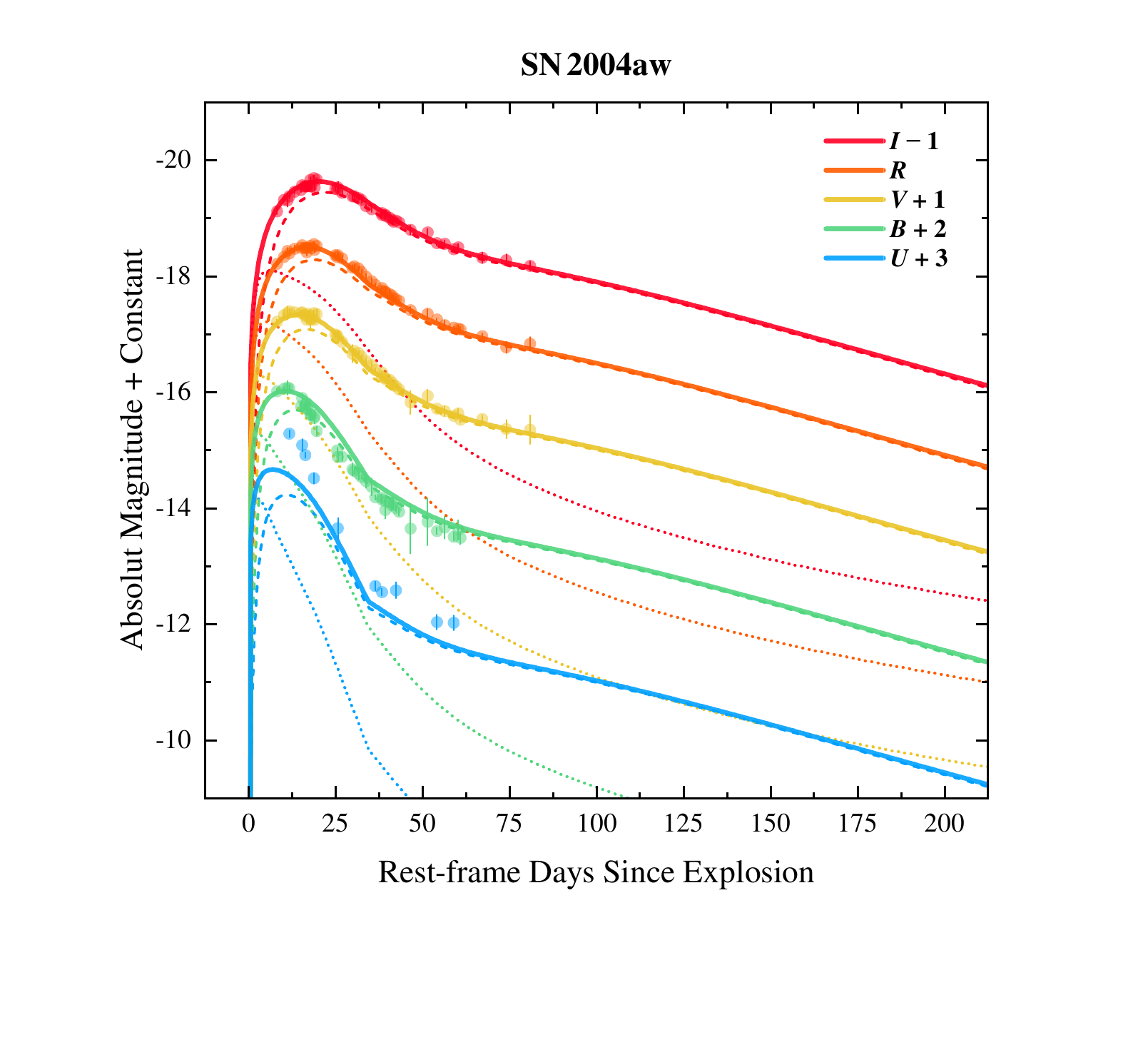}
    \includegraphics[width = 0.32\linewidth , trim = 80 65 93 35, clip]{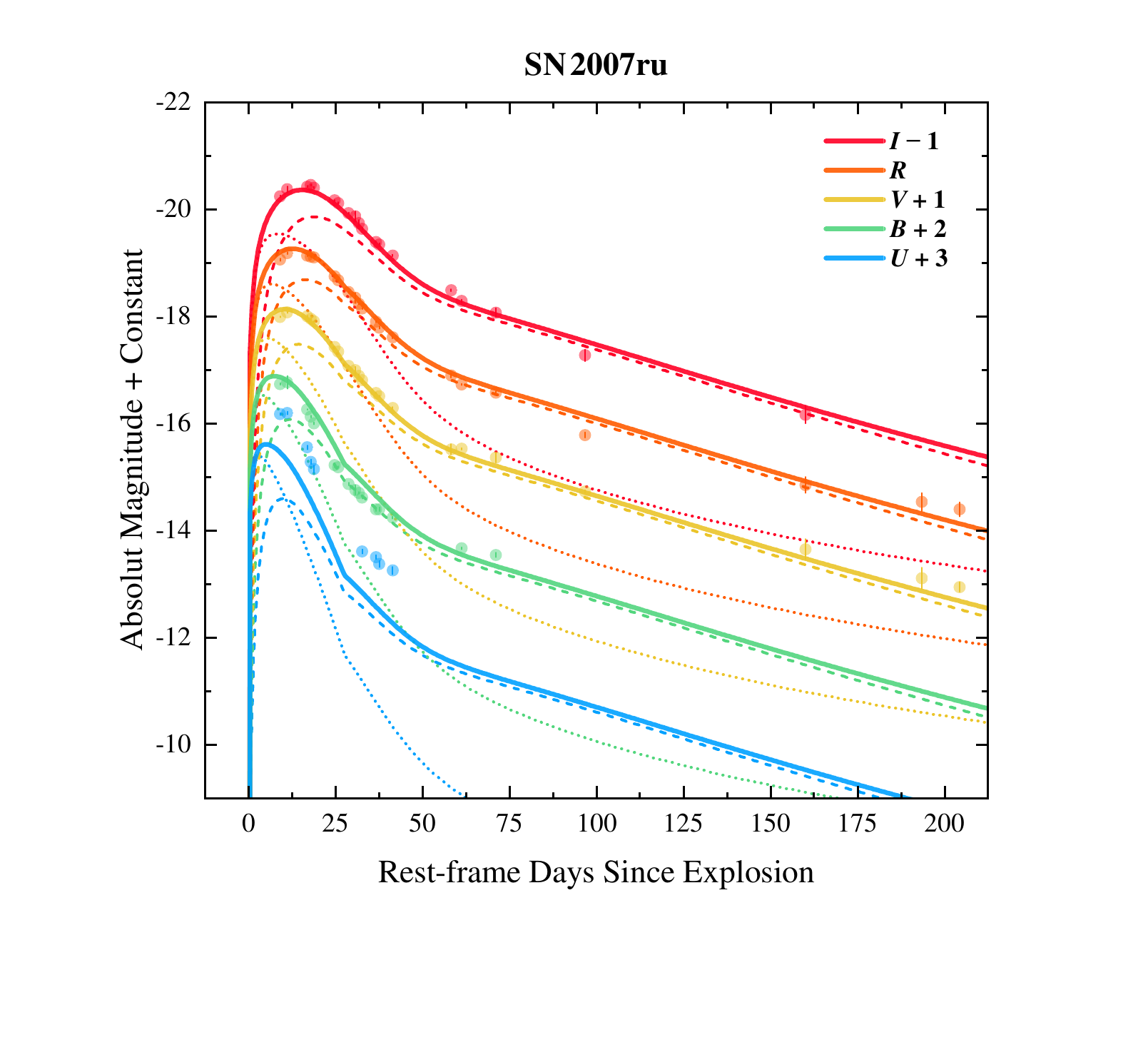}
    \includegraphics[width = 0.32\linewidth , trim = 80 65 93 35, clip]{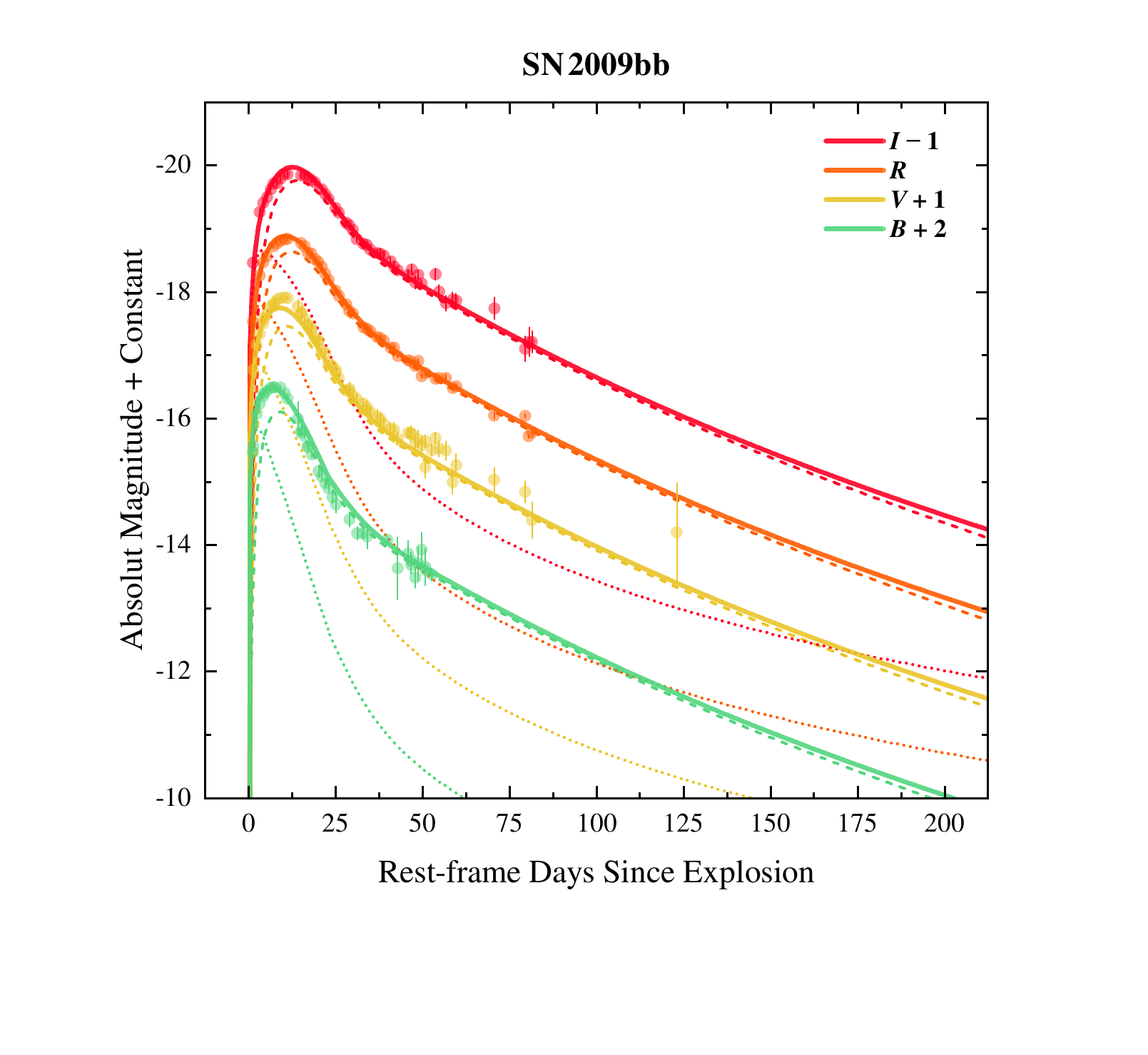}
    \includegraphics[width = 0.32\linewidth , trim = 80 65 93 35, clip]{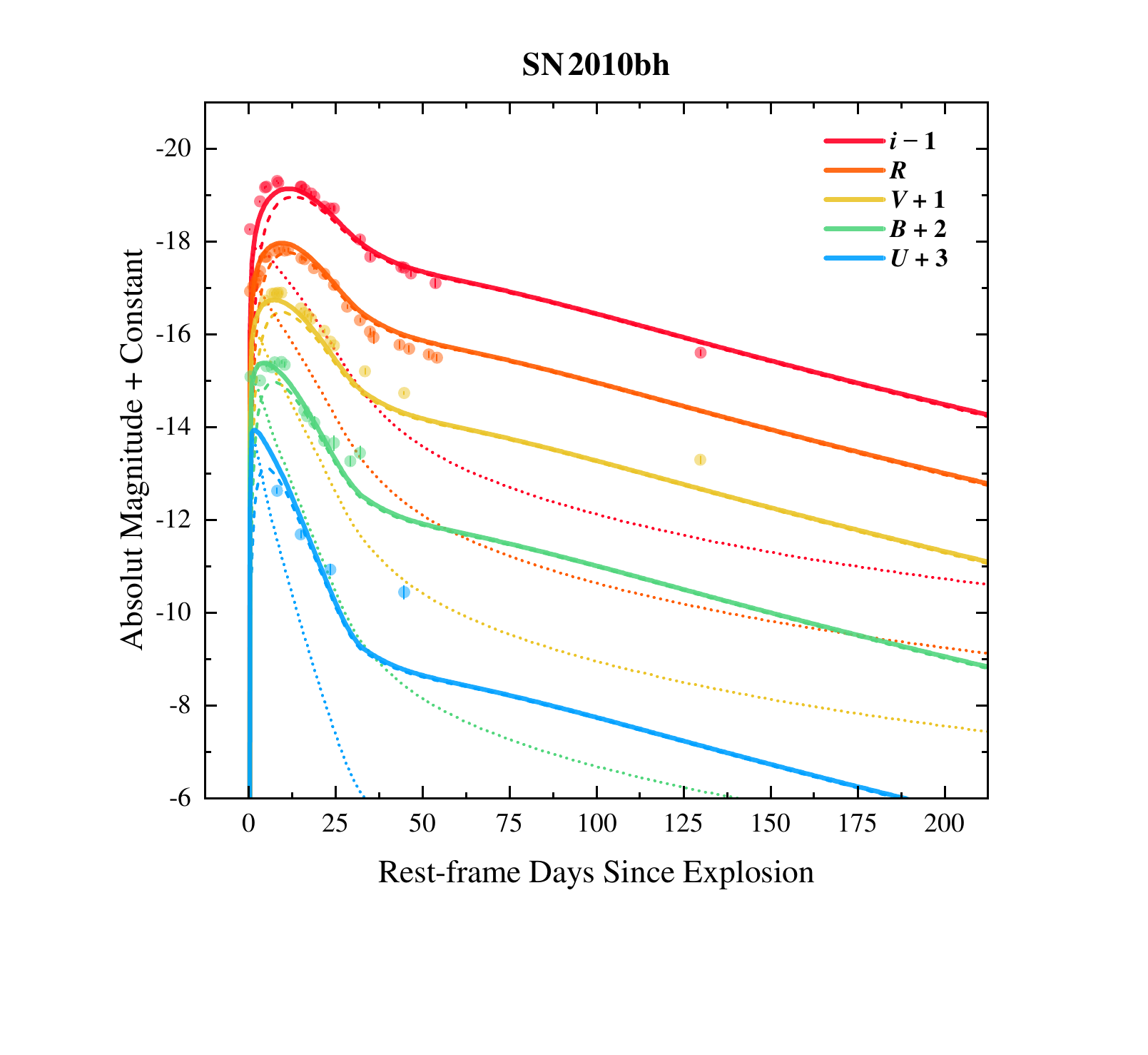}
    \includegraphics[width = 0.32\linewidth , trim = 80 65 93 35, clip]{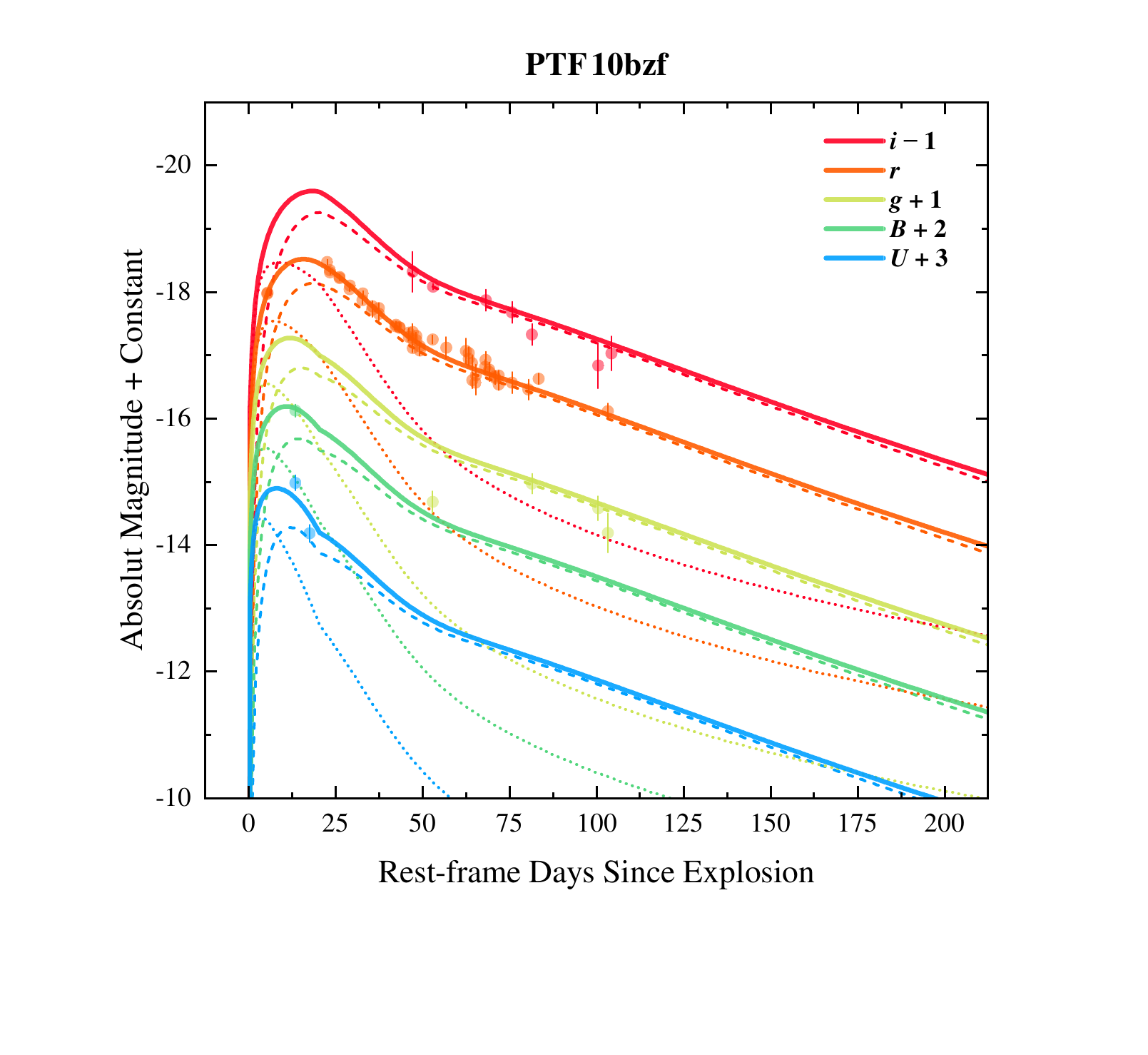}
    \includegraphics[width = 0.32\linewidth , trim = 80 65 93 35, clip]{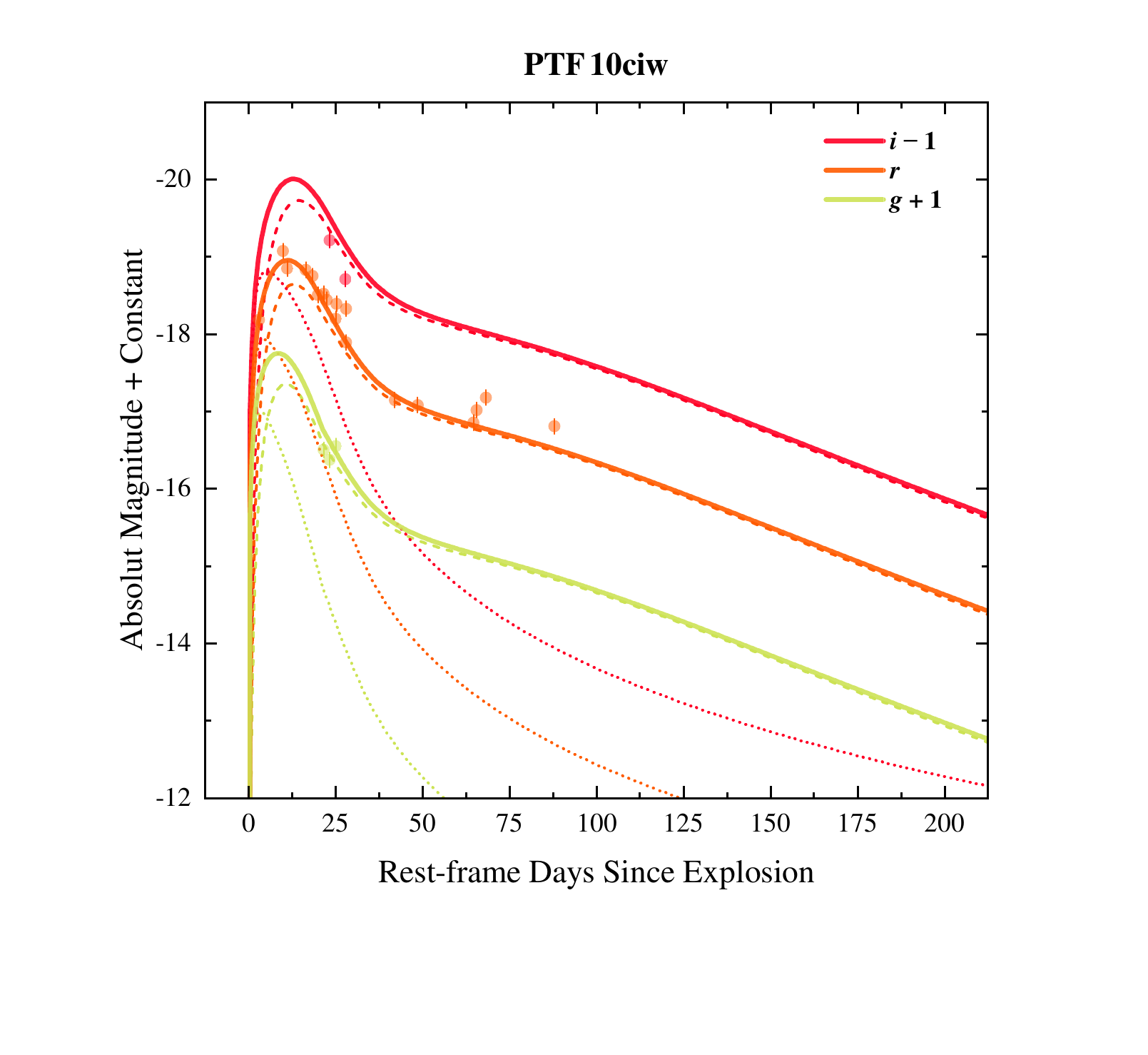}
    \includegraphics[width = 0.32\linewidth , trim = 80 65 93 35, clip]{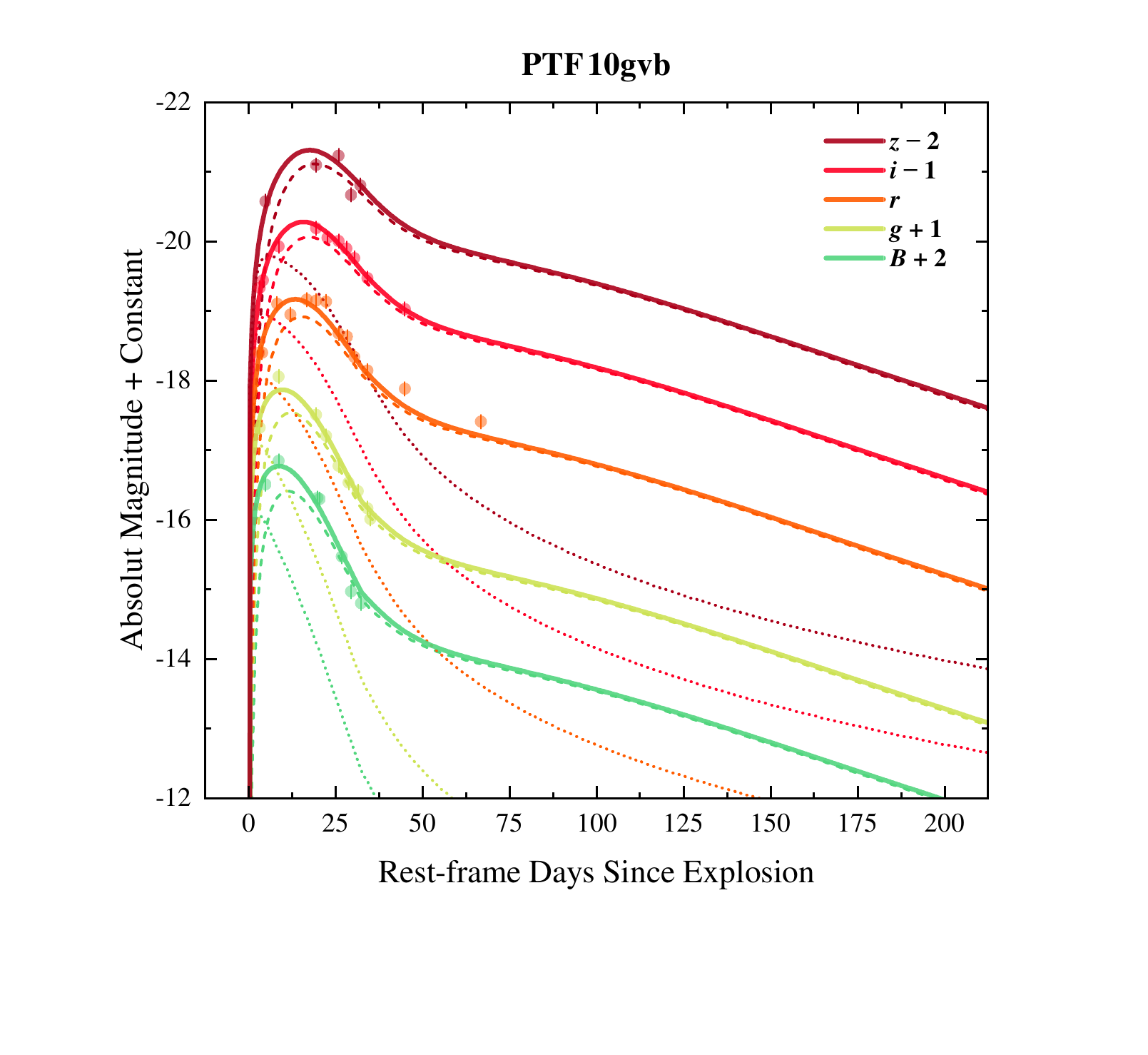}
    \caption{Multi-band observations and fittings for the SNe Ic-BL by the magnetar-powered model. The solid lines represent the best fits for the multi-band lightcurves, while the contributions from the magnetar, $^{56}$Ni, and afterglow emissions are marked as the dotted, dashed, and dotted-dashed lines, respectively. The colors corresponding to different bands are shown in the top-right legend of each panel. }
    \label{fig:Fits}
\end{figure*}

\begin{figure*}
    \ContinuedFloat
    \centering
    \includegraphics[width = 0.32\linewidth , trim = 80 65 93 35, clip]{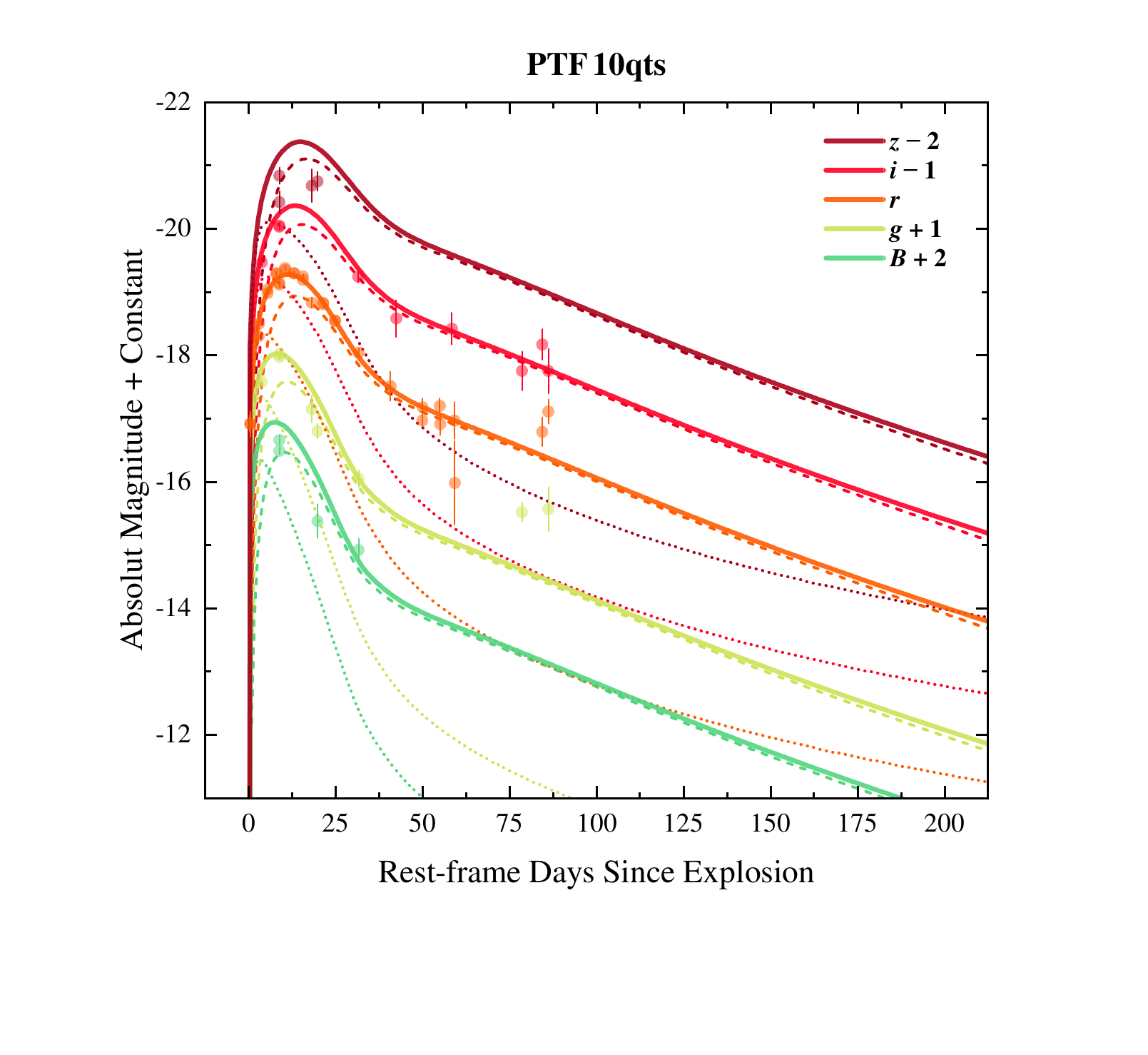}
    \includegraphics[width = 0.32\linewidth , trim = 80 65 93 35, clip]{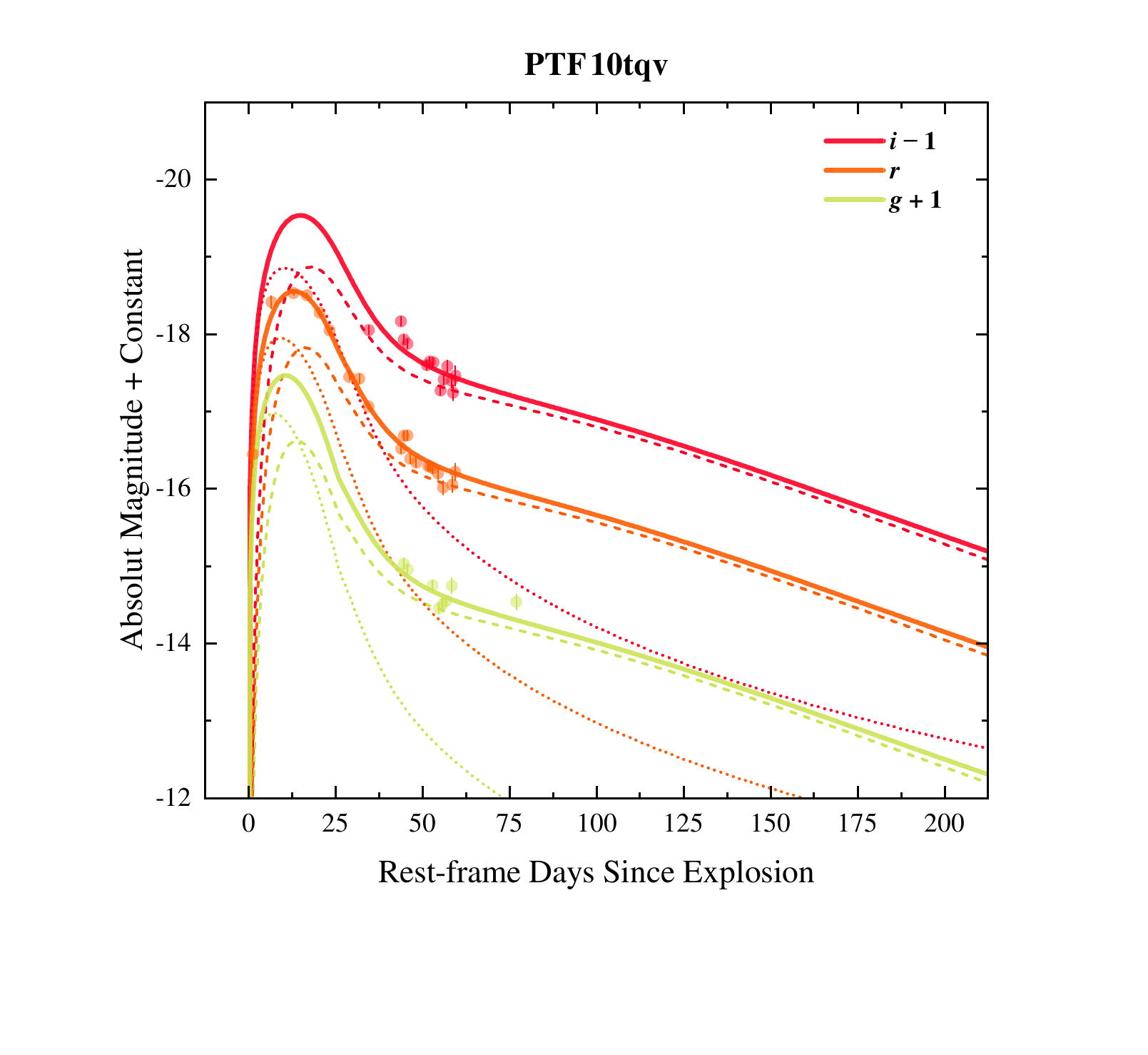}
    \includegraphics[width = 0.32\linewidth , trim = 80 65 93 35, clip]{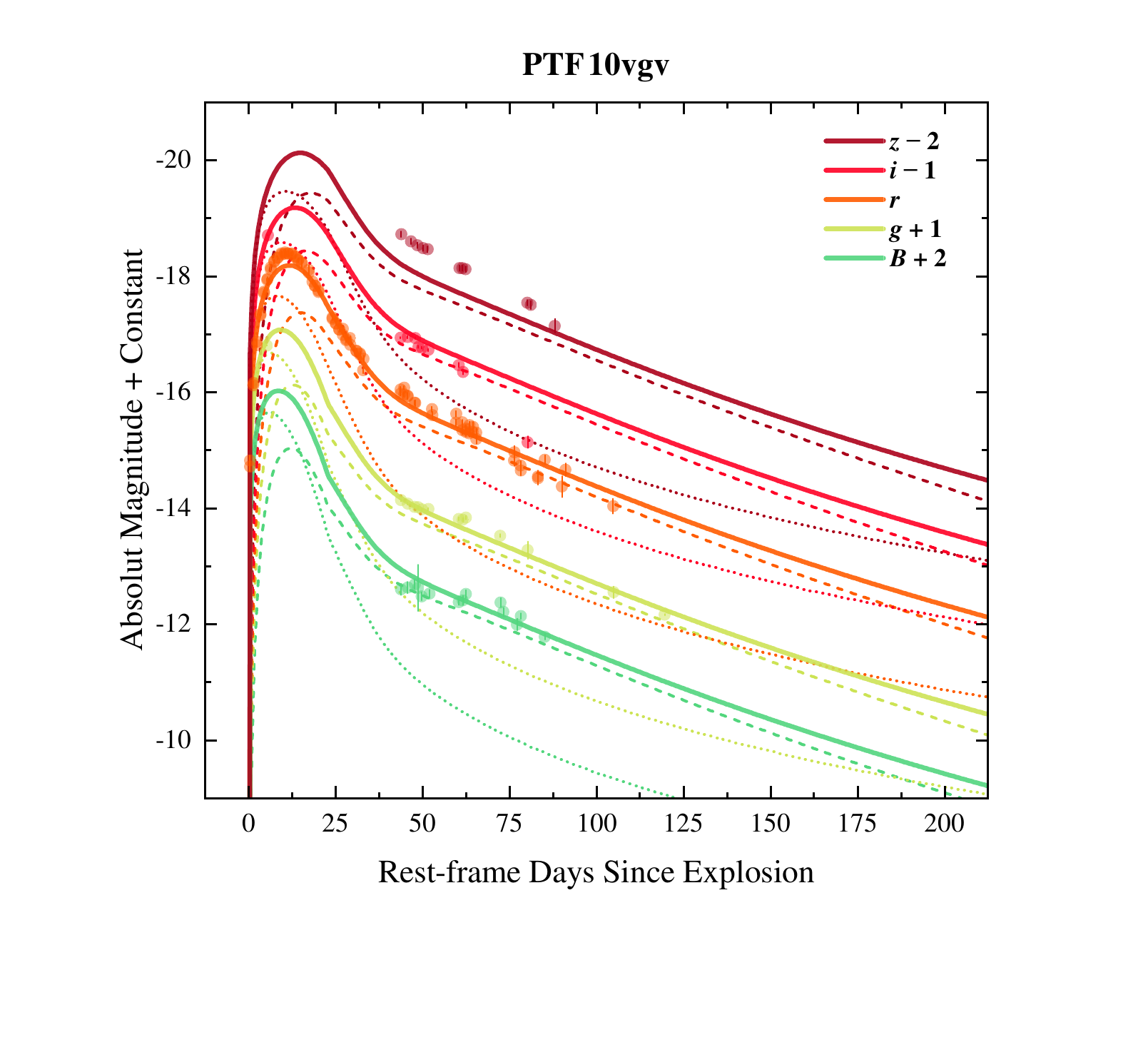}
    \includegraphics[width = 0.32\linewidth , trim = 80 65 93 35, clip]{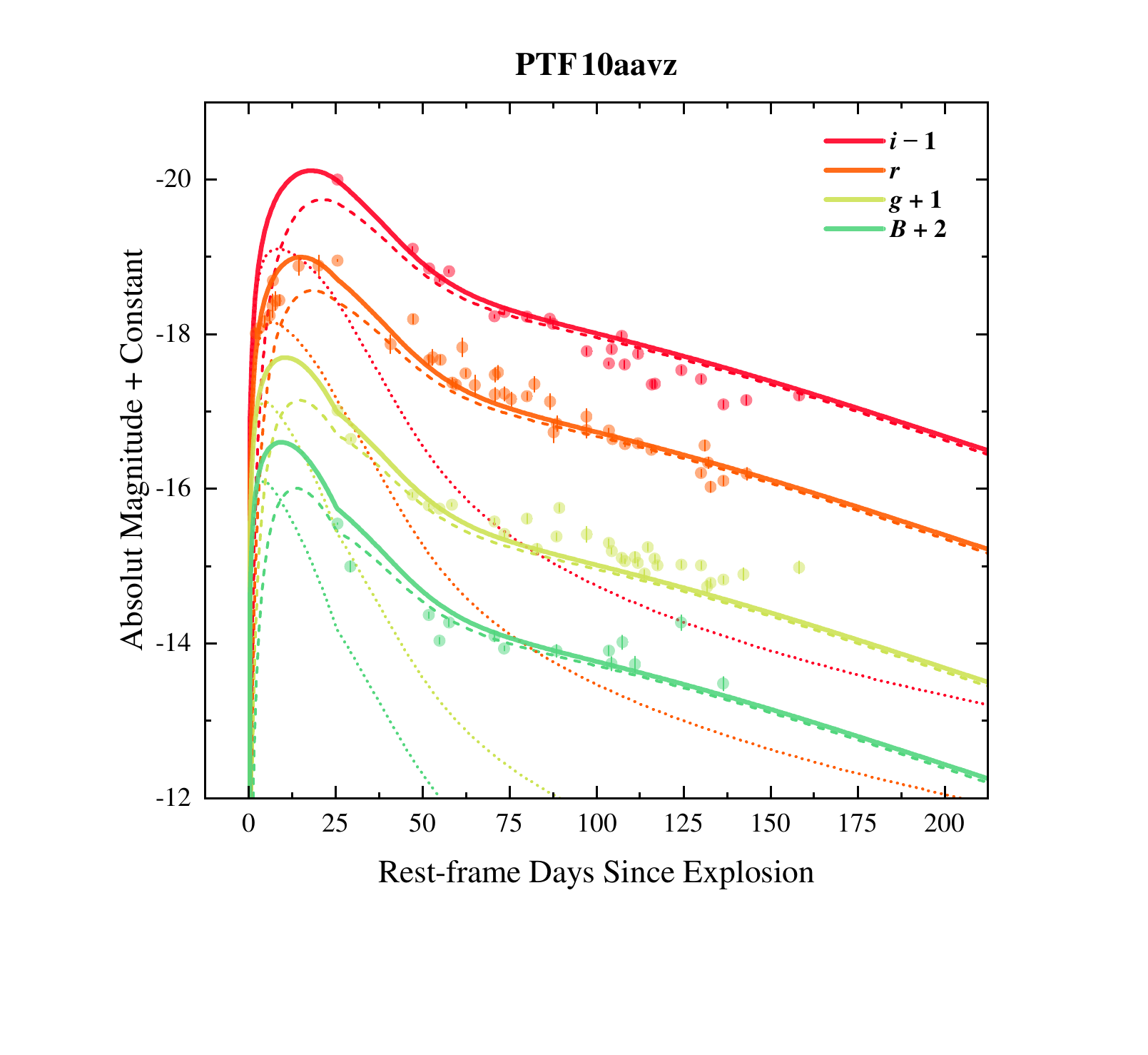}
    \includegraphics[width = 0.32\linewidth , trim = 80 65 93 35, clip]{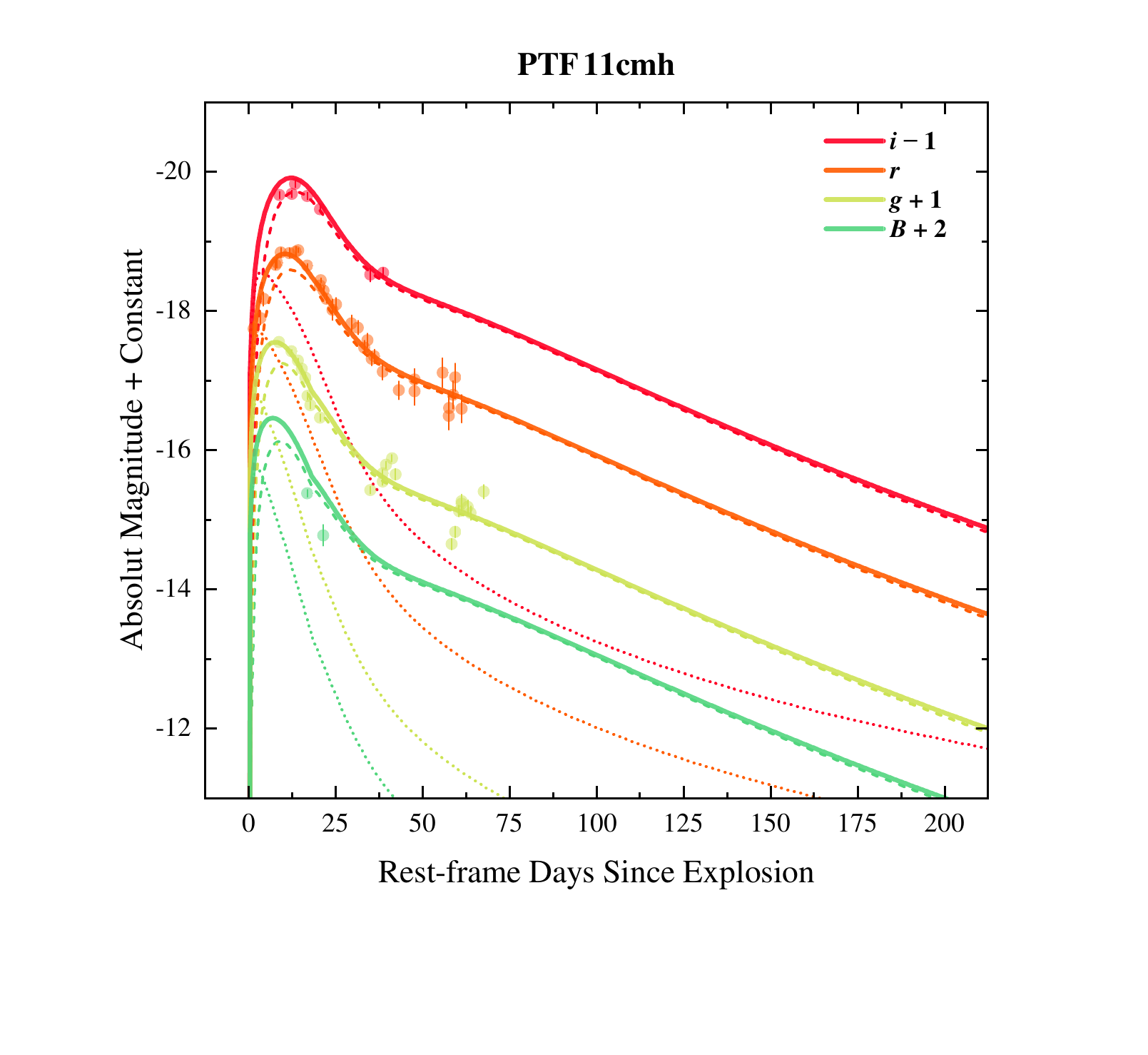}
    \includegraphics[width = 0.32\linewidth , trim = 80 65 93 35, clip]{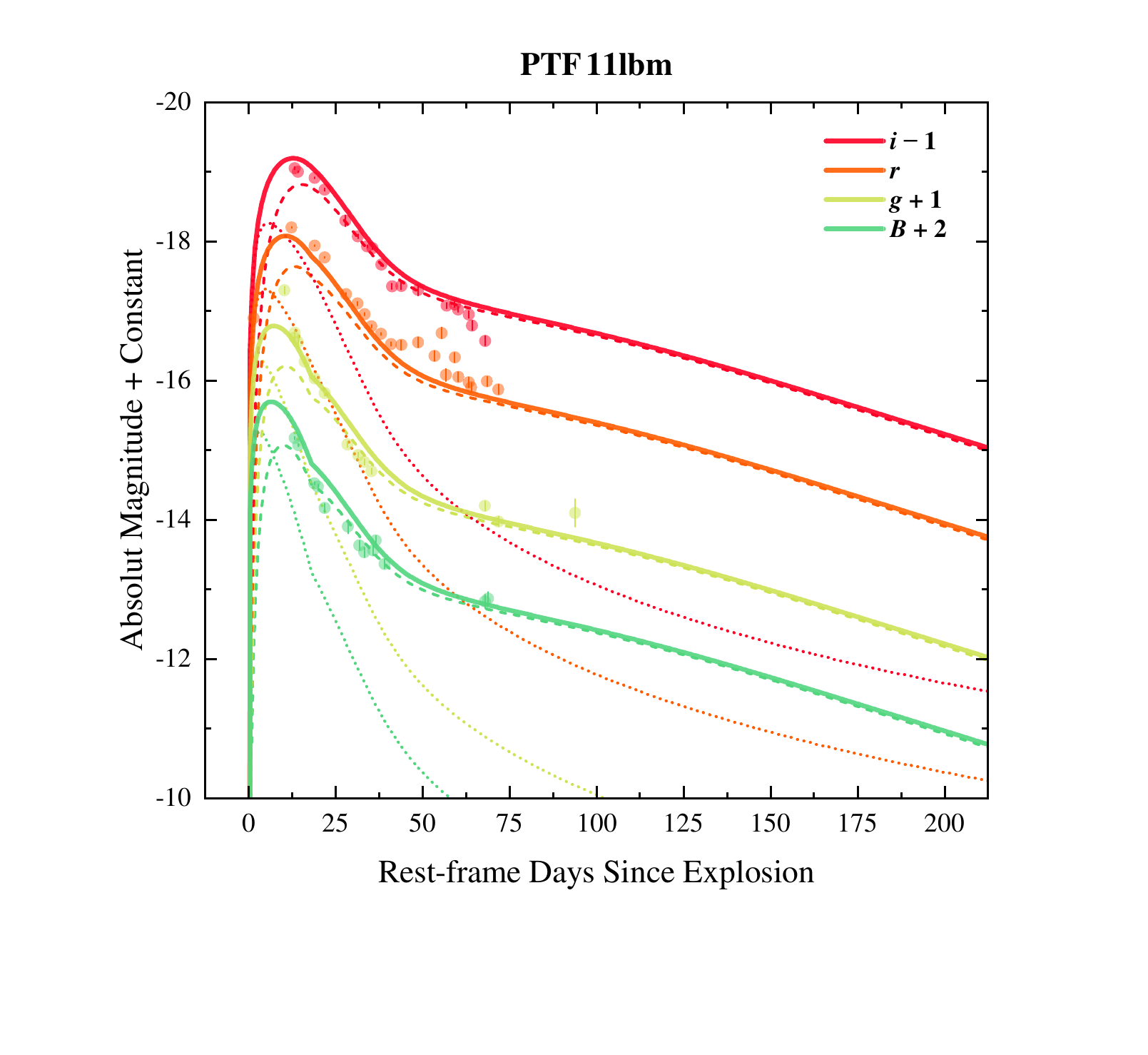}
    \includegraphics[width = 0.32\linewidth , trim = 80 65 93 35, clip]{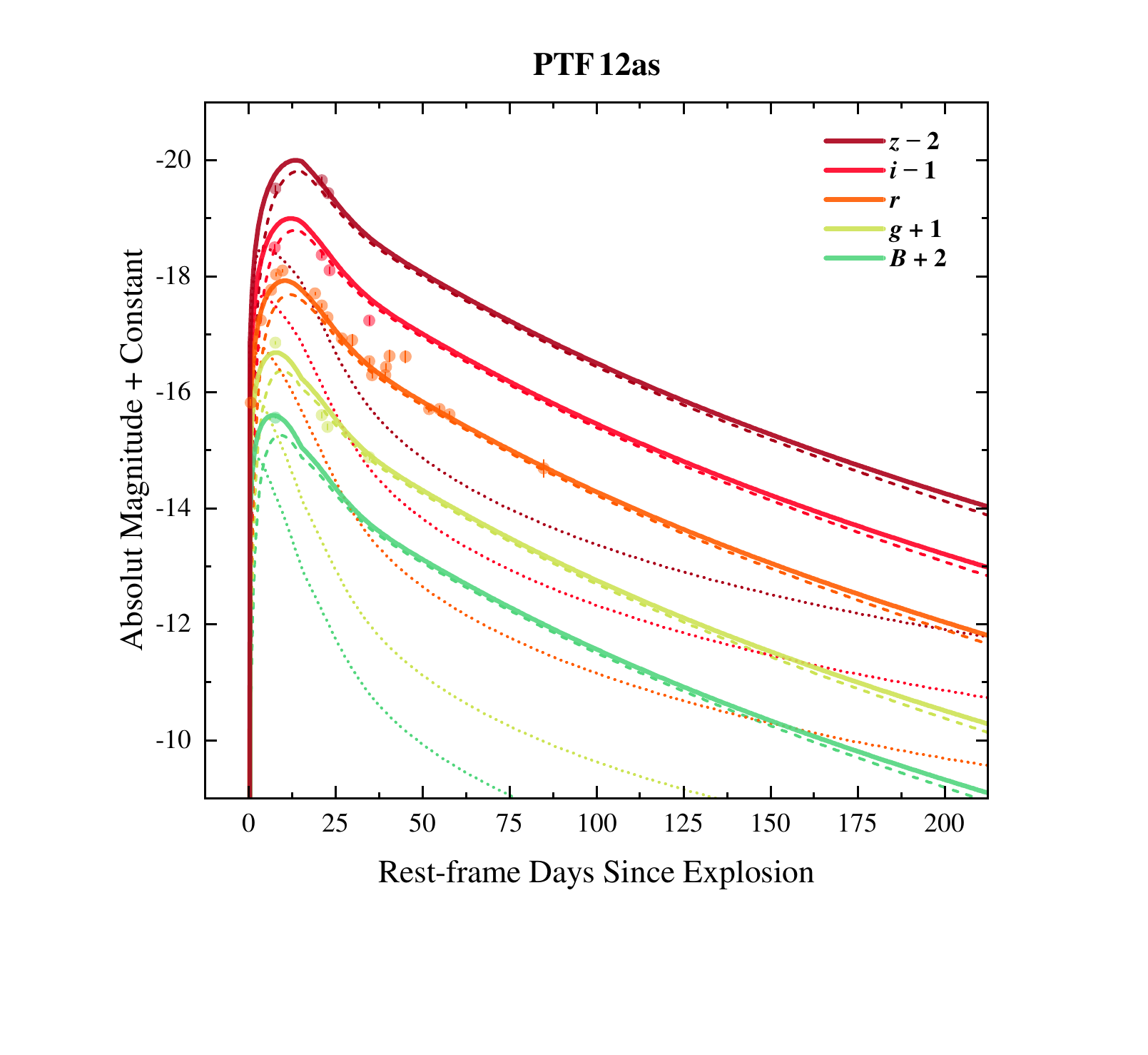}
    \includegraphics[width = 0.32\linewidth , trim = 80 65 93 35, clip]{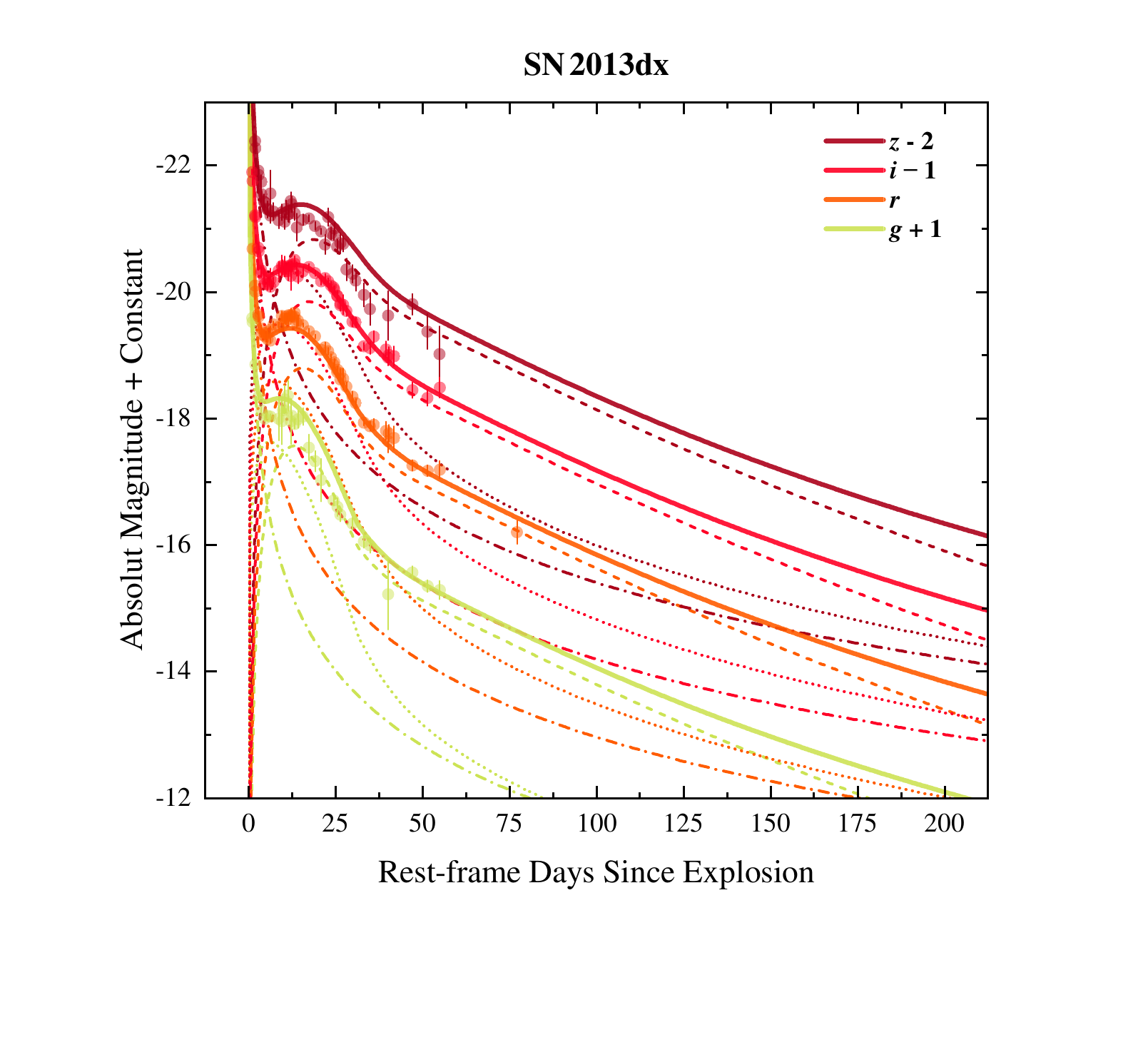}
    \includegraphics[width = 0.32\linewidth , trim = 80 65 93 35, clip]{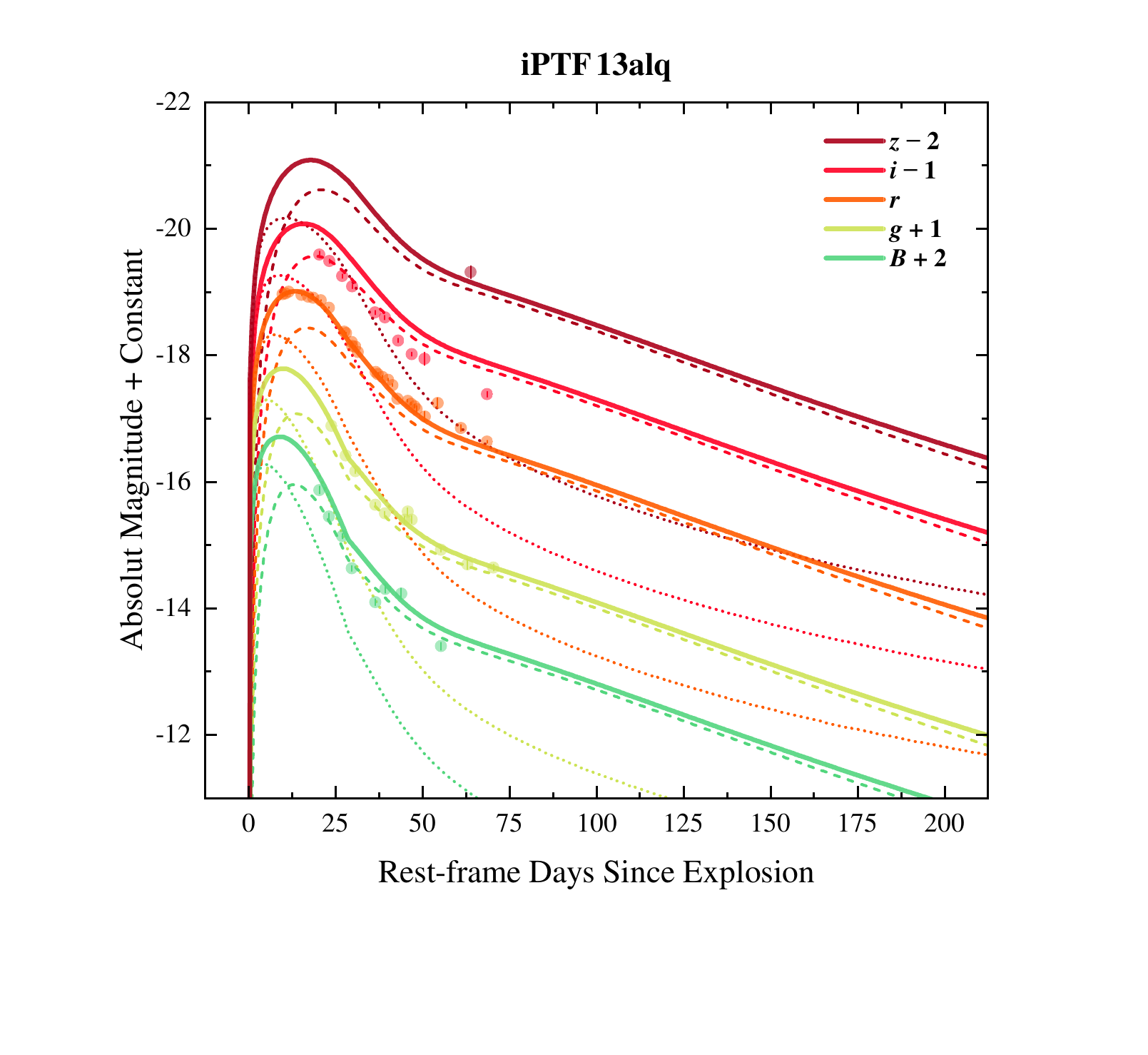}
    \includegraphics[width = 0.32\linewidth , trim = 80 65 93 35, clip]{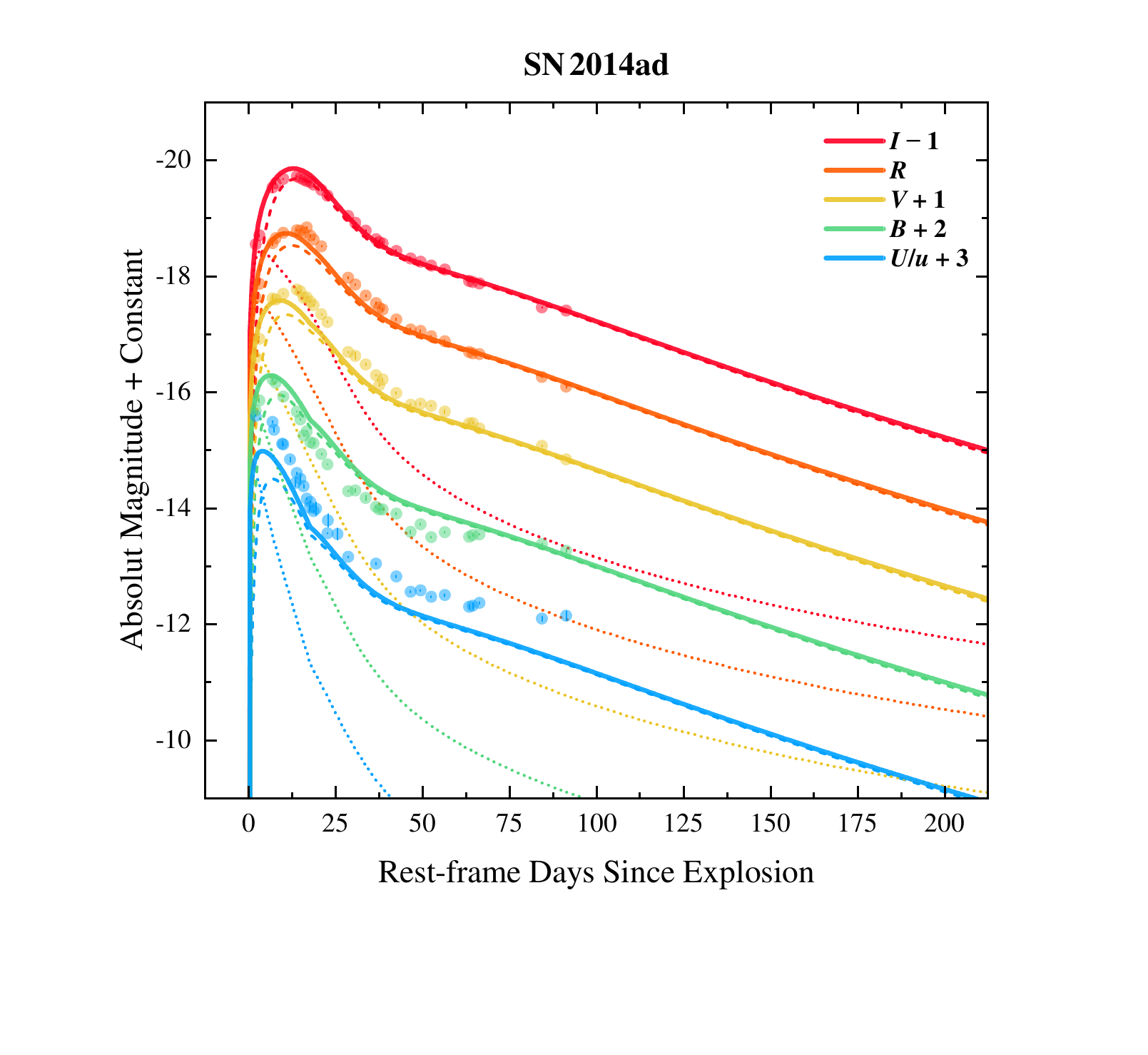}
    \includegraphics[width = 0.32\linewidth , trim = 80 65 93 35, clip]{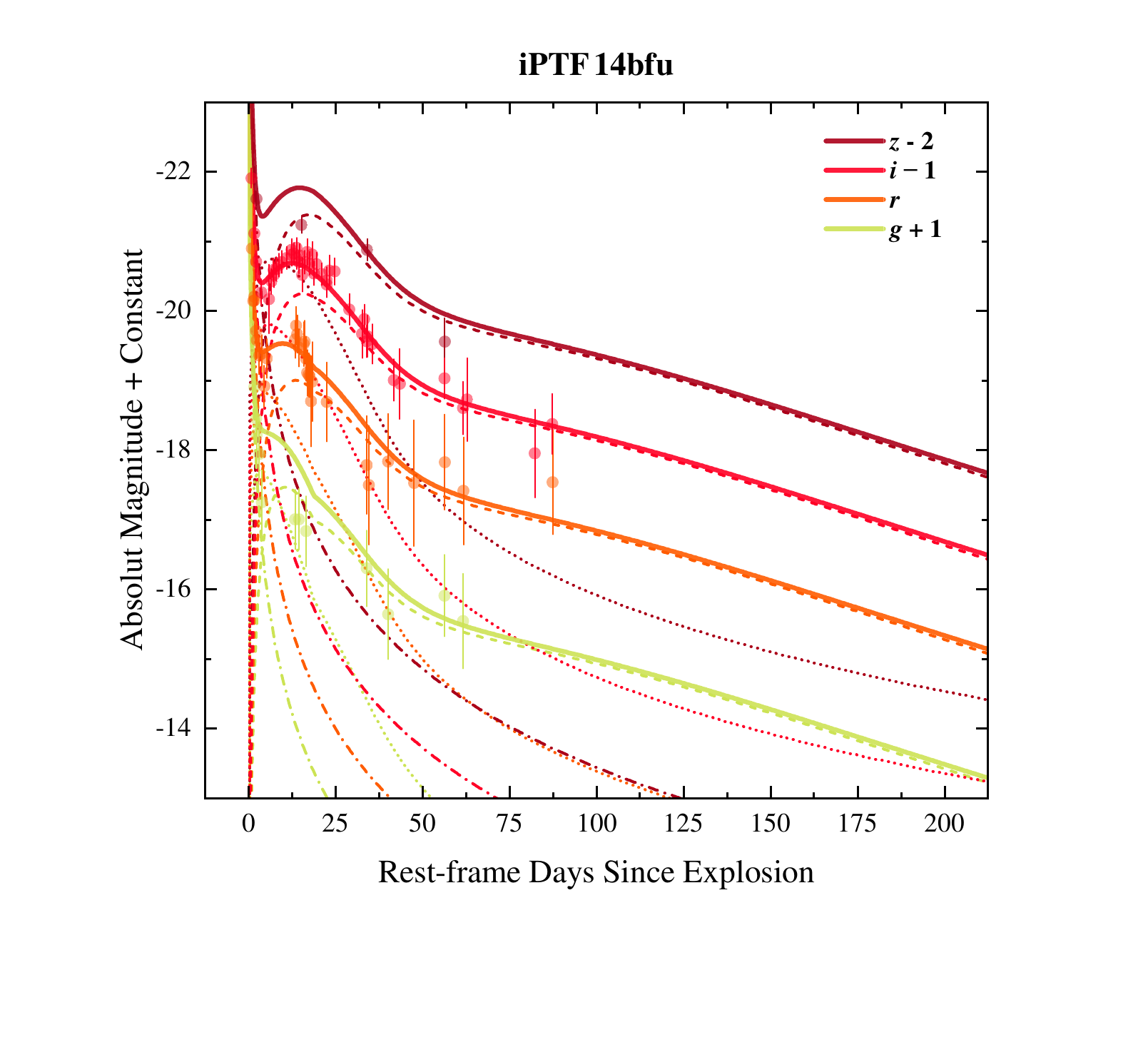}
    \includegraphics[width = 0.32\linewidth , trim = 80 65 93 35, clip]{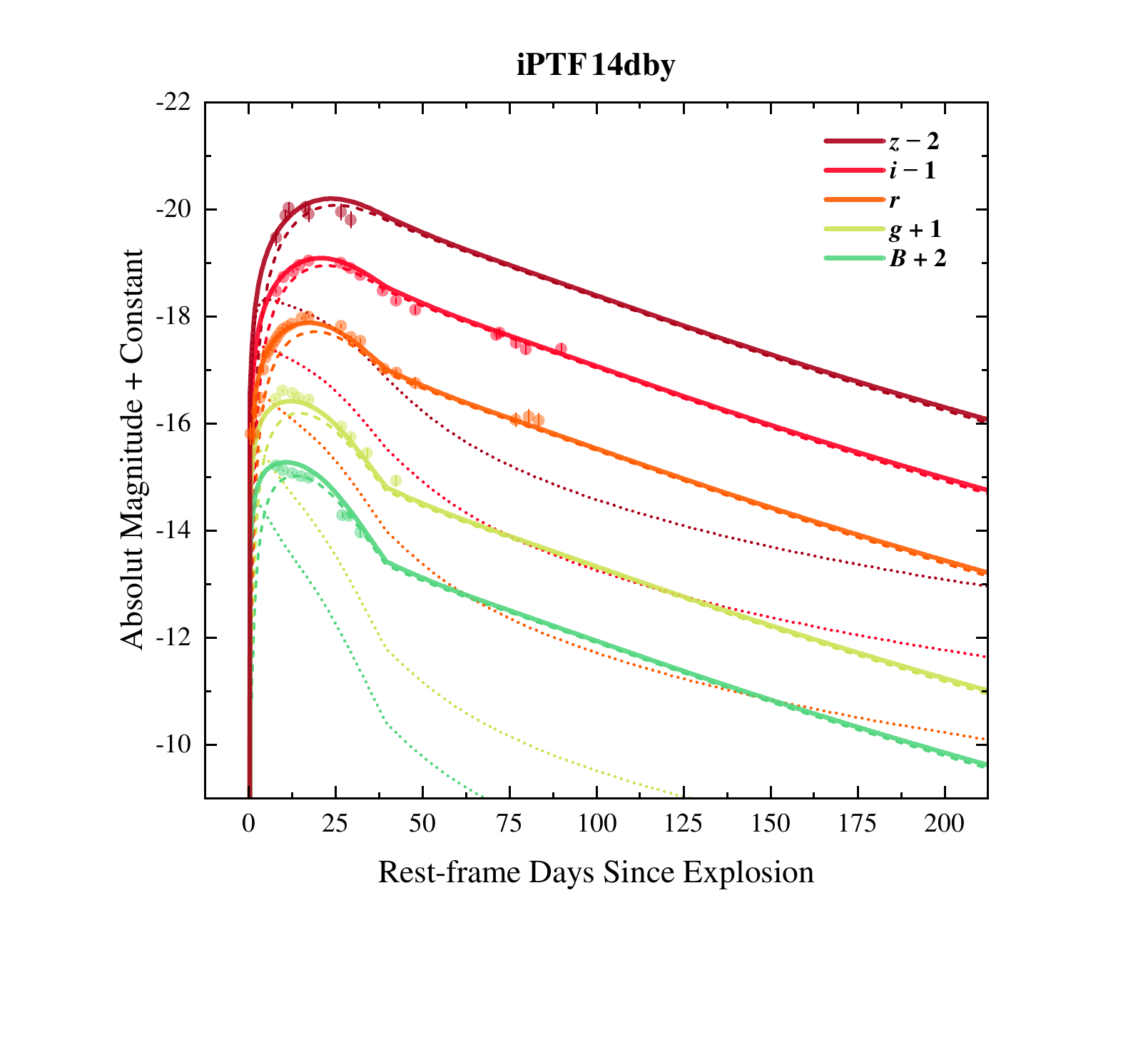}
    \caption{(Continued.)}
\end{figure*}

\begin{figure*}
    \ContinuedFloat
    \centering
    \includegraphics[width = 0.32\linewidth , trim = 80 65 93 35, clip]{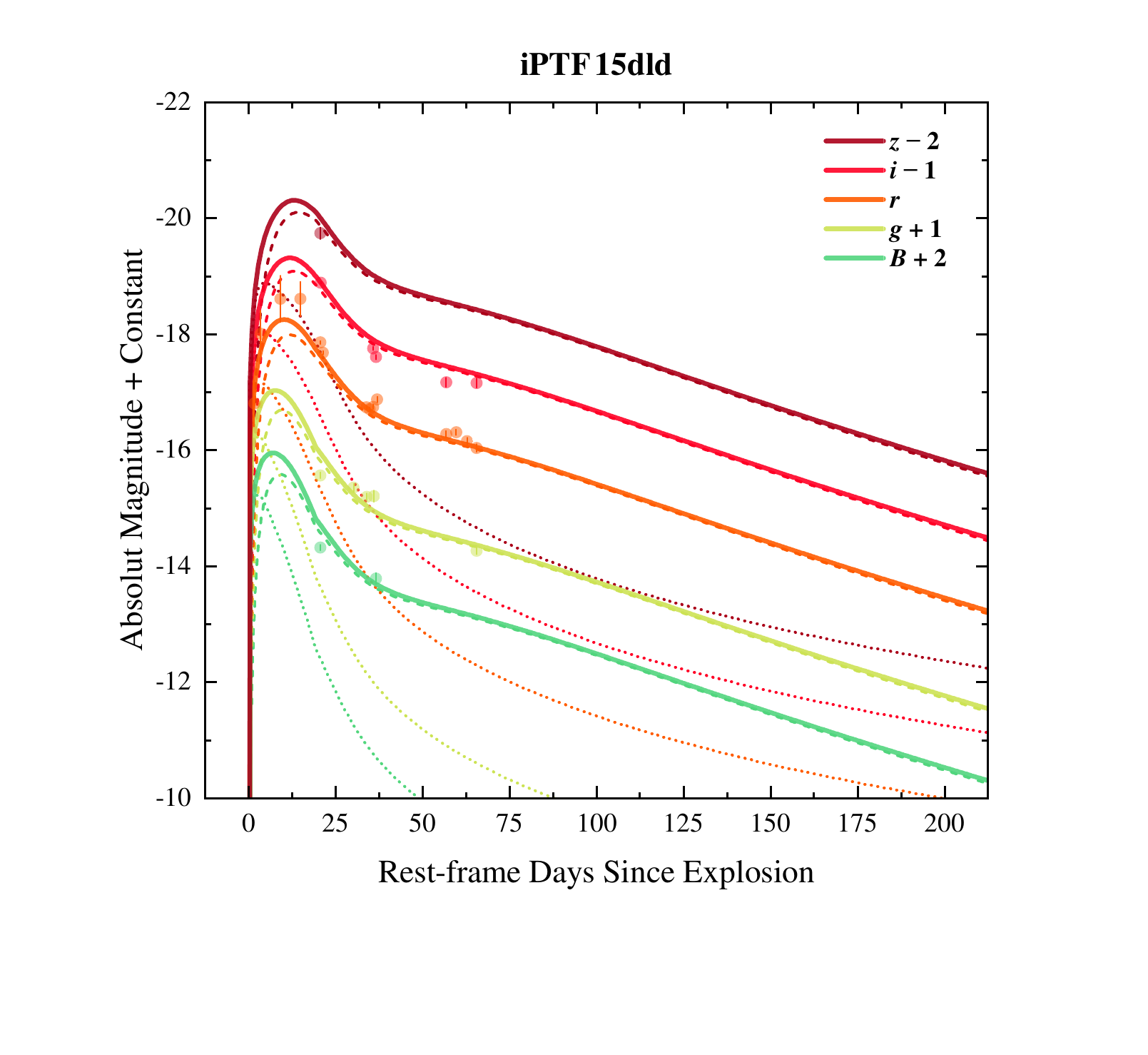}
    \includegraphics[width = 0.32\linewidth , trim = 80 65 93 35, clip]{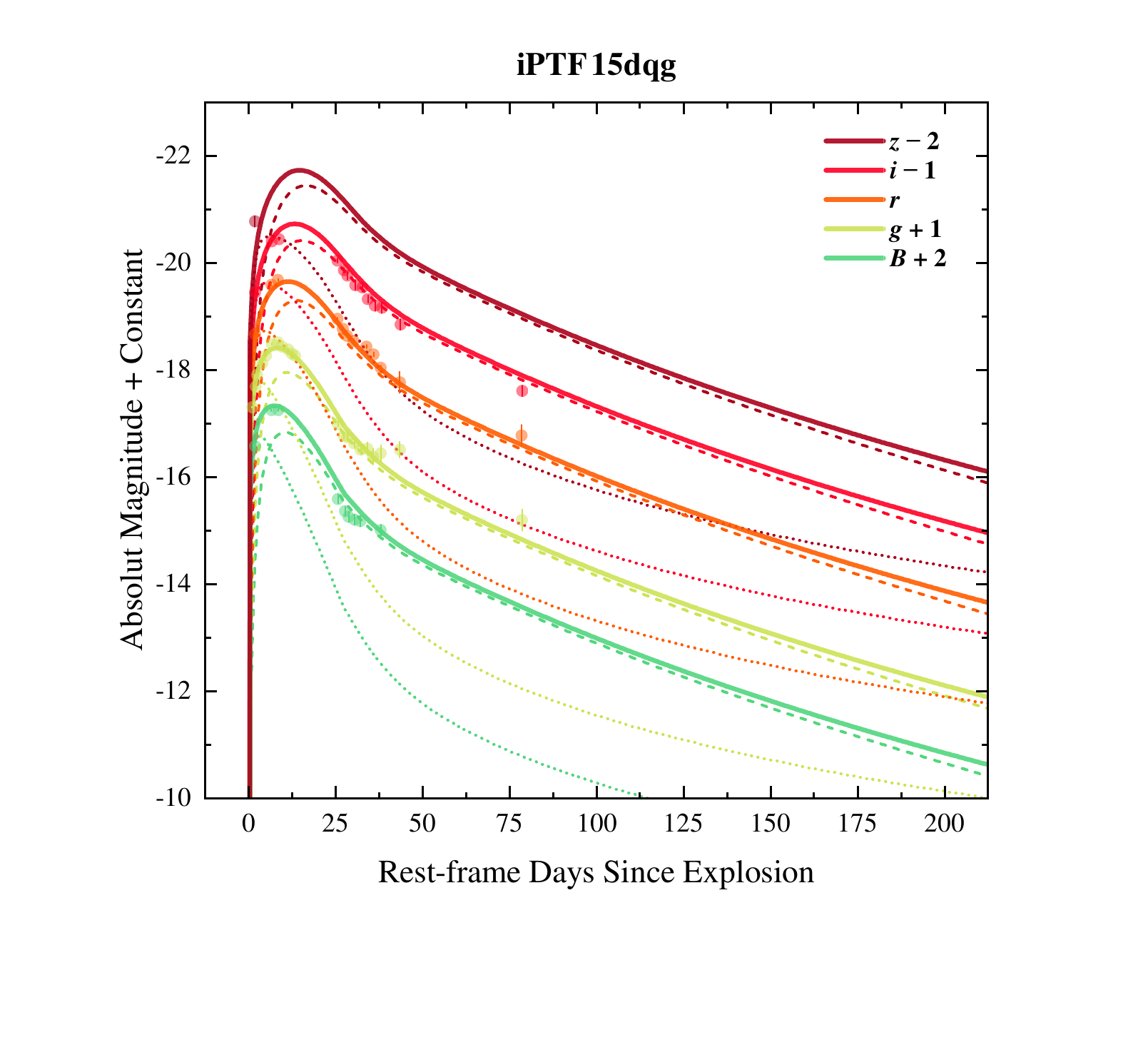}
    \includegraphics[width = 0.32\linewidth , trim = 80 65 93 35, clip]{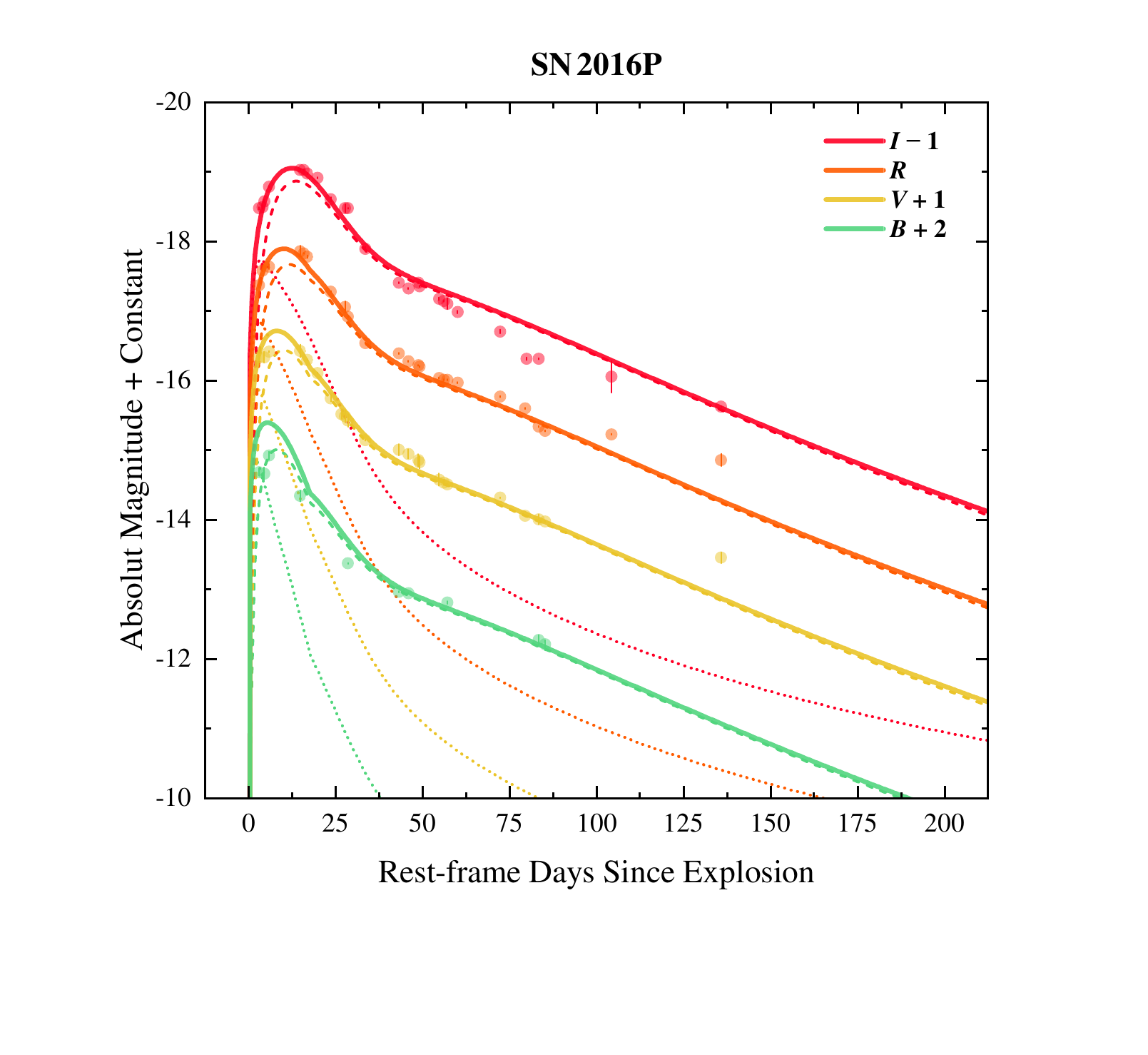}
    \includegraphics[width = 0.32\linewidth , trim = 80 65 93 35, clip]{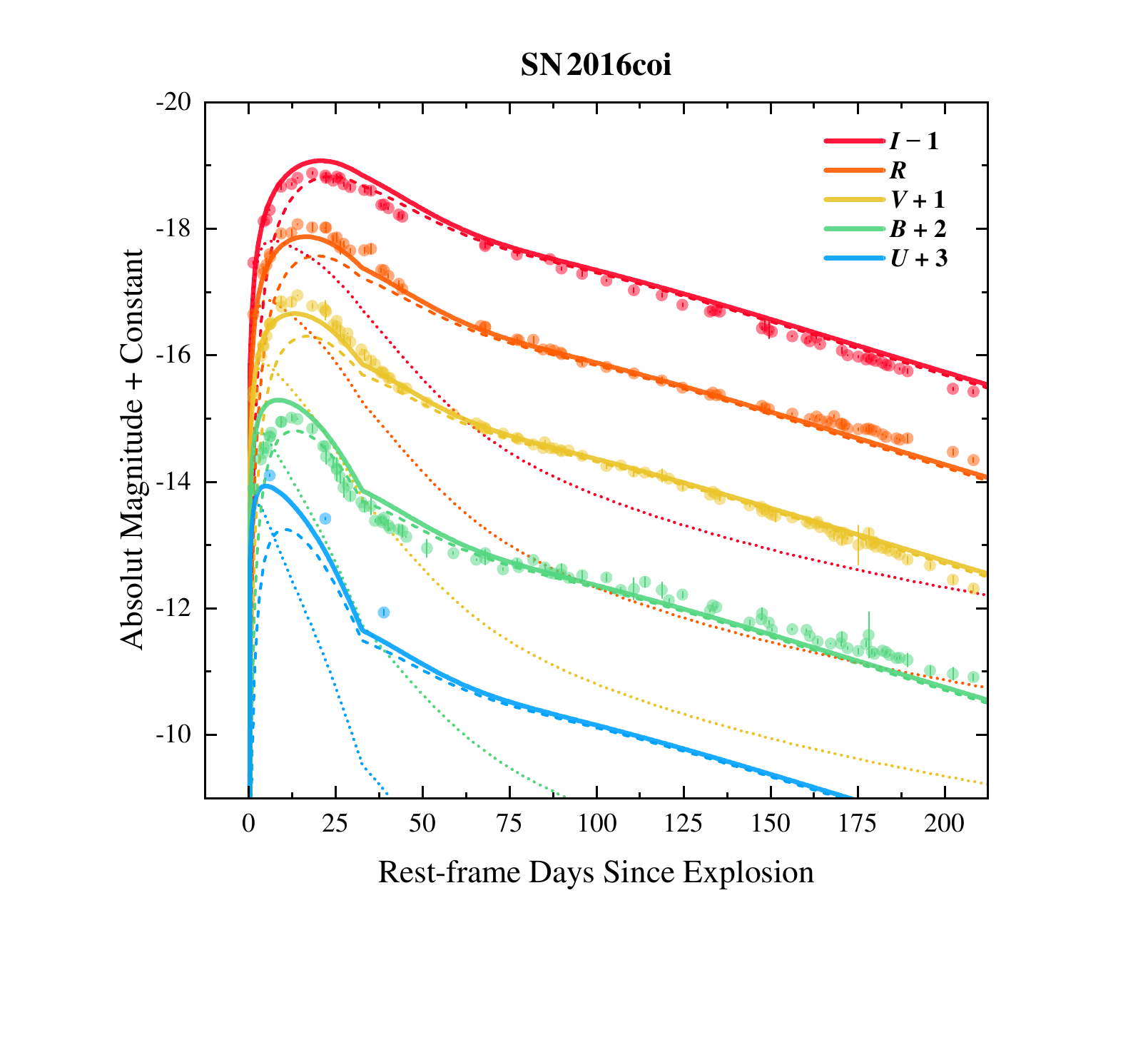}
    \includegraphics[width = 0.32\linewidth , trim = 80 65 93 35, clip]{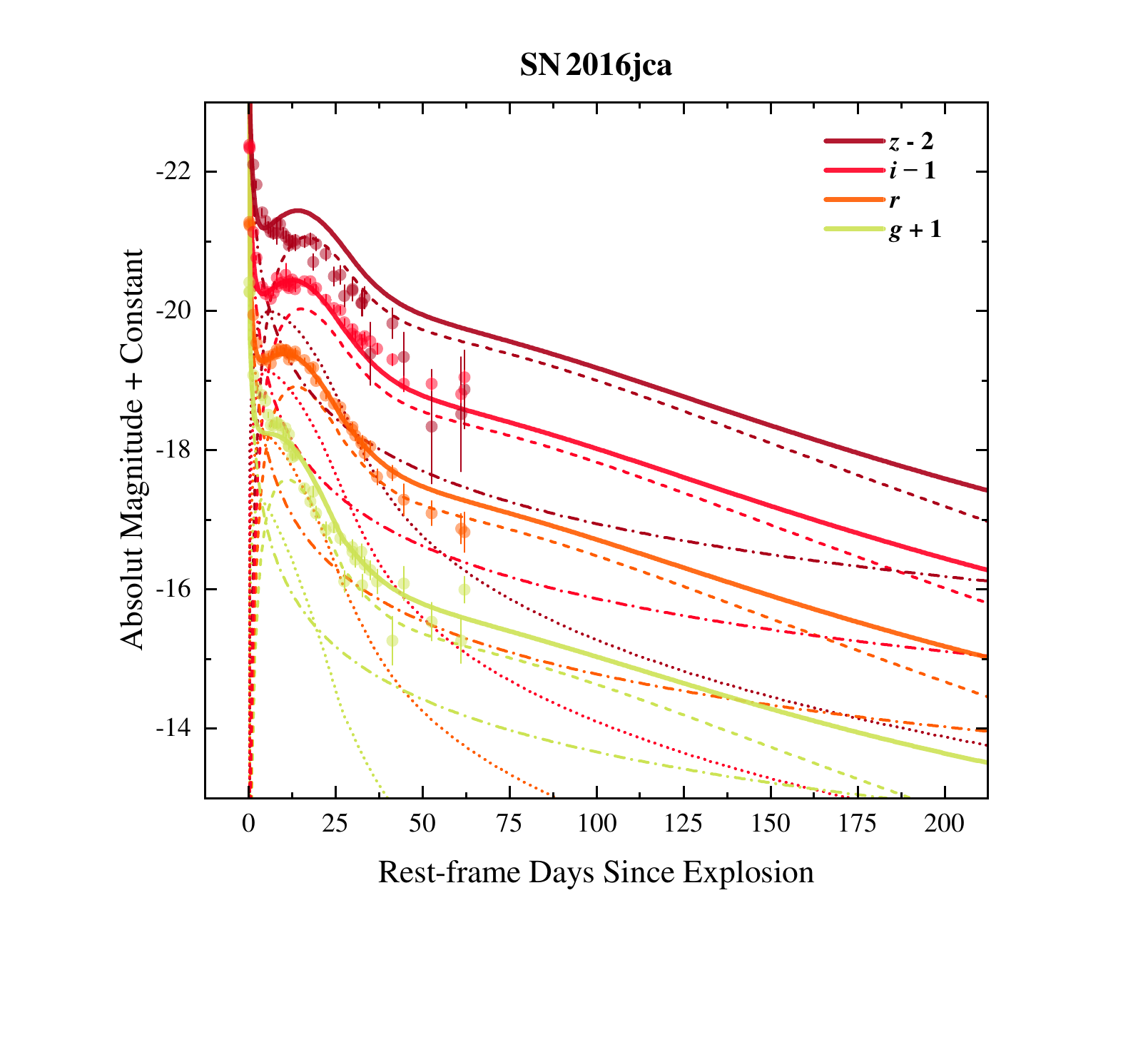}
    \includegraphics[width = 0.32\linewidth , trim = 80 65 93 35, clip]{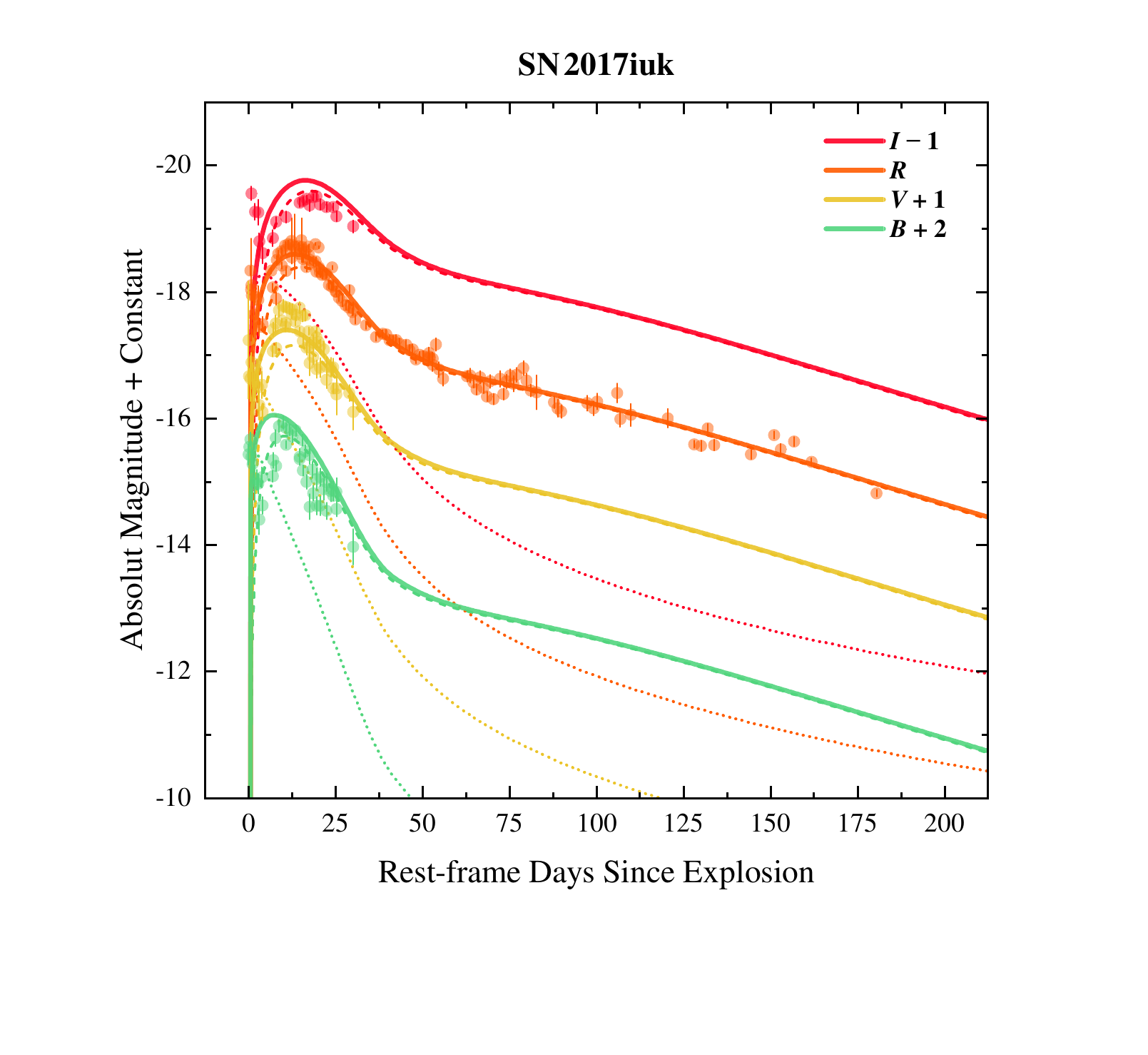}
    \includegraphics[width = 0.32\linewidth , trim = 80 65 93 35, clip]{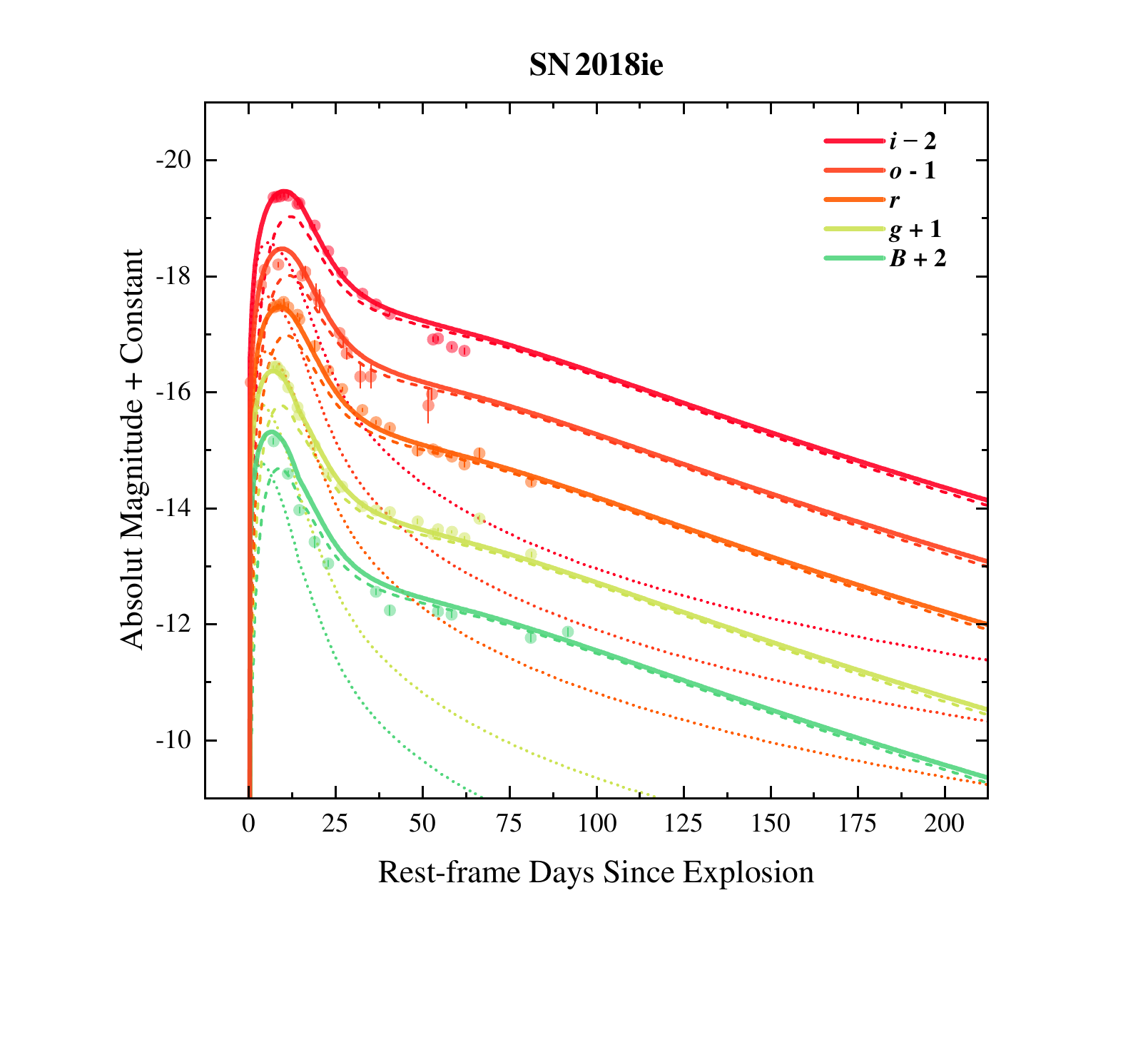}
    \includegraphics[width = 0.32\linewidth , trim = 80 65 93 35, clip]{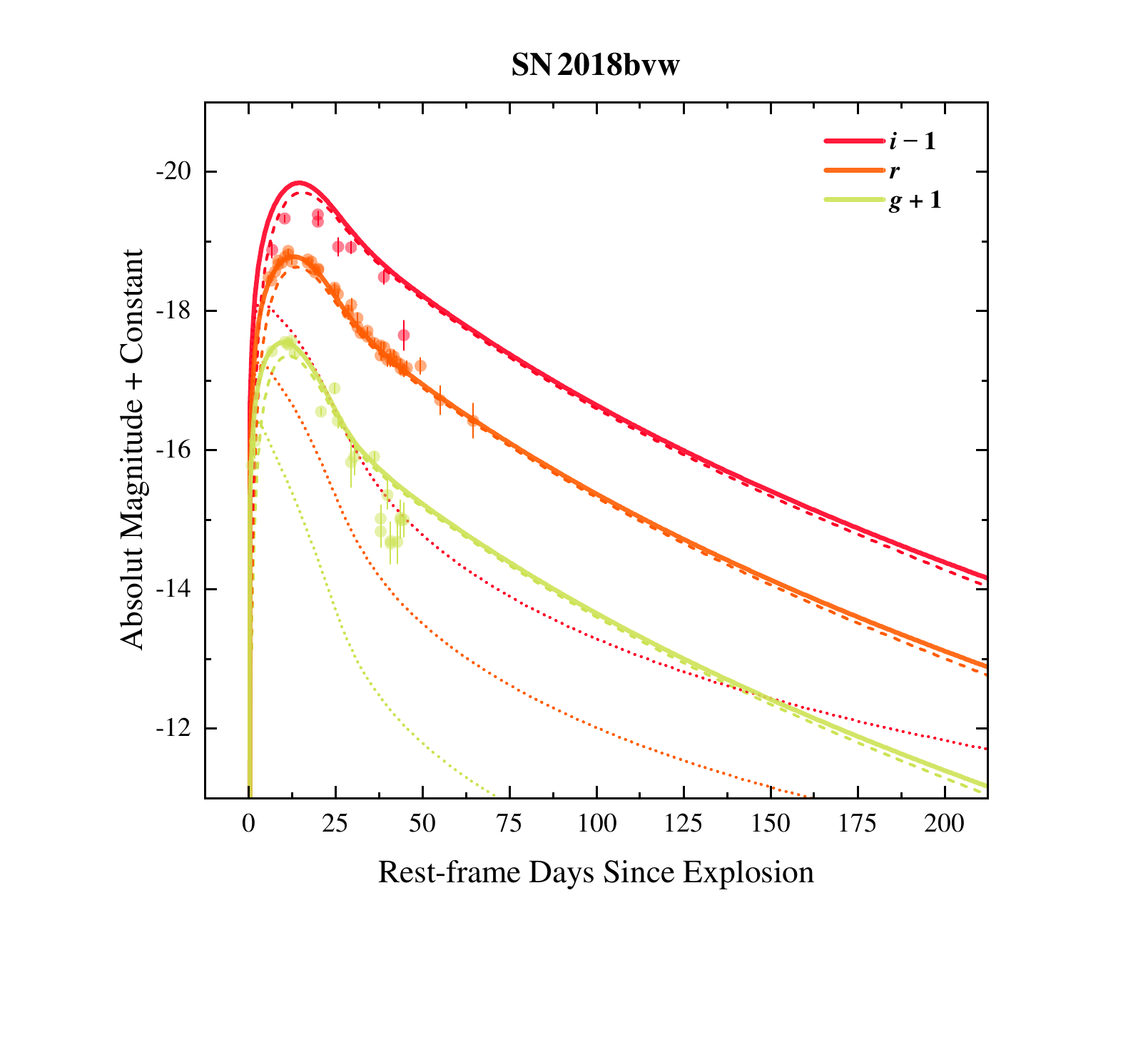}
    \includegraphics[width = 0.32\linewidth , trim = 80 65 93 35, clip]{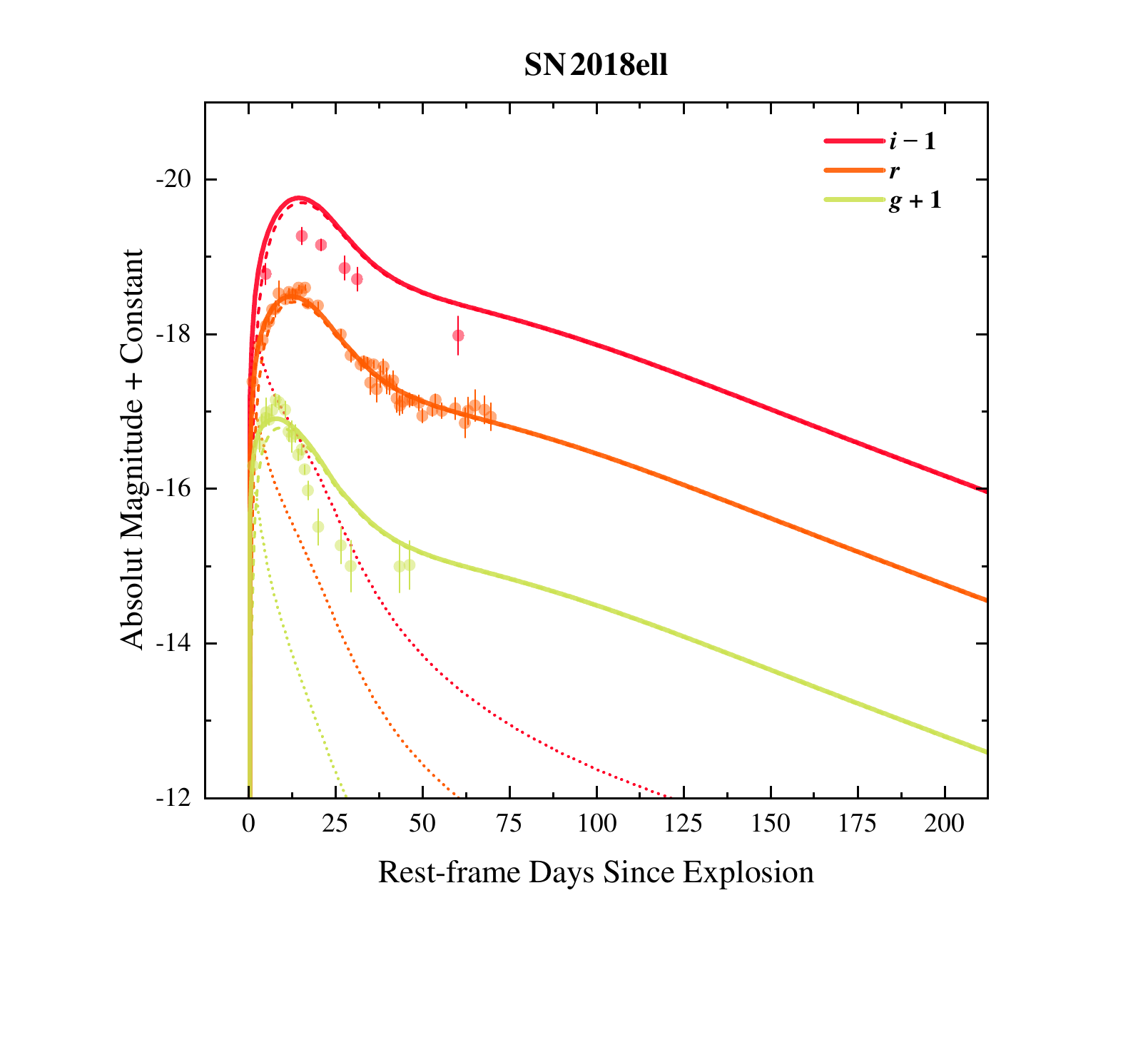}
    \includegraphics[width = 0.32\linewidth , trim = 80 65 93 35, clip]{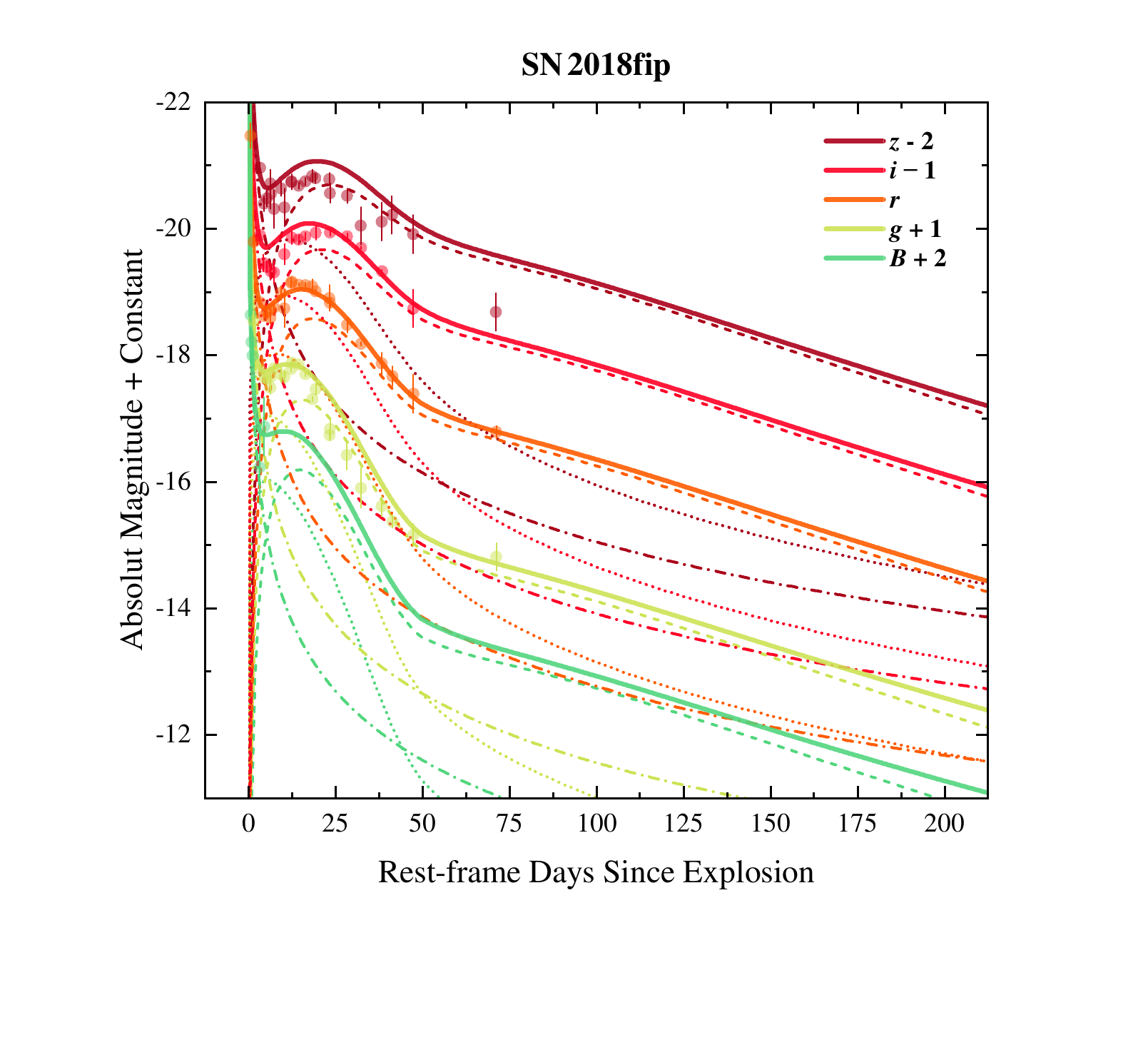}
    \includegraphics[width = 0.32\linewidth , trim = 80 65 93 35, clip]{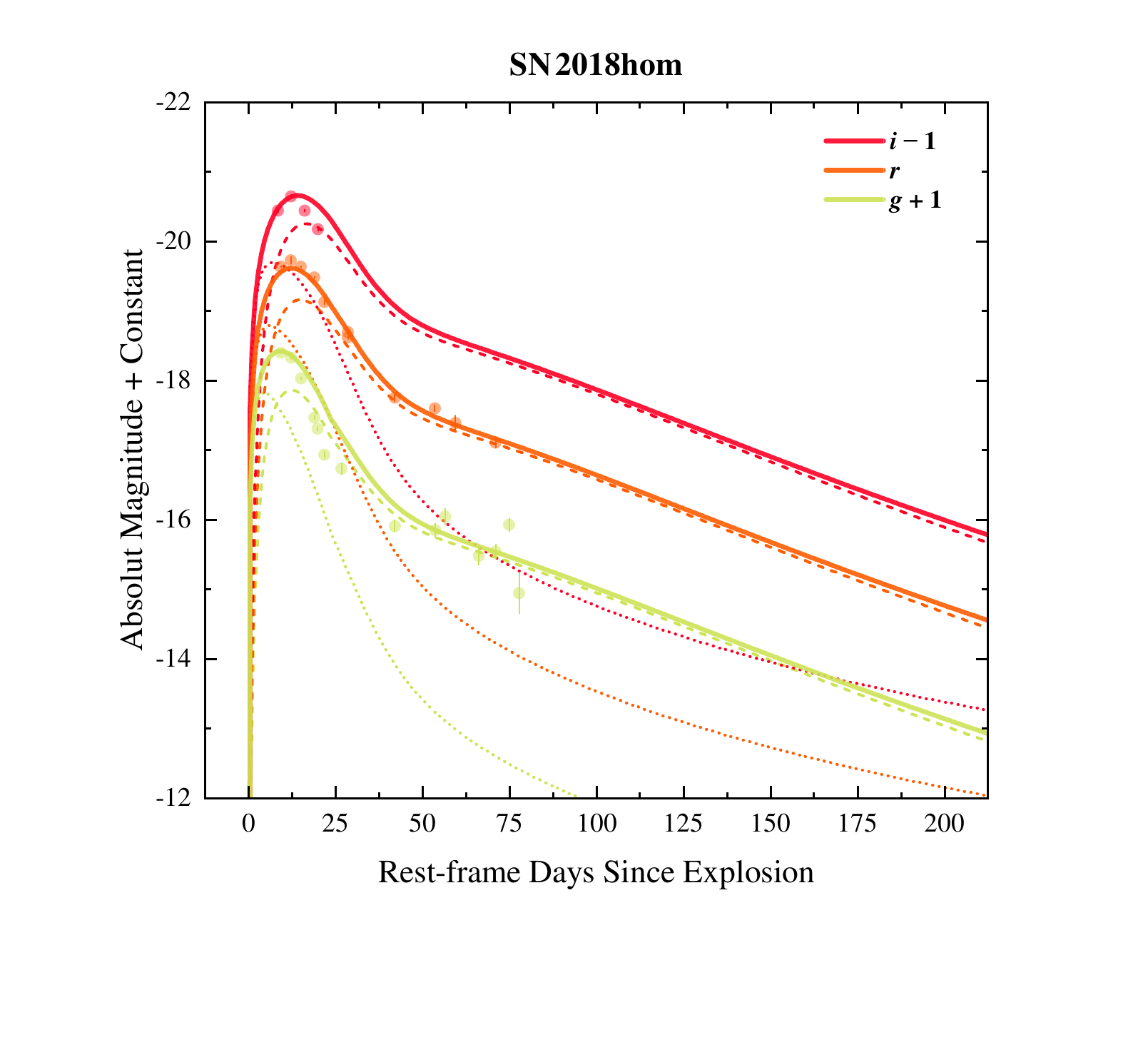}
    \includegraphics[width = 0.32\linewidth , trim = 80 65 93 35, clip]{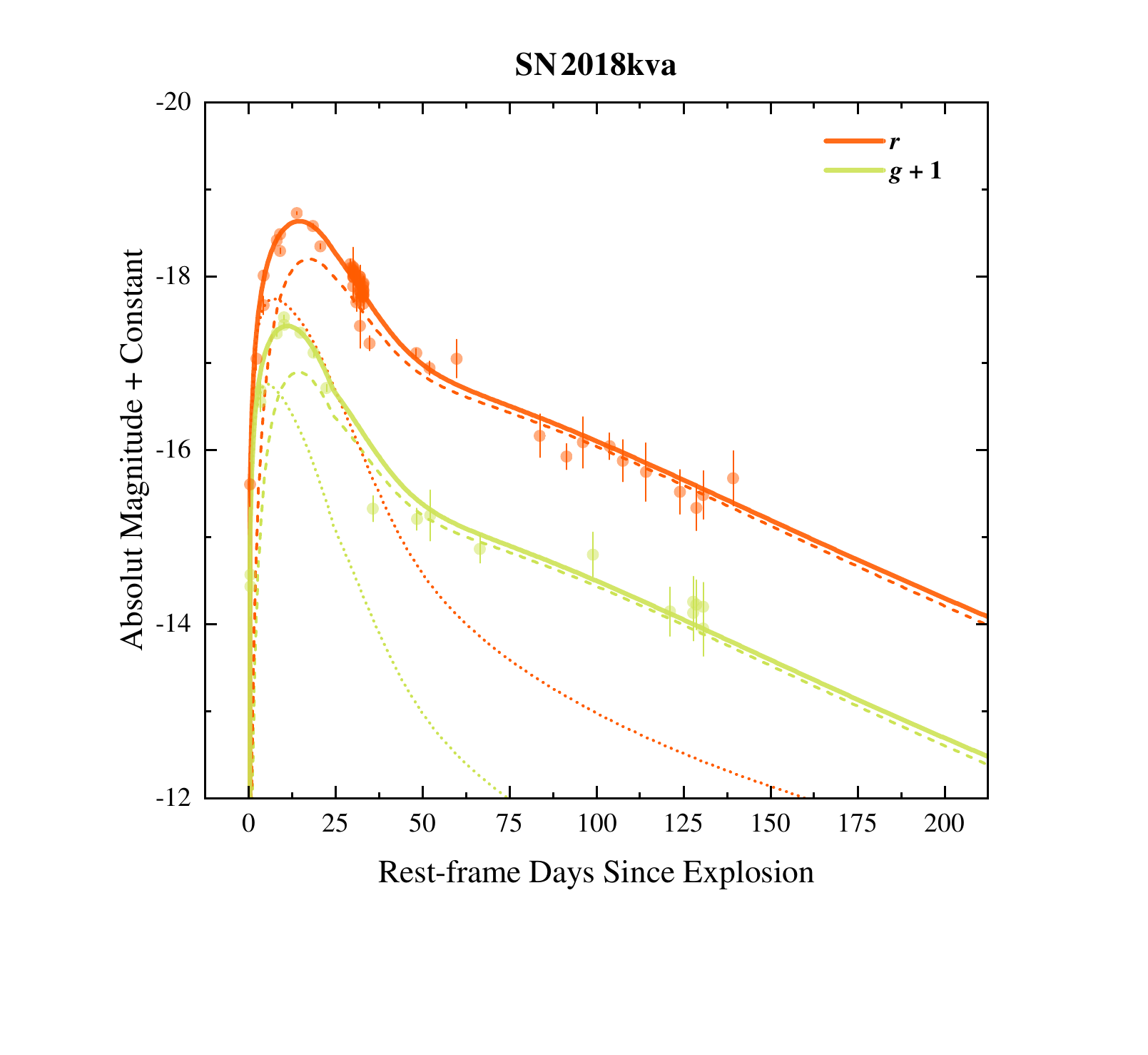}
    \caption{(Continued.)}
\end{figure*}

\begin{figure*}
    \ContinuedFloat
    \centering
    \includegraphics[width = 0.32\linewidth , trim = 80 65 93 35, clip]{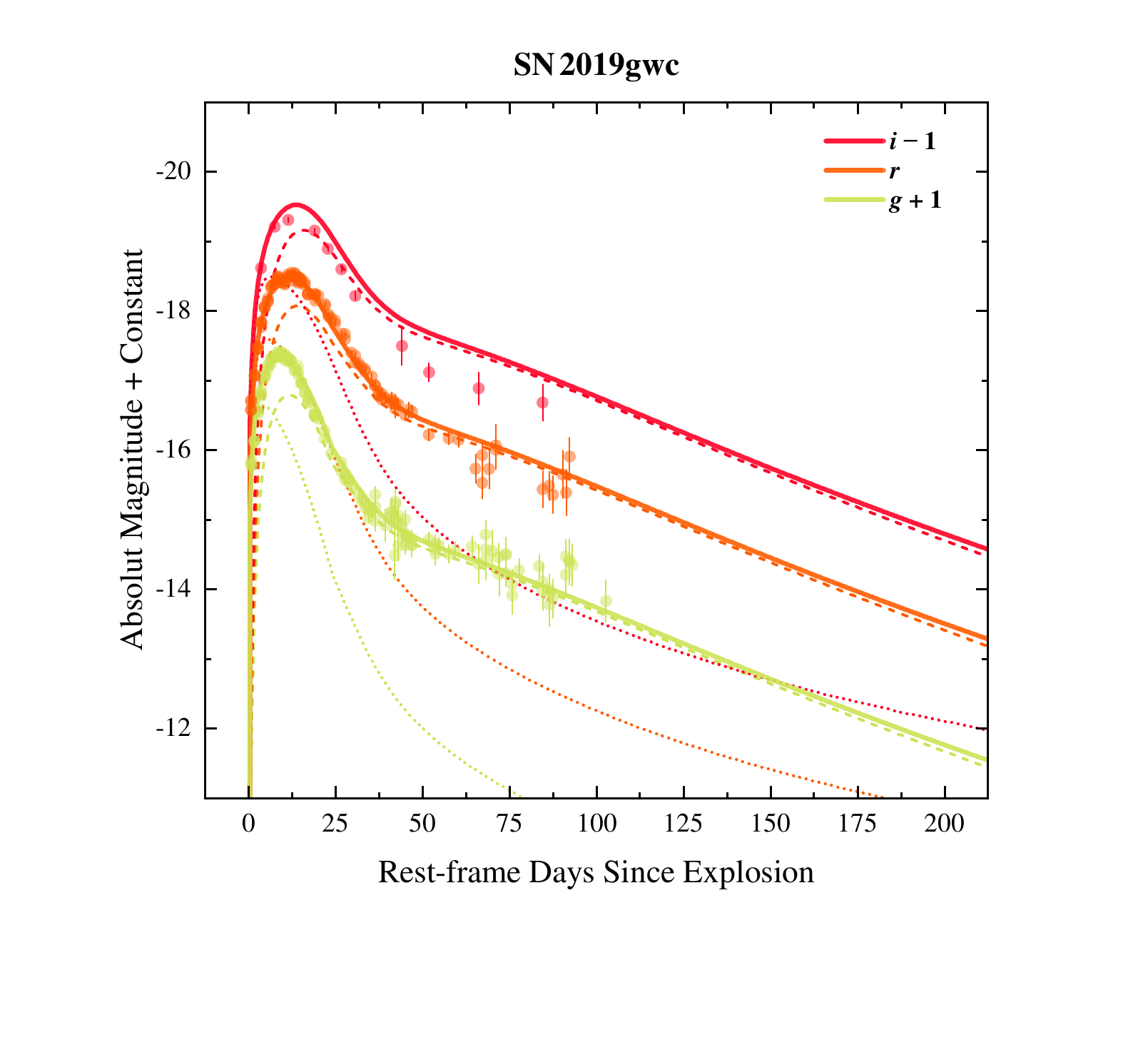}
    \includegraphics[width = 0.32\linewidth , trim = 80 65 93 35, clip]{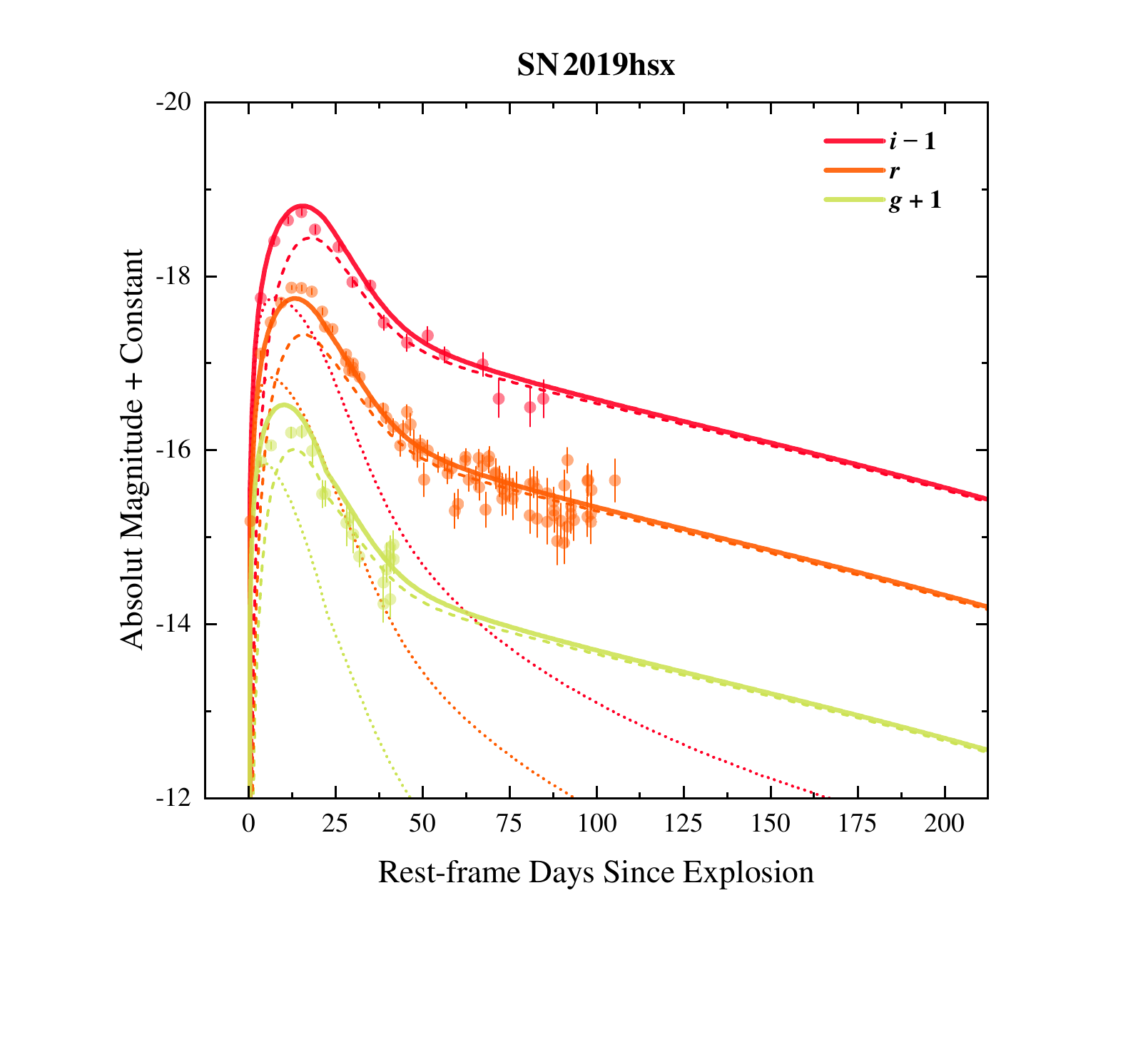}
    \includegraphics[width = 0.32\linewidth , trim = 80 65 93 35, clip]{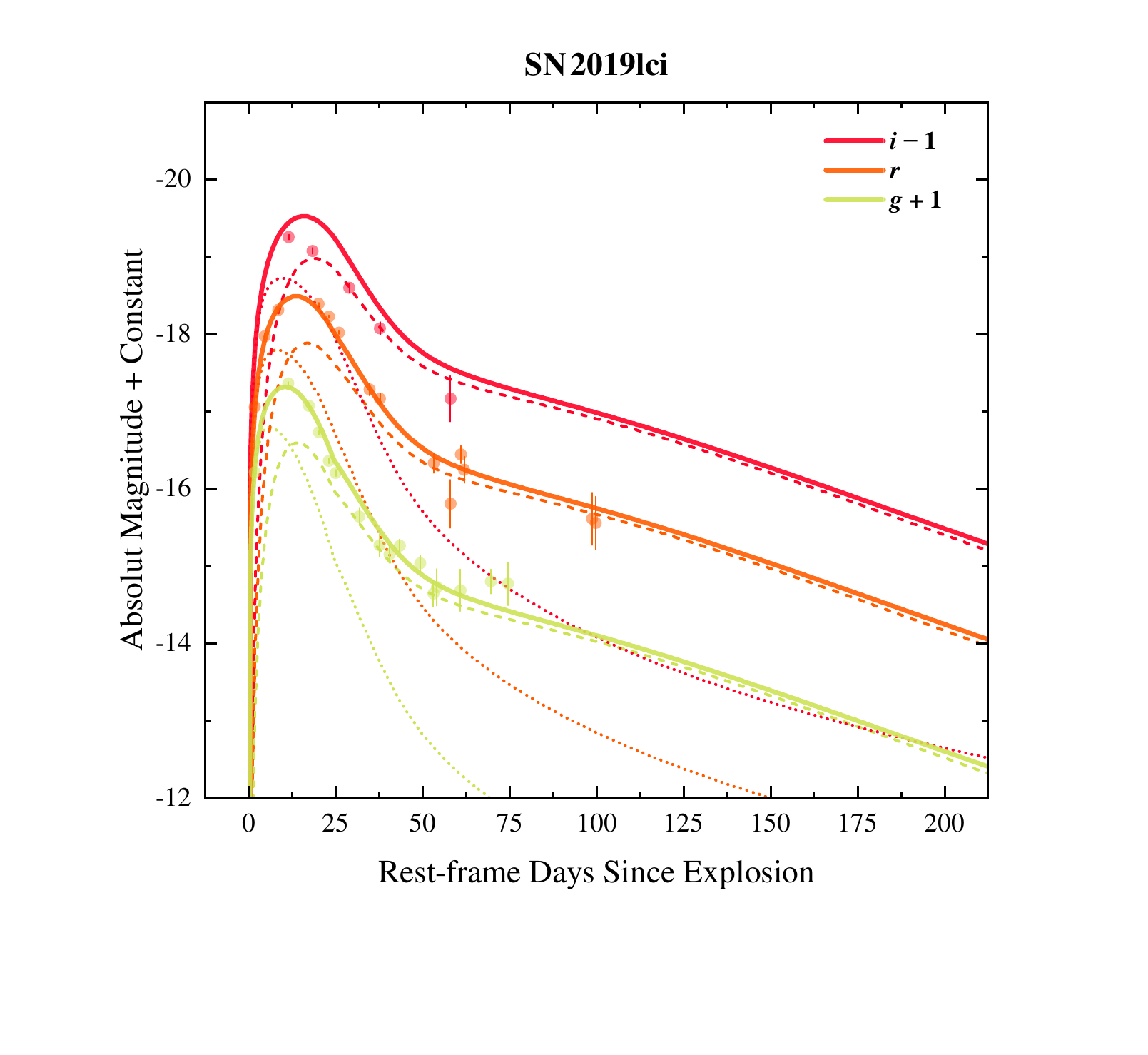}
    \includegraphics[width = 0.32\linewidth , trim = 80 65 93 35, clip]{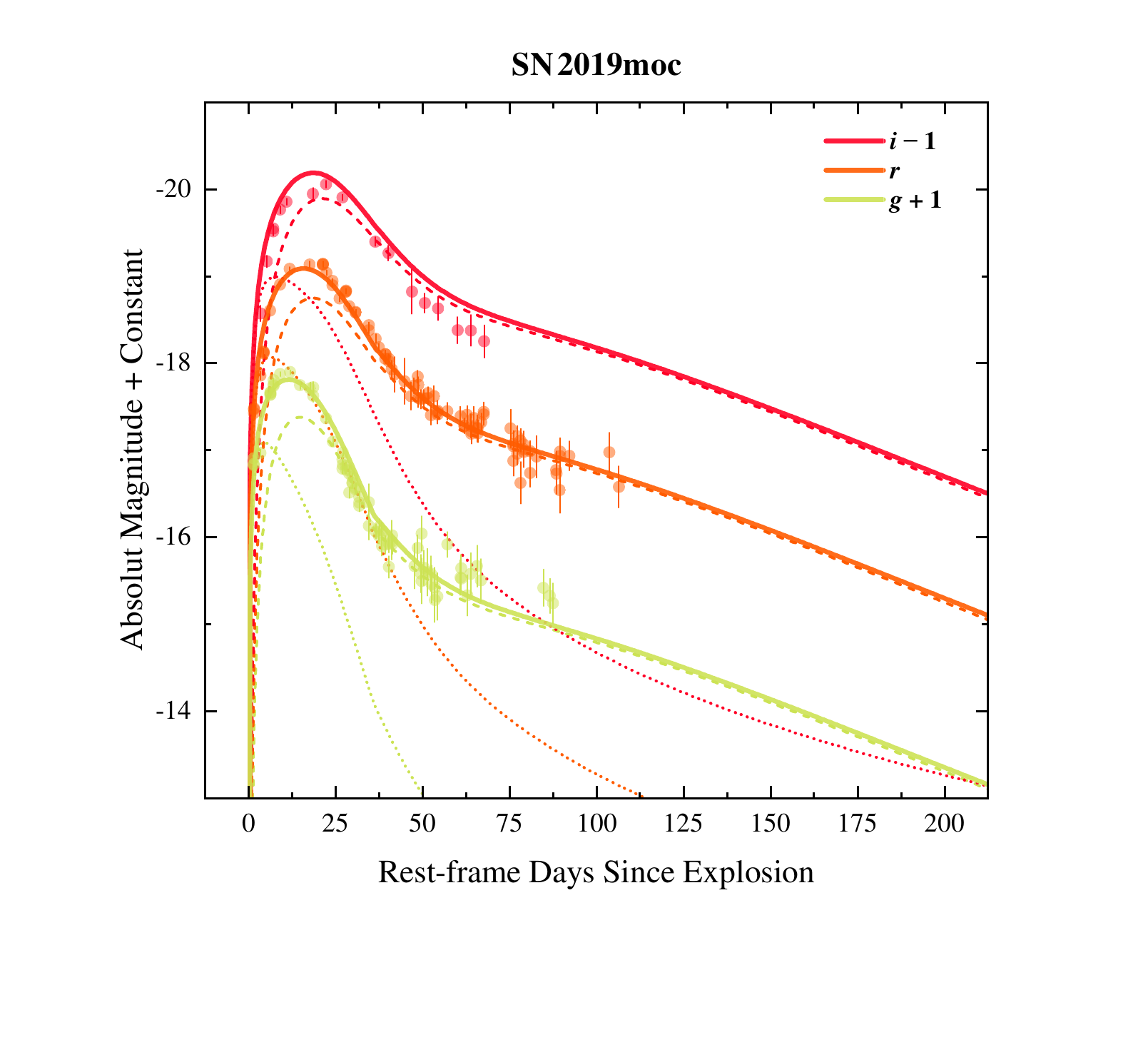}
    \includegraphics[width = 0.32\linewidth , trim = 80 65 93 35, clip]{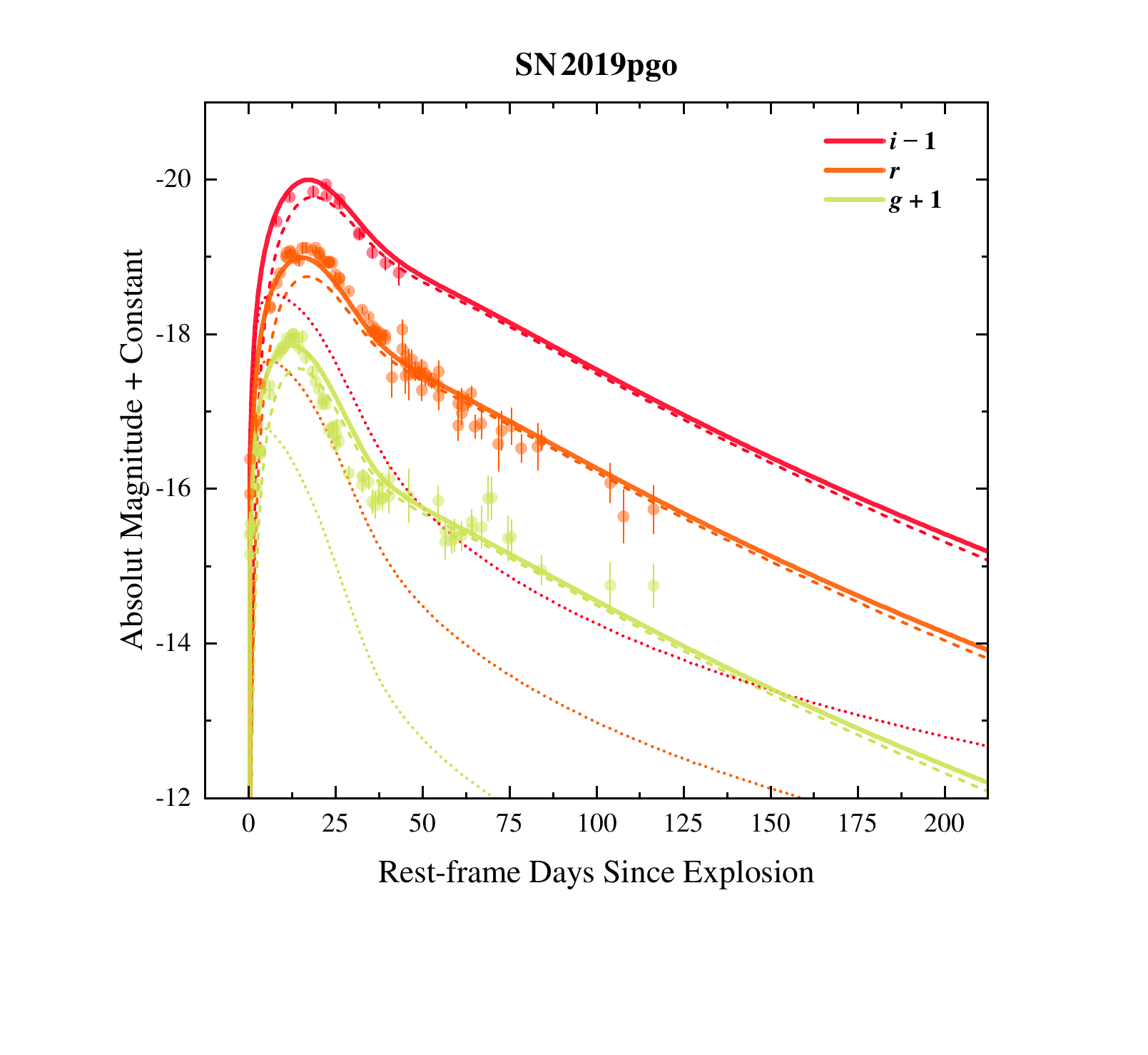}
    \includegraphics[width = 0.32\linewidth , trim = 80 65 93 35, clip]{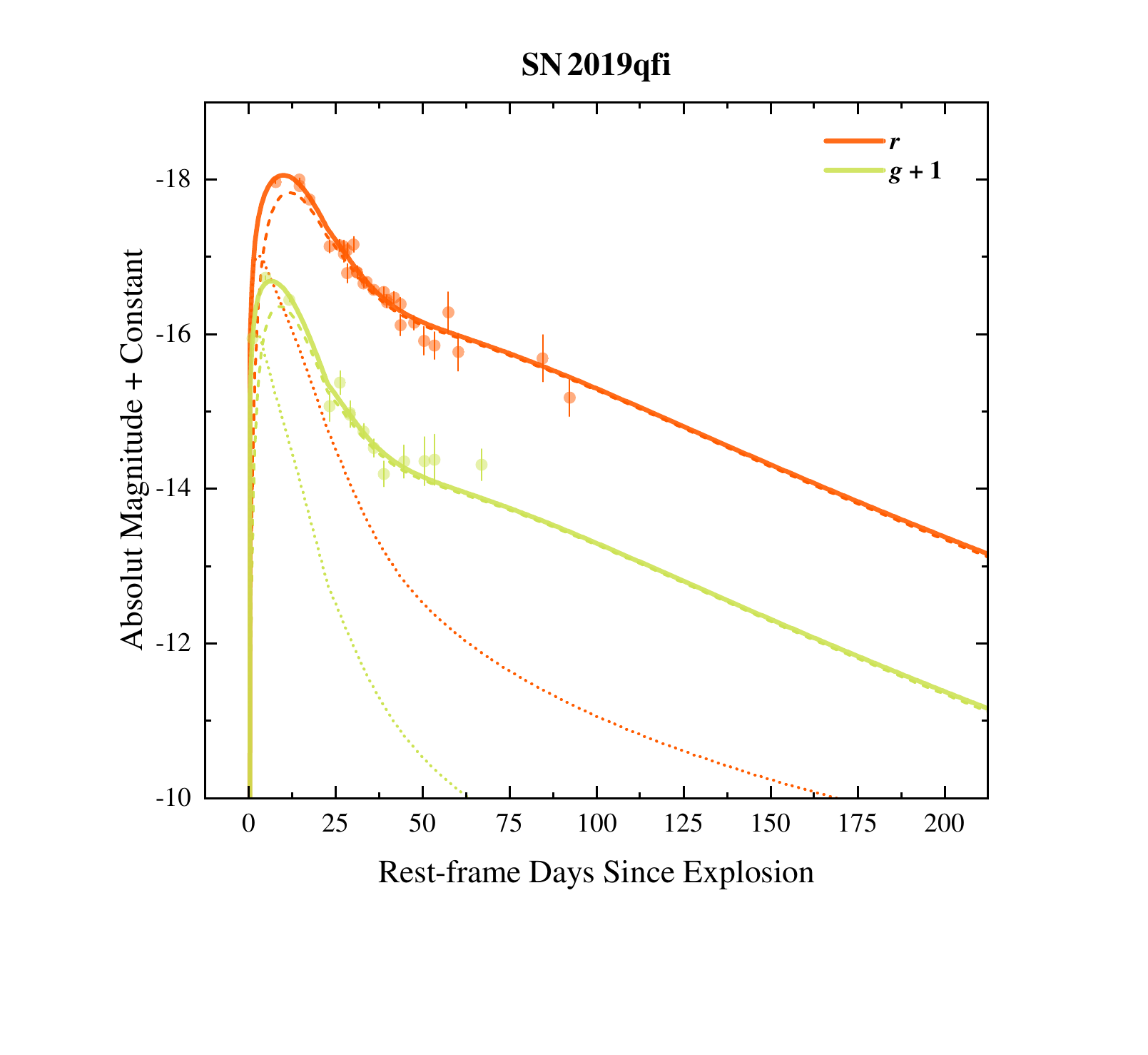}
    \includegraphics[width = 0.32\linewidth , trim = 80 65 93 35, clip]{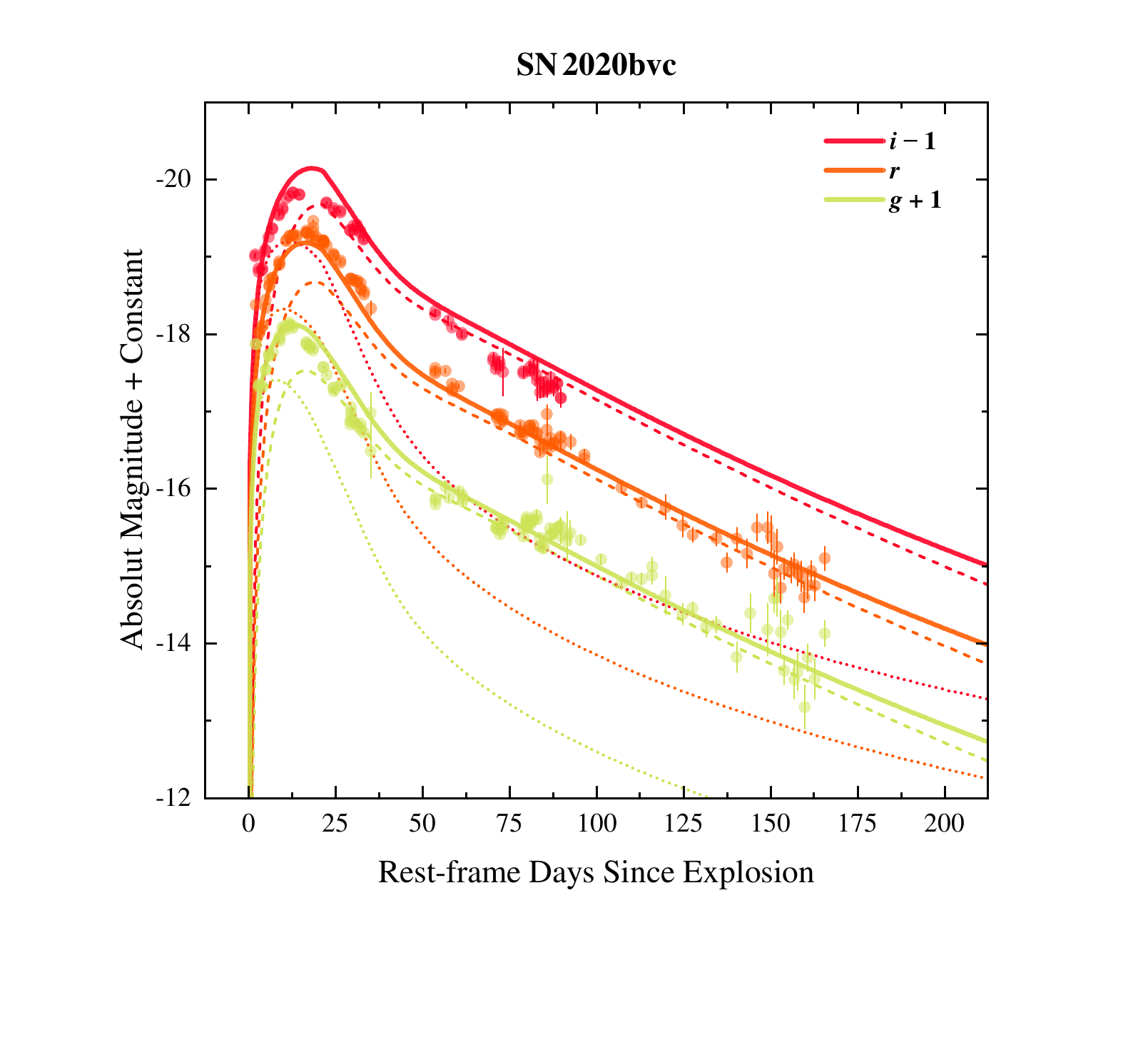}
    \includegraphics[width = 0.32\linewidth , trim = 80 65 93 35, clip]{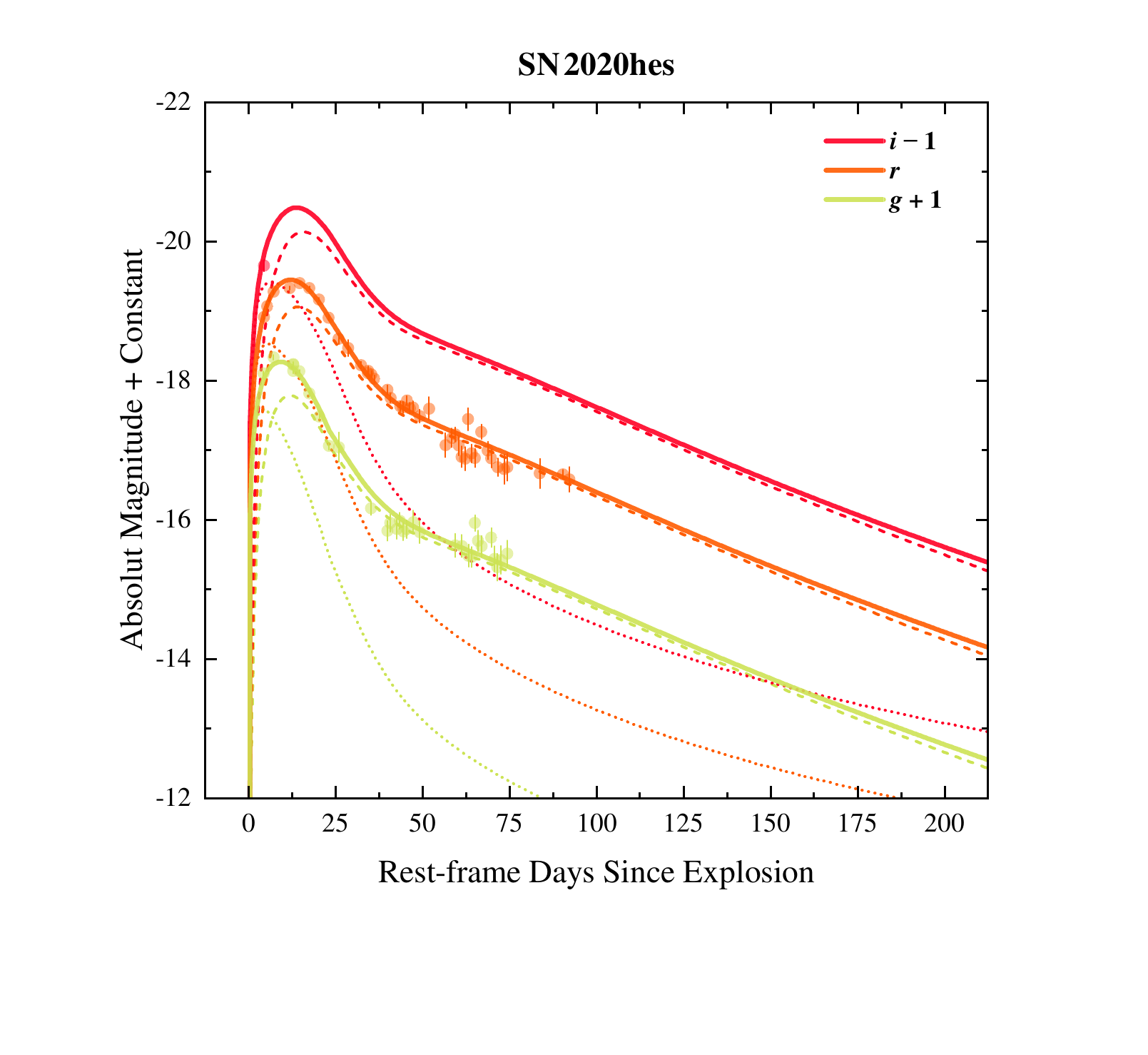}
    \includegraphics[width = 0.32\linewidth , trim = 80 65 93 35, clip]{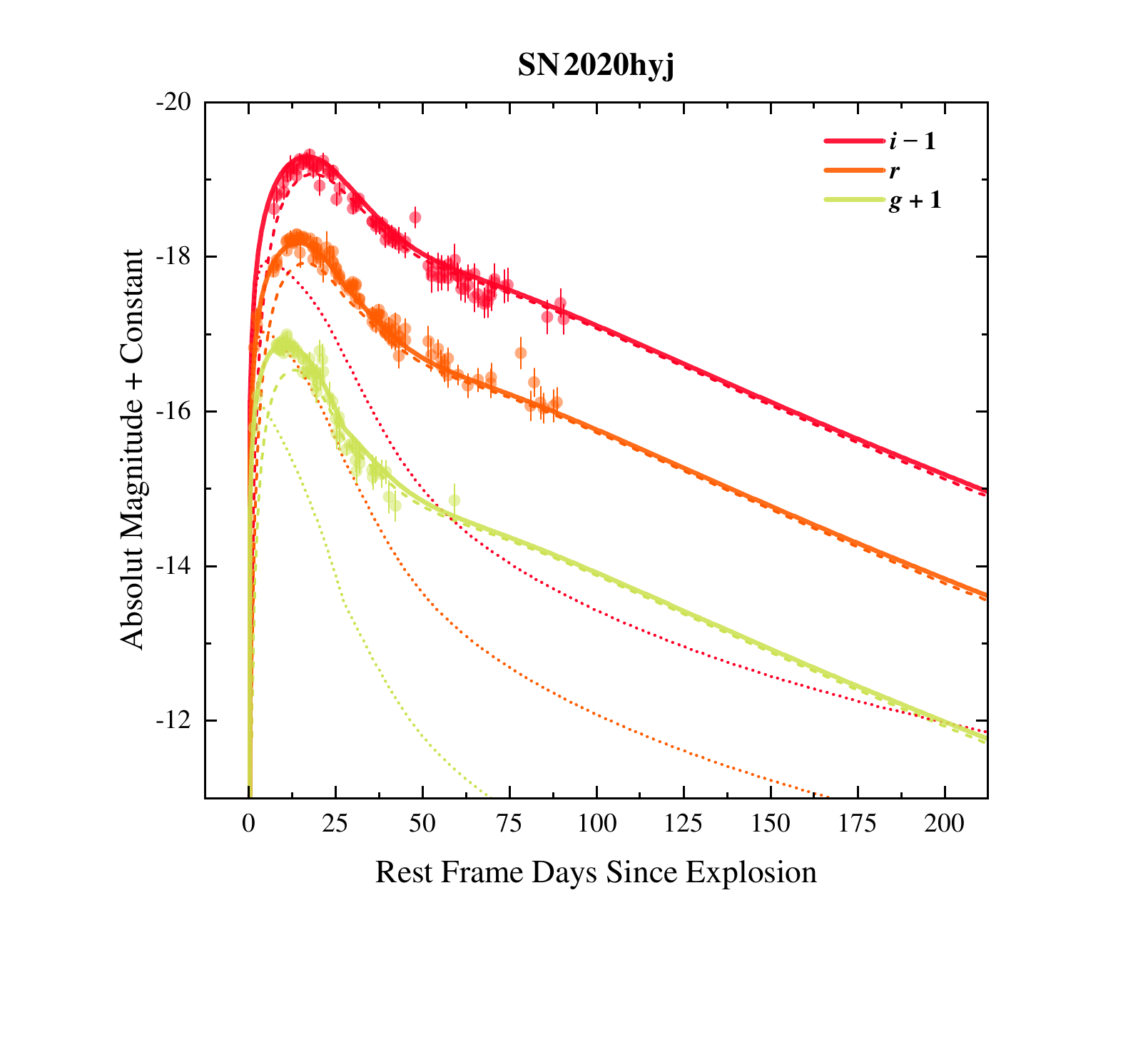}
    \includegraphics[width = 0.32\linewidth , trim = 80 65 93 35, clip]{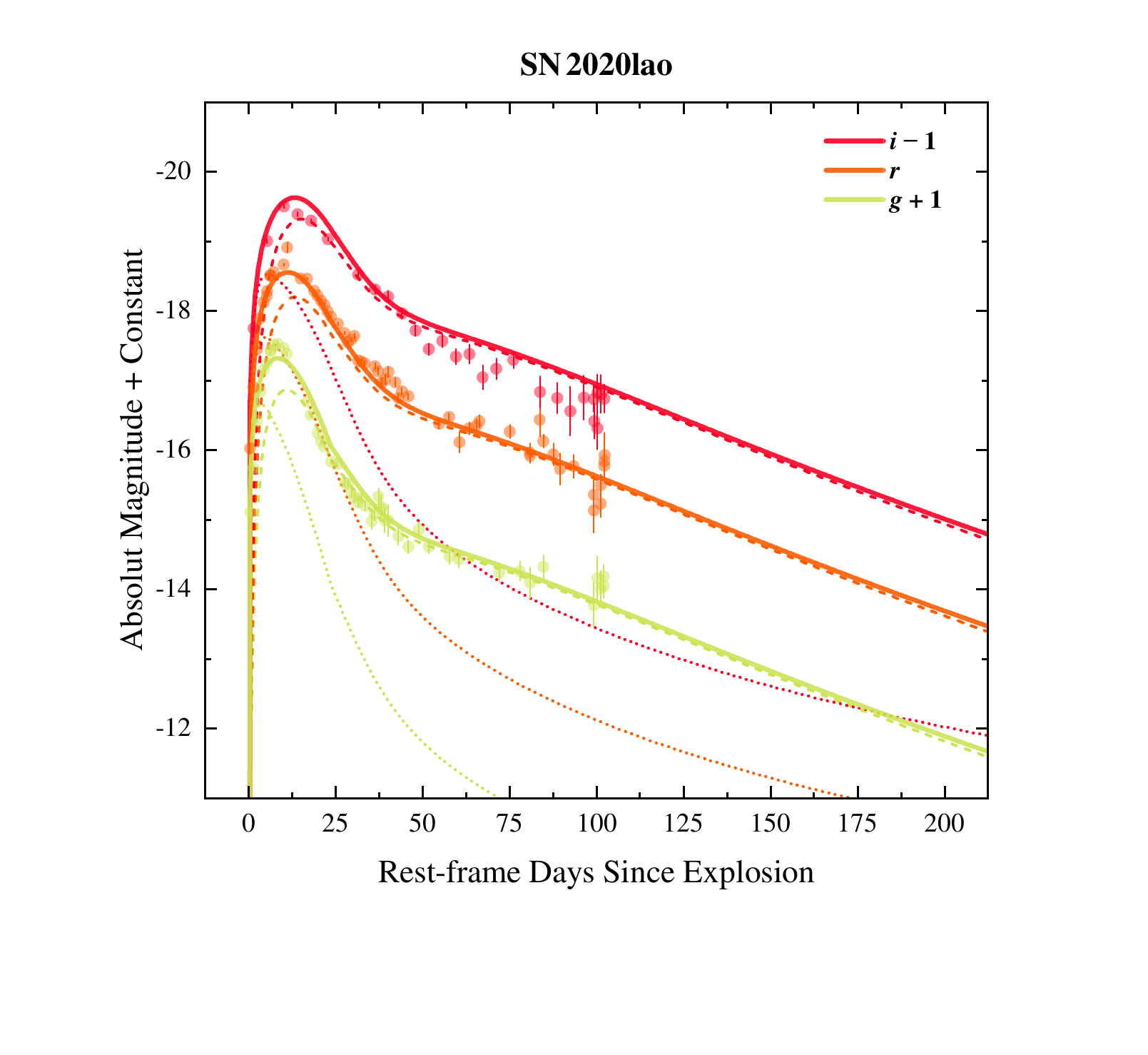}
    \includegraphics[width = 0.32\linewidth , trim = 80 65 93 35, clip]{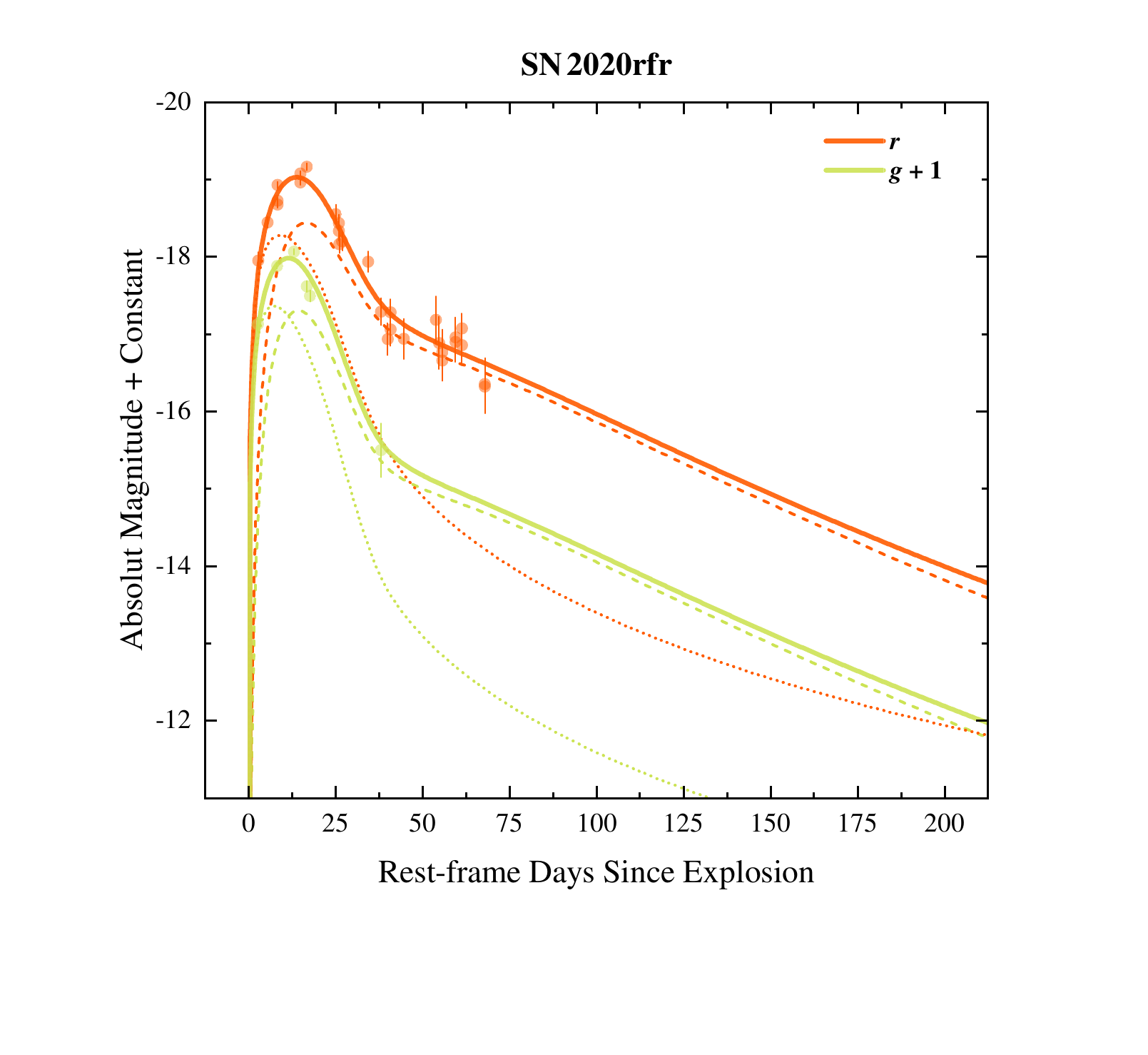}
    \includegraphics[width = 0.32\linewidth , trim = 80 65 93 35, clip]{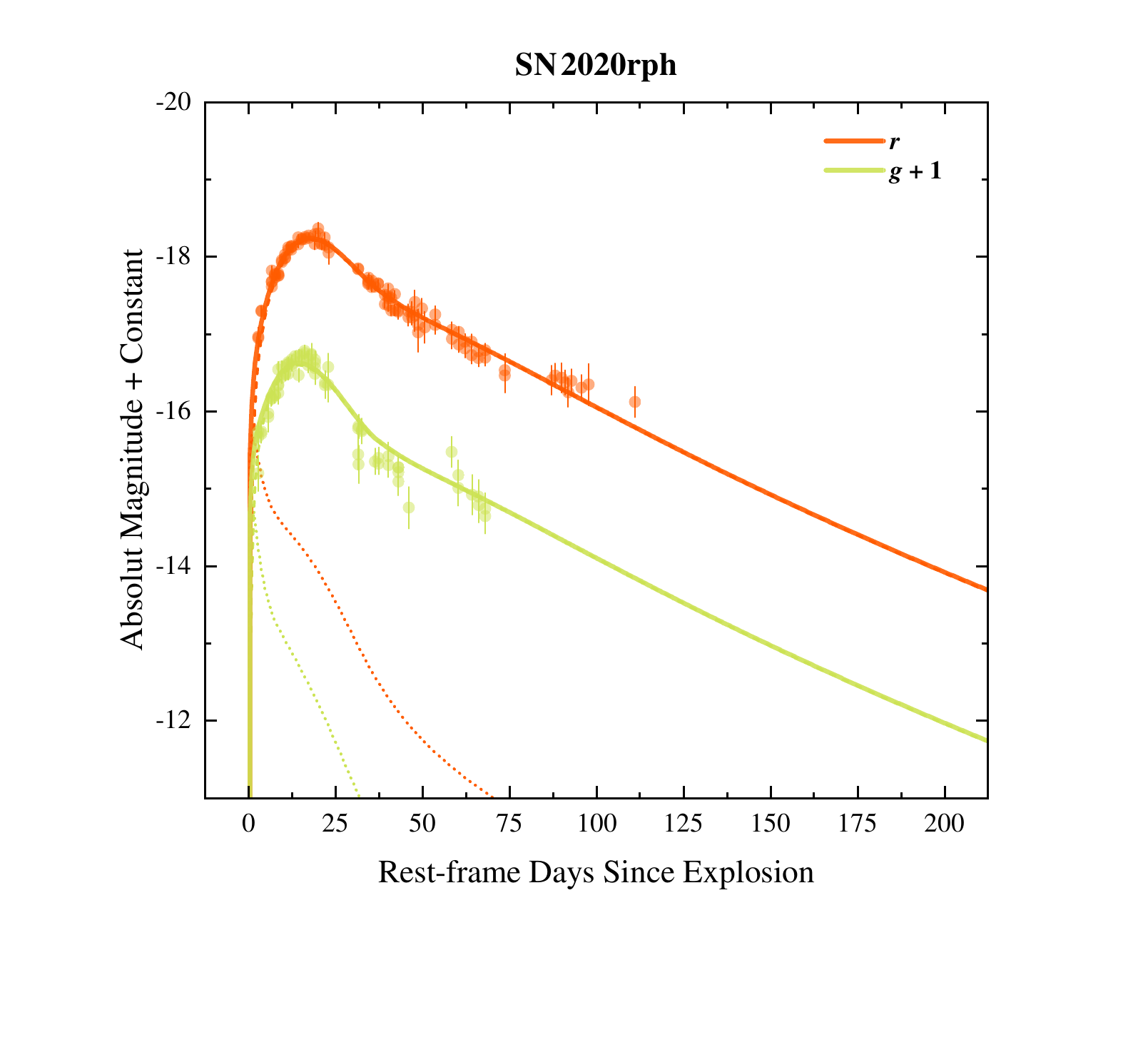}
    \caption{(Continued.)}
\end{figure*}

\begin{figure*}
    \ContinuedFloat
    \centering
    \includegraphics[width = 0.32\linewidth , trim = 80 65 93 35, clip]{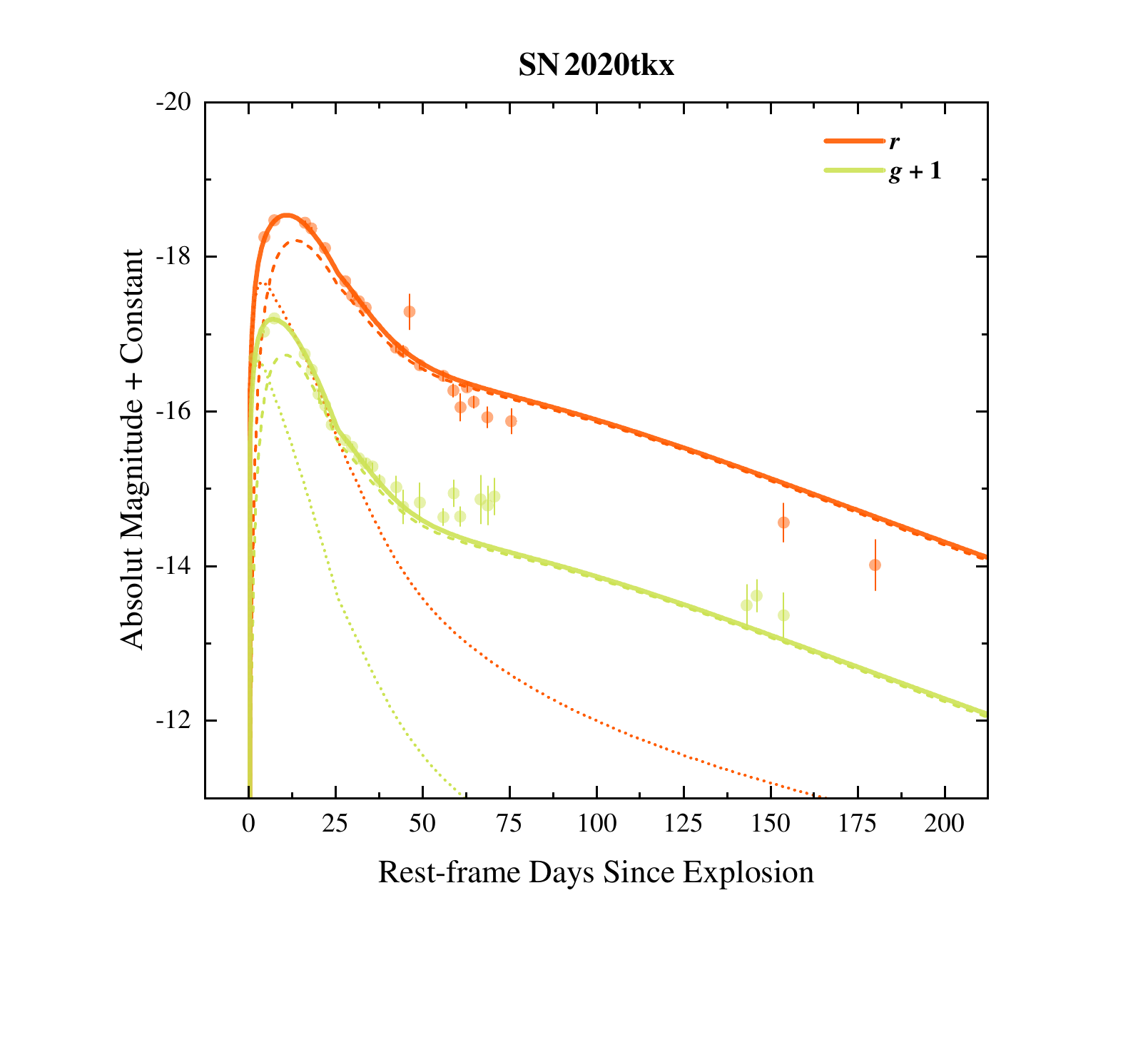}
    \includegraphics[width = 0.32\linewidth , trim = 80 65 93 35, clip]{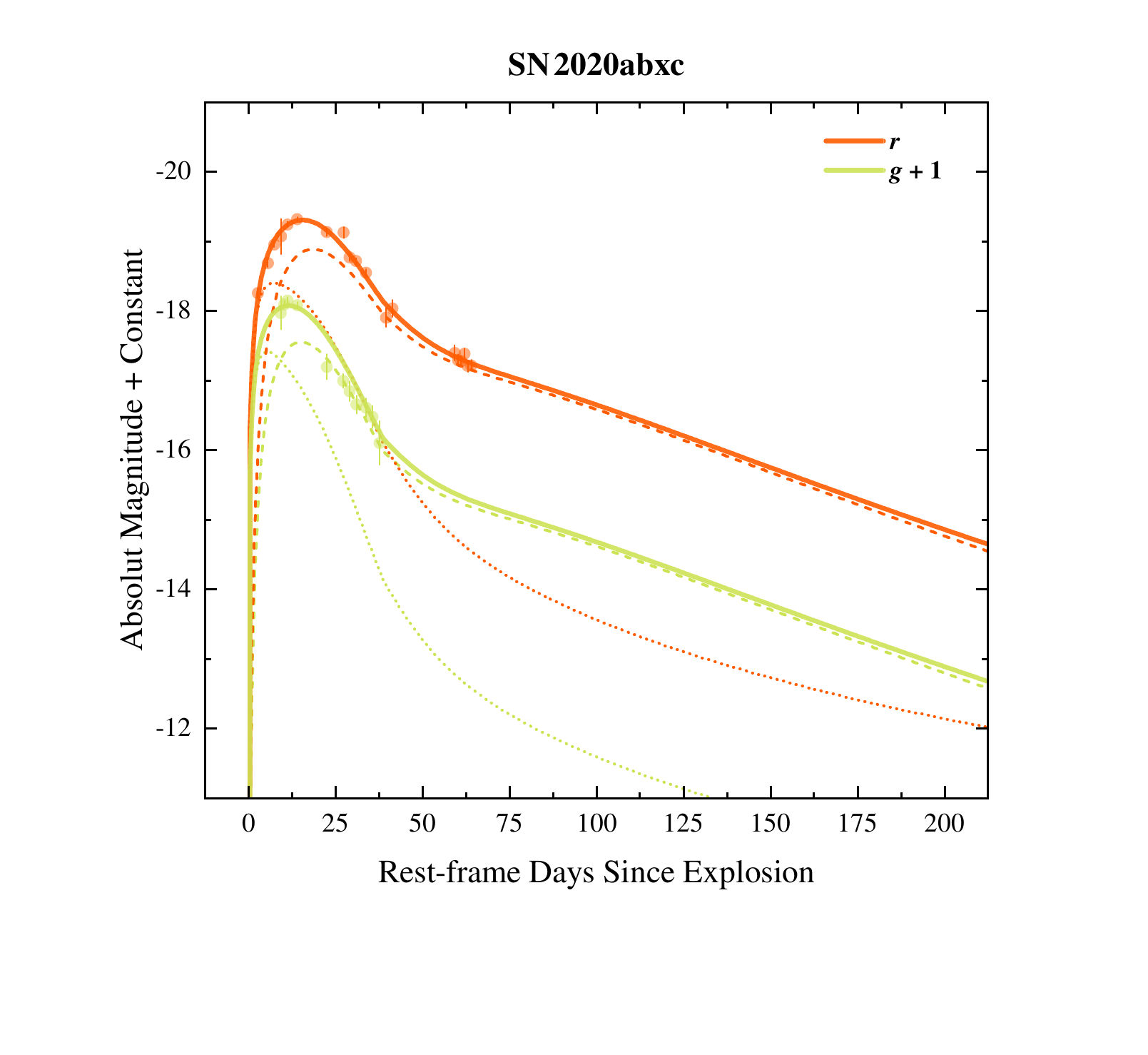}
    \includegraphics[width = 0.32\linewidth , trim = 80 65 93 35, clip]{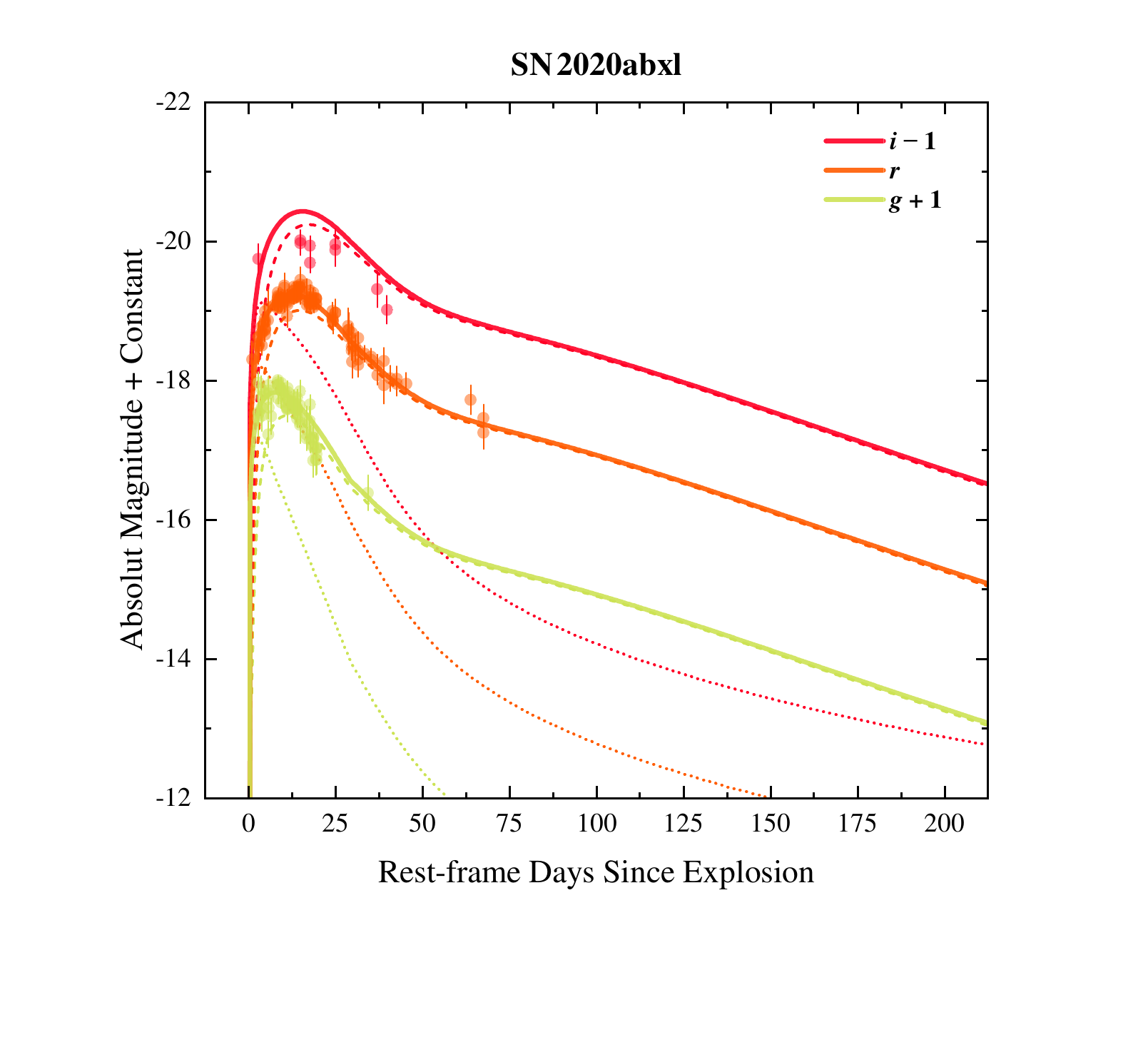}
    \includegraphics[width = 0.32\linewidth , trim = 80 65 93 35, clip]{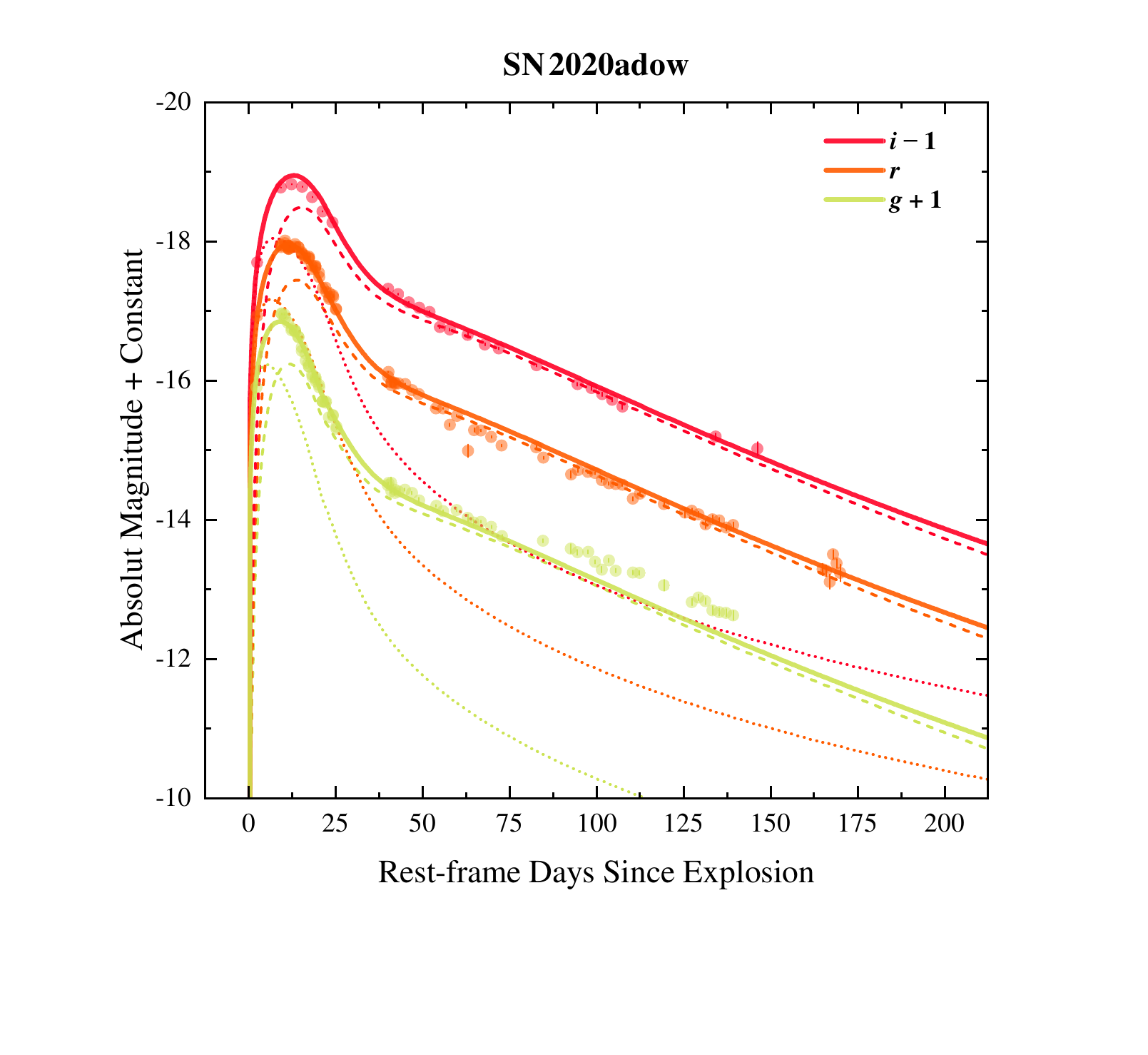}
    \includegraphics[width = 0.32\linewidth , trim = 80 65 93 35, clip]{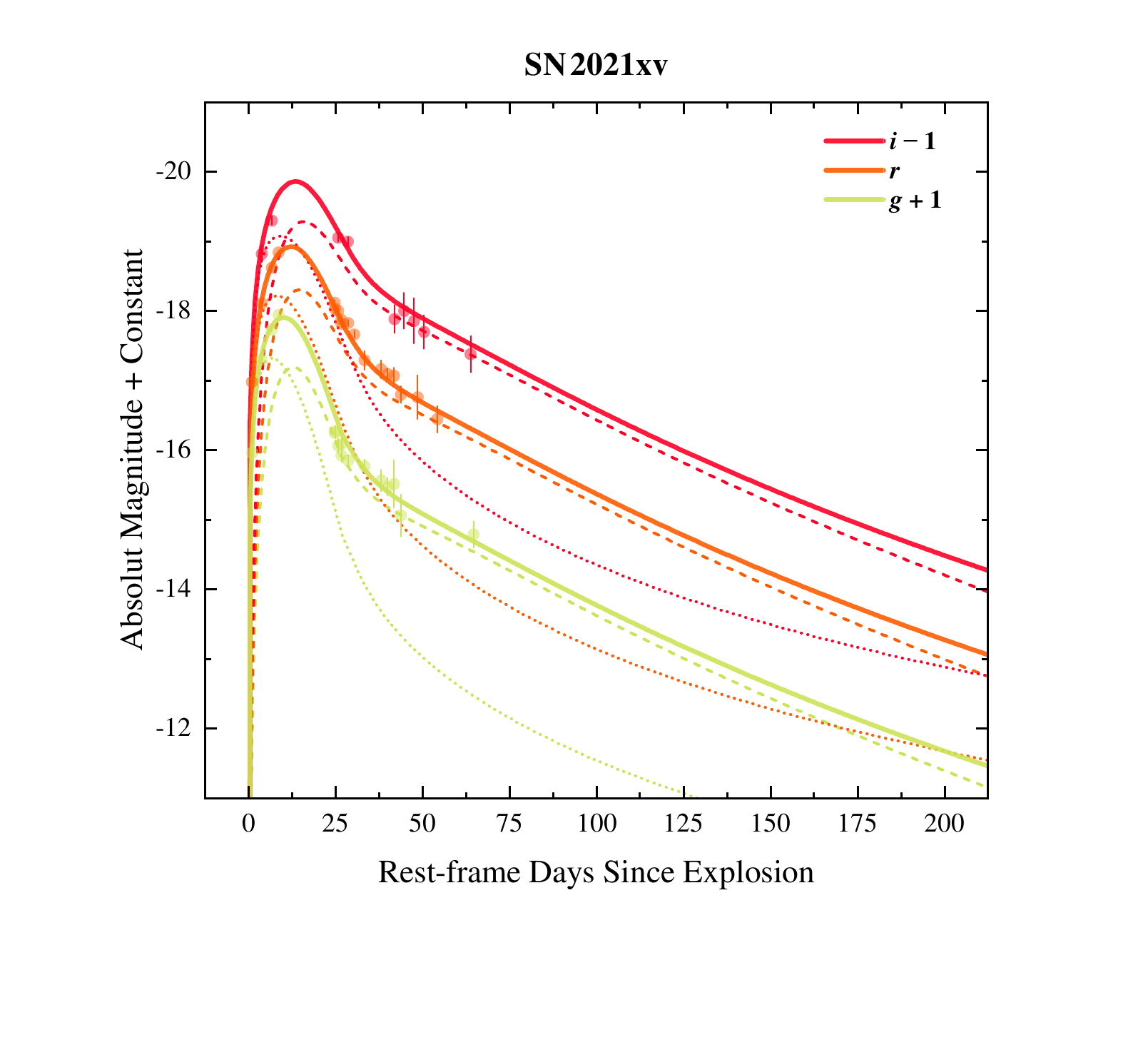}
    \includegraphics[width = 0.32\linewidth , trim = 80 65 93 35, clip]{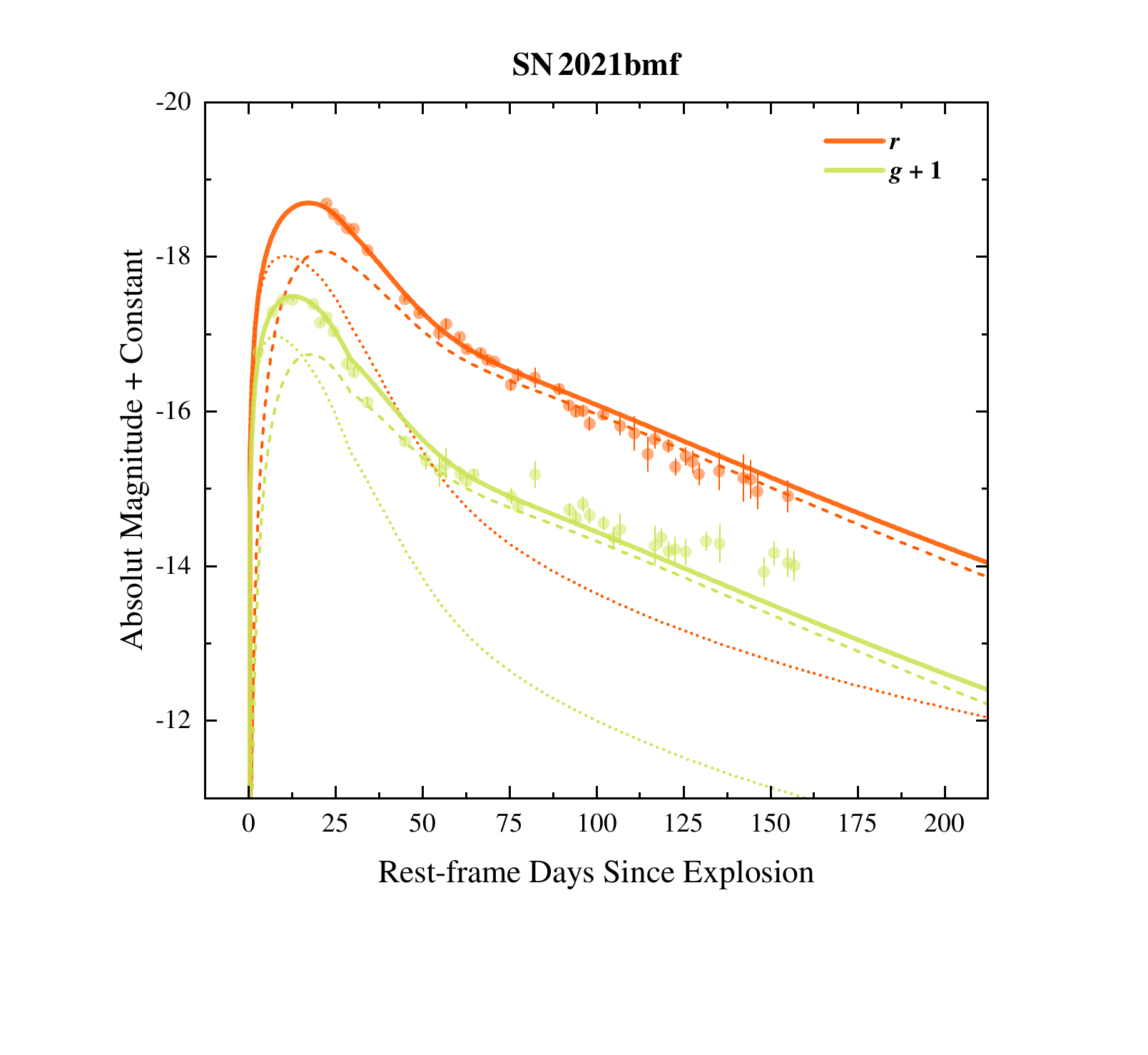}
    \includegraphics[width = 0.32\linewidth , trim = 80 65 93 35, clip]{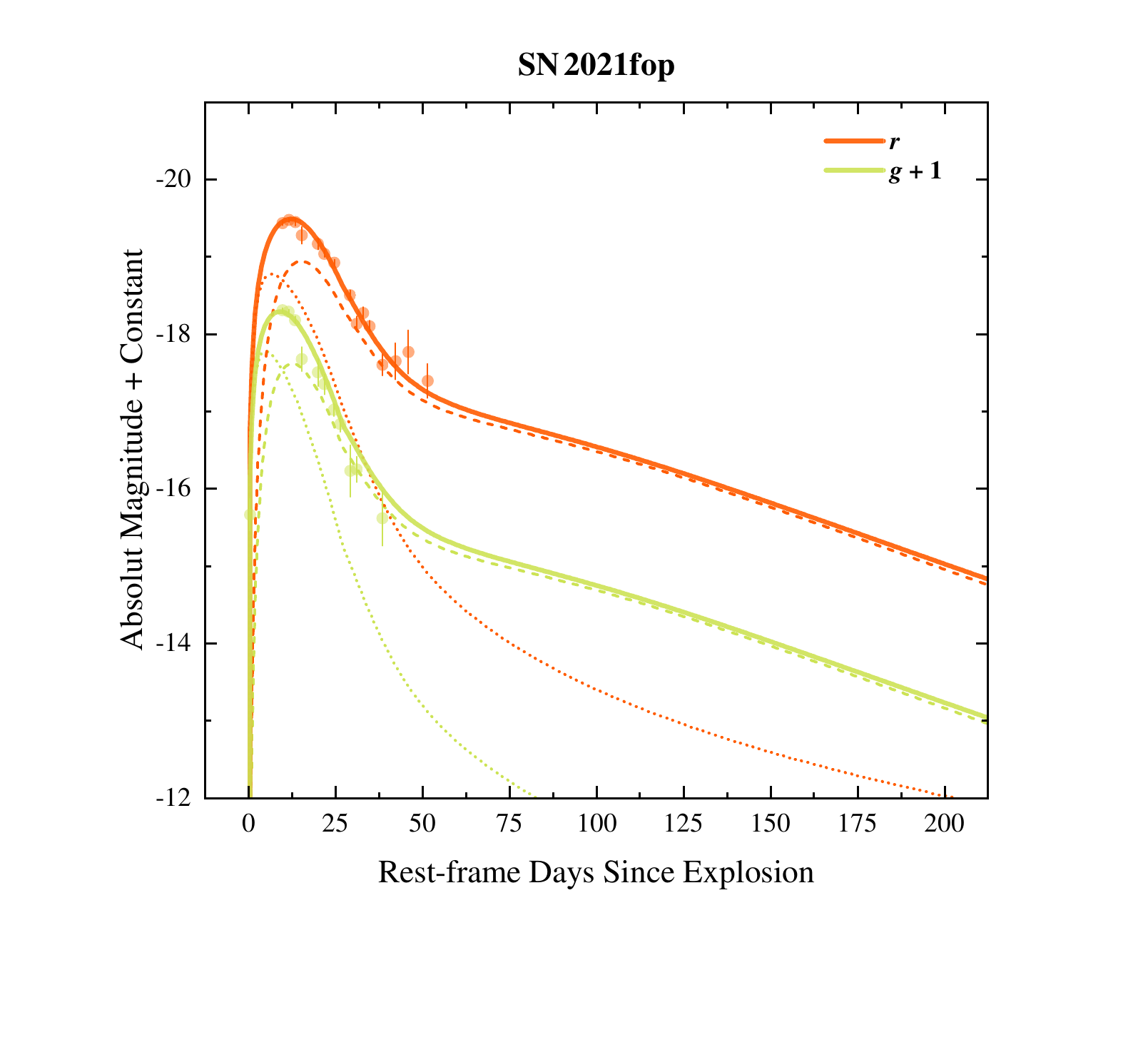}
    \includegraphics[width = 0.32\linewidth , trim = 80 65 93 35, clip]{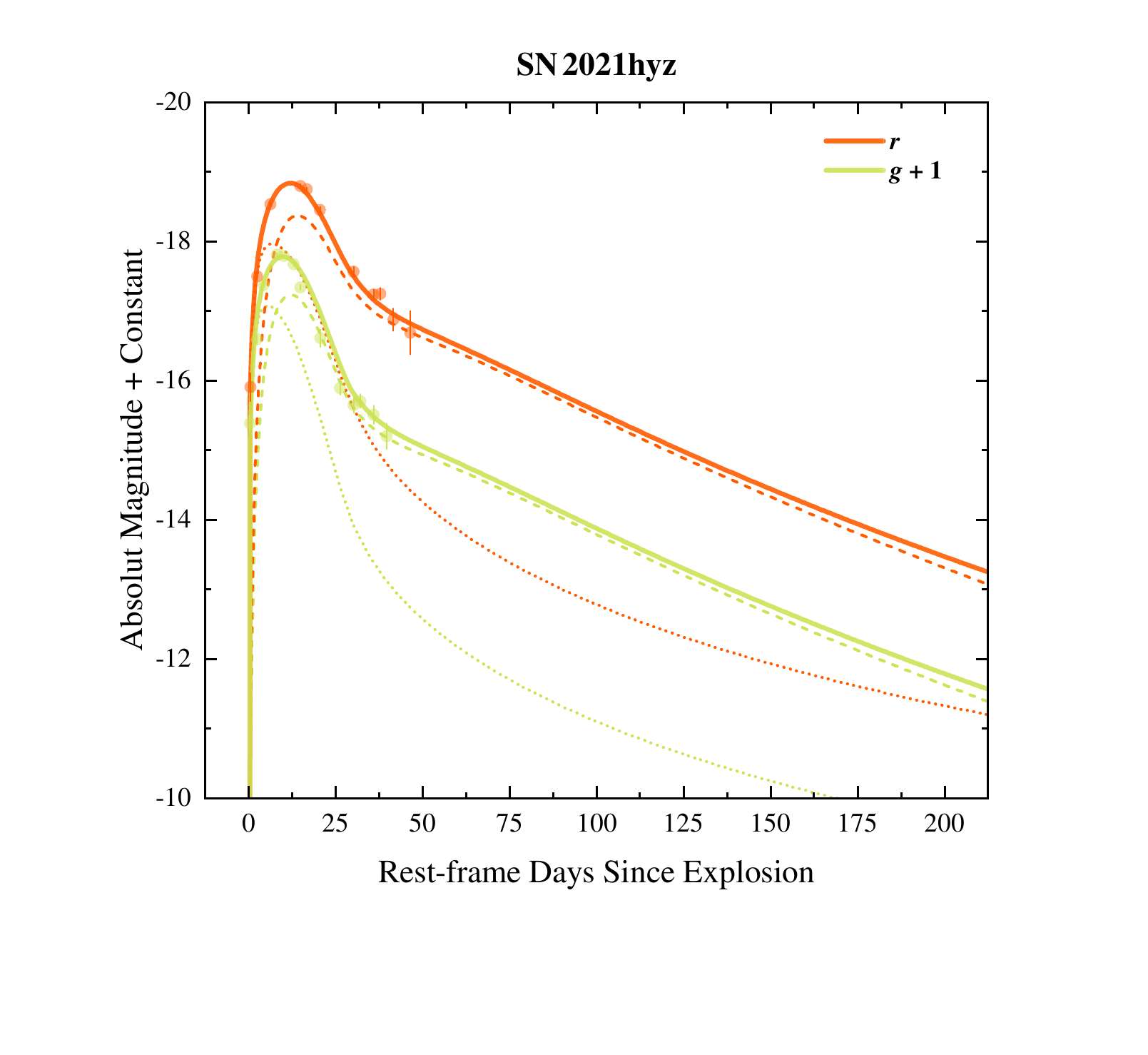}
    \includegraphics[width = 0.32\linewidth , trim = 80 65 93 35, clip]{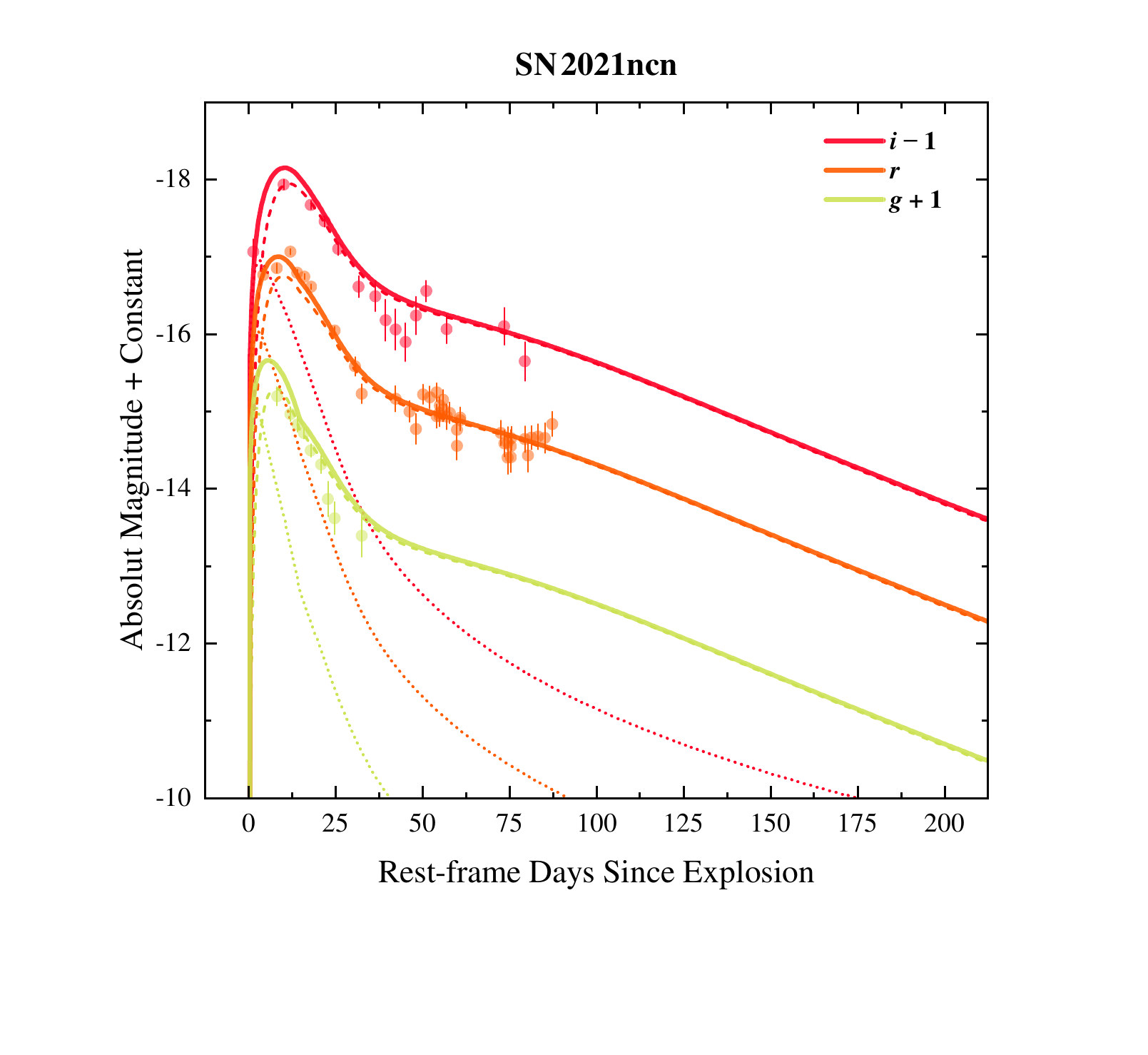}
    \includegraphics[width = 0.32\linewidth , trim = 80 65 93 35, clip]{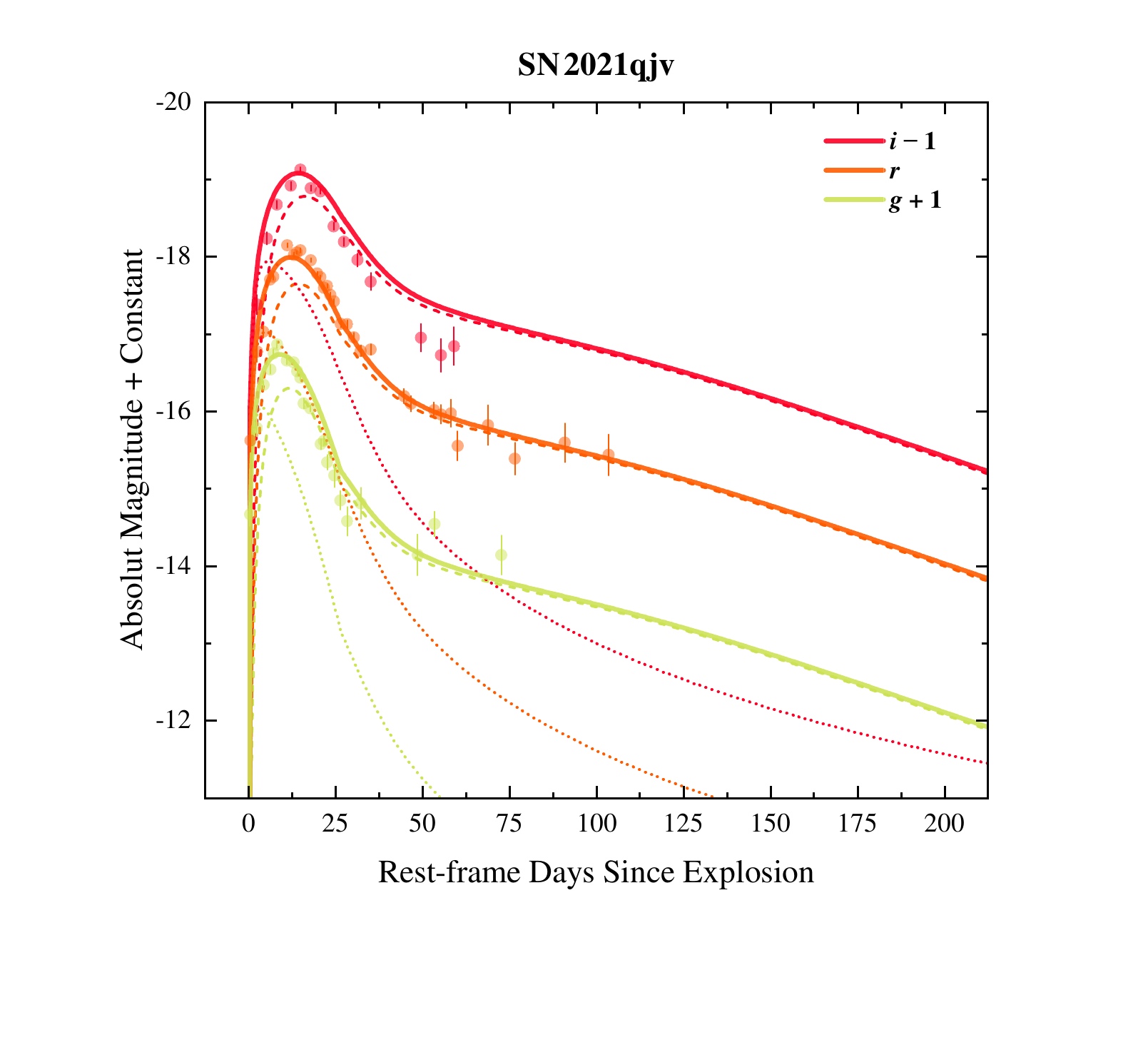}
    \includegraphics[width = 0.32\linewidth , trim = 80 65 93 35, clip]{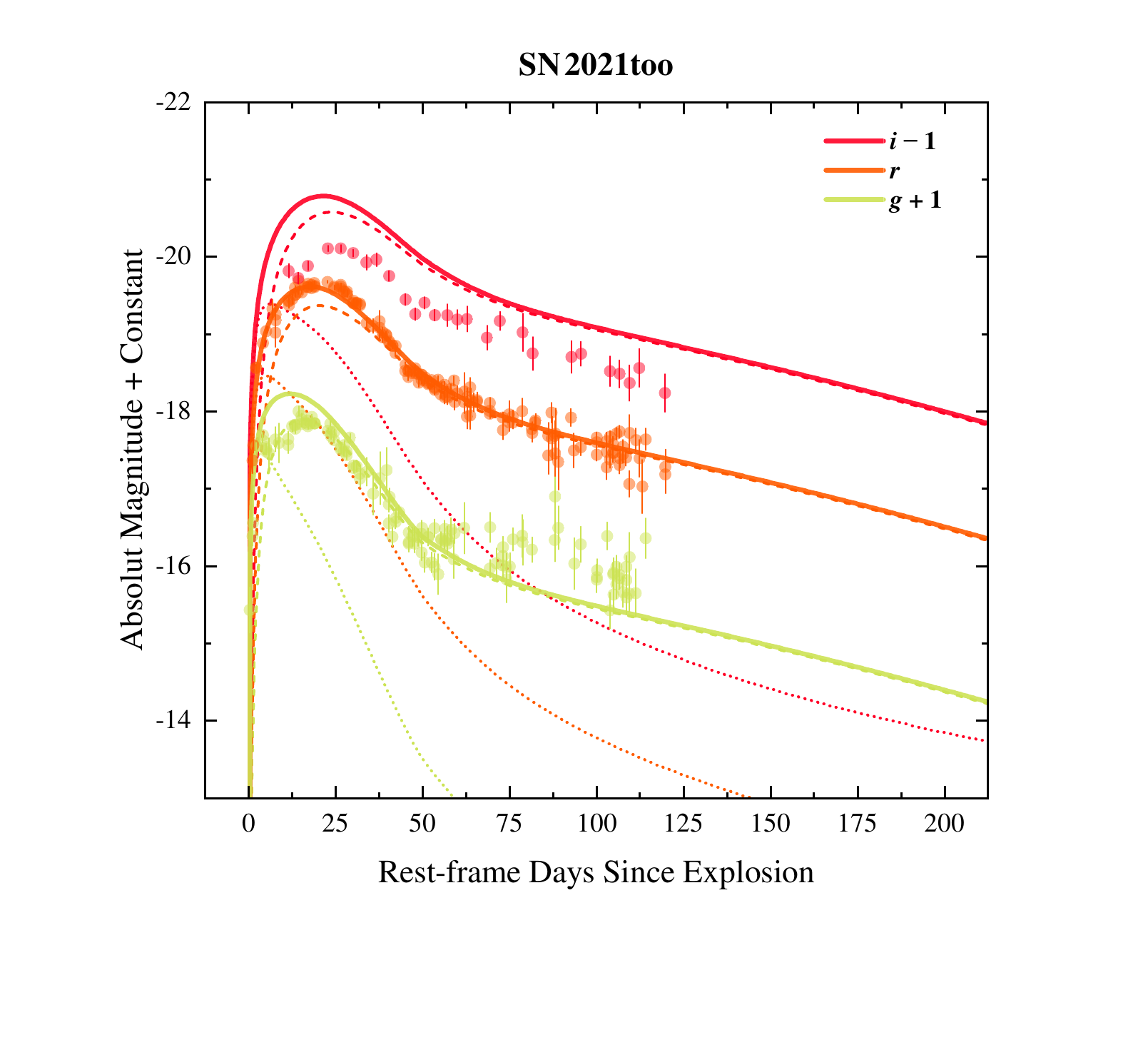}
    \includegraphics[width = 0.32\linewidth , trim = 80 65 93 35, clip]{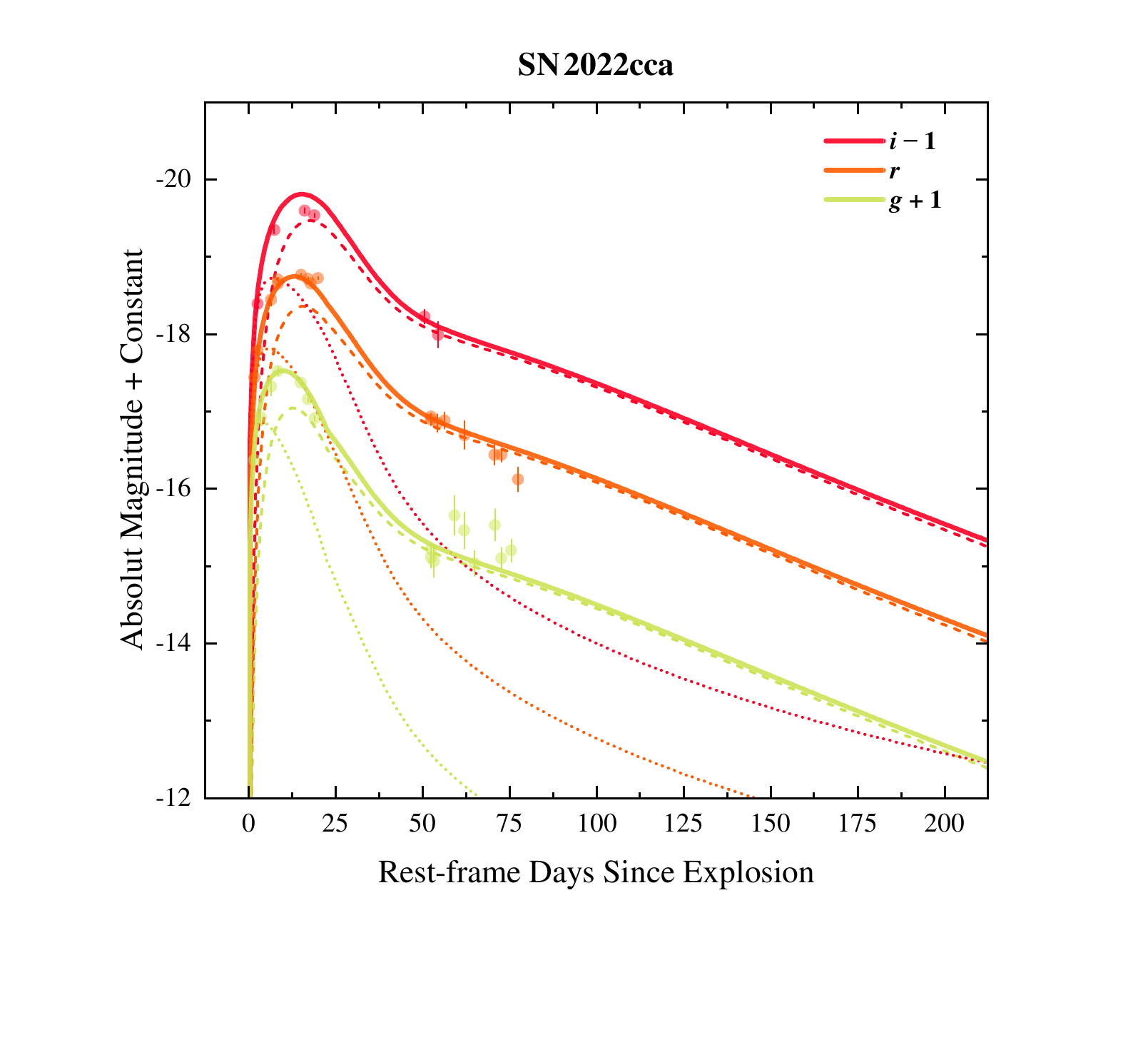}
    \caption{(Continued.)}
\end{figure*}

\begin{figure*}
    \ContinuedFloat
    \centering
    \includegraphics[width = 0.32\linewidth , trim = 80 65 93 35, clip]{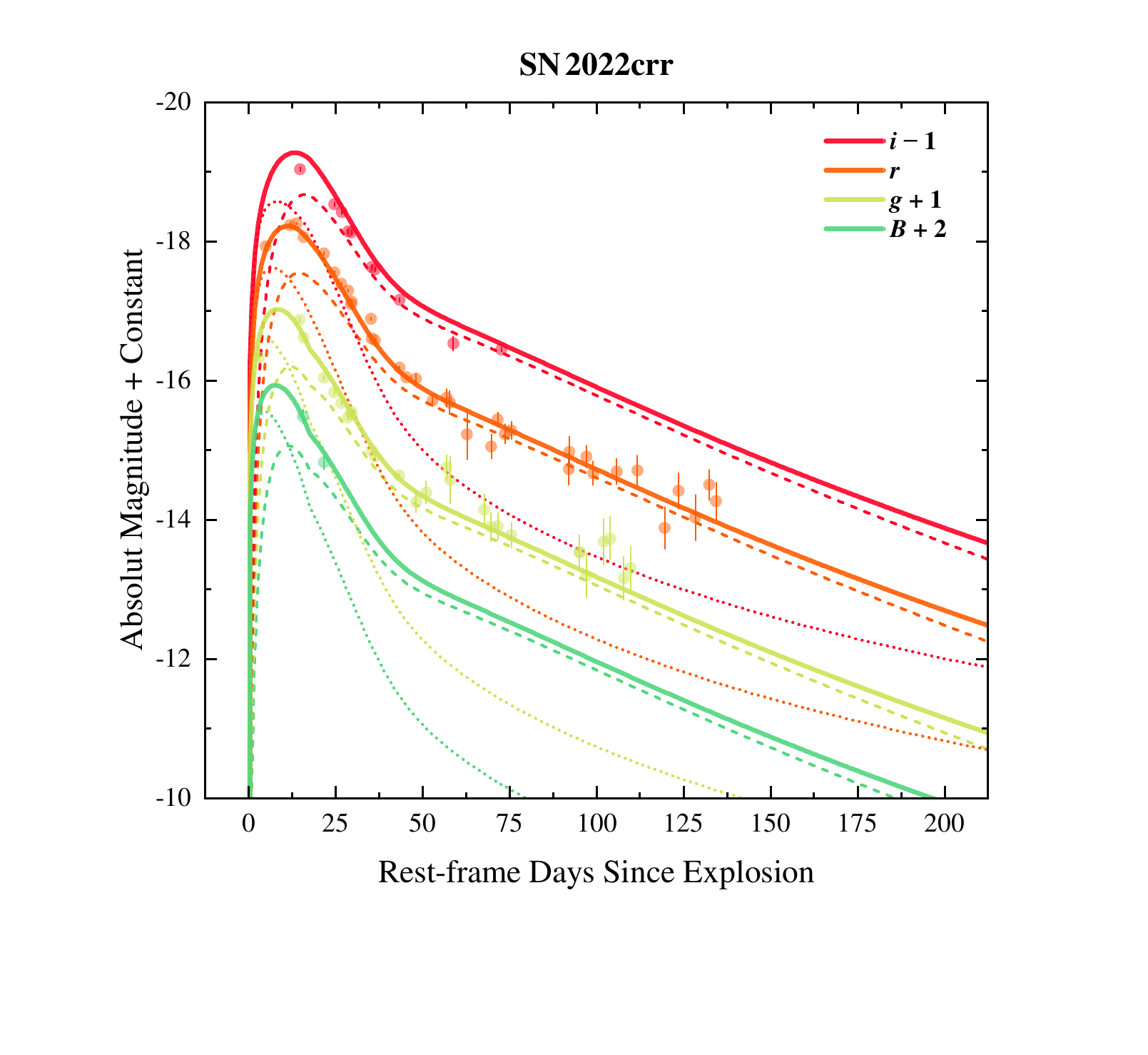}
    \includegraphics[width = 0.32\linewidth , trim = 80 65 93 35, clip]{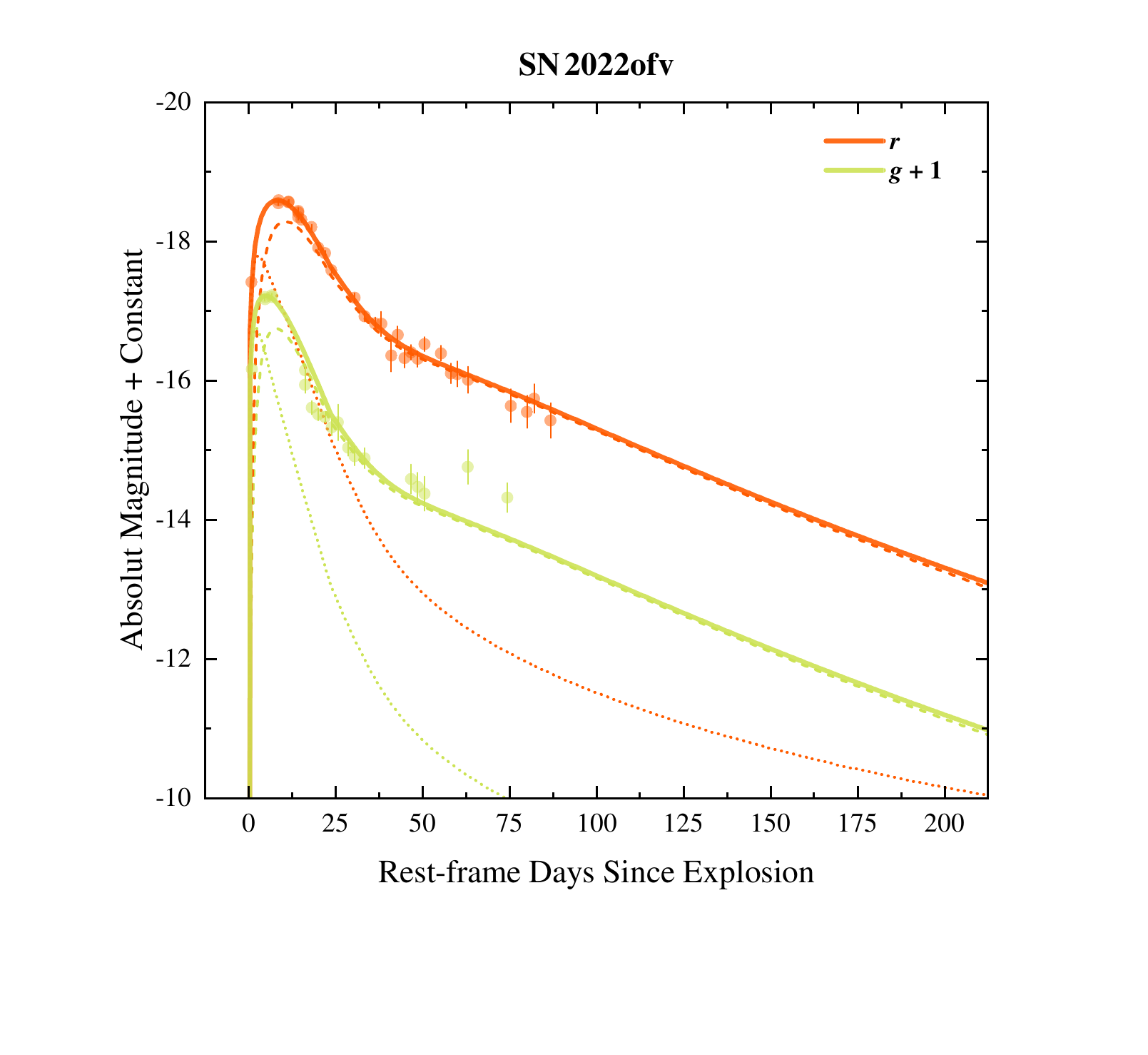}
    \includegraphics[width = 0.32\linewidth , trim = 80 65 93 35, clip]{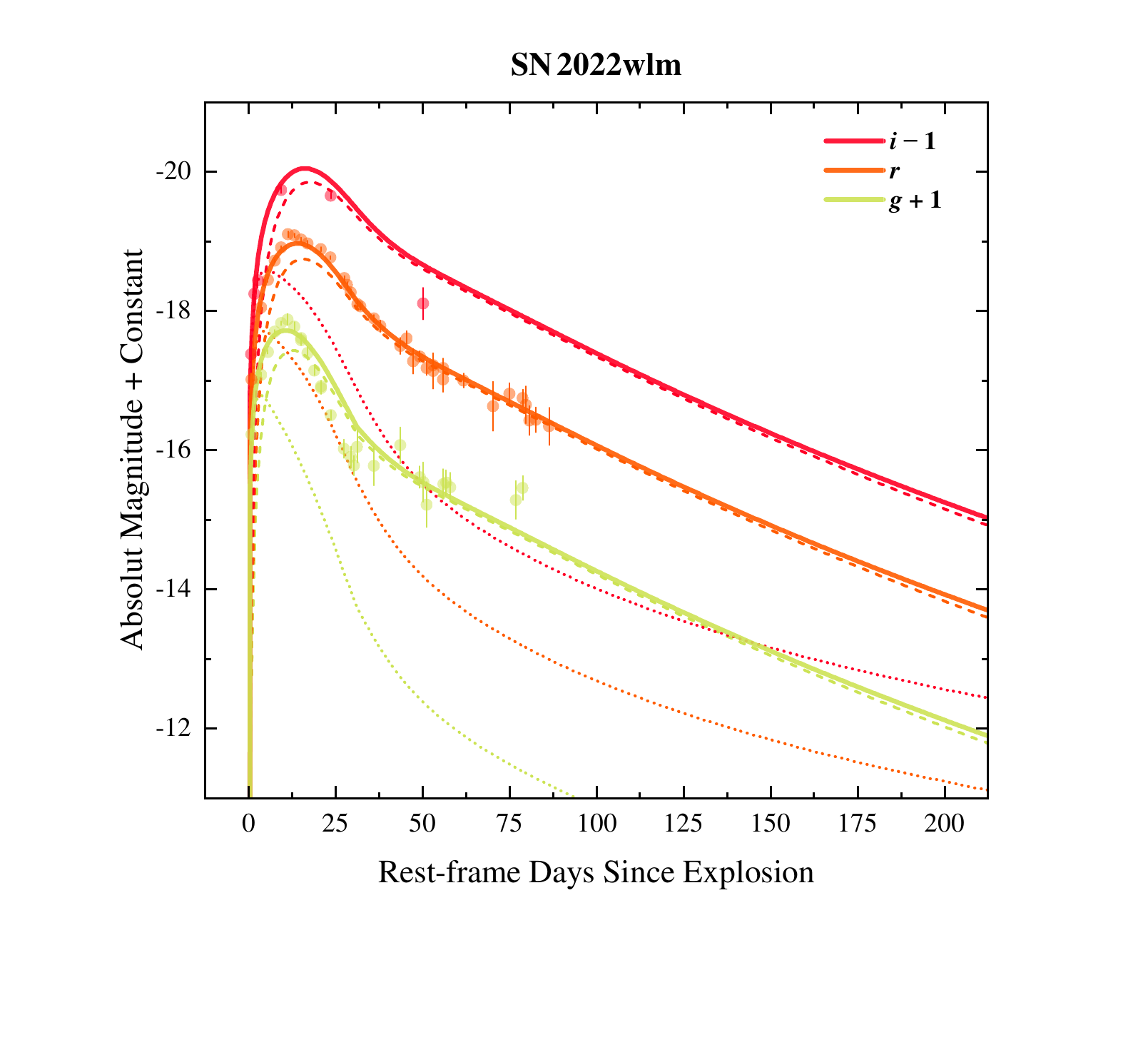}
    \includegraphics[width = 0.32\linewidth , trim = 80 65 93 35, clip]{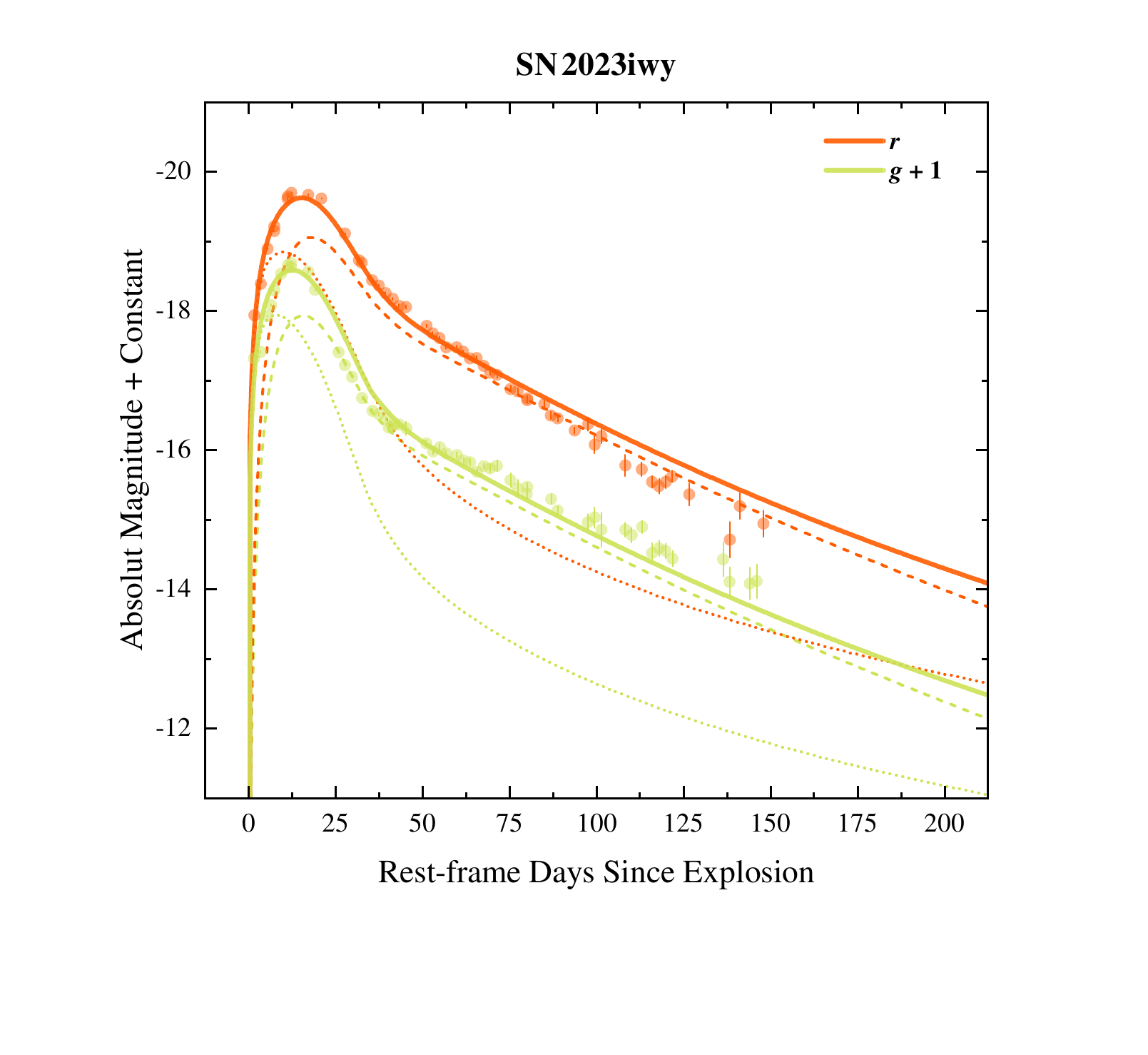}
    \includegraphics[width = 0.32\linewidth , trim = 80 65 93 35, clip]{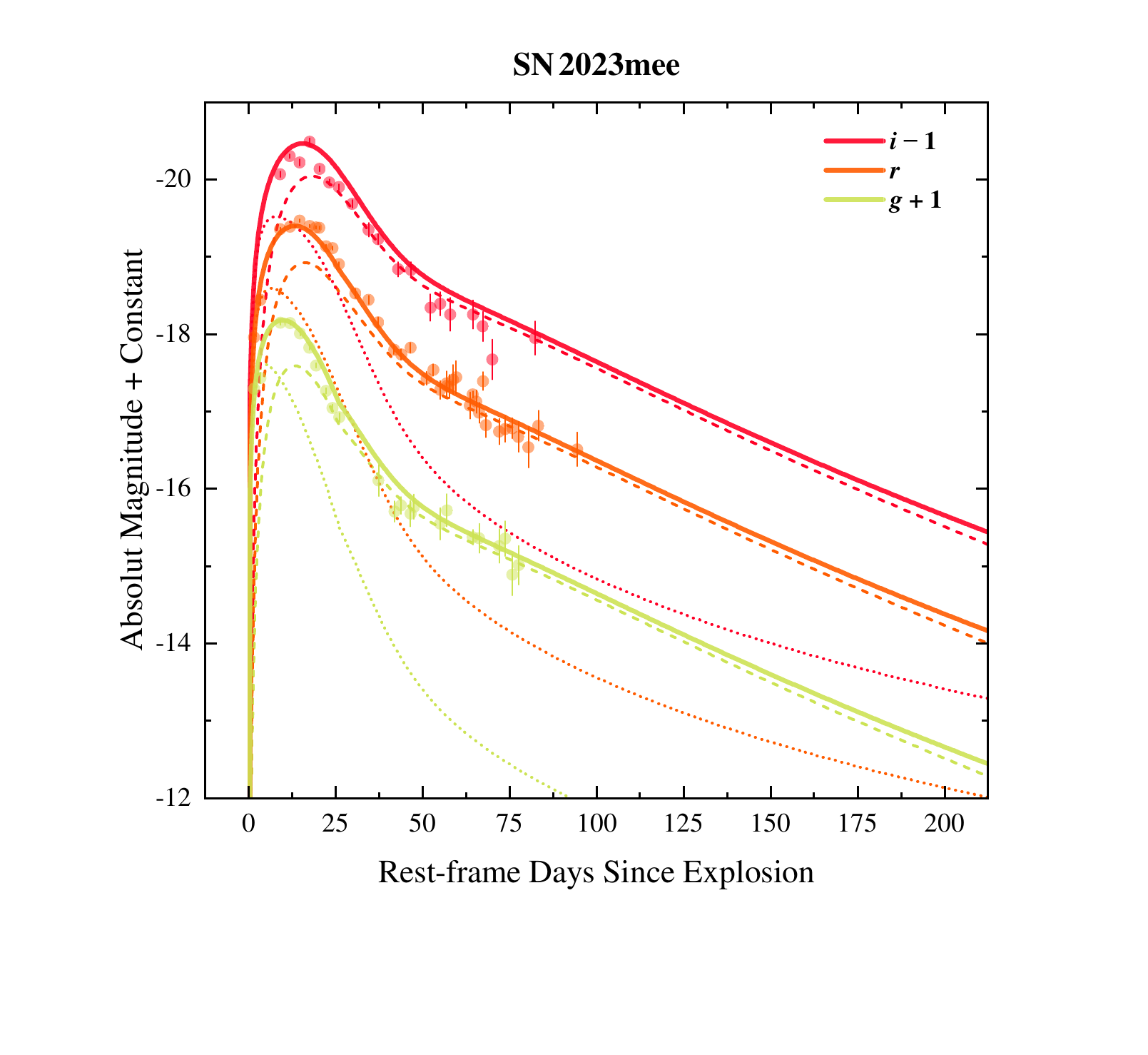}
    \includegraphics[width = 0.32\linewidth , trim = 80 65 93 35, clip]{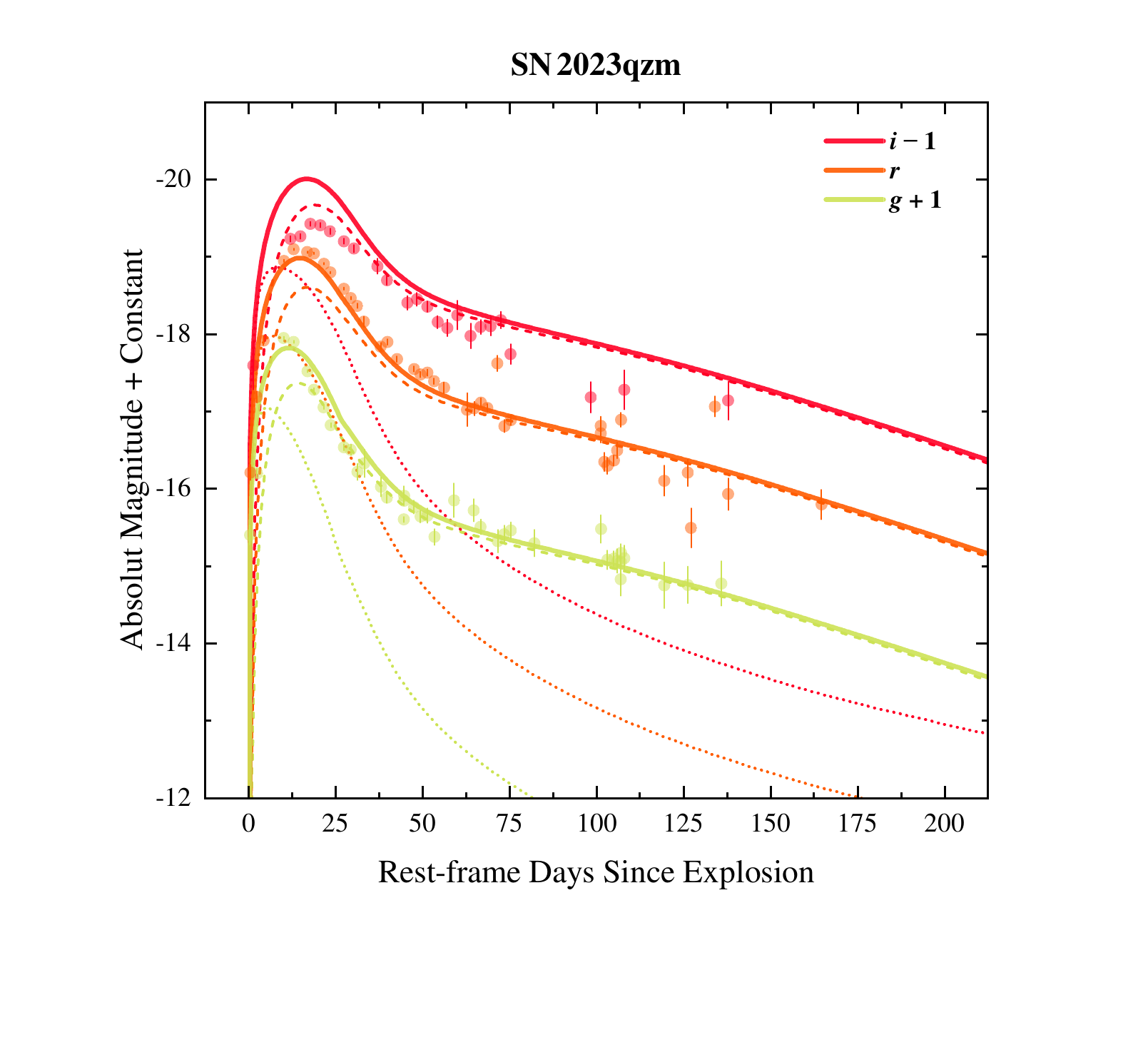}
    \includegraphics[width = 0.32\linewidth , trim = 80 65 93 35, clip]{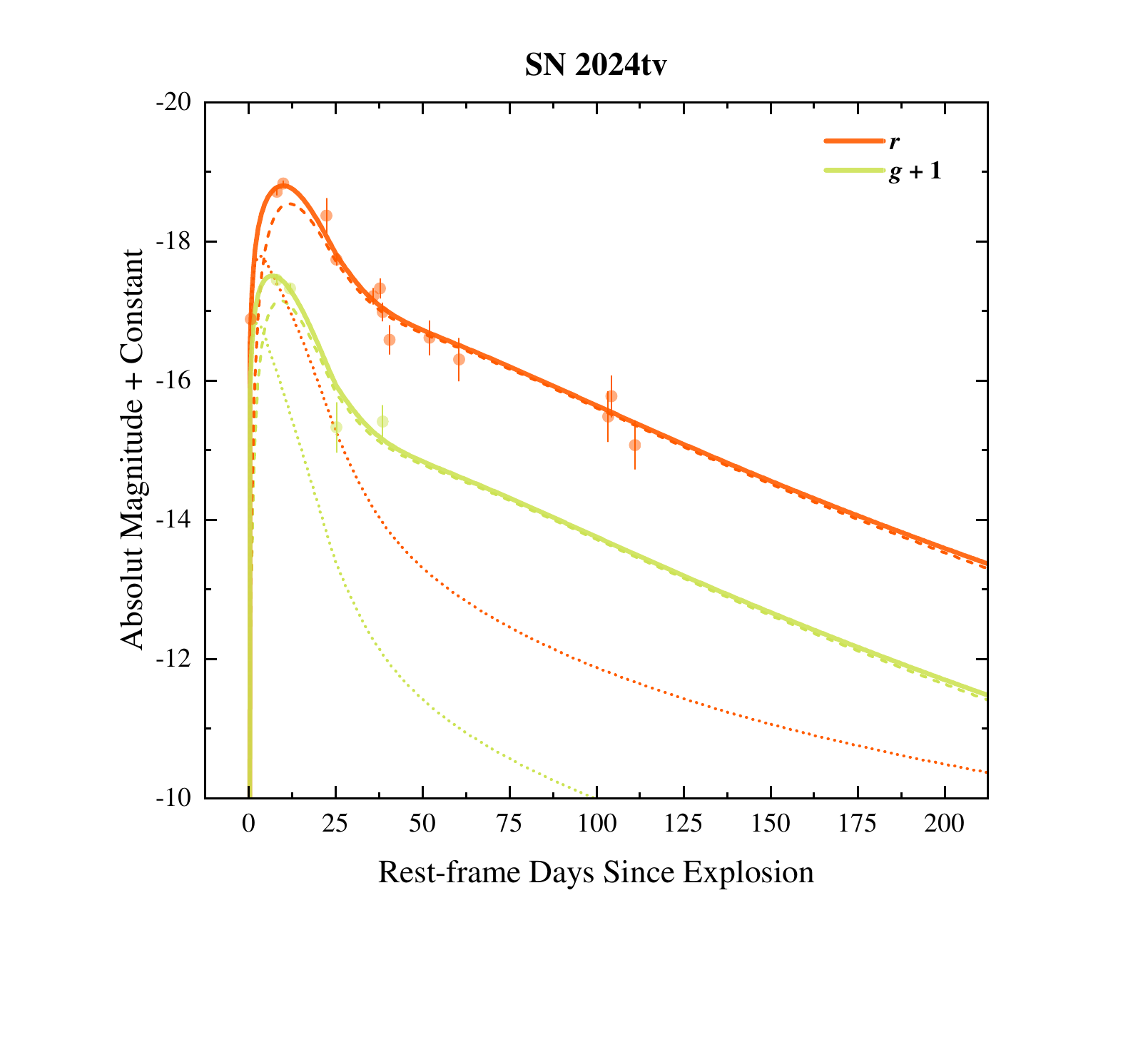}
    \includegraphics[width = 0.32\linewidth , trim = 80 65 93 35, clip]{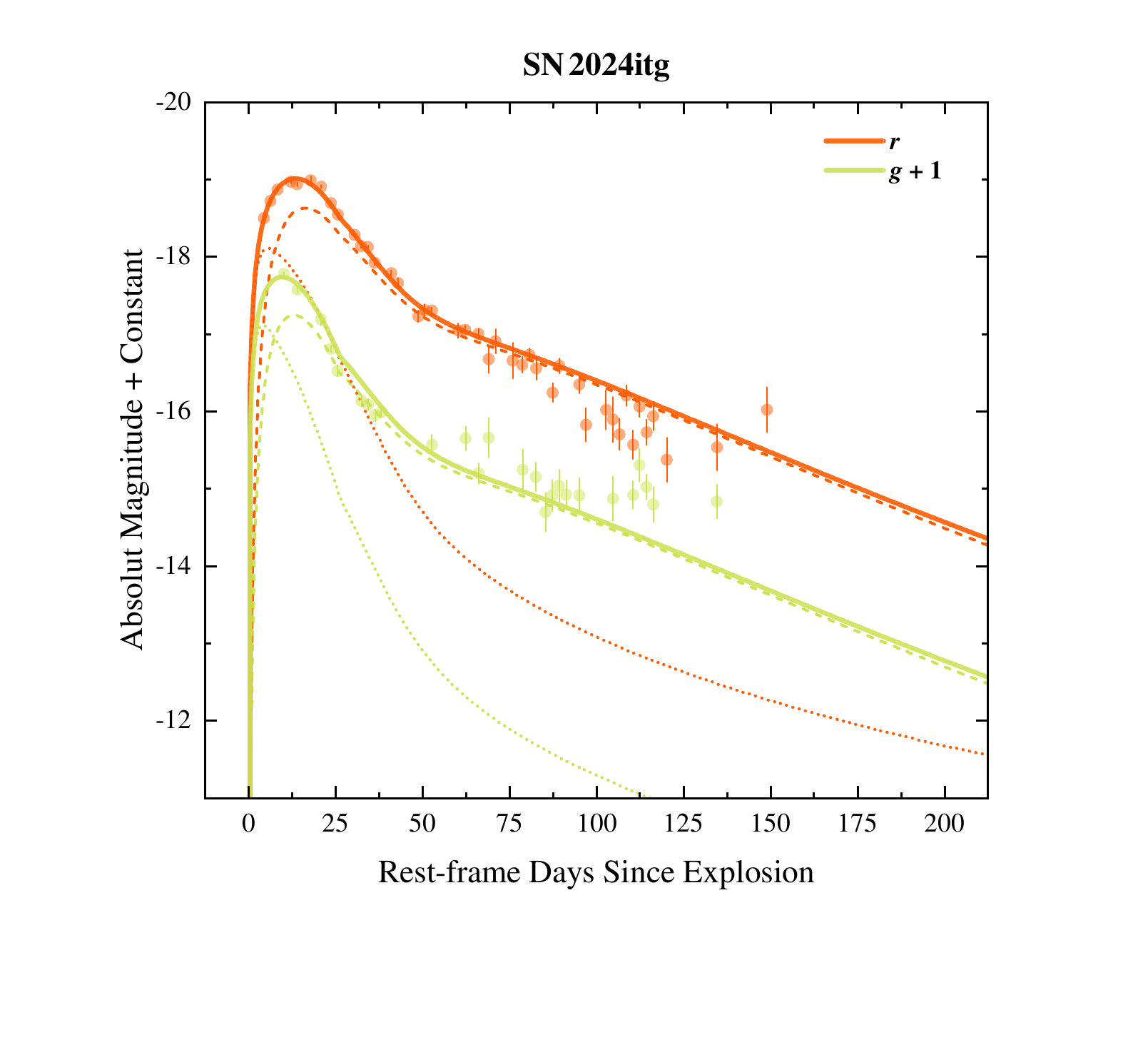}
    \includegraphics[width = 0.32\linewidth , trim = 80 65 93 35, clip]{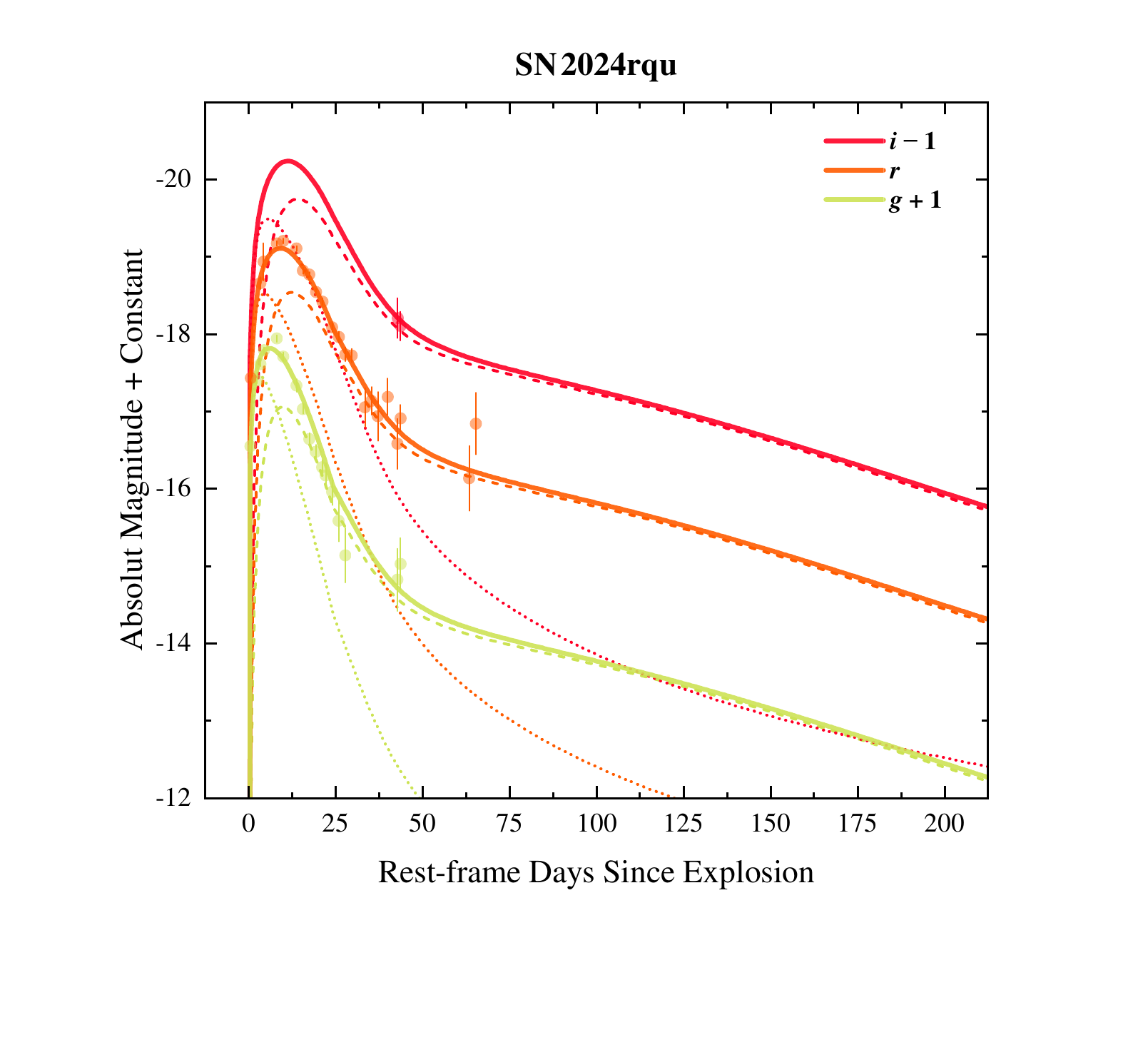}
    \includegraphics[width = 0.32\linewidth , trim = 80 65 93 35, clip]{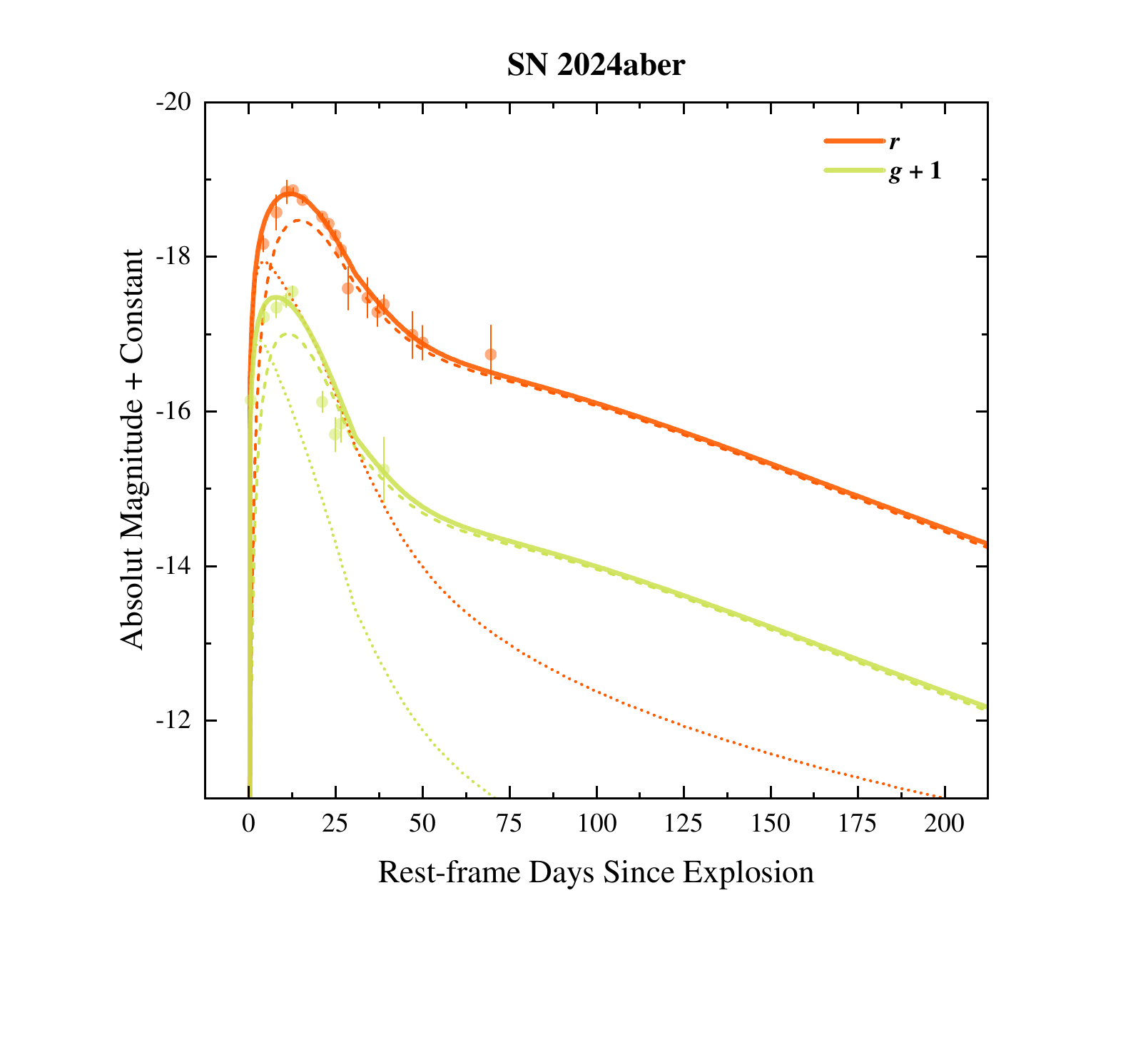}
    \includegraphics[width = 0.32\linewidth , trim = 80 65 93 35, clip]{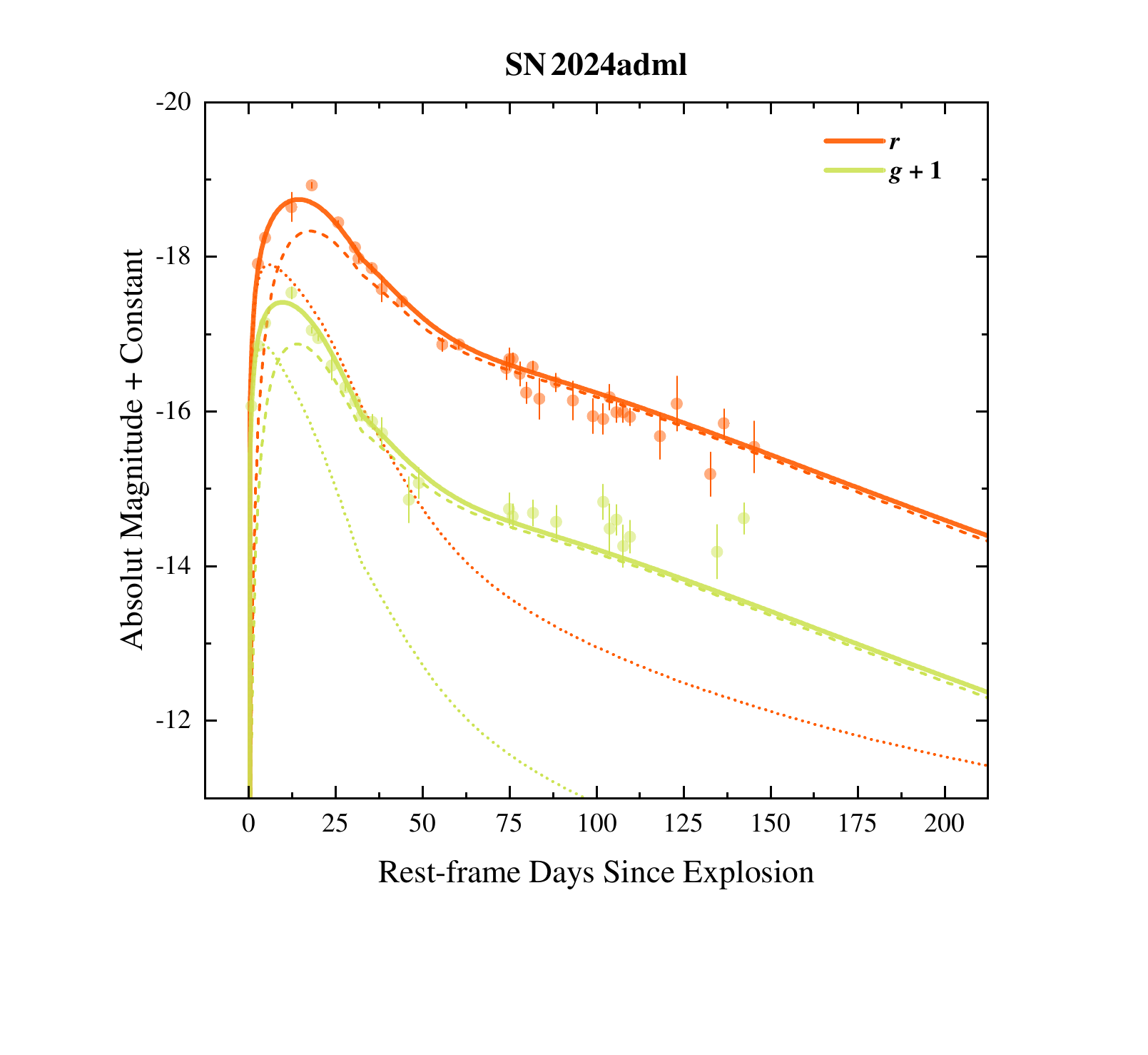}
    \includegraphics[width = 0.32\linewidth , trim = 80 65 93 35, clip]{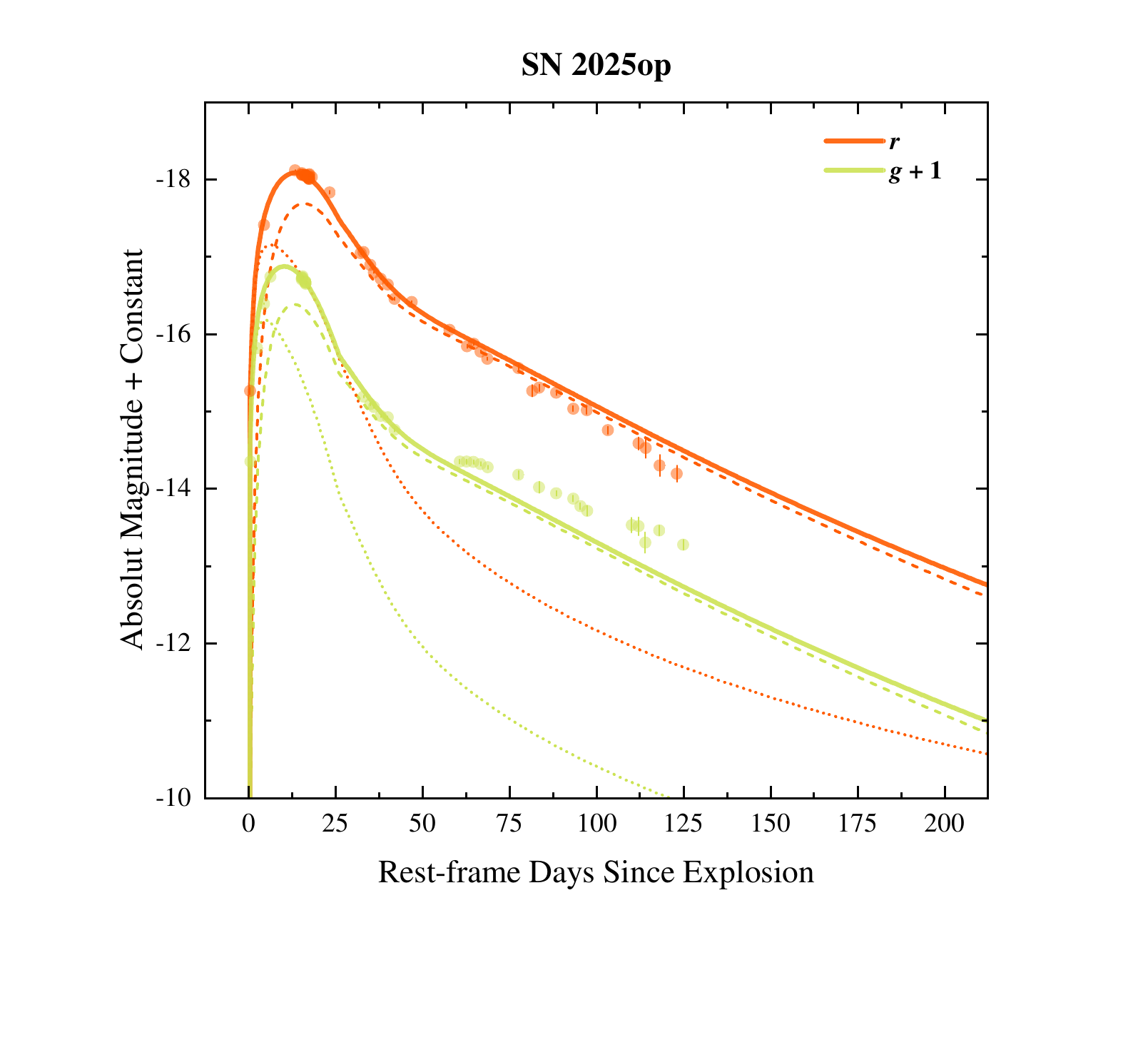}
    \caption{(Continued.)}
\end{figure*}

\begin{figure*}
    \ContinuedFloat
    \centering   
    \includegraphics[width = 0.32\linewidth , trim = 80 65 93 35, clip]{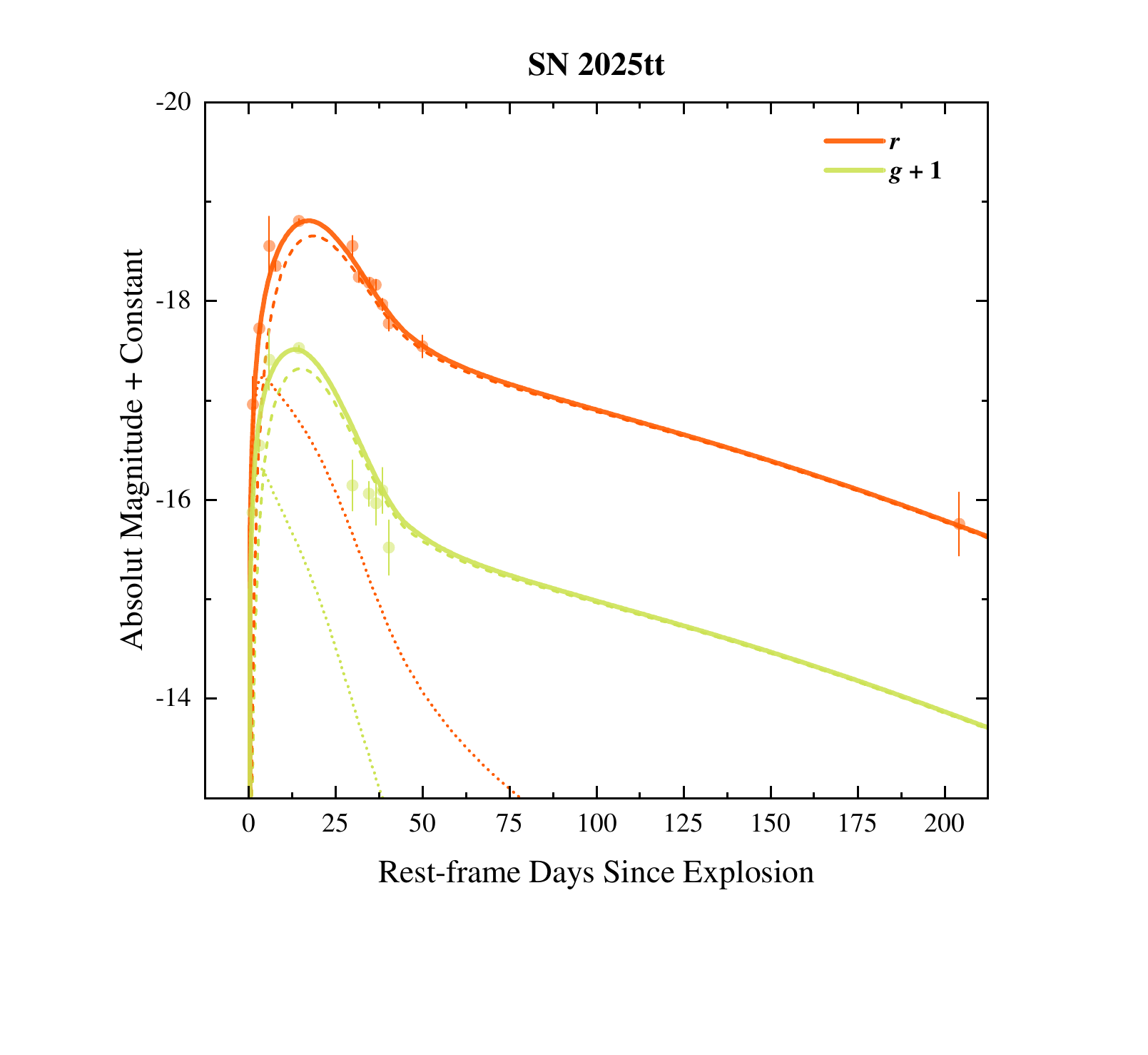}
    \includegraphics[width = 0.32\linewidth , trim = 80 65 93 35, clip]{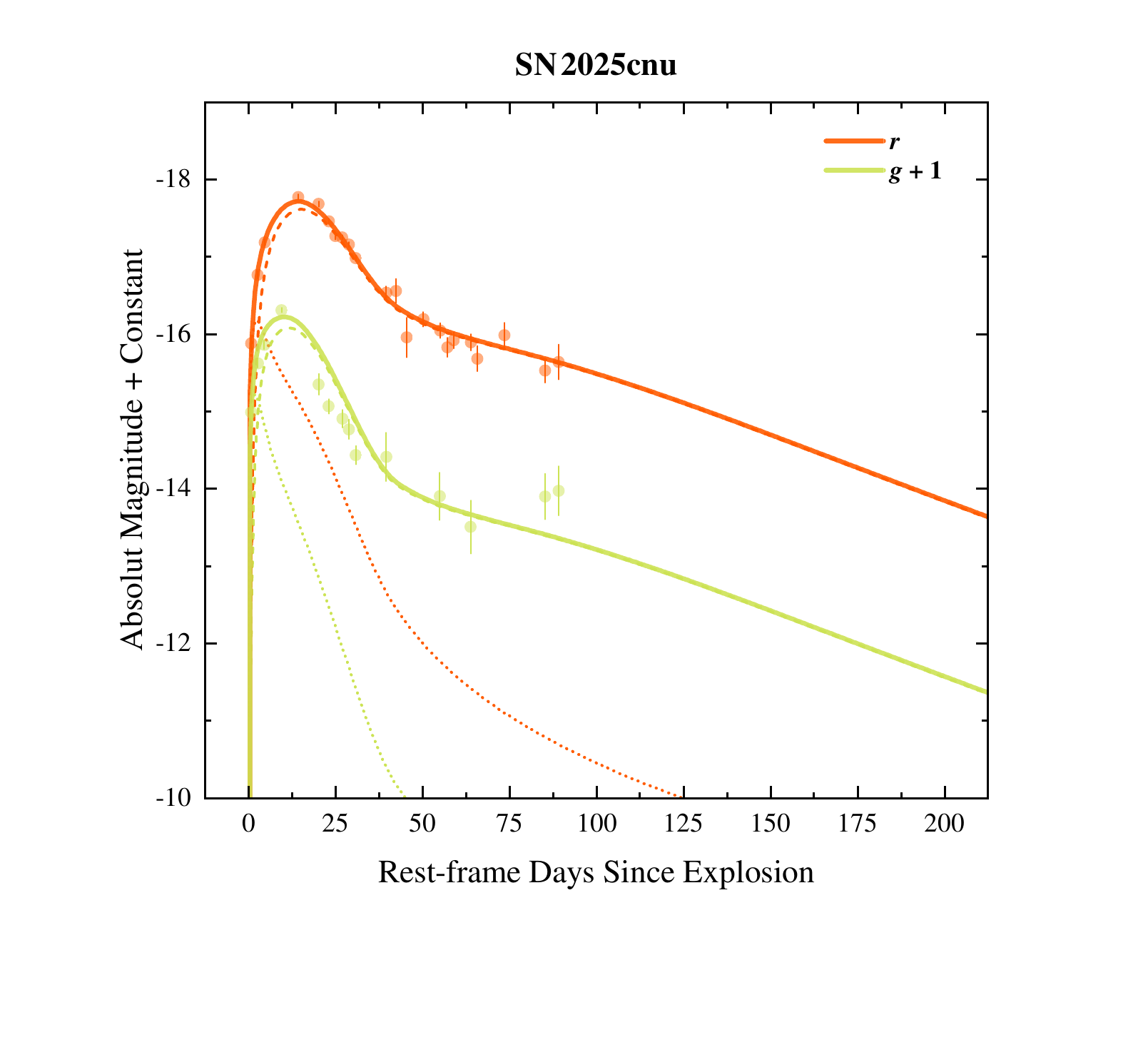}
    \includegraphics[width = 0.32\linewidth , trim = 80 65 93 35, clip]{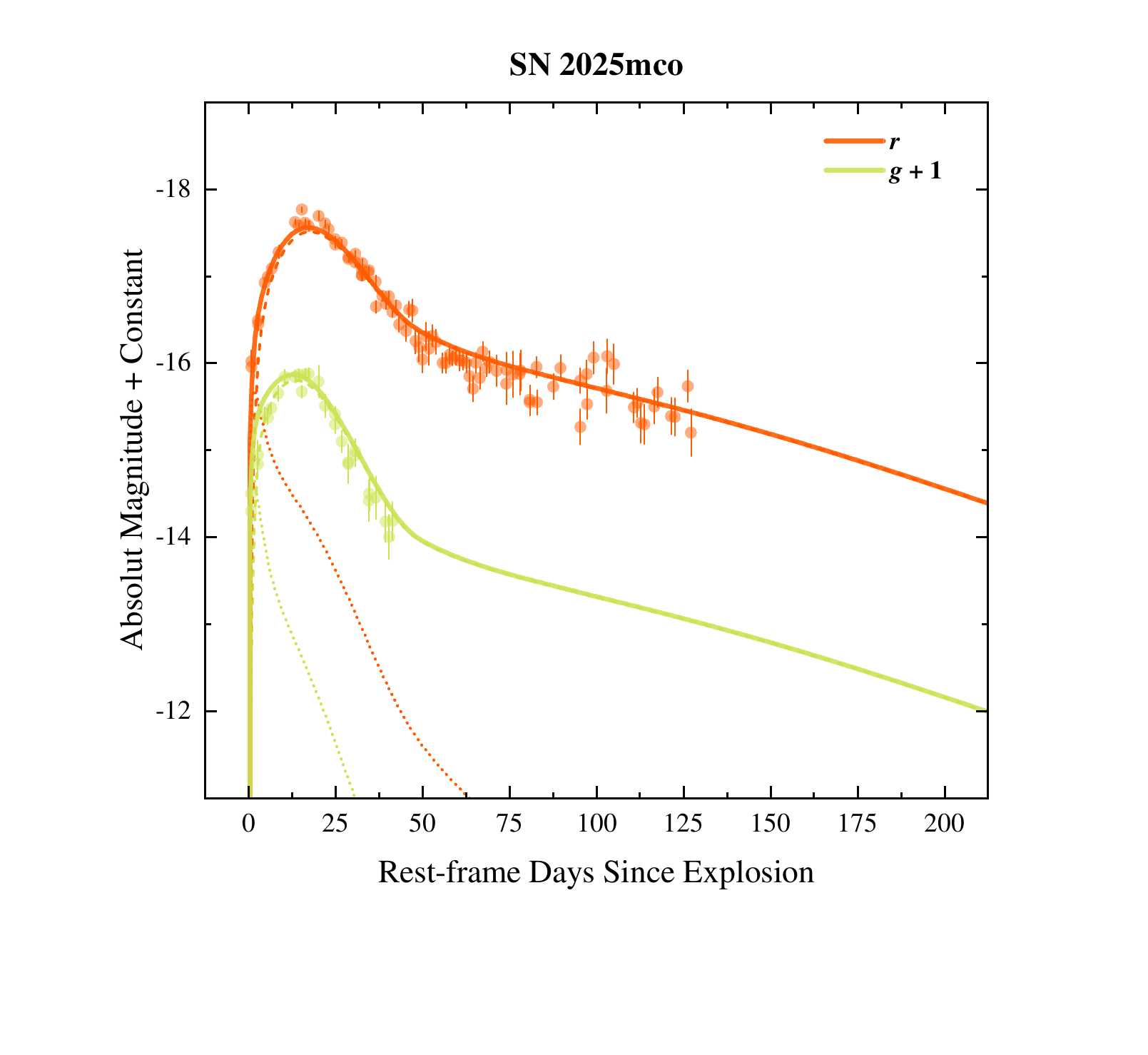}
    \includegraphics[width = 0.32\linewidth , trim = 80 65 93 35, clip]{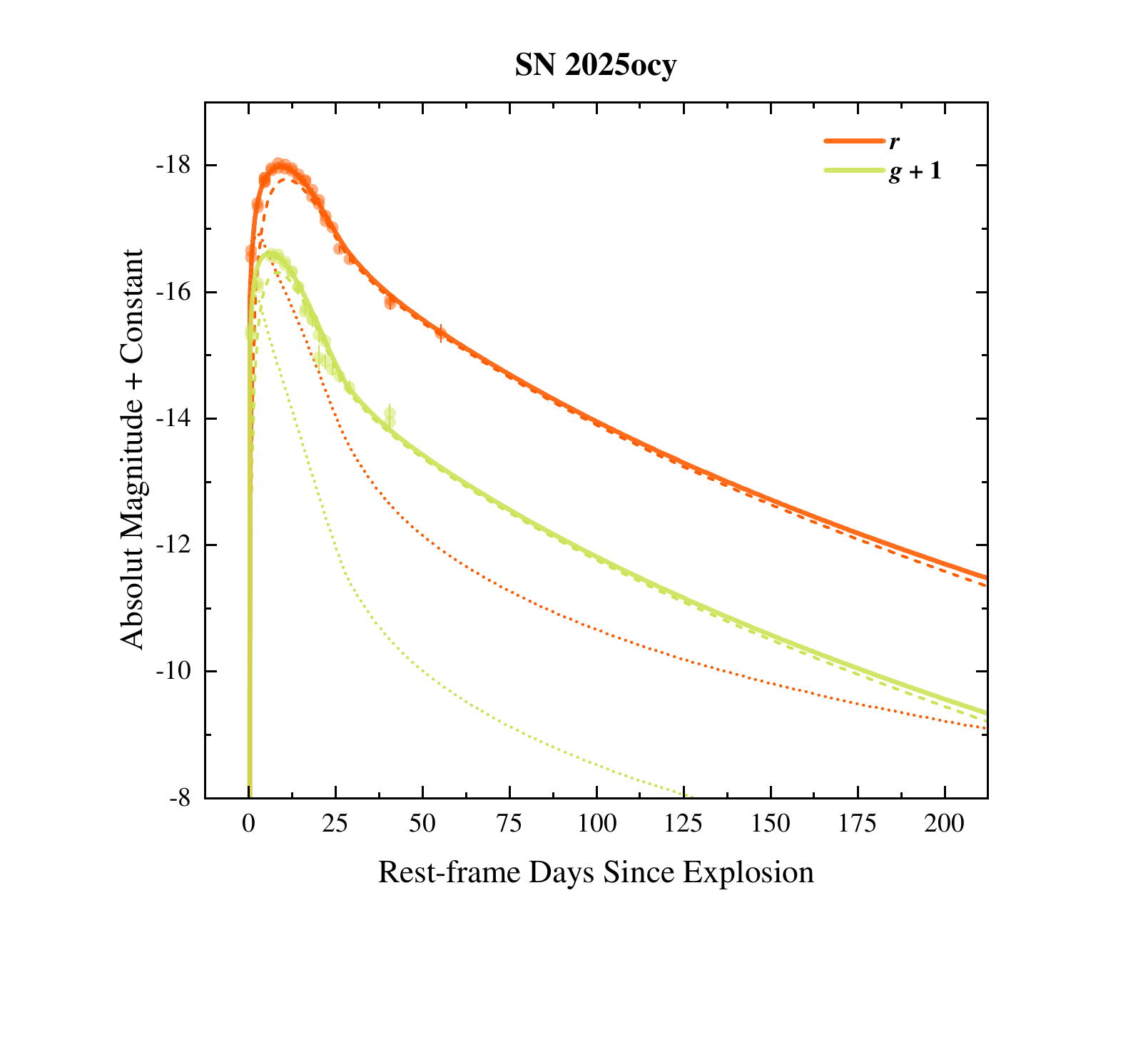}
    \includegraphics[width = 0.32\linewidth , trim = 80 65 93 35, clip]{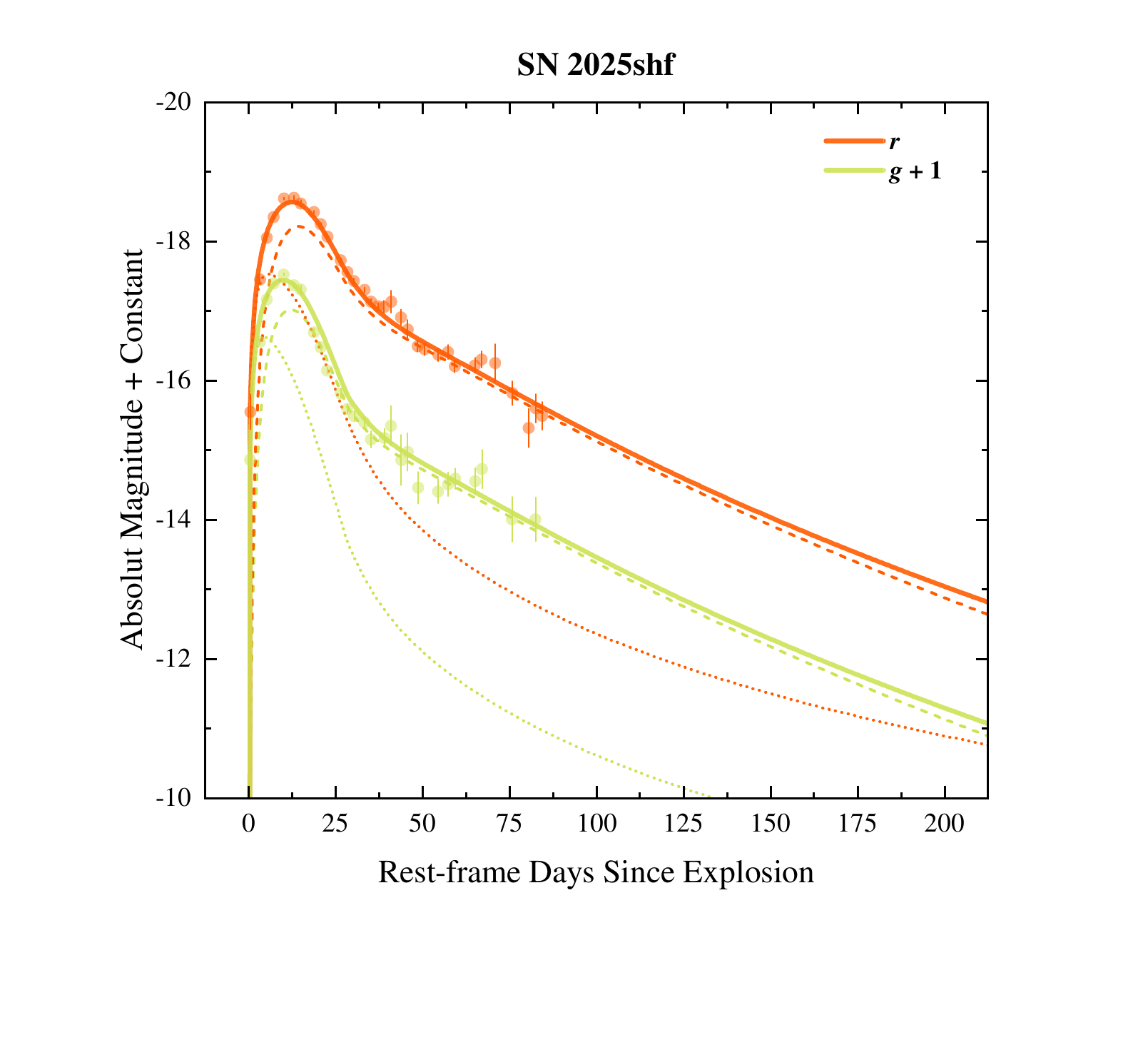}
    \includegraphics[width = 0.32\linewidth , trim = 80 65 93 35, clip]{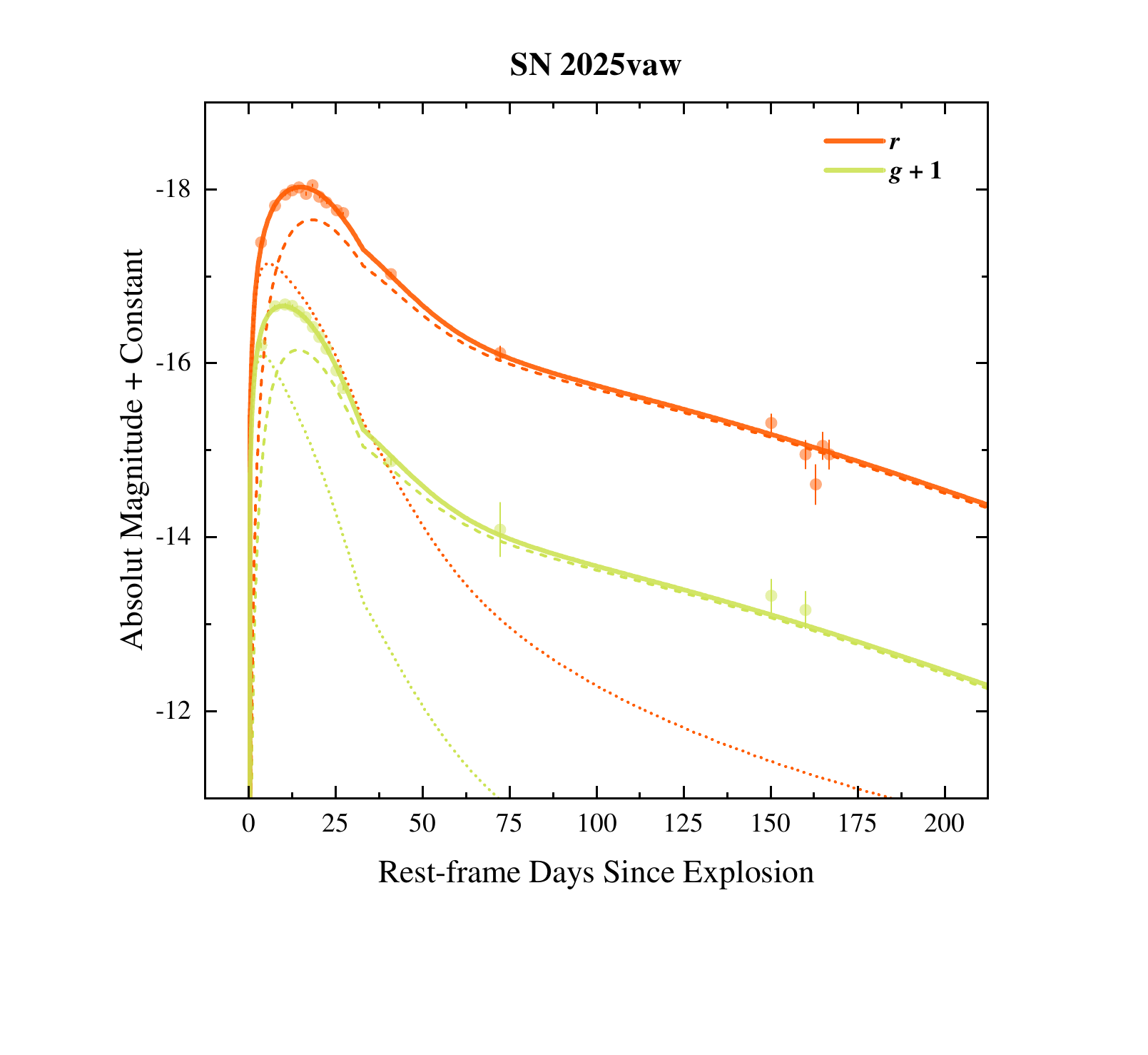}
    \includegraphics[width = 0.32\linewidth , trim = 80 65 93 35, clip]{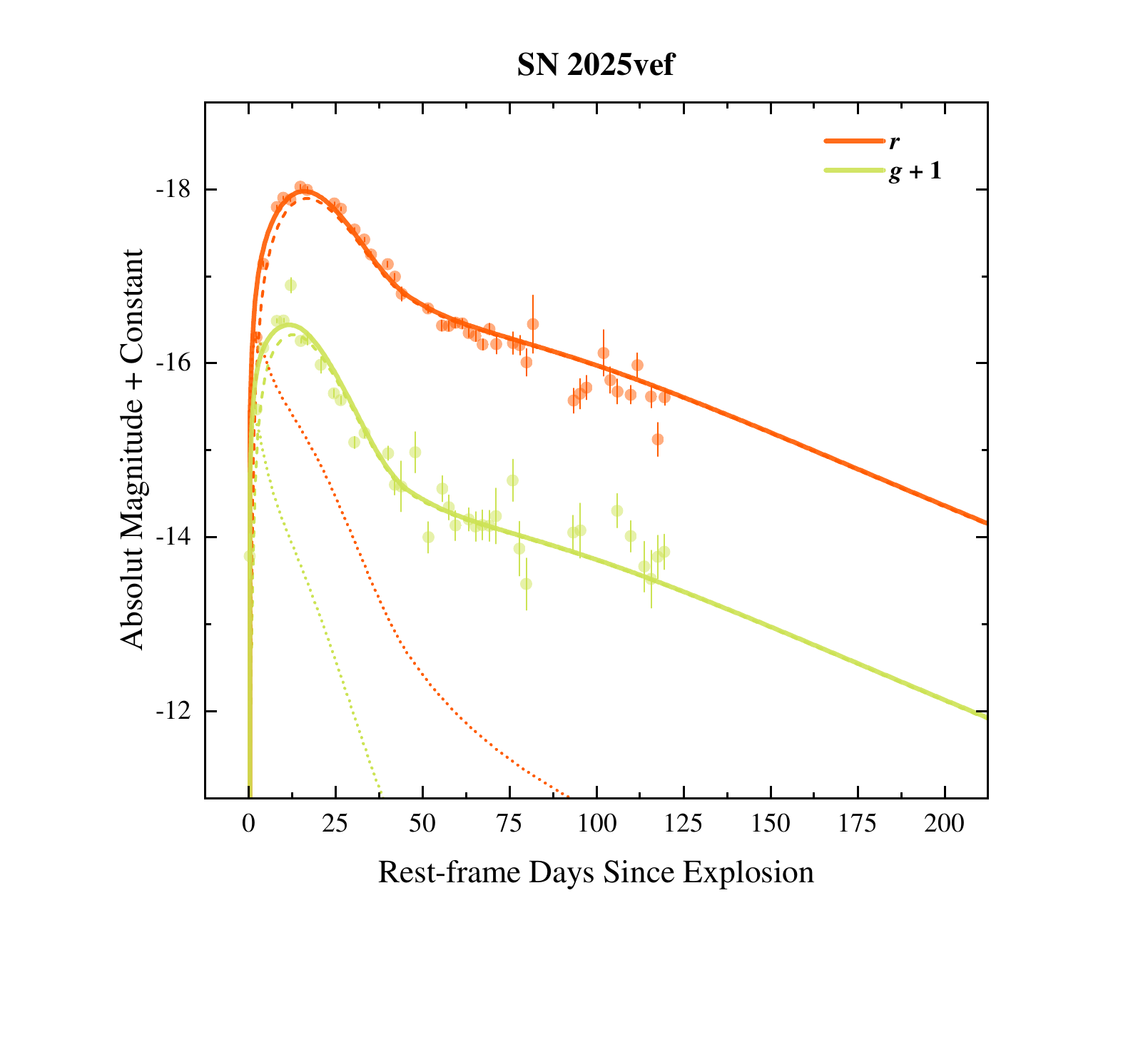}
    \includegraphics[width = 0.32\linewidth , trim = 80 65 93 35, clip]{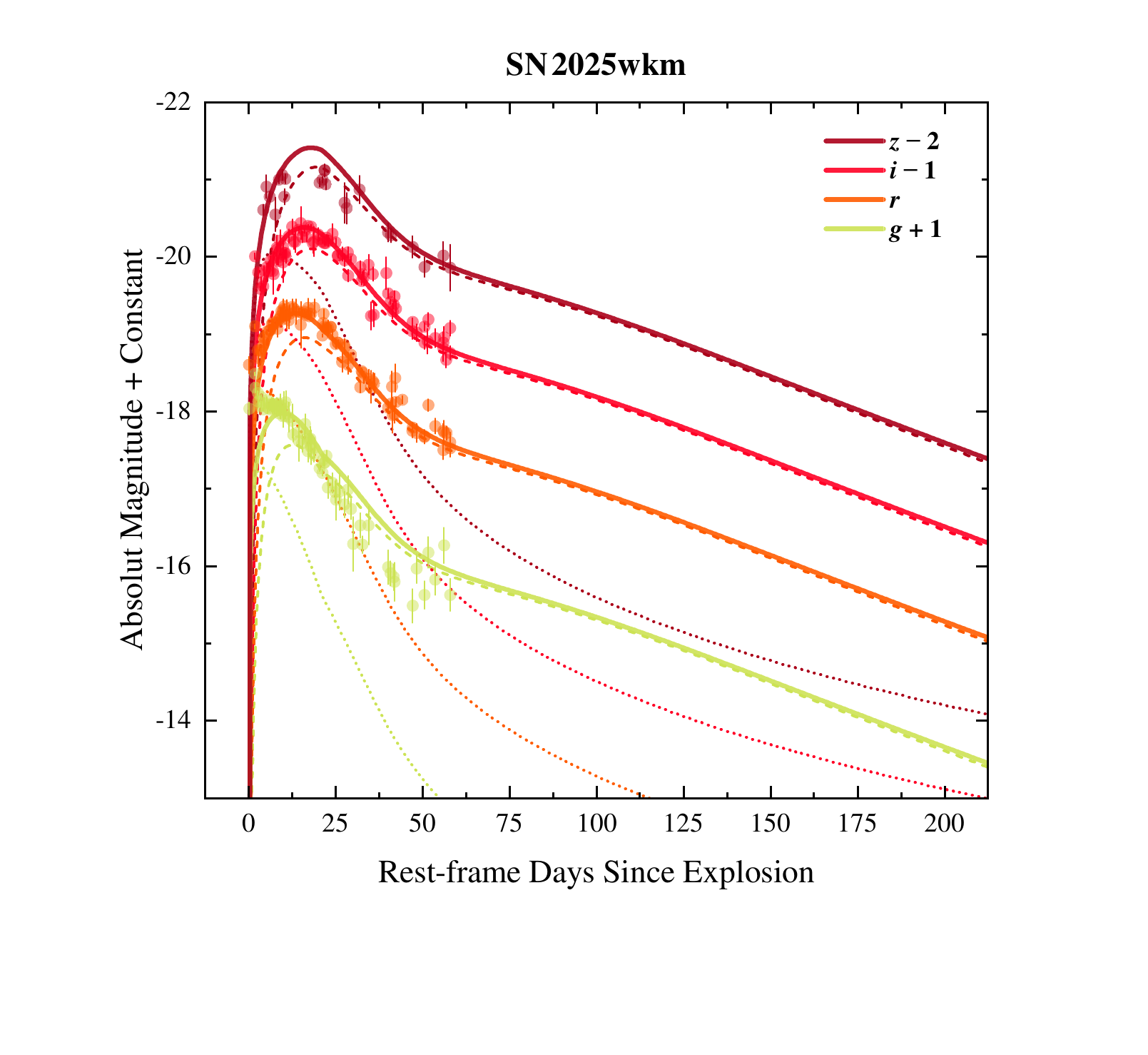}
    \caption{(Continued.)}
\end{figure*}

\clearpage
\bibliography{ms}{}
\bibliographystyle{aasjournalv7}

\end{document}